\documentclass[12pt]{report}
\usepackage[a4paper,left=40mm,right=20mm,top=20mm,bottom=20mm]{geometry}
\usepackage{masthesis}
\usepackage{bm}
\usepackage{times}
\usepackage{float}
\usepackage{color}
\usepackage{fixltx2e} %to display {figure*}s in right order
\usepackage{hyperref}

\graphicspath{{./fig/}{./png/}{./ps/}}

\newcommand{\mathbfss}[1]{\textbf{\textsf{#1}}}
\newcommand{\vect}[1]{{{\mbox{\boldmath $#1$}}}}%also makes bold Greek letters
\newcommand\sfrac[2]{{\textstyle{\frac{#1}{#2}}}}
\newcommand\deriv[2]{\displaystyle\frac{\upartial #1}{\upartial #2} }
\newcommand\average[1]{{\langle #1 \rangle}} %volume average
\newcommand\averaged[1]{{\langle #1 \rangle}} 		%volume average
\newcommand\mean[1]{\overline{#1}}         %ensemnle average
  \newcommand{\phm}{{\phantom{0}}} %empty pace, 0-wide, useful to align table entries
  \newcommand{\phd}{{\phantom{0.}}} %empty pace, 0-wide, useful to align table entries
  \newcommand{\cool}{{_{\rm cool}}}  %radiative cooling
  \newcommand{\mesh}{{_{\Delta}}}
  \newcommand{\tphi}{{\Phi}}  %ensemble-averaged philling factor
  \newcommand{\Mesh}{{\Delta}}
  
  \newcommand{\rrms}{{_{\rm rms}}}
  \newcommand{\sound}{{_{\mathrm{s}}}}  %sound
  \newcommand{\st}{^{\rm st}}     %first
  \newcommand{\nd}{^{\rm nd}}     %second
  \newcommand{\rd}{^{\rm rd}}     %second
  \newcommand{\uth}{^{\rm th}}     %second
  \newcommand{\SN}{_{\rm SN}}     %supernova
  \newcommand{\SNk}{_{\rm SN,kin}}     %supernova
  \newcommand{\SNt}{_{\rm SN,th}}     %supernova
  \newcommand{\dd}{{\rm d}}       %for derivatives
  \newcommand{\HI}{{\rm H}{\sc i}}
  \newcommand{\HII}{{\rm H}{\sc ii}}
  \newcommand{\eB}{\,{e_{\rm B}}}
  \newcommand{\eK}{\,{e_{\rm kin}}}
  \newcommand{\eT}{\,{e_{\rm th}}}
  \newcommand{\cv}{\,{c_{\rm v}}}
  \newcommand{\cp}{\,{c_{\rm p}}}

  \newcommand{\Rm}{\mathrm{Rm}}
  \newcommand{\Op}{{WSWa}}
  \newcommand{\OpH}{{WSWah}}
  \newcommand{\WSWa}{{WSWb}}
  \newcommand{\Omp}{{\Omega}}
  \newcommand{\Ompa}{{\mathrm{B}1\rm{\Omega} }}
  \newcommand{\Ompb}{{\mathrm{B}1\rm{\Omega}\rm{O}}}
  \newcommand{\Ompc}{{\mathrm{B}1\rm{\Omega}\rm{SN}}}
  \newcommand{\Ompd}{{\mathrm{B}2\rm{\Omega}}}
  \newcommand{\Ompe}{{\mathrm{B}1\rm{\Omega}\rm{Sh}}}
  \newcommand{\Omph}{{\mathrm{H}1\rm{\Omega}}}
  \newcommand{\str}{\mathcal{D}}
  \newcommand{\corr}{\mathcal{C}}
  \newcommand{\lcorr}{\mathcal{L}}
  \newcommand{\turb}{_{\rm turb}}     			%turbulent
  \newcommand{\cm}{\,{\rm cm}}
  
  \newcommand{\cmcube}{\,{\rm cm^{-3}}}
  \newcommand{\dyn}{\,{\rm dyn}}
  \newcommand{\erg}{\,{\rm erg}}
  \newcommand{\g}{\,{\rm g}}

  \newcommand{\km}{\,{\rm km}}
  \newcommand{\kms}{\,{\rm km\,s^{-1}}}

  \newcommand{\K}{\,{\rm K}}
  \newcommand{\kpc}{\,{\rm kpc}}
  \newcommand{\pc}{\,{\rm pc}}
  
  \newcommand{\Myr}{\,{\rm Myr}}
  \newcommand{\Gyr}{\,{\rm Gyr}}
  \newcommand{\G}{\,{\rm G}}
  
  \newcommand{\mkG}{\,\umu{\rm G}}
  
  \newcommand{\Msol}{{M_{\odot}}}
  \newcommand{\p}{\,{\rm pc}}
  
  \newcommand{\s}{\,{\rm s}}
  \newcommand{\yr}{\,{\rm yr}}
  \newcommand{\Pe}{{\rm Pe}}								%Peclet
  \newcommand{\Ry}{{\rm Re}}								%Peclet
  \newcommand{\nG}{\,{\rm nG}}

     \newcommand{\la}{\,\rlap{\raise 0.5ex\hbox{$<$}}{\lower 1.0ex\hbox{$\sim$}}\,}
     \newcommand{\ga}{\,\rlap{\raise 0.5ex\hbox{$>$}}{\lower 1.0ex\hbox{$\sim$}}\,}

  \definecolor{burntorange}{RGB}{255,97,0}
  \definecolor{violet}{RGB}{105,20,200}
  \definecolor{royalblue}{RGB}{0,137,255}
  \definecolor{mygreen}{RGB}{0,120,0}
  \newenvironment{freply}
    {\color{black}}{}
  \newenvironment{frcorr}
    {\color{black}}{}
\begin{document}

  \phdtitle{Supernovae Driven Turbulence in the Interstellar Medium} % Title.
           {Frederick Armstrong Gent}          % Author.
           {fig/shield}                 % Graphic for the title page.
           {November $2012$}                % Date.

  \thispagestyle{empty}
  \cleardoublepage

  \begin{dedication}
  My son, Declan. Wherever you're going may you always enjoy the journey \ldots
\end{dedication}
\thispagestyle{empty}
\cleardoublepage
                    % Optional dedication.

  \begin{acknowledgements}
  I would like to thank my supervisors -- Prof. Anvar Shukurov, Dr. Graeme
  Sarson and Dr. Andrew Fletcher -- who have provided a collaborative and 
  stimulating environment in which to work. 
  They have been constructive in their criticism and very supportive both
  with resources and advice. 
  They have provided opportunities for me to participate in the broader
  research community in this country and internationally.
  I have always felt that they valued and were generally interested in our
  project and have enjoyed the freedom to think and work autonomously.

  Further I thank my host in Helsinki -- Dr. Maarit Mantere, her family and
  her colleagues, amongst them Dr. Petri K\"aply\"a, Dr. Thomas Hackmann and
  Dr. Jorma Harju -- who has provided invaluable guidance in the application and
  understanding of the use of the pencil-code and in particular principles of
  numerical modelling and specifically modelling the ISM.
  She, and they, also made my visits to Finland for work an extremely 
  pleasurable social and cultural experience.

  In addition I feel privileged to have shared offices with a fabulous group
  of friends, who have filled my experience of post graduate life with 
  amusement, drama and insight. 
  In particular I thank my ever present room mates Timothy Yeomans, Donatello
  Gallucci, Dr. Joy Allen and Anisah Mohammed.
  Other graduate students and Post Docs who have made memorable contributions 
  to my experience during my graduate studies are: 
  Dr. Sam James, 
  Dr. Pete Milner,
  Dr. Jill Johnson, 
  Dr. Andrew Baggaley,
  Dr. James Pickering, 
  Dr. Kevin Wilson,
  Dr. Daniel Maycock, 
  Dr. David Elliot, 
  Dr. Drew Smith, 
  Dr. Paul McKay, 
  Dr. Angela White, 
  Nathan Barker, 
  Nuri Badi,
  Matt Buckley, 
  Christian Perfect, 
  Alix Leboucq, 
  Rob Pattinson, 
  Holly Ainsworth, 
  Rute Vieira, 
  Kavita Gangal, 
  Nina Wilkinson,
  Asghar Ali,
  Gavin Whitaker, 
  Stacey Aston, 
  Lucy Sherwin, 
  Jamie Owen, 
  David Cushing, 
  Tom Fisher.

  I wish to thank the school computing support officers, Dr. Anthony Youd and
  Dr. Michael Beaty, and administrative staff Jackie Williams, Adele Fleck, 
  Jackie Martin and Helen Green, plus Andrea Carling in MathsAid and
  Gail de-Blaquiere of the SAgE Faculty Office. I also thank
  Prof. Carlo Barenghi, 
  Prof. Ian Moss, 
  Prof. David Toms, 
  Prof. Robin Johnson,
  Dr. Paul Bushby, 
  Dr. Nikolaos Proukakis, 
  Dr. Nicholas Parker, 
  Prof. John Matthews, 
  Prof. Robin Henderson, 
  Dr. Peter Avery,
  Dr. Colin Gillespie,
  Dr. Phil Ansel, 
  Dr. David Walshaw, 
  Dr. Jian Shi,
  Dr. Zinaida Lykova,
  Prof. Peter Jorgensen,
  Dr. Rafael Bocklandt and
  Dr. Alina Vdovina
  for their general practical and intellectual assistance during my research
  and previously.

  I owe a special debt of gratitude to Dr. James Ford and Dr. Bill Foster, who 
  enabled me to successfully apply for a place on the degree programme in 2004,
  leading me towards a new scientific vocation.

  I wish to express my gratitude and respect for my hosts at the Inter
  University Centre for Astronomy and Astrophysics, in Pune, India, where I
  spent a stimulating, effective and enjoyable month, writing up my thesis,
  and investigating some new material: 
  Prof. Kandu Subramanian, 
  Luke Charmandy,
  Pallavi Bhatt, 
  Nishant Singh and
  Dr. Pranjal Trivedi.

  I acknowledge the support of the staff and resources from;
  the Center for Science and Computing, Espoo, Finland, where the bulk of my
  simulations were computed; 
  the HPC-Europa~II programme, which funded my
  research visits to Finland; 
  UKMHD, who provided computing on the UK MHD
  Cluster, St. Andrews, Scotland and support for attendance at UKMHD
  conferences; Nordita, Stockholm, Sweden, and in particular 
  Prof. Axel Brandenburg,
  who provided financial and practical support for attendance at conferences 
  and development of the simulation code; 
  the International Space Science Institute, Berne, Switzerland support for
  attending their workshop and publishing;
  my funding research council the Engineering and Physical Sciences Research
  Council;
  the Science and Technology Facilities Council for additional support;
  and the pencil-code developers,
  among them Prof. Axel Brandenburg, Dr. Antony Mee, Dr. Wolfgang Dobler,
  Dr. Boris Dintrans, Dr. Dhrubaditya Mitra, Dr. Matthias Rheinhardt.

  In addition to those who have contributed directly to my research I would
  like to thank many, who have communicated informally, either by email or in 
  person, and in particular: Prof. Miguel de Avillez and Dr. Oliver Gressel for 
  advice arising from their previous experiences of similar modelling;
  Dr. Greg Eyink, Dr. Eric Blackman and Prof. Russell Kulsrud for their
  insights into volume averaging, magnetic helicity and cosmic rays;
  the anonymous reviewers of my submitted articles to Monthly Notices for
  their conscientious and constructive assessment of the work;
  for informal international discussions Dr. Marijke Haverkorn, Dr. Dominik
  Schleicher, Dr. Reiner Beck, Dr. Simon Calderesi, Dr. Gustavo
  Guerrero, Dr. J\"orn Warnecke, Dr. Sharanya Sur, Anne Liljstr\"ohm, 
  Nadya Chesnok, Prof. Sridhar Seshadri;
  and visitors to my department Dr. Michele Sciacco, Luca Galantucci,
  Alessandra Spagniolli, Carl Schneider, Dr. Mike Garrett, Brendan Mulkerin and
  Dr. Rodion Stepanov.

  In conclusion I thank my internal examiner, Dr. Paul Bushby, and my external
  examiner Prof. James Pringle for an interesting, intelligent and rigorous
  viva. 
  
  Finally I thank my parents and family for their patience and support 
  throughout my academic studies.

  \end{acknowledgements}
\thispagestyle{empty}
\cleardoublepage
                   % Acknowledgements.

  \begin{abstract}

  I model the multi-phase interstellar medium (ISM) randomly heated and 
  shocked by supernovae (SN), with gravity, differential rotation and other
  parameters we understand to be typical of the solar neighbourhood. 
  The simulations are in a 3D domain extending horizontally $1\times1\kpc^2$
  and vertically $2\kpc$, symmetric about the galactic mid-plane.
  They routinely span gas number densities  $10^{-5}$--$10^2\cmcube$,
  temperatures 10--$10^8\K$, speeds up to $10^3\kms$ and Mach number up to 25.  
  Radiative cooling is applied from two widely adopted parameterizations, 
  and compared directly to assess the sensitivity of the results to cooling.

  There is strong evidence to describe the ISM as comprising well defined cold,
  warm and hot regions, typified by $T\sim10^2\,,10^4$ and $10^6\K$, which
  are statistically close to thermal and total pressure equilibrium.
  This result is not sensitive to the choice of parameters considered here.
  The distribution of the gas density within each can be robustly modelled
  as lognormal.
  Appropriate distinction is required between the properties of the
  gases in the supernova active mid-plane and the more homogeneous phases 
  outside this region. 
  The connection between the fractional volume of a phase and its various
  proxies is clarified.
  An exact relation is then derived between the fractional volume and the
  filling factors defined in terms of the volume and probabilistic averages. 
  These results are discussed in both observational and computational contexts. 
 
  The correlation scale of the random flows is calculated from the velocity
  autocorrelation function; it is of order 100\,pc and tends to grow with
  distance from the mid-plane. 
  The origin and structure of the magnetic fields in the ISM is also 
  investigated in non-ideal MHD simulations.
  A seed magnetic field, with volume average of roughly ${4\nG}$, grows
  exponentially to reach
  a statistically steady state within 1.6\,Gyr.
  Following \citet{G92}, volume averaging is applied with a Gaussian kernel to 
  separate magnetic field into a mean field and fluctuations.
  Such averaging does not satisfy all Reynolds rules, yet allows a formulation
  of mean-field theory.
  The mean field thus obtained varies in both space and time. 
  Growth rates differ for the mean-field and fluctuating field and there is
  clear scale separation between the two elements, whose integral scales are
  about $0.7\kpc$ and $0.3\kpc$, respectively.
  
  Analysis of the dependence of the dynamo on rotation, shear and SN rate is
  used to clarify its mean and fluctuating contributions. 
  The resulting magnetic field is quadrupolar, symmetric about the mid-plane,
  with strong positive azimuthal and weak negative radial orientation.
  Contrary to conventional wisdom, the mean field strength increases away from
  the mid-plane, peaking outside the SN active region at $|z|\simeq300\p$.
  The strength of the field is strongly dependent on density, and in particular
  the mean field is mainly organised in the warm gas, locally very strong in the
  cold gas, but almost absent in the hot gas.
  The field in the hot gas is weak and dominated by fluctuations. 

\end{abstract}
\thispagestyle{empty}
\cleardoublepage
                      % Abstract.

  \pagenumbering{roman}                 % Change the page numbering style for
                                        % Table of Contents and Lists.

  \tableofcontents

  \listoffigures

  \listoftables

  \clearpage                            % End the current page making sure all
  \thispagestyle{empty}                 % tables/figures are printed.
  \cleardoublepage                      % Necessary for correct page numbering.

  \pagenumbering{arabic}                % Reset the page numbering style.

  \part{Motivation and outline: magnetism and \\the interstellar medium 
        in galaxies}\label{part:intro}

%  \clearpage                            % End the current page making sure all
%  \thispagestyle{empty}                 % tables/figures are printed.
%  \cleardoublepage                      % Necessary for correct page numbering.

  \begin{chapter}{How to shed some light on galaxies? \label{chap:intro}}
\section{\label{sect:whatsinit} Outline of the content}

\subsection{\label{subsect:motivation}Motivation behind this work}

  As we look into the sky, for generations mankind has been captivated by its
  beauty, awestruck by the spectacular (aurora, comets), and reassured by
  its familiarity and constancy over the centuries. 
  The combination of 
  improved observations and measurements, mathematics and validated theory
  have transformed our
  understanding of the objects populating our own solar system and the most
  distant of galaxies.
  In doing so the universe has grown to fill spaces and
  time so vast our minds cannot easily conceive them. We have discovered
  new entities (black holes, supernovae, jets and pulsars, etc) that have
  been so
  vast or alien, that they are a challenge to describe. To do so has required
  combining theories about the very large (general relativity) and the very 
  small (bosons, cosmic rays, quantum mechanics, etc). 
  Hence, in contrast to the constant and familiar, astronomy and 
  astrophysics have instead
  continually challenged our preconceptions and uncovered shocking and 
  surprising discoveries. 

  In more recent years it has even become possible to visit space and
  to devise
  machines that enable us to peer further into space, and also to view the 
  universe in wavelengths invisible to the human eye, the radio spectrum, 
  x-rays, infrared, etc. 
  The rate
  of discovery and understanding has accelerated. 
  It has become apparent that magnetic fields are ubiquitous, with
  many planets and stars generating their own magnetic fields. Of interest to
  this study, it has also been discovered that the tenuous gas between the
  stars is magnetized and in many galaxies, including our own, these magnetic
  fields are organized on a galactic scale. 
  Scientific controversy surrounds these structures, which cannot easily be 
  explained with what we know so far about magnetism. 

  Despite all the advances in technology, even with instruments in orbit,
  observational data is effectively constrained to line of
  site measurements. 
  On galactic time scales, we are also in effect viewing a 
  freeze frame of the sky, and any understanding of motion and evolution has 
  to be inferred by comparing differences between galaxies at different stages 
  of development. Assumptions and estimates must be invoked about the 
  composition and distribution of the gas along these lines of measurement. If
  we can accurately model astrophysical objects in three
  dimensions with numerical simulations, these can be used to assess the best 
  methods of estimating the 3D composition. The movement and structure of the 
  astrophysical models can be investigated more easily and inexpensively than 
  the galaxies directly, and be used to motivate useful and effective targets
  for future observations and investigation.

\subsection{\label{subsect:order} Structure and contents}

  In the rest of the Introduction, I provide a
  brief description of galaxy structure, some of the primary mysteries of 
  interest here and how this research might address these.
  Section~\ref{chap:review} reviews some of the progress made to date
  with similar models and discusses the significant differences between 
  what is and is not included within the various models. 

  The most substantial result of my research has been the construction of a 
  robust and versatile numerical model capable of simulating a 3D section of
  a spiral galaxy. The system is extremely complicated physically and also,
  inevitably, numerically. The simulations need to evolve over weeks or months
  and overcoming numerical problems has taken a large proportion of my research 
  time. However the most interesting outcomes are not numerical, but the 
  results and analysis obtained from the simulations. I therefore 
  reserve my comments on some of the critical numerical insights to the 
  Appendices. These may be of interest to others interested in numerical 
  modelling, but less so to the wider scientific community. However, I think
  it is important not to brush these issues under the carpet for at least two
  reasons. First, it is important to be clear and honest about the deficits
  in one's model, so that the reader may be able to understand how robustly
  the model may capture various features of the interstellar medium.
  Secondly, sharing the 
  experiences may help to improve the models of others.
  
  However, so that the reader may reasonably understand the outcomes, in 
  Part~\ref{part:model} I fully describe the model. I state what
  physical components are included, and discuss the 
  implications of any necessary numerical limitations or exclusions.    
  In Parts~\ref{part:mphase}--\ref{part:dynamo} I describe and explain in 
  detail the new
  results I have obtained. These include data from two sets of simulations,
  conducted over a period exceeding 12 months. They differ in that the 
  second set include the addition of a seed magnetic field. Rather than 
  discuss each simulation or set separately, I describe various 
  characteristics of the ISM and consider to what extent these depend on 
  any model parameters.
  Part~\ref{part:sum} contains a summary of my work, conclusions and 
  discussion of the future of this research approach.  

  \subsubsection{Collaborative and previously available content}
  Some of the work presented in this thesis is the result of collaboration,
  which has been submitted for publication. A review
  of previous related numerical work has appeared in a published chapter
  \citep{FG11}, which will form part of a
  book from the International Space Science Institute (ISSI). 
  Chapter~\ref{chap:review} covers similar subject matter, but has been
  substantially revised.
  The thesis contains material, submitted
  to MNRAS \citep{FG12,FG12a}, in which I am lead author with 
  co-authors Prof.~A.~Shukurov, Dr.~A.~Fletcher, Dr.~G.~R.~Sarson and
  Dr.~M.~Mantere. The latter
  article forms the basis of Chapter~\ref{chap:meanB}, with some additional 
  new work in Section~\ref{sect:eval}. Material from the former article is 
  included in Chapters~\ref{chap:NI}, \ref{chap:HDref} and \ref{chap:flow}, 
  plus Sections~\ref{subsect:COOL} and \ref{sect:summulti}, and 
  Appendices~\ref{chap:EISNR} and \ref{chap:stab}. 
  All numerical simulation code refinement, testing and running was performed 
  by myself.

%--------------------------------------------------------------------------
\section{\label{sect:gal} A brief description of galaxies}
%--------------------------------------------------------------------------

  The largest dynamically bound structures in the universe so far 
  observed are galaxy clusters. These collections of multiple galaxies of
  various types interact gravitationally and also contain an 
  intergalactic magnetic field among other features. They occupy
  regions of space spanning order of 10 -- 100 mega-parsecs (Mpc).
  Within clusters
  galaxies may collide, merge or accrete material from one to another. 
  Groups of galaxies, typically spanning order 1\,Mpc, contain a few dozen 
  gravitationally bound galaxies. Our own galaxy, the Milky Way, belongs to 
  the Local Group containing about 40 galaxies. We refer to the Galaxy 
  (capitalised), to 
  denote the Milky Way, to distinguish it from galaxies in general 
  (lower case).

\subsection{\label{subsect:class}Classification}

  The next largest category in the hierarchy of such structures are the
  galaxies. All galaxies will interact with the intergalactic 
  material surrounding them to a greater or lesser extent. Nevertheless they
  generally
  contain powerful internal forces and structure, which may govern the internal
  dynamics irrespective of the external environment. An early attempt to 
  identify the different categories of galaxies was made by Edwin Hubble in
  the 1920s, illustrated in Fig.~\ref{fig:hubble}.  
   
     \begin{figure}[htb]
      \begin{center}
       \includegraphics[width=0.5\linewidth]{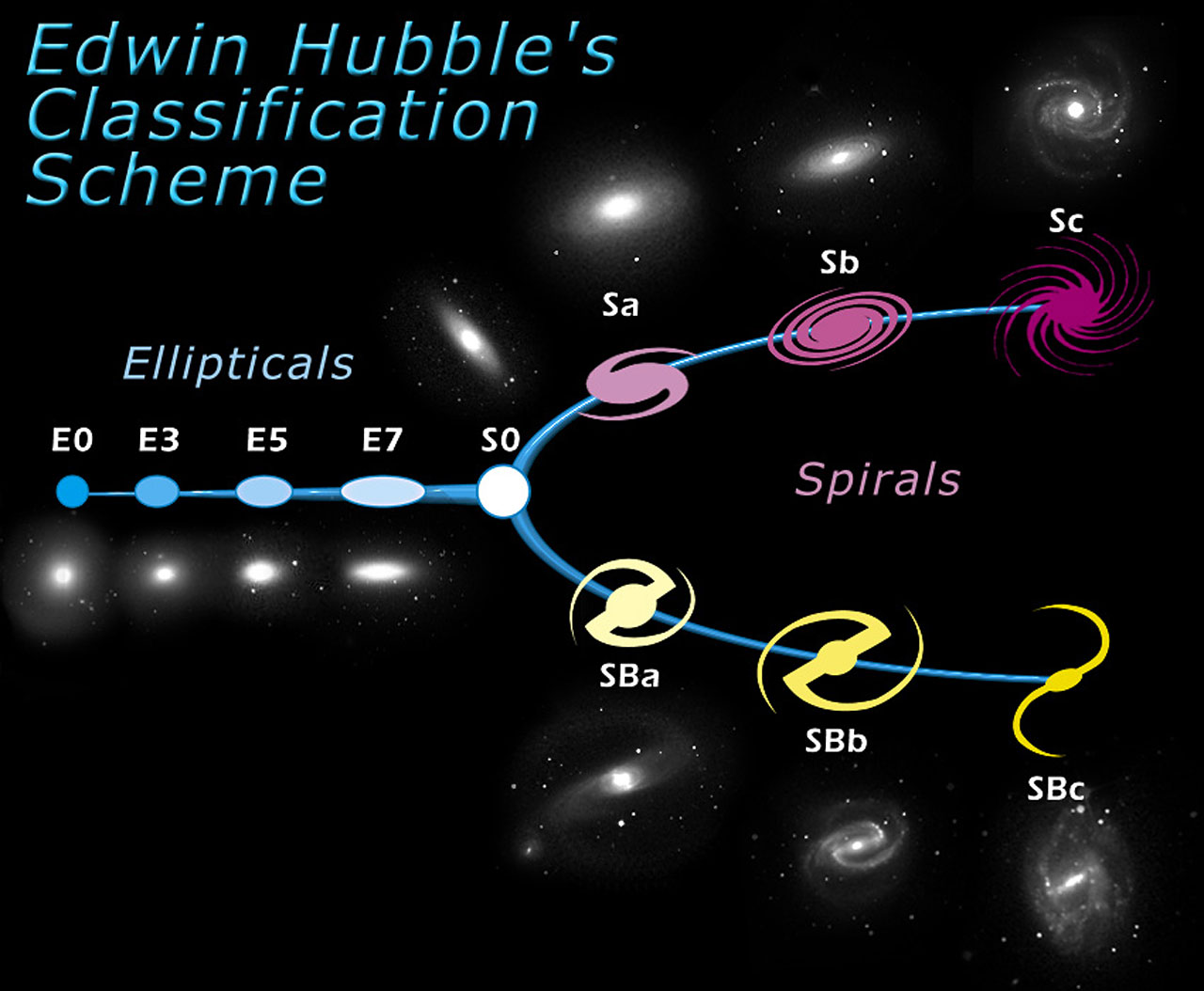}
       \caption[Hubble classification diagram]{Hubble's  traditional tuning fork classification diagram for 
  galaxies. Courtesy of the European Space Agency (ESA) 
  \url{http://www.spacetelescope.org/images/heic9902o/}
  Galaxies are defined as elliptical (E) or spiral (s) with further division of
  spirals into normal or barred (Sb)
       \label{fig:hubble}}
      \end{center}
     \end{figure}

  The classification of galaxies turned out to be more complicated than 
  Hubble could have envisaged. 
  In the Hubble classification galaxies are 
  elliptical (E) or spiral (S). 
  Spirals also include barred galaxies 
  where the spiral arms emerge from the tips of an elongated central layer, 
  rather than the central bulge. 
  {\freply
  {It may not always be certain whether a galaxy, which appears elliptical as 
    viewed from our solar system, is not in fact a spiral, viewed face on.}}
  Subsequently \citet{vandenBergh76}  
  revised Hubble's morphological classification of galaxies, 
  but more recently \citet{MNR18600}, using the ATLAS$^{3\rm D}$ 
  astronomical survey, proposed a more physical classification based on 
  the kinematic properties of galaxies. Their revised classification diagram
  is shown in Fig.~\ref{fig:atlas}. 

     \begin{figure}[htb]
      \begin{center}
       \includegraphics[width=0.66\linewidth]{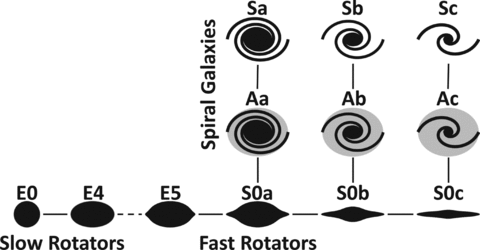}
       \caption[Revised kinematic galaxy classification diagram]{ The revised galaxy classification 
  diagram of \cite{MNR18600} using the ATLAS$^{3\rm D}$ astronomical   
  survey. Galaxies are categorized as slow rotators, corresponding to
  elliptical shapes and fast rotators, which become increasingly flattened
  with smaller central bulge as a result of higher angular momentum.
       \label{fig:atlas}}
      \end{center}
     \end{figure}
   
  Considering the kinematic properties of the galaxies improves the 
  consistency of classification and reduces the reliance for identification
  on our viewing angle. 
  It also helps us to understand the underlying cause 
  for the variation in morphology. Disc galaxies rotate faster and, 
  significantly,
  about a single axis of rotation. In some cases elliptical galaxies exhibit
  very weak rotation. In others the shape is the result of multiple axes of 
  rotation, with angular momentum acting in two or three transverse
  directions. This action counteracts the tendency to flatten perpendicular   
  to an axis of rotation.

  \citet{MNR18600} found only a very small proportion of galaxies are
  actually elliptical (about 4\% in most clusters) and the rate of rotation 
  as well as stellar density is very important to the morphology 
  the strength of the magnetic field.  
  It is my long term aim to model the ISM in any class of galaxy, although
  the current model is constructed to simulate spiral galaxies. The most 
  comprehensive data pertains to the Galaxy, so as a first iteration with
  which we can verify the relevance of the model, it is convenient to use
  parameters matching the Galaxy in the solar neighbourhood and compare the
  outcomes with what we understand of the real Galaxy. 

  The Galaxy is estimated to {\freply{have a stellar disc of radius 
  approximately 16 kiloparsecs (kpc) with atomic hydrogen \HI\ up to 40\kpc}}, 
  and the Sun, is estimated to be 7 -- 9\,kpc from the Galactic centre. 
  (Here capitalised Sun denotes our own star.) Most of the mass in the 
  Galaxy is in the form of non-interacting dark matter and stars, with
  stars accounting for about 90\% of the visible mass with gas and dust the
  other 10\%. 
  There is a large central
  bulge and although hard to identify from our position inside the Galaxy,
  it is likely that we live in a barred galaxy. Away from the centre 
  most of the mass is contained within a thin disc $\pm200$ parsecs (pc) of the 
  mid-plane \citep{Kulkarni87,Clemens88,Bronfman88,F01} over a Galactocentric 
  radius of about 4 -- 12\kpc. Either side of the disc is a region 
  referred to as the galactic halo with height of order 10\,kpc. 
  Halo gas is generally more diffuse and hotter gas in the disc.

  \begin{figure}[htb]
    \begin{center}
  \includegraphics[width=0.45\linewidth]{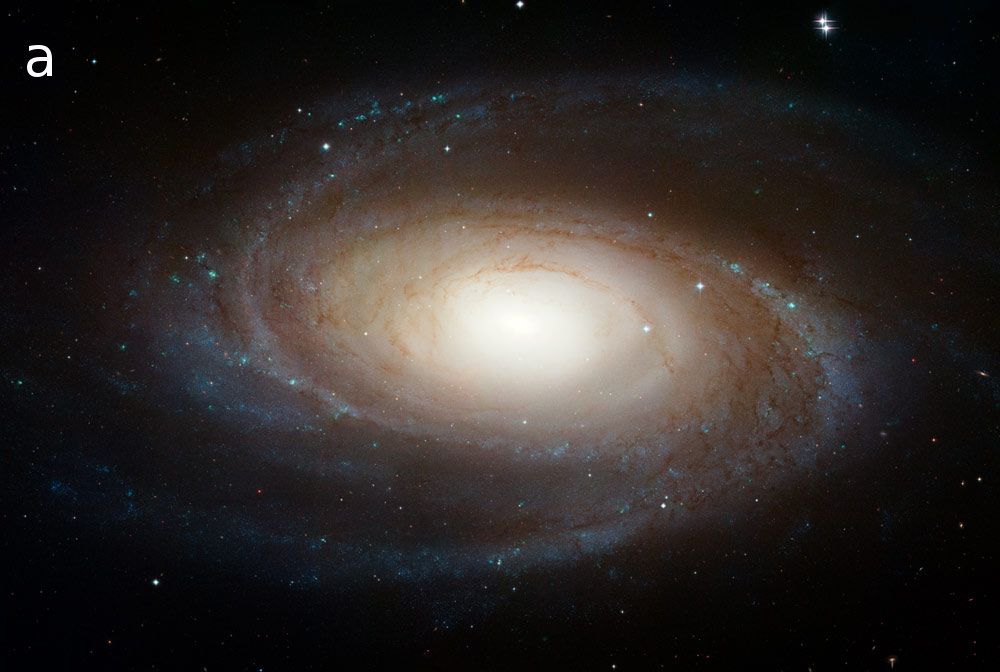}\vspace{0.5cm}\\
  \hspace{-0.75cm}
  \includegraphics[width=0.4175\linewidth]{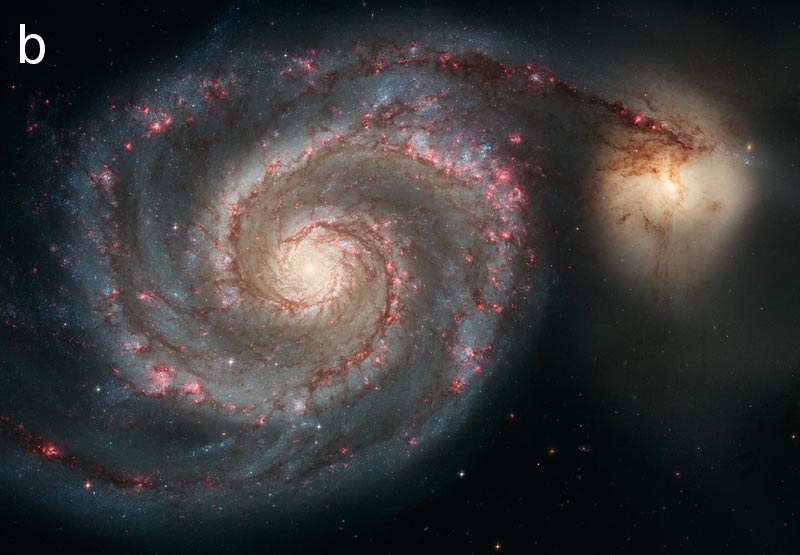}
  \hspace{0.35cm}
  \includegraphics[width=0.51\linewidth]{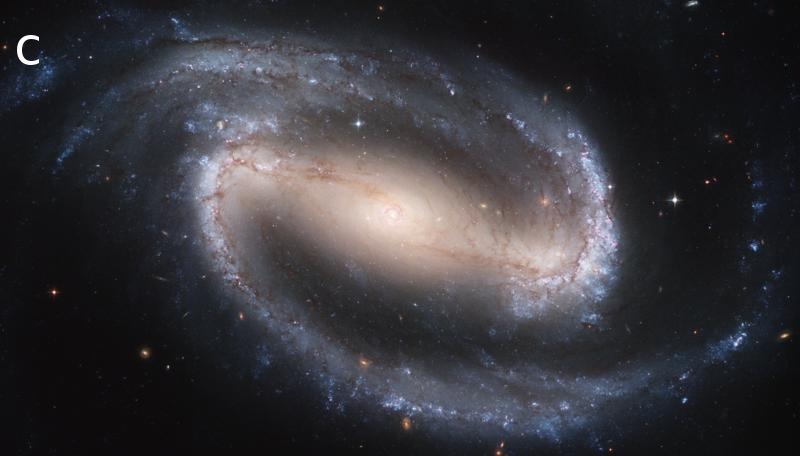}
  \hspace{-0.75cm}\\
  \vspace{0.5cm}
  \includegraphics[width=0.465\linewidth]{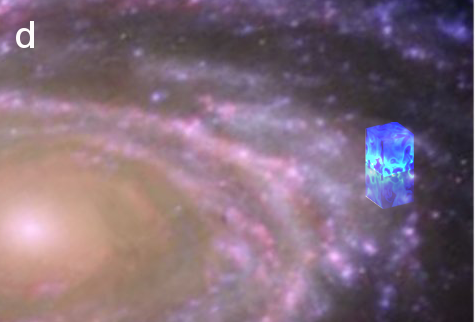}\hspace{3.95cm}
    \caption[Images of galaxies]{
  Images of spiral galaxies M81 (a), Whirlpool (b), and  the
       barred spiral galaxy NGC 1300 (c)
       from \url{http://hubblesite.org/gallery/album/galaxy/} The image in
       panel (d) represents the typical location of the simulation region,
       superimposed as a blue box, 
       orbiting the galaxy within the mid-plane of the disc. 
       It extends above and below the mid-plane. 
    \label{fig:galaxy}}
    \end{center}
  \end{figure}
   
  Figure~\ref{fig:galaxy} shows examples of disc galaxies, in
  panels (a -- c). Panel d indicates the typical location and size
  of the simulation 
  region in a galaxy, which will be described in Part~\ref{part:model}.
  Here the disc is rotating around the axis at
  about $230\kms$ (for the Galaxy in the solar neighbourhood). As the ISM
  nearer to the centre of a galaxy travels a shorter orbit, differential 
   rotation induces a shear through the ISM in the azimuthal direction, with
  the ISM trailing further away from the galactic centre. Rotation and shear
  vary between different galaxies, and also over different radii within each
  galaxy. These features can affect the classification of a galaxy and appear
  to be
  important to the strength and organization of the magnetic field.

\subsection{\label{subsect:ism}The interstellar medium}

  The space between the stars is filled with a tenuous gas, the interstellar 
  medium (ISM), which is mainly hydrogen (90.8\% by number [70.4\% by mass])
  and helium (9.1\% [28.1\%]), with much smaller amounts of heavier elements 
  also present and a very small proportion of the hydrogen in molecular
  form \citep{F01}.
  The density of the ISM in the thin disc is on average below 1 atom
  $\cmcube$. (In air at sea level the gas number density is about
  $10^{27}\cmcube$.) The ISM contains gas with a huge range
  in density and temperature, and can be described as a set of phases each with 
  distinct characteristics.
  
  The multi-phase nature of the ISM affects all of its properties, including
  its evolution, star formation rate, galactic winds and fountains, and 
  behaviour of the magnetic fields, {\freply{which in turn feed back into the}}
  cosmic rays. 
  In a widely accepted picture \citep{CS74,MO77}, most of the volume is
  occupied by the hot ($T\simeq10^6\K$), warm ($T\simeq10^4\K$) and cold
  ($T\simeq10^2\K$) phases. 
  The concept of the multi-phase ISM in pressure equilibrium has endured with 
  modest refinement \citep{Cox05}. 
  Perturbed cold gas is quick to return to equilibrium due to short cooling 
  times, while warm diffuse gas with longer cooling times has persistent 
  transient states significantly out of thermal pressure balance 
  \citep[][ and references therein]{KKRev09}. 
  Dense molecular clouds, while containing most of the total mass of the
  interstellar gas, occupy a negligible fraction of the total volume and are of 
  key importance for star formation 
  \citep[e.g.][]{Kulkarni87, Kulkarni88, Spitzer90, McKee95}. 
  Gas that does not form one of the three main, stable phases but is in a 
  transient state, may also be important in some processes. 
  The main sources of energy maintaining this complex structure are supernova explosions (SNe) and stellar
winds \citep[][and references therein]{MLK04}. The clustering of SNe in OB
associations facilitates the escape of the hot gas into the halo thus reducing
the volume filling factor of the hot gas in the disc, perhaps down to 10\% at
the mid-plane \citep{NI89}. The energy injected by the SNe not only produces
the hot gas but also drives compressible turbulence in all phases,
as well as driving outflows from the disc associated with the galactic wind
or fountain, as first suggested by \citet{Bregman80}. Thus turbulence, the
multi-phase structure, and the disc-halo connection are intrinsically related
features of the ISM.
  
  Although accounting for only about 10\% of the total visible mass in 
  galaxies, the turbulent ISM is nevertheless important to their dynamics and
  structure. 
  {\freply{Heavy elements are forged by successive generations of stars.
  These are recycled into the ISM by SN feedback and transported by the ISM 
  around the galaxy to form new stars. 
  The ISM supports the amplification and regulation of a magnetic field, which
  also affects the movement of gas and subsequent location of new stars.
  Heating and shock waves within the ISM, primarily driven by supernovae, 
  results in thermal and turbulent pressure alongside magnetic and cosmic ray 
  pressure, which supports the disc against gravitational collapse.}}
  
\end{chapter}
                     % A chapter.
  \begin{chapter}{\label{chap:review} A brief review of interstellar modelling}
  \section{\label{sect:review}Introduction}

Work to produce a comprehensive description of the complex dynamics of the multi-phase ISM has
been significantly advanced by numerical simulations in the last three decades,
starting with \citet{CP85}, followed by many others including
\citet{Rosen93,RB95,V-SPP95,PV-SP95,RBK96,Korpi99,GP99,WN99,Avillez00,WN01,AB01,
AM-L02,WMN02,AB04,Balsara04,AB05a,AB05b,Slyz05,M-LBKA05,Joung06,AB07,WN07,
Gressel08,HJMBM12}.

Numerical simulations of this type are demanding even with the best computers
and numerical methods available. The self-regulation cycle of the ISM includes
physical processes spanning enormous ranges of gas temperature and density,
as well as requiring a broad range of spatial and temporal scales. It involves star formation in the
cores of molecular clouds, assisted by gravitational and thermal instabilities
at larger scales, which evolve against the global background of transonic
turbulence, in turn, driven by star formation and subsequent SNe \citep{MLK04}. 
It is understandable that none of the existing numerical models cover the whole
range of parameters, scales and physical processes known to be important.

Two major approaches in earlier work focus either on the dynamics of diffuse
gas or on dense molecular clouds. In this chapter I will review a range of
  previous and current models, relating them to the physical features they
  include or omit. As with the general structure of the thesis, it is the 
  extent to which physical processes are investigated that is of most 
  interest, therefore I shall consider the physical problems and detail how
  various numerical approaches have been applied to these.

\section{\label{sect:sform}Star formation and condensation of molecular clouds}

  I shall not explicitly consider 
  star formation. This occurs in the most dense clouds, which form from 
  compressions in the turbulent ISM. As densities and temperatures attain
  critical levels in the clouds the effects of self-gravity and thermal 
  instability may accelerate the collapse of regions within the clouds to
  form massive objects, which eventually ignite under high pressures to form
  stars. 

  The clouds themselves typically occupy regions spanning only a few parsecs. 
  The final stages of star formation occurs on scales many magnitudes smaller  
  than that, which 
  may be regarded as infinitesimal by comparison to the dynamics of the 
  galactic rotation and disc-halo interaction.
  \citet{BDRP06} model star formation by identifying regions of self 
  gravitating gas of number density above $10^5\cmcube$ and size$\le0.5\p$. 
  To track the evolution of even one star, requires very powerful computing
  facilities, and a time
  resolution well below 1 year. Currently this excludes the possibility of
  simultaneously modelling larger scale interstellar dynamics.
  
  Modelling on the scale of molecular clouds, incorporating subsonic and 
  supersonic turbulence and self-gravity is reported in \citet{Klessen00}
  and \citet{Heitsch01} with and without magnetic field. These authors 
  investigated
  the effect of various regimes of turbulence on the formation of multiple
  high density regions, which may then be expected to form stars. The 
  subsequent collapse into stars was beyond the resolution of these models.
   Magnetic
  tension inhibited star formation, but this was subordinate to the effect
  of supersonic turbulence, which enhanced the clustering of gas within
  the clouds. These clusters provide the seed for star formation. 
  In these models, supersonic turbulence was imposed by a numerical 
  prescription. Physically the primary drivers of turbulence are SNe, and
  these evolve on scales significantly larger than the computational domain,
  so it would be difficult to include SNe and the fine structure of molecular 
  clouds in the same model. 

  In addition to compression and self-gravity, thermal instability may also
  accelerate the formation of high density structures within the ISM. 
  \citet{BKM07} investigated the effects of turbulence and thermal instability,
  including scales below 1\,pc. Their analysis concluded that  
  turbulent compression dominated that of thermal instability in the formation 
  of the densest regions. 
  The
  minimum cooling times were far longer than the typical turnover time of
  the turbulence.  
    
  \citet{Slyz05} included star formation and a thermally unstable cooling 
  function with a numerical domain spanning 1.28\,kpc. The typical resolution 
  was 10 or 20\,pc. As previously stated, the dynamics of compression at 
  higher resolution substantially dominate those of self-gravity and 
  thermal instability. At this resolution, maximum density was about 
  $10^2\cmcube$ and minimum temperatures were above 300. They found a 
  sensitivity in their results to the inclusion or exclusion of 
  self-gravity, which \citet{{Heitsch01}}, with higher resolution, found to be
  subordinate to the effects of turbulence on the density profile. I would 
  argue that this indicates that including self-gravity at such large 
  scales may therefore produce artificial numerical rather than physical
  effects. The gravitational effect of the ISM on scales greater than 1\,pc
  are negligible, compared to stellar gravity and other dynamical effects.
  
  In summary, modelling star formation and molecular clouds in the ISM 
  necessitates a resolution significantly less than 1\,pc. Self-gravity 
  within the ISM is significant only for dense structures inside molecular
  clouds. Such gravitation is proportional to $l^{-2}$, where $l$ denotes the
  distance from some mass sink, so is a highly localised phenomenon.  
  Star formation in large scale modelling can therefore at best be
  parametrized by the removal of mass from the ISM based on the 
  distribution of dense mass structures and self-gravity neglected.

  \section{\label{sect:disc} Discs and spiral arms}

  To model a whole spiral galaxy requires a domain radius of order  
  $10\kpc$ and potentially double that in height to include the halo. Often
  to avoid excessive computation, when modelling scales spanning the whole
  disc some authors either adopt a
  2D horizontal approach \citep{Slyz03}, or adopt a very low vertical extent
  in 3D \citep{Dobbs08,Hanasz09,KZ09,DP10}. Given the
  mass, energy and turbulence are predominantly located within a few hundred
  parsecs of the mid-plane,
  some progress can be made with this approach.

  A number of important physical features of the ISM must be neglected at this
  resolution. An important factor in the self-regulation of the ISM is the 
  galactic fountain, in which over-pressured hot gas in the disc
  is convected into the halo,
  where it cools and rains back
  to the disc. 
  How this affects star-formation and SNe rates is poorly understood. 

  The multi-phase ISM cannot currently be resolved at these scales. The 
  cold diffuse gas exists in clumpy patches of at most a few parsecs. Even the
  hot bubbles of gas, generated at the disc by SNe and clusters of SNe are
  typically less than a few hundred parsecs across. The resolution of these 
  global models is of order 100\,pc, which is insufficient to include the separation  
  of scales of about $10^3$ between the hot and the warm gas. 
  Generally these models apply an isothermal ISM.
  \citet{Dobbs08} include a multiphase medium, by 
  defining cold and warm particles, with fixed temperature and density, but
  must exclude any interaction between the phases. 
   
  The random turbulence driven by SNe must also be neglected or weakly
  parametrized. \citet{Dobbs08} exclude SNe with the effect of significantly
  reducing the disc height between the spiral arms. \citet{Hanasz09} neglect 
  the thermal and kinetic energy and inject energy only through current rings
  of radius 50\,pc and 1\% of the total SN energy. \citet{Slyz03} use forced
  supersonic random turbulence to generate density perturbations. 

  These models can reproduce structures similar to spiral arms and, where 
  a magnetic field is included, an ordered field on the scale of the spiral
  arms. These can be related to different rates of rotation and for various
  density profiles. Even in 3D however these only describe the flat 2D 
  structure of the disc and spiral arms. Both the velocity and the magnetic 
  fields are vectors and subject to non-trivial transverse effects. The 
  effect of asymmetry on these vectors in the vertical direction cannot be 
  well represented with this geometry. 

  Features such as the vertical density distribution of the ISM, the thermal 
  properties and turbulent velocities cannot be understood from these models.
  They must be introduced as model parameters, derived from observational 
  measurements or smaller scale simulations. On the other hand structures 
  such as spiral arms are too large to generate in more
  localised simulations. The global simulations may help to identify suitable
  parameters for introducing spiral arms into local models, such as the scale 
  of the density
  fluctuations, orientation of the arms, typical width of the arms, how the
  strength or orientation of the magnetic field varies between the arm and
  interarm regions. 
  
  \section{\label{sect:local} The multi-SNe environment}

  This thesis concerns modelling the ISM in a domain larger than the size
  of molecular clouds (Sect~\ref{sect:sform}) and much smaller than a
  galaxy, or even just the disc (Sect~\ref{sect:disc}). Models on
  this scale were introduced by \citet{Rosen93}, in 2D with 
  much lower resolution, and without SNe. Nonetheless these already were 
  able to identify features of the galactic fountain and the multiphase 
  structure of the ISM. Individual SNe were investigated numerically by
  \citet{Cowie81}. Extensive high resolution simulations led to the 
  refinement of the classic Sedov-Taylor solution \citep{Sedov59,Taylor50} 
  and subsequent snowplough describing the evolution of an expanding blast wave 
  \citep{Ostriker88,Cioffi98}. The modelling of multiple SNe in the form of
  superbubbles within a stratified ISM was advanced by \citet{Tomisaka98}.
  The first 3D simulations of SNe driven turbulence in the ISM were by 
  \citet{Korpi99} and, independently, \citet{Avillez00}. In this section I 
  consider the key physical and numerical elements of these and
  subsequent similar models, and the extent to which their inclusion or 
  omission improves or hinders the results.

  \subsection{\label{subsect:strat} The stratified interstellar medium}

  From the images in Fig.\ref{fig:galaxy} and the profiles illustrated in 
  Fig.~\ref{fig:atlas} it can be seen that 
  the matter in spiral galaxies is substantially flattened into a disc. 
  This morphology
  strongly depends on the rate and form of rotation of the galaxy. 
  Towards the centre of these galaxies is a bulge, where the density
  structure and dynamics differ substantially from those in the disc.
  In elliptical galaxies the structure may involve asymmetries that 
  are complicated to model.
  I shall defer consideration of the central region and 
  elliptical galaxies to elsewhere and confine this thesis to the investigation 
  of the disc region of fast rotating galaxies.

  Along the extended plane of the disc in fast rotators 
  the dense region of gas may reasonably be approximated as unstratified
  to within about a few hundred parsecs of the mid-plane depending on the   
  galaxy. The thickness of the disc away from the central bulge can, in 
  many cases, to first approximation be regarded as constant, although 
  it does usually increase exponentially away from the galactic centre. Thus,
  towards the outer galaxy, even over relatively short distances the flat disc 
  approximation breaks down.
  Providing the model domain remains sufficiently within the thickness of the 
  disc, some progress can be made by ignoring stratification, 
  \citep[eg][]{Balsara04,Balsara05}. 

  Important to the dynamics of the interstellar medium is its separation
  into characteristic phases in apparent pressure equilibrium. Is this 
  separation a real physical effect, or just a statistical noise? What are
  the thermodynamical properties of the gas, what are the filling factors, 
  typical motions and, if the phases do in fact differ qualitatively, how 
  do they interact? The answers to these questions are critically different
  when considering the ISM to be stratified or not. \citet{Balsara04} 
  investigated the effect of increasing rates of supernovae on the 
  composition of the ISM. With initial density set to match the mid-plane
  of the Galaxy, increasing the rate of energy injection resulted in increased 
  proportions of cold dense and hot diffuse gas and correspondingly less warm
  gas. The typical radius of 
  SN remnant shells in such a dense medium remained below 50\,pc before being
  absorbed into the surrounding turbulence. The result of increased SNe is
  higher pressures and eventually overheating of the ISM. 

  However when a stratified ISM is included, the increased pressure from 
  SNe near the mid-plane is released away from the disc, by a combination of
  diffuse convection and blow outs of hot high pressure bubbles. 
  \citet{Korpi99a,Avillez00,AB01} included stratification. 
  They found superbubbles, which
  combine multiple SNe merging or exploding inside existing remnants in the
  disc and then breaking out into the more diffuse layers, where they 
  transport the hot gas away. They also had individual SNe exploding at 
  distances $\ga 500\p$ from the mid-plane, where the ambient ISM is very
  diffuse and the remnants of these SNe extend to a radius of a few hundred
  parsecs. This also has 
  an effect on the typical pressures, turbulent velocities and mixing
  scales. Instead of the cold gas and hot gas becoming more abundant at the
  mid-plane, the hot gas is transported away and the thickness of the disc 
  expands reducing the mean density and mean temperature at the mid-plane. 

  Due to the gravitational potential, dominated by stellar mass near the 
  mid-plane, there is a vertical density and pressure gradient. This will
  vary depending on the galaxy, but in most simulations of this nature it is
  useful to adopt the local parameters, since data relating to the solar neighbourhood
  is generally better understood, and therefore is convenient for benchmarking
  the results of numerical simulations. 

  There is a 
  natural tendency for cold dense gas to be attracted to the mid-plane and 
  hot diffuse gas to rise and cool. Estimates from observations in the 
  solar vicinity place the Gaussian scale height of the cold molecular gas at
  about  $80\p$ and the neutral atomic hydrogen about $180\p$ 
  \citep{F01}. The warm gas is estimated to have an
  exponential scale height of $390-1000\p$. Observationally hot gas is more
  difficult to isolate and estimates vary for the exponential scale height; 
  ranging from as low as 1\,kpc to over 5\,kpc. Hence to reasonably produce the 
  interaction between the cold and warm gas, we at least need to extend 
  $\pm 1\kpc$ vertically. \citet{Avillez00} with a vertical extent of 
  $\pm4\kpc$ found the ISM above about 2.5\,kpc to be 100\% comprised of
  hot gas, and this vertical extent was not 
  high enough to observe the cooling and recycling of the hot gas back to the
  disc. Subsequent models 
  \citep[including][]{Avillez01,AB01,AM-L02,Joung06,Joung09,HJMBM12} use an
  extended range in $z$ of $\pm 10\kpc$ and find the volume 
  above about $\pm2.5\kpc$ almost exclusively occupied by the hot gas for
  column densities 
  and SNe rates comparable to the solar vicinity. 
  Many of these models exclude the magnetic field and all exclude cosmic rays
  so the models have somewhat thinner discs than expected from observations. 
  With the inclusion of magnetic and cosmic ray pressure, we expect the
  disc to be somewhat thicker and the scale height of the hot gas to increase.
  
  The build up of thermal and turbulent pressure by SNe near the mid-plane 
  generates strong vertical flows of the hot gas towards the halo. With 
  vertical domain $|z|=10\kpc$  \citet{AB04} are able to attain an equilibrium with
  hot gas cooling sufficiently to be recycled back to the mid-plane, 
  replenishing the star forming regions and subsequently SNe. Such recycling 
  may be expected to occur in the form of a galactic fountain. 
  However  
  \citet{Korpi99a,Korpi99} found the correlation scale in the warm gas of the
  velocity
  field remained quite consistently about 30\,pc independent of $z$ location,
  but in the hot gas it increased from about 20\,pc at the mid-plane to over 
  150\,pc at $|z|\simeq300\p$ and increasing with height. 
  Other authors report correlation scales similar near the mid-plane,   
  but the vertical dependence of correlation lengths requires further
  investigation.
  If there is an increase in the correlation length
  of the velocity field with height, as $|z|\rightarrow1\kpc$ the scale of the
  turbulent structures of the ISM can become comparable to the typical size of
  the numerical domain. Consequently in extending the $z$-range to 
  include the vertical dimension of the hot gas, its 
  horizontal extent may exceed the width of too narrow a 
  numerical box. With a vertical extent of only $\pm1\kpc$, or with 
  \cite{Gressel08} $\pm2\kpc$, there is a net outflow of gas. Without a 
  mechanism for recycling the hot gas, net losses eventually exhaust the disc
  and the model parameters cease to be useful. \citet{Korpi99} found this
  restriction limited the useful time frame to a few hundred Myr. 
   It is difficult to combine in one model the detailed dynamics
  of the SNe driven turbulence about the mid-plane with a realistic global
  mechanism for the recycling of the hot gas. 

  There has also been some discussion on the effect of magnetic tension 
  inhibiting these outflows. Using 3D simulations of a stratified but  
  non-turbulent ISM with a purely horizontal magnetic field \citet{Tomisaka98}
  placed an upper bound on this confinement. When the ISM is turbulent the 
  magnetic field becomes disordered, the ISM contains pockets of
  diffuse gas, and the opportunity for hot gas blow outs increases 
  substantially \citep[see e.g.][]{Korpi99a,AB04,AB05b}.

%  \citet{AB04,AB05b} Compare the filling factors for the gas they
%  obtain including stratification with results by \citet{Balsara04}.  

  \subsection{\label{subsect:shocks} Compressible flows and shock handling} 
  
  The ISM is highly compressible, with much of the gas moving at
  supersonic velocities.
  The most powerful shocks are driven by SNe. Accurate modelling of a single 
  SNe in 3D requires high resolution and short time steps \citep{Ostriker88}.
  On the scales of interest here, hundreds or thousands of SNe are required 
  and this constrains the time and resolution available to model the SNe.
  Approximations need to be made, which retain the essential characteristics
  of the blast waves and structure of the remnants at the minimum 
  physical scales resolved by the model. 

  Shock handling is optimised 
  \citep[by e.g.][]{Avillez00,Balsara04,MLK04,Joung06,Gressel08} through adaptive mesh
  refinement (AMR),
  where increased resolution is applied locally for regions containing the
  strongest contrasts in density, temperature or flow. With increased spatial
  resolution the time step must also be much smaller, so there are 
  considerable computational overheads to the procedure. Many of the codes 
  utilising AMR do not currently include modelling of differential rotation, 
  although \citet{Gressel08} does apply shear using the Nirvana
  \footnote{http://nirvana-code.aip.de/} code.  An alternative 
  approach is to apply enhanced viscosities where there are strong convergent
  flows \citep[by e.g.][]{Korpi99}. This broadens the shock profiles and removes
  discontinuities, so care needs
  to be taken that essential physical properties are not compromised.   
   
  \subsection{\label{subsect:SNpop} Distribution and modelling of SNe}
  
  There are a number of different types of supernovae, with different
  properties and origin. Type~Ia SNe arise from white dwarfs, which are older stars that have used up most of 
  their hydrogen and comprise mainly of heavier elements oxygen and carbon. 
  If, through accretion or other mechanisms, their mass exceeds the
  Chandrasekhar limit \citep{C31} of approximately $1.38M\odot$ they become
  unstable and explode. Type~Ia are observed in all categories of galaxy and
  can have locations isolated from other types of SNe. 

  Type~II SNe are produced by massive, typically more than $10M\odot$, relatively young
  hot stars, which rapidly exhaust their fuel and collapse under their own
  gravity before exploding. Type~Ib and Ic are characterized by the absence of
  hydrogen lines in their spectrum and, for Ic, also their helium, which has been stripped by either
  stellar winds or accretion, before the more dense residual elements collapse
  to form SNe. Otherwise Type~II, Ib and Ic SNe are very similar, located only
  in fast rotating galaxies and populating mainly the star forming dense
  gas clouds near the mid-plane of the disc. For the remainder of this thesis
  I shall collectively denote these as Type~CC (core collapsing)
  and Type~I shall refer only to 
  Type~Ia. Type~CC SNe are more 
  strongly correlated in space and time than Type~I, because they form in
  clusters, evolve over a few million years and explode in rapid succession.

  In many disc galaxies, including the Milky Way, Type~CC SNe
  are much more prolific than Type~I. Both types effectively inject
  $10^{51}\erg$ into the ISM (equivalent to $10^{40}$ modern thermo-nuclear 
  warheads exploded simultaneously). Type~CC also contribute $4-20M\odot$   
  as ejecta at supersonic speeds of a few thousand $\kms$. Mainly
  located in the most dense region of the disc,
  Type~CC SNe energy can be more rapidly absorbed and explosion sites reach
  a radius of typically 50\,pc before becoming subsonic, while some Type~I 
  explode in more diffuse gas away from the mid-plane and can expand to 
  a radius of a few hundred parsecs. Because Type~CC are correlated in
  location and time, many explode into or close to existing remnants and form
  superbubbles of hot gas, which help to break up the dense gas shells of 
  SN remants and increase 
  the turbulence. Heating outside the mid-plane by Type~I SNe helps to
  disrupt the thick disc and increase the circulation of gas from within.

  A common prescription for location of SNe in numerical models, is to locate 
  them randomly, but uniformly in the horizontal plane and to apply a
  Gaussian or exponential random probability distribution in the vertical
  direction, with its peak at the mid-plane $z=0$. 
  Having determined the 
  location, an explosion is then modelled by injecting a roughly spherical
  region containing mass, energy and/or a divergent velocity profile. The 
  sphere needs to be large enough, that for the given resolution, the 
  resulting thermal and velocity gradients are numerically resolved. 
  This means that for typical resolution of a few pc the radius of the injection site will be several parsecs and 
  indicative of the late Sedov-Taylor or early snowplough phase (See 
  Section~\ref{sect:MSN} and Appendix~\ref{sect:SNPL}) some 
  thousands of years after the explosion. Subsequently gas is expelled by  
  thermal pressure or kinetic energy from the interior to form an expanding
  supersonic shell, which drives turbulent motions and heats up the ambient
  ISM. Problems have been encountered with modelling the subsequent
  evolution of these remnants. Cooling can dissipate the energy before the
  remnant shell is formed. Using kinetic rather than thermal energy to 
  drive the SNe evolution is hampered by the lack of resolution.

  As well as being physically consistent, clustering of SNe mitigate against 
  energy losses by ensuring they explode in the more diffuse, hot gas where they 
  are subject to lower radiative losses. To achieve clustering for
  the Type~CC SNe \citet{Korpi99a} selected sites with a random exponential
  distribution in $z$,
  excluded sites in the horizontal plane with density below average for the
  layer and then applied uniform random selection. \citet{Avillez00}
  apply a more rigorous prescription to locate 60\% of 
  Type~CC in superbubbles, in line with observational estimates 
  \citep{CSY79}. In addition to a simplified clustering scheme 
  \citet{Joung06} and \citet{Joung09} also adjust the radii of their
  SNe injection sites to enclose $60M\odot$ of ambient gas. This appears to
  optimise the efficiency of the SNe modelling; avoiding high energy losses
  at $T>10^8\K$ with a smaller injection site or avoiding too large an
  injection site, which dissipates the energy before the snowplough phase can
  be properly reproduced.

  The critical features of the SNe modelling must ensure that the SNe are 
  sufficiently energetic and robust to stir and heat the ambient ISM. It is 
  unlikely that the individual remnants can be accurate in every detail, 
  given the constraints on time and space resolution, but the 
  growth of the remnants should be physically consistent, producing remnants and 
  bubbles of multiple remnants at the appropriate scale. In particular we can
  expect
  about 10\% of the energy to be output into the ISM in the form of kinetic
  energy when the remnant merges with the ambient ISM. \citet{Spitzer78}
  estimates this to be 3\% and \citet{Dyson97} to be 30\%. 

  \subsection{\label{subsect:rotation} Solid body and differential rotation }
   
  To understand the density and temperature composition of the ISM the 
  inclusion of the rotation of the galaxy is not essential. Even an 
  appreciation of the characteristic velocities and Mach numbers 
  associated with the regions of varying density or temperature can be 
  obtained without rotation. 
  \citet{Balsara04,Balsara05,M-LBKA05} and \citet{AB05a} investigate 
  the magnetic field in the ISM without rotation or shear, the latter in the vertically
  extended stratified ISM, otherwise in an unstratified $(200\p)^3$ section. 
  The latter two models impose a uniform $\vect{B}$ of between $2-5\mkG$. The
  turbulence breaks up the ordered field (mean denoted $\vect{B}_0$) and with
  modest amplification of the
  total magnitude $B$, although the simulations last only a few hundred Myr. 
  Over a longer period, it would be expected that $\vect{B}_0$ would continue
  to dissipate in the absence of any restoring mechanism (rotation). 
  They analyze the distribution with respect to density and temperature of the
  fluctuations, $\vect{b}$, and mean parts of the field, and their 
  comparative strengths. 

  \citet{Balsara04} and \citet{Balsara05} investigate the 
  fluctuation dynamo, introducing a seed field of order $10^{-3}\mkG$
  and increasing it to $10^{-1}\mkG$ over 40\,Myr. This is insufficiently long 
  to saturate and there is no mean field, but they do examine the 
  turbulent structure of $\vect{b}$ and the dynamo process.

  However galactic differential rotation is a defining feature of the global
  characteristics
  of disc galaxies and is felt locally by the Coriolis effect and shearing
  in the azimuthal direction. There is 
  observational evidence that the magnetic field in disc galaxies, which is
  heavily randomised by turbulence, is also strongly organized along the
  direction of the spiral arms \citep[e.g.][]{Pat06,TKFB08,MNR18065}. 
  Differential rotation is responsible for the spiral structure, so  must be
  included to understand the source of the mean field in the galaxy. 

  There is also considerable controversy over the nature and even the 
  existence of the the galactic dynamo. A fluctuation dynamo can be generated
  by turbulence alone, but it is not clear that it can be sustained 
  indefinitely. The mean field dynamo cannot be generated without anisotropic
  turbulence and large scale systematic flows. 

  \citet{Gressel08,Gressel08a} include differential rotation and by 
  introducing a small seed field, are able to derive a mean field dynamo for
  a stratified ISM.
  The dynamo is only sustained in their model for rotation $4\Omega_0$,
  where $\Omega_0\approx230\kms$ is the angular velocity of the Galaxy in the
  solar vicinity. Otherwise they apply parameters the same as or similar to
  the local Galaxy. \citet{Gressel08b} also considered solid body rotation and
  found no dynamo, but concluded that shear, without rotation, cannot support
  the galactic dynamo. Without rotation diamagnetic pumping is too weak to 
  balance the galactic wind. The resolution of the model at $(8\p)^3$ limits
  the magnetic Reynolds numbers available, which
  may explain the failure to find a dynamo with rotation of $\Omega_0$.
  For a comprehensive understanding of the galactic magnetic field 
  differential rotation should not be neglected.

%  \subsection{\label{subsect:resolution}} 
%  \subsection{\label{subsect:boxsize}} 
%  \subsection{\label{subsect:shear}} 
  \subsection{\label{subsect:cooling} Radiative cooling and diffuse heating } 
         
  The multi-phase structure of the ISM would appear to be the result of a
  combination of the complex effects of supersonic turbulent compression,
  randomized heating by SNe and
  the differential cooling rates of the ISM in its various states. 
  Although excluding SNe, and having a low resolution, 
  \citet{Rosen93} generated a multi-phase ISM with just diffuse stellar 
  heating from the mid-plane and non-uniform radiative cooling. 

  Radiative cooling depends on the density and temperature of the gas and 
  different absorption properties of various chemicals within the gas on the
  molecular level. For models such as the ones discussed here detailed 
  analysis of these effects and estimates of the
  relative abundances of these elements in different regions of the ISM
  \citep[see e.g.][]{Wolfire95} need to be adapted to fit a monatomic gas
  approximation. Stellar heating varies locally, on scales far below the 
  resolution of these models, due to variations in interstellar cloud density 
  and composition as well ambient ISM temperature sensitivities. 

  Subtle variations in the application of these cooling and heating 
  approximations may have very unpredictable effects, due to the complex 
  interactions of the density and temperature perturbations and shock waves.
  In particular thermal instability is understood to be significant in 
  accelerating the gravitational collapse of dense clouds,
  by cooling the cold dense gas more quickly than the ambient diffuse warm
  gas. 

  Different models have used radiative cooling functions
  which are qualitatively as well as quantitatively distinct, making direct
  comparison uncertain. 
  \citet{VS00} compared their thermally unstable model to a different model
  by \citet{Scalo98} who used a thermally stable cooling function. They 
  concluded that thermal instability on $1\p$ scales is insufficient to 
  account for the phase separation of the ISM, but in the presence of other 
  instabilities increases the tendency towards thermally stable temperatures.
  Similarly, \citet{AB04} and \citet{Joung06} compared results obtained
  with different cooling functions, but comparing models with different 
  algorithms for the SN distribution and control of the explosions. 
  In the absence of SNe driven turbulence the ISM does separate into two 
  phases with reasonable pressure parity and in the presence of background
  turbulence this separation persists \citep{Sanchez02,BKM07}.
  Although the pressure distribution broadens, the bulk of both warm and
   cold gases broadly retain pressure parity.

  An additional complexity is advocated by \citet{ABnei10}. Generally it is 
  assumed that ionization through heating and recombination of atoms by 
  cooling are in equilibrium. However recombination time scales lag 
  considerably, so that cooling is not just dependent on temperature and 
  density. Cooler gas which had been previously heated to $T>10^6\K$ will 
  remain more ionized than gas of equivalent density and
  temperature which has not been previously heated. 
  \citet{ABnei10} show that this significantly 
  affects the radiation spectra of the gas, therefore they use a cooling
  function which varies locally,
  depending on the thermal history of the gas.  
 
  The choice of cooling and heating functions has a quantitative effect on the 
  results of ISM models and there is evidence that the inclusion or exclusion
  of thermal instability in various temperature ranges has a qualitative impact 
  on the structure of the ISM. However uncertainty remains over whether
  the adopted cooling functions are indeed faithfully reproducing the physical
  effects. There is also uncertainty over how critical the 
  differences between various parameterizations are to the results, once
  SNe, magnetic fields and other processes are considered. It is 
  certain, however, that differential radiative cooling is an essential 
  feature of the ISM. 
   
  \subsection{\label{subsect:mhd} Magnetism} 

  Estimates of the strength of magnetic fields indicate that 
  the magnetic energy density has the same order of
  magnitude as the kinetic and thermal energy densities of the ISM. Including
  magnetic pressure substantially increases the thickness of the galactic 
  disc compared to the purely hydrodynamic regime. Magnetic tension and the 
  Lorentz force will affect
  the velocities and density perturbations in the ISM. Ohmic heating and 
  electrical conductivity will affect the thermal composition of the ISM.
  Hydrodynamical models provide an excellent benchmark, but to 
  accurately model the ISM and be able to make direct comparisons with
  observations the magnetic field should be included.

  Comparisons between MHD and HD models have been made
  \citep{Korpi99a, Balsara04,  Balsara05,M-LBKA05,AB05a}. These either
  impose a 
  relatively strong initial mean field or contain only a random field, 
  and are evolved over a comparatively short time frame. So the composition
  of the field cannot reliably be considered authentic. 
  Ideally it would be useful to understand the dynamo mechanism; are there
  minimum and maximum rates of rotation conducive to magnetic fields? How
  does the SNe rate or distribution affect the magnetic field? What is the
  structure of the mean field, how do the mean and fluctuating parts
  of the field compare and how is the magnetic field related to the 
  multi-phase composition of the ISM? To ensure the field is the product of the
  ISM dynamics the mean field needs to be generated consistently with the key
  ingredients of differential rotation and turbulence, from a very small 
  seed field. The time over which the model evolves must be sufficiently long
  that no trace of the seed field properties persist. 
  
  \subsection{\label{subsect:cosmic} Cosmic Rays} 
 
  Cosmic rays are high energy charged particles, travelling at relativistic 
  speeds. Their trajectories are strongly aligned to the magnetic field lines
  \citep{Kulsrud78}. It is speculated that, amongst other potential sources,
  they are generated and accelerated in shock fronts around supernovae.
  They also occur in solar mass ejections, some of which are caught in the
  Earth's magnetosphere. The polar aurora result from these travelling along
  the field lines and heating the upper atmosphere as they collide.

  Cosmic rays are of interest in themselves, but also are estimated to make a 
  similar contribution to the energy density and pressures in the ISM as each 
  of the magnetic field, the kinetic and the thermal energies 
  \citep{Parker69}. As such their
  inclusion can be expected to significantly affect the global properties of
  the ISM.

  Although a substantial component of the ISM is not charged the highly 
  energized cosmic rays strongly affect the bulk motion of the gas. 
  The interaction between cosmic rays and the ISM is highly non-linear, but
  can be considered in simplified form by a diffusion-advection equation,
  \begin{align}\label{eq:cosmic}
   \frac{\partial e_{\rm cr}}{\partial t}+\nabla(e_{\rm cr}\vect{u})&=
   -p_{\rm cr}\nabla\vect{u}+\nabla(\hat{K}\nabla e_{\rm cr})+Q_{\rm cr}\,,&\\
    p_{\rm cr}&=(\gamma_{\rm cr}-1)e_{\rm cr}\,,&
  \end{align}  
  where $e_{\rm cr}$ and $p_{\rm cr}$ are the cosmic ray energy and pressure
  respectively, and $\hat{K}$ the cosmic ray diffusion tensor,
  $\gamma_{\rm cr}$ the cosmic ray 
  ratio of specific heats, and $Q_{\rm cr}$ a cosmic ray source term such as
  SNe. $\vect{u}$ is the ISM gas velocity. 

  \citet{Hanasz04,Hanasz05,Hanasz06,Hanasz11} use cosmic rays in a magnetized
  isothermal simulation to perturb the magnetic field and drive the dynamo in
  their global disc galaxy simulations. 
  None of the thermodynamic, stratified models of the ISM have yet included 
  cosmic rays. 

  \subsection{\label{subsect:ideal} Diffusivities }

  The dimensionless parameters characteristic of the ISM, such as the kinetic
  and magnetic Reynolds numbers (reflecting the relative importance of gas
  viscosity and electrical resistivity) and the Prandtl number (quantifying
  thermal conductivity) are too large to be obtainable with current computers.
  \citet{Lequeux05} estimates the Reynolds number $\Ry\simeq10^7$ for the cold
  neutral medium, with viscosity $\nu\simeq3\times10^{17}\cm^2\s^{-1}$.
  Uncertainty applies to these estimates in the various phases and also to 
  thermal and magnetic diffusivities, but physical $\Ry$ and $\Rm$
  (magnetic Reynolds number) are far
  larger and diffusivities far smaller than can be resolved numerically. 
  Given the very low physical diffusive co-efficients in the ISM some 
  models approximate them as zero. Applying ideal MHD where electrical 
  resistivity is ignored \citep[e.g.][and their subsequent models]{Avillez00,
  M-LBKA05,Joung06,Li05}
  the magnetic field is modelled as 
  frozen in to the fluid. 
  Although magnetic diffusion is not explicitly included in their equations, it 
  exists through numerical diffusion which is determined by the grid scale.
  In models with adaptive mesh refinement, the diffusion varies locally with
  the variations in resolution. 

  In fact, within the turbulent environment of the ISM, although the 
  diffusivity is very low, shocks and compressions can create conditions, 
  in which the time scales of the diffusion are comparable with other 
  time scales such as the diffusion of heat, charge and momentum.
  \citet{Korpi99} and \citet{Gressel08} include bulk
  diffusivity in their equations, which are some orders of magnitude larger
  than those typical of the ISM. These are constrained by the minimum that can 
  be resolved with the numerical resolution available, but ensure the level 
  of diffusion is consistent throughout. 
%  In addition \citet{Korpi99} include
%  hyper-diffusion, . The advantage is that lower bulk diffusion can be applied, 
%  but hyper-diffusion applies enhanced diffusion to prevent the growth of 
%  instabilities due to the higher derivatives with inconsistent results. 

  Although, the maximum effective Reynolds numbers are much lower than
  estimates for the real ISM, there is considerable consistency in 
  structures derived by models on the meso-scales, e.g., thickness of the disc, size of 
  remnant structures and super bubbles, typical velocities, densities and 
  temperatures. Uncertainties are greater in describing the fine 
  structures; remnant shells, condensing clouds, and dispersions of velocity,
  density and temperature. 

  \subsection{\label{subsect:fountain} The galactic fountain} 

  It is reasonable to conclude from what we understand from observations and
  the work of \citet{AM-L02} and their subsequent models, that the modelling
  of the galactic fountain requires a scale height of order $\pm10\kpc$. 
  They found $z=\pm5\kpc$ was insufficient to establish a {{duty}} cycle
  with hot gas condensing and recycling back to the disc.
  \cite{Gressel08} model the ISM to $z=\pm2\kpc$ and throughout their model 
  there is a net outflow across the outer horizontal surfaces. 
  
  Less clear is how large a horizontal span is required.
  \citet{Korpi99} found the correlation length of the velocity of the hot gas
  increased from about 20\,pc at $z=0$ to about 150\,pc at $|z|=300\p$ and 
  the scale of the box at  $|z|=850\p$. This indicates that as height increases
  above 1\,kpc the size of the physical structures of the hot gas may well
  exceed the model. As such reliable modelling of the galactic fountain may
  well require a substantially larger computational domain in all directions.
  \citet{AM-L02} model large $|z|$ with much lower resolution than the 
  mid-plane, to manage the computational demands. It may be that the scale of
  the task may necessitate sacrificing resolution throughout to expand the
  range of the models to investigate the galactic fountain.

  \subsection{\label{subsect:spiral} Spiral arms} 

  Only models on the larger scales previously described 
  \citep[][and subsequent models]{Slyz03,Dobbs08,Hanasz04} have reproduced
  features similar to the spiral arms in rotating galaxies. It is unlikely
  that the dynamics of the spiral arms can be understood from models in the
  scale of 1--2\,kpc in the horizontal, because they are structures typical
  of the full size of the galaxy. 
  Rather than an outcome of such models, they could perhaps be included as
  an input in the form of a density wave through the box at appropriate 
  intervals, with other associated effects, such as fluctuations in SNe 
  rates. 
  
  \section{Summary}

  Many models have been used, some to describe different astrohpysical features
  or alternative approaches to the same environment. Direct comparison of
  results is consequently difficult. 
  All models must omit ingredients and approximate or parameterize some of the
  challenging dynamics. 
  Computational power restricts resolution, domain size and physical 
  components, so each model must be adapted to match the requirements of the
  particular astrophysics under investigation; e.g. star formation permits only
  a small domain, enables high resolution, requires thermal instability; 
  spiral arms impose the requirement of a global domain, restrict resolution 
  or turbulence. 
  A comprehensive understanding requires a heirarchy of models, spanning an
  overlapping range of scales, encompassing the all the critical physical
  features.

\end{chapter}    
                     % A chapter.
%
%%  \part{Critical elements affecting the model}\label{part:critics}
%
%  \clearpage                            % End the current page making sure all
%  \thispagestyle{empty}                 % tables/figures are printed.
%  \cleardoublepage                      % Necessary for correct page numbering.
  \part{Modelling the interstellar medium}\label{part:model}
%  \clearpage                            % End the current page making sure all
%  \thispagestyle{empty}                 % tables/figures are printed.
%  \cleardoublepage                      % Necessary for correct page numbering.
  %--------------------------------------------------------------------------
\begin{chapter}{\label{chap:NI}Basic equations and their numerical implementation}
%--------------------------------------------------------------------------

%----------------------------------------------------------------------------
\section{\label{sect:eq}Basic equations}

%----------------------------------------------------------------------------
  I solve numerically a system of equations using the {\sc Pencil Code} 
  \footnote{http://code.google.com/p/pencil-code}, which is designed for fully
  nonlinear, compressible magnetohydrodynamic (MHD) simulations. 
  I do not currently include cosmic rays, which will be considered elsewhere
  subsequently. 
  
  The basic equations include the mass conservation equation, 
  the Navier--Stokes equation (written here in the rotating frame), the heat
  equation and the induction equation:
%----------------------------------------------------------------------------
%
%----------------------------------------------------------------------------
  \begin{align}
  \label{eq:mass}
  \frac{D\rho}{Dt}\, &=
      \,- \nabla \cdot (\rho \vect{u})\,+\dot{\rho}\SN,\\
%----------------------------------------------------------------------------
  \label{eq:mom}
  \frac{D\vect{u}}{Dt} \,&=
      \,-\rho^{-1}\nabla{\sigma}\SNk
        \,-c\sound^2\nabla\left({s}/{\cp}+\ln\rho\right)
        \,-\nabla\Phi
        \,-Su_x\bm{\hat{y}}
        \,-2\vect{\Omega}\times\vect{u}\nonumber\\
      &\,+\,\rho^{-1}\vect{j}\times\vect{B}
        \,+\nu \left( \nabla^{2} \vect{u}+\sfrac{1}{3}\nabla \nabla 
          \cdot \vect{u}+2{\mathbfss W}\cdot\nabla \ln\rho\right)
        \,+\zeta_{\nu}\left(\nabla\nabla \cdot \vect{u} \right),\\
%----------------------------------------------------------------------------
  \label{eq:ent}
  \rho T\frac{D s}{Dt}\, &=
      \,\dot{\sigma}\SNt
        \,+\rho\Gamma
        \,-\rho^2\Lambda
        \,+\nabla\cdot\left(\cp\rho\chi\nabla T\right)
        \,+\eta\mu_0\vect{j}^2  \nonumber\\
      &\,+\,2 \rho \nu\left|{\mathbfss W}\right|^{2}
    +\zeta_\chi\left(\rho T\nabla^2s+\nabla\ln\rho T\cdot\nabla s\right)
        +\rho T\nabla\zeta_\chi\cdot\nabla s,\\
%----------------------------------------------------------------------------
  \label{eq:ind}
%    \frac{D \vect{A}}{Dt}=&\mbox{}\phm(\vect{U}+\vect{u})\times\vect{B}
    {\frcorr{\frac{\partial \vect{A}}{\partial t}}}\,   &=
        \,{\frcorr{\vect{u}\times\vect{B}
        \,-S A_y\vect{\hat{x}}
        \,-S x \frac{\partial \vect{A}}{\partial y}
        }} \nonumber\\
      &\,+ \,(\eta+\zeta_\eta)\nabla^2\vect{A}
        %       +(\nabla\cdot\vect{A})\nabla\zeta_\eta,
       \,+\nabla\cdot\vect{A}({\frcorr{\nabla\eta+}}\nabla\zeta_\eta),
  \end{align}
%----------------------------------------------------------------------------
  where $\rho$, $T$ and $s$ are the gas density, temperature and specific
  entropy, respectively, $\vect{u}$ is the deviation of the gas velocity from
  the background rotation profile 
  (here called the \textit{velocity perturbation\/}), and
  $\vect{B}$ and $\vect{A}$ are the magnetic field and magnetic potential,
  respectively, such that $\vect{B}=\nabla\times\vect{A}$.
  Also $\vect{j}$ is the current density, $c\sound$ is the adiabatic speed of
  sound, $\cp$ is the heat capacity at constant pressure, $S$ is the velocity
  shear rate associated with the Galactic differential 
  rotation at the angular velocity $\vect{\Omega}$ (see below), assumed to be
  aligned with the $z$-axis.
  The Navier--Stokes equation includes the viscous term with the viscosity   
  $\nu$ and the rate of strain tensor $\mathbfss W$ whose components are given
  by
  \[
  \label{eq:str}
    2 W_{ij}= \frac{\upartial u_{i}}{\upartial x_{j}}+
        \frac{\upartial u_{j}}{\upartial x_{i}}
        -\frac{2}{3}\delta_{ij}\nabla \cdot\vect{u},
  \]
  as well as the shock-capturing viscosity $\zeta_\nu$ and the Lorentz force,
  $\rho^{-1}\vect{j}\times\vect{B}$. The implementation of shock-capturing is 
  discussed in Section~\ref{sect:NS}

  The system is driven by SN energy injection per unit volume, at rates
  $\dot{\sigma}\SNk$ in the form of kinetic energy in Eq.~(\ref{eq:mom}) and 
  thermal energy ($\dot{\sigma}\SNt$) in Eq.~(\ref{eq:ent}). 
  Energy injection is confined to the interiors of SN remnants, and
  the total energy injected per supernova is denoted $E\SN$. The mass of the SN
  ejecta is included in Eq.~(\ref{eq:mass}) via the source $\dot{\rho}\SN$. The
  forms of these terms are specified and further details are given in
  Section~\ref{sect:MSN}. 

  The heat equation also contains a thermal energy source due
  to photoelectric heating $\rho\Gamma$, energy loss due to optically thin 
  radiative cooling $\rho^2\Lambda$, heat conduction with the thermal 
  diffusivity $\chi$,
  viscous heating (with $|{\mathbfss W}|$ the determinant of ${\mathbfss W}$),
  Ohmic heating $\eta\mu_0\vect{j}^2$, 
  and the shock-capturing thermal diffusivity $\zeta_\chi$. $K=\cp\rho\chi$ is 
  the thermal conductivity and $\mu_0$ is vacuum permeability.
 
  The induction equation includes magnetic diffusivity $\eta$ and
  shock-capturing magnetic diffusivity $\zeta_\eta$. 
  
  The advective derivative,
  \begin{equation}
  \frac{D}{Dt}= \frac{\upartial}{\upartial t} + \left( \vect{U} +
      \vect{u} \right) \cdot \nabla,
  \label{eq:advection}
  \end{equation}
  includes transport by an imposed shear flow $\vect{U}=(0,Sx,0)$ 
  in the local Cartesian coordinates (taken to be linear across the local 
  simulation box), with the   velocity $\vect{u}$ representing a deviation 
  from the overall rotational  velocity $\vect{U}$.
  As discussed later, the perturbation velocity $\vect{u}$ consists of two
  parts, a mean flow and a turbulent velocity.
  A mean flow is considered using kernel averaging techniques
   \citep[e.g.][]{G92}
  applying a Gaussian kernel:
    \begin{align}\label{eq:ugauss}
    \average{ \vect{u}}_\ell(\vect{x})
  	&=\int_{V}\vect{u}(\vect{x}')G_\ell(\vect{x}-\vect{x}')\,\dd^3\vect{x}',\\
      G_\ell(\vect{x})&=\left(2\upi \ell^2\right)^{-{3}/{2}}
  	\exp\left[-{\vect{x}^2}/({2\,\ell^2})\right],\nonumber
    \end{align}
  where, as discussed in Chapter~\ref{chap:meanB}, $\ell\simeq50\p$ is 
  determined to be an appropriate smoothing scale. 
  The random flow $\vect{u}\turb$ is then $\vect{u}-\average{\vect{u}}_\ell$. 
  The differential rotation of the galaxy is modelled with a background
  shear flow along the local azimuthal ($y$) direction, $U_y=Sx$. The
  shear rate is $S=r\upartial\Omega/\upartial r$ in terms of galactocentric
  distance $r$, which translates into the $x$-coordinate for the local Cartesian
  frame. In this thesis I consider models with rotation and shear
  relative to those in the solar neighbourhood, $\Omega_0=-S=25\kms\kpc^{-1}$.
  
  The ISM is considered an ideal gas, with thermal pressure given by
  \[
      \label{eq:eos}
      p = \frac{k_\mathrm{B}}{\mu m_\mathrm{p}}\rho T,
  \]
  where $k_\mathrm{B}$  is the Boltzmann constant, $m_\mathrm{p}$ is the proton
  mass. 
  I have assumed 
  the
  gas to have the Solar chemical composition and the level of ionization to be
  uniform\footnote{I thank Prof. J. Pringle 
  for helping me during my viva to derive an appreciation for the determination
  of $\mu$ appropriately for the distinct ionized states of different ISM
  phases.} adopting a value of $\mu=0.62$ for the mean molecular weight.
  The calculation does not explicitly solve for temperature nor pressure, but
  for entropy.
  Due to its complexity ions and electrons are not modelled directly, however
  even without this it is reasonable to consider different states of ionization
  for gas depending on temperature.
  Expressions including $T$ in Eq.~\ref{eq:ent} are reformulated in terms
  of $s$ and $\rho$ for the calculation.
  However the equation of state and 
  the effect of $\mu$ is applied indirectly through the specific heat 
  capacities $\cv$ and $\cp$ and
  via the temperature dependent radiative cooling.
  In future simulations and analysis we could consider applying $\mu$, which
  varies according to the temperature
  and its corresponding level of ionization.
  As a result the thermal pressure might reduce for the cold and warm gas,
  and increase for the hot gas, when compared to the analysis presented in this
  thesis. 
  Further clarification is included in Appendix~\ref{chap:units}.  

  In Eq.~(\ref{eq:mom}), $\Phi$ is the gravitational potential produced by
  stars and dark matter. For the Solar vicinity of the Milky Way,
  \citet{Kuijken89} suggest the following form of the vertical gravitational
  acceleration \citep[see also][]{F01}:
  \begin{equation}
  \label{eq:grav}
    g_{z}=-\deriv{\Phi}{z}
        =-\frac{a_1z}{\sqrt{z_1^{2}+z^{2}}}-a_2\frac{z}{z_2},
  \end{equation}
  with $a_1=4.4\times10^{-14}\km\s^{-2}$, $a_2=1.7\times10^{-14} \km\s^{-2}$,
  $z_1=200\p$ and $z_2=1\kpc$. Self-gravity of the interstellar gas is 
  neglected because it is subdominant at the scales of interest.

  If self-gravity were included it applies on scales below the Jeans length 
  \begin{equation}
    \lambda_{\rm J}\approx\sqrt{\frac{k_{\rm B} T r^3}{GM\mu}},
  \end{equation}
  where the gravitational energy $GM\mu/r$ of the total mass $M$ 
  enclosed within a radius of $r$ exceeds the thermal energy per particle
  $k_{\rm B} T$ within $r$.
  If the density of cold gas increases sufficiently such that locally
  $\lambda_{\rm J}$ becomes less than the grid length $\Delta$ then it can 
  no longer be resolved. 
  \citet{DBP11} exclude this in their models by locating SNe where the
  the ISM becomes sufficiently self-gravitating, thus preventing the
  process exceeding the grid resolution.
  This is also used to regulate the SN rate.

%------------------------------------------------------------------------
\section{\label{sect:MSN}Modelling supernova activity}

  I include both Type~CC and Type~I SNe in these simulations, distinguished only
  by their frequency and vertical distribution. The SNe frequencies are those in
  the Solar neighbourhood \citep*[e.g.][]{TLS94}. Type~CC SNe are introduced at
  a rate, per unit surface area, of $\nu_\mathrm{II}=25\kpc^{-2}\Myr^{-1}$
  ($0.02\yr^{-1}$ in the whole Galaxy), with fluctuations of the order of
  $10^{-4}\yr^{-1}$ at a time scale of order $10\Myr$. Such fluctuations
  in the SN~II rate are natural to introduce; there is some evidence
  that they can enhance dynamo action in MHD models
  \citep{Hanasz04,Balsara04}. The surface density rate of Type~I SNe is
  $\nu_\mathrm{I}=4\kpc^{-2}\Myr^{-1}$ (an interval of 290 years between
  SN~I explosions in the Galaxy).
  
  Unlike most other ISM models of this type, the SN energy in the injection 
  site is split between thermal and kinetic parts, in order to further reduce
  temperature and energy losses at early stages of the SN remnant evolution.
  Thermal energy density is distributed within the injection site as 
  $\exp[-(r/r\SN)^6]$, with $r$ the local spherical radius and $r\SN$ the 
  nominal location of the remnant shell (i.e.\ the radius of the SN bubble) 
  at the time of injection. Kinetic energy is injected by adding
  a spherically symmetric velocity field $u_r\propto\exp[-(r/r\SN)^6]$;
  subsequently, this rapidly redistributes matter into a shell. 
  To avoid a discontinuity in $\vect{u}$ at the centre of the injection site, 
  the centre is simply placed midway between grid points.
  I also inject $4\Msol$ as stellar ejecta, with density profile 
  $\exp[-(r/r\SN)^6]$.
  Given the turbulent environment, there are significant random motions
  and density inhomogeneities within the
  injection regions. Thus, the initial kinetic energy is not the same in each
  region, and, injecting part of the SN energy in the kinetic form results 
  in the total kinetic energy varying between SN remnants. 
  I therefore record the energy added for every
  remnant so I can fully account for the rate of energy injection. For example,
  in Model~{\Op} I obtain the energy per SN in the range
  \[
    0.5 < E\SN< 1.5\times10^{51}\erg,
  \]
  with the average of $0.9\times10^{51}\erg$.
  
  The SN sites are randomly distributed in the horizontal coordinates $(x,y)$.
  Their vertical positions are drawn from normal distributions with 
  scale heights of $h_\mathrm{II}=0.09\kpc$ for SN~II and
  $h_\mathrm{I}=0.325\kpc$ for Type~I SNe. Thus, Eq.~(\ref{eq:mass}) contains
  the mass source of $4\Msol$ per SN,
  \[
    \dot\rho\SN\simeq4\Msol\left(\frac{\nu_\mathrm{II}}{2 h_\mathrm{II}}
        +\frac{\nu_\mathrm{I}}{2 h_\mathrm{I}}\right) 
        [\Msol \kpc^{-3} \Myr^{-1}],
  \]
  whereas Eqs.~(\ref{eq:mom}) and (\ref{eq:ent}) include kinetic and thermal
  energy sources of similar strength adding up to approximately $E\SN$ per SN:
  \[
    {\dot\sigma}\SNk \simeq {\dot\sigma}\SNt = \tfrac12 
        E\SN\left(\frac{\nu_\mathrm{II}}{2 h_\mathrm{II}}
        +\frac{\nu_\mathrm{I}}{2 h_\mathrm{I}}\right)
        [\erg \kpc^{-3} \Myr^{-1}].
  \]
  The only other constraints applied when choosing SN sites are to
  reject a site if an SN explosion would result in a local temperature
  above $10^{10}\K$ or if the local gas number density exceeds
  $2\cmcube$. The latter requirement ensures that the thermal energy
  injected is not lost to radiative cooling before it can be converted
  into kinetic energy in the ambient gas. More elaborate prescriptions
  can be suggested to select SN sites
  \citep{Korpi99a,Avillez00,Joung06,Gressel08}; I found this
  unnecessary for the present purposes.
  
  Arguably the most important feature of SN activity, in the present
  context, is the efficiency of evolution of the SNe energy from thermal to
  kinetic energy in the ISM, a transfer that occurs via the
  shocked, dense shells of SN remnants. Given the relatively low resolution of
  this model (and most, if not all, other models of this kind), it is essential
  to verify that the dynamics of expanding SN shells are captured %correctly.
  correctly:
  inaccuracies in the SN remnant evolution would indicate that
  the modelling of the thermal and kinetic energy processes was unreliable.
  Therefore, I present in Appendix~\ref{chap:EISNR} detailed numerical
  simulations
  of the dynamical evolution of an individual SN remnant at spatial grid 
  resolutions
  in the range $\Delta=1$--$4\p$. The SN remnant is allowed to evolve from the 
  Sedov--Taylor stage
  (at which SN remnants are introduced in these simulations) for 
  $t\approx3.5\Myr$.
  The remnant enters the snowplough regime,
  with a final shell radius exceeding $100\p$, and 
  the numerical results are compared with the analytical solution of
  \citet{Cioffi98}. The
  accuracy of the numerical results depends on the ambient gas density $n_0$:
  larger $n_0$ requires higher resolution to reproduce the
  analytical results. I show that agreement with
  \citet{Cioffi98} in terms of the shell radius and speed is very good at
  resolutions $\Delta\leq2\p$ for $n_0\simeq1\cmcube$, and excellent
  at $\Delta=4\p$, for $n_0\approx0.1$ and $0.01\cmcube$. 
  
  Since shock waves in the immediate vicinity of an SN site are usually
  stronger than anywhere else in the ISM, these tests also confirm that this
  handling of shock fronts is sufficiently accurate and that the shock-capturing
  diffusivities that are employed
  do not unreasonably affect the shock evolution.
  
  The standard resolution is 4\,pc. To be minimally resolved, the initial radius
  of an SN remnant must span at least two grid points. Because the origin is set
  between grid points, a minimum radius of 7\,pc for
  the energy injection volume is sufficient.
   The size of the energy
  injection region in the model must be such that the gas temperature is above
  $10^6\K$ and below $10^8\K$: at both higher and lower temperatures, energy
  losses to radiation are excessive and adiabatic expansion cannot be
  established. Following \citet{Joung06}, I adjust the radius of the energy
  injection volume to be such that it contains $60\Msol$ of gas. 
  For example, in Model~{\Op} this results in a mean $r\SN$ of $35\p$, 
  with a standard deviation of $25\p$ and a maximum of $200\p$.
  The distribution of radii appears approximately lognormal,
  so $r\SN>75\p$ is very infrequent and the modal value is about $10\p$; 
  this corresponds 
  to the middle of the Sedov--Taylor phase of the SN expansion.  
  Unlike \citet{Joung06}, I found that mass redistribution within the 
  injection site was not necessary. Therefore I do not impose uniform site 
  density, particularly as
  it may lead to unexpected consequences in the presence of magnetic fields.

%------------------------------------------------------------------------
\section{\label{sect:ssvrb}Radiative cooling and photoelectric heating}
%-----------------------------------------------------------------------------
  
%-----------------------------------------------------------------------------
  \begin{table}[ht]
  \centering
    \caption[The RBN cooling function]{The cooling function of \citet{Rosen93}, labelled RBN in the text
    (and in the labels of the numerical models),
    with $\Lambda=0$ for $T<10\K$.
    \label{table:coolRB}}
    \begin{tabular}{ccc}
    \hline
    $T_k$       &$\Lambda_k$                              &$\beta_k$        \\
    $[\K]$      &$[\erg\g^{-2}\s^{-1}\cm^3\K^{-\beta_k}]$ &                 \\ 
    \hline
    10        &$9.88\times10^5\phantom{^5}$             &\phantom{$-$}6.000 \\
    300       &$8.36\times10^{15}$                      &\phantom{$-$}2.000 \\
    2000      &$3.80\times10^{17}$                      &\phantom{$-$}1.500 \\
    8000      &$1.76\times10^{12}$                      &\phantom{$-$}2.867 \\
    $10^{5}$  &$6.76\times10^{29}$                      &          $-0.650$ \\
    $10^{6}$  &$8.51\times10^{22}$                      &\phantom{$-$}0.500 \\
    \hline
    \end{tabular}
  \end{table}
%-----------------------------------------------------------------------------

  I consider two different parameterizations of the optically thin radiative
  cooling appearing in Eq.~(\ref{eq:ent}), both of the piecewise form
  $\Lambda=\Lambda_{k}T^{\beta_{k}}$ within a number of temperature ranges 
  $T_{k}\le T<T_{k+1}$, with $T_k$ and $\Lambda_k$ given in
  Tables~\ref{table:coolSS} and \ref{table:coolRB}. Since this is just a crude
  (but convenient) parameterization of numerous processes of recombination and
  ionization of various species in the ISM, there are several approximations
  designed to describe the variety of physical conditions in the ISM. Each of
  the
  earlier models of the SN-driven ISM adopts a specific cooling curve, often
  without explaining the reason for the particular choice or assessing its
  consequences. In Section~\ref{subsect:COOL} I discuss the sensitivity of 
  results to the choice of the cooling function.
  
%-----------------------------------------------------------------------------
  \begin{table}[htb]
  \centering
    \caption[The WSW cooling function]{The cooling function of \citet{Wolfire95} at $T<10^5\K$, joined to 
  that of \citet{Sarazin87} at higher temperatures, with $\Lambda=0$ for 
  $T<10\K$. This cooling function is denoted WSW in the text (and in the labels
  of the numerical models).  \label{table:coolSS}}
    \begin{tabular}{ccc}
    \hline
    $T_k$                 &$\Lambda_k$           &$\beta_k$           \\
    $[\K]$                &$[\erg\g^{-2}\s^{-1}\cm^3\K^{-\beta_k}]$&  \\ 
    \hline
    10                    & $3.70\times10^{16}$ &   \phantom{$-$}2.12 \\
    141                   & $9.46\times10^{18}$ &   \phantom{$-$}1.00 \\
    313                   & $1.18\times10^{20}$ &   \phantom{$-$}0.56 \\
    6102                  & $1.10\times10^{10}$ &   \phantom{$-$}3.21 \\
    $10^{5}$              & $1.24\times10^{27}$ &              $-0.20$\\
    $2.88\times 10^{5}$   & $2.39\times10^{42}$ &              $-3.00$\\
    $ 4.73\times 10^{5}$  & $4.00\times10^{26}$ &              $-0.22$\\
    $2.11\times 10^{6}$   & $1.53\times10^{44}$ &              $-3.00$\\
    $3.98\times 10^{6}$   & $1.61\times10^{22}$ &   \phantom{$-$}0.33 \\
    $ 2.00\times 10^{7}$  & $9.23\times10^{20}$ &   \phantom{$-$}0.50 \\
    \hline
    \end{tabular}
  \end{table}
%-----------------------------------------------------------------------------
  
  One parameterization of radiative cooling, labelled WSW and shown in
  Table~\ref{table:coolSS}, consists of two parts. For $T<10^5\K$, the cooling
  function fitted by \citet{Sanchez02} to the `standard' equilibrium 
  pressure--density relation of \citet[][Fig.~3b therein]{Wolfire95} is 
  used. For higher temperatures, the cooling function of \citet{Sarazin87} is
  adopted. This part of the cooling function (but extended differently to 
  lower temperatures) was used by \citet{Slyz05} to study star formation in 
  the ISM. The WSW cooling function was also used by \cite{Gressel08}. It has 
  two thermally unstable ranges: at $313<T<6102\K$, the gas is isobarically
  unstable ($\beta_k<1$); at $T>10^5\K$, some gas is isochorically or
  isentropically unstable ($\beta_k<0$ and $\beta_k<-1.5$, respectively).

  Results obtained with the WSW cooling function are compared with those
  using the cooling function of \citet{Rosen93}, labelled RBN, whose parameters 
  are shown in Table~\ref{table:coolRB}. This cooling function has a thermally 
  unstable part only above $10^5\K$. \citet{Rosen93} truncated their cooling
  function at  $T=300\K$. Instead of an abrupt truncation, I have smoothly 
  extended the cooling function down to $10\K$. This has no palpable physical 
  consequences as the radiative cooling time at these low temperatures 
  becomes longer than other time scales in the model, so that adiabatic cooling
  dominates. The minimum temperature reported in the model of \citet{Rosen93} 
  is about $100\K$. Here, with better spatial resolution, the lowest 
  temperature gas is at times below $10\K$, with some gas at $50\K$ present most
  of the time.
  
  I took special care to accurately ensure the continuity of the cooling
  functions, as small discontinuities may affect the performance of the
  code; hence the values of $\Lambda_k$ in Table~\ref{table:coolSS} differ
  slightly from those given by \citet{Sanchez02}. The two cooling functions are
  shown in Fig.~\ref{fig:cool}. The cooling function used in each numerical
  model is identified with a prefix RBN or WSW in the model label
  (see Table~\ref{table:models}). The purpose of
  Models {RBN} and {\WSWa} is to assess the impact of the choice of the
  cooling function on the results (Section~\ref{subsect:COOL}). Other models
  employ the WSW cooling function.

  %-----------------------------------------------------------------------------
  \begin{figure}[htb]
  \begin{center}
    \includegraphics[width=0.75\linewidth]{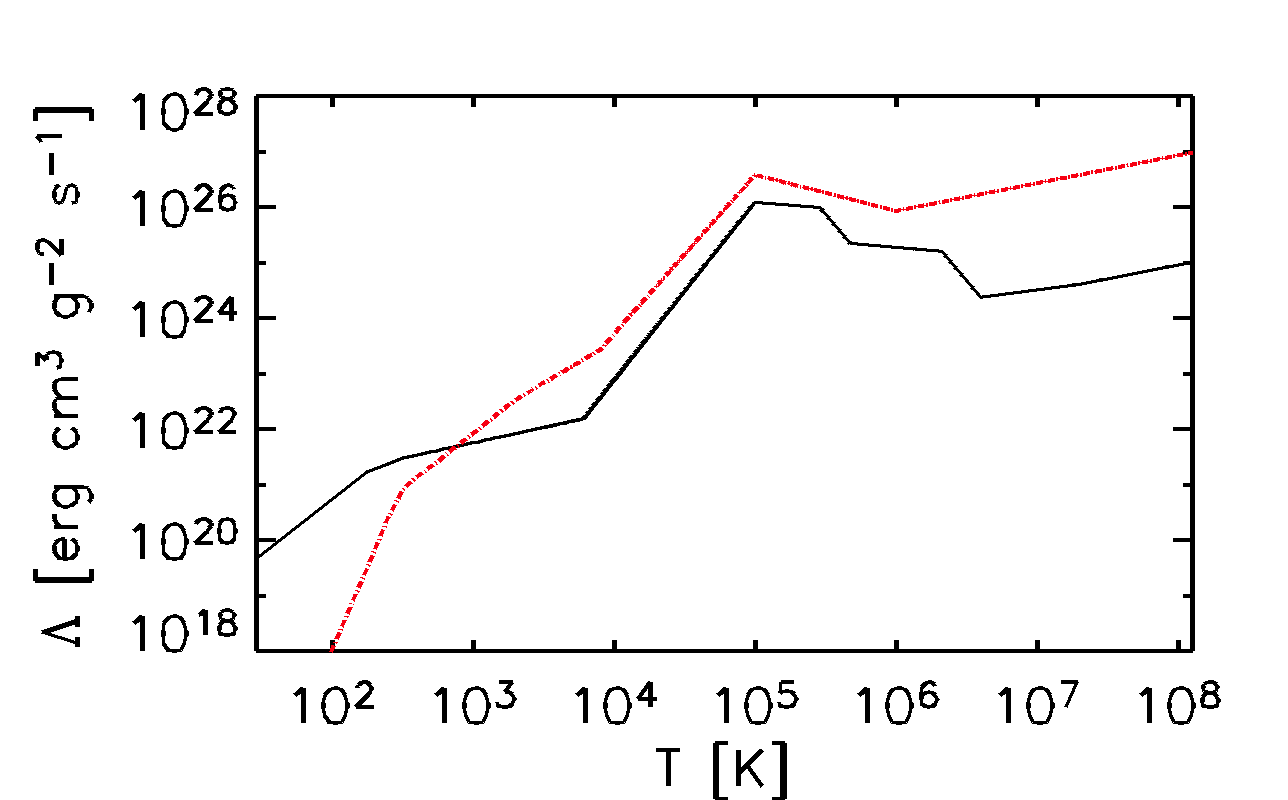}
    \caption[The cooling functions]{The cooling functions WSW (solid, black) and RB (red, 
    dash-dotted), with parameters given in Tables~\ref{table:coolSS} 
    and \ref{table:coolRB}, respectively.\label{fig:cool} }
    \end{center}
  \end{figure}
  %-----------------------------------------------------------------------------
  
  I also include photoelectric heating in Eq.~(\ref{eq:ent}) via the stellar
  far-ultraviolet (UV) radiation $\Gamma$, following \citet{Wolfire95}, and
  allowing for its decline away from the Galactic mid-plane with a length scale
  comparable to the scale height of the stellar disc near the Sun 
  \cite[cf.][]{Joung06}:
  \[
    \label{eq:zheat}
    \Gamma(z)=\Gamma_0\exp\left(-|z|/300\p\right),
  \quad
  \Gamma_0=0.0147\erg\g^{-1}\s^{-1}.
  \]
  This heating mechanism is smoothly suppressed at $T>2\times10^4\K$, since the
  photoelectric effect due to UV photon impact on PAHs (Polycyclic Aromatic
  Hydrocarbons) and small dust grains is impeded at high temperatures
  \citep{Wolfire95}.
  
%------------------------------------------------------------------------
\section{\label{sect:NS}Numerical methods}

  The model occupies a relatively small region within the galactic disc and 
  lower halo with parameters typical of the solar neighbourhood.
  Using a three-dimensional Cartesian grid, the results have been obtained for
  a region $1.024\times1.024\times 2.240\kpc^{3}$ in size 
  ($1.024\times1.024\times 2.176\kpc^{3}$ for the MHD models), with $1.024\kpc$ 
  in the radial and azimuthal directions and $1.120\kpc$ ($1.088\kpc$)
  vertically on either side of the galactic mid-plane.
  Assuming that the correlation length of the interstellar turbulence is 
  $l_0\simeq0.1\kpc$, it might be expected that the computational domain
  encompass about 2,000 turbulent cells, so the statistical properties of the
  ISM should be reliably captured.
  Testing individual SNe in diffuse ISM and snapshot sampling over 
  various runs confirms the computational domain is 
  sufficiently broad to accommodate comfortably even the largest SN remnants at
  large heights, so as to exclude any self-interaction of expanding remnants
  through the periodic boundaries.
  
  Vertically, the reference model accommodates ten scale heights of the cold
  \HI\ gas, two scale heights of diffuse \HI\ (the Lockman layer), and one scale
  height of ionized hydrogen (the Reynolds layer).
  The vertical size of the domain in the reference model is insufficient to
  include the scale height of the hot gas and the Galactic halo.
  Therefore, it may be desirable to increase its vertical extent in future
  work.
  Note, however, that the size of SN remnants, and the correlation scale of the
  flow above a few hundred parsec height may approach the horizontal size of the
  numerical domain and may even exceed it at greater heights \citep{Korpi99}.
  The periodic boundary conditions exclude divergent flows at scales comparable
  to the horizontal size of the box, suggesting that an increase in height may
  necessitate an appropriate increase in the horizontal dimensions to reliably
  capture the dynamics of the flow, and remain consistent with the assumptions
  of the (sheared) periodic boundary conditions in $x$ and $y$.
  
  The standard resolution (numerical grid spacing) is $\Delta x = \Delta y =
  \Delta z=\Delta= 4\p$.
  The grid is $256\times256\times560$ (excluding 'ghost' boundary zones) for
  the HD models, slightly reduced to $256\times256\times544$ for the MHD models.
  One model at doubled resolution, $\Delta=2\p$, has a grid 
  $512\times512\times1120$ in size.
  A sixth-order finite difference scheme for spatial vector operations and a
  third-order Runge--Kutta scheme for time stepping are applied.
  
  The spatial and temporal resolutions attainable impose lower limits on the
  kinematic viscosity $\nu$ and thermal conductivity $K$, which are,
  unavoidably, much higher than any realistic values. These limits result from
  the Courant--Friedrichs--Lewy (CFL) condition which requires that the
  numerical time step must be shorter than the crossing time over the mesh
  length $\Delta$ for each of the transport processes involved. It is desirable 
  to avoid unnecessarily high viscosity and thermal diffusivity. The cold and 
  warm phases have relatively small perturbation gas speeds (of order 
  $10\kms$), so $\nu$ and $\chi$ are prescribed to be proportional to the local
  speed of sound, $\nu=\nu_1 c\sound/c_1$ and $\chi=\chi_1 c\sound/c_1$. To ensure the
  maximum Reynolds and P\'eclet numbers based on the mesh separation $\Delta$
  are always less than unity throughout the computational domain (see
  Appendix~\ref{sect:BCND}), $\nu_1\approx4.2\times10^{-3}\kms\kpc$, 
  $\chi_1\approx4.1\times10^{-4}\kms\kpc$ and $c_1=1\kms$. This gives, for
  example, $\chi=0.019\kms\kpc$ at $T=10^5\K$ and $0.6\kms\kpc$ at $T=10^8\K$.
  
  Numerical handling of the strong shocks widespread in the ISM needs special
  care. To ensure that they are always resolved, shock-capturing diffusion
  of heat, momentum and induction are included, with the diffusivities 
  $\zeta_{\chi}$, $\zeta_{\nu}$ and $\zeta_{\eta}$, respectively, defined as
  \begin{equation}
    \zeta_\chi=
        \begin{cases}
          c_{\chi}\Delta x^2\max_5|\nabla\cdot\vect{u}|, 
              &\text{if }\nabla\cdot\vect{u}<0,\\
          0,  &\text{otherwise},
        \end{cases}\label{shockdiff}
  \end{equation}
  (and similarly for $\zeta_{\nu}$ or $\zeta_{\eta}$, but with coefficients
   $c_\nu$ or $c_\eta$), where ${\rm max_5}$ denotes the maximum value
  occurring at any of the five nearest mesh points (in each coordinate). Thus, 
  shock-capturing diffusivity is proportional to the maximum divergence 
  of the velocity in the local neighbourhood, and are confined to the regions
  of convergent flow. Here, $c_{\chi}=c_\nu=c_\eta$ is a dimensionless
  coefficient which has been adjusted empirically to 10. This prescription 
  spreads a shock front over sufficiently many (usually, four) grid points. 
  Detailed test simulations of an isolated expanding SN remnant in
  Appendix~\ref{chap:EISNR} confirm that this prescription produces quite 
  accurate results, particularly in terms of the most important
  goal: the conversion of thermal to kinetic energy in SN remnants.
  
  With a cooling function susceptible to thermal instability, thermal
  diffusivity $\chi$ has to be large enough as to allow us to resolve the most
  unstable normal modes:
  \[
  \chi\geq\frac{1-\beta }{\gamma~\tau\cool}\left( \frac{\Delta}{2 \upi}\right)^2,
  \]
  where $\beta$ is the cooling function exponent in the thermally unstable
  range ($\beta=0.56$ in the WSW model),
  $\tau\cool$ is the radiative cooling time, and $\gamma=5/3$ is the adiabatic
  index.
  From the contours of constant cooling time plotted in Fig.~\ref{fig:pdf2dop} 
  of Chapter~\ref{chap:SMP}, it is evident that for these models
  $\tau\cool$ is typically in excess of 1\,Myr in the thermally unstable regime.
  Further details can be found in Appendix~\ref{sect:ti} where it is
  demonstrated that, with the parameters chosen in my models, thermal
  instability is well resolved by the numerical grid.
  
  The shock-capturing diffusion broadens the shocks and increases the spread of 
  density around them.
  An undesirable effect of this is that the gas inside SN remnants cools faster
  than it should, thus reducing the maximum temperature and affecting the
  abundance of the hot phase.
  Having considered various approaches while modelling individual SN remnants
  in  Appendix~\ref{chap:EISNR}, a prescription is adopted which is numerically 
  stable, reduces gas cooling within SN remnants, and confines extreme cooling
  to the shock fronts. 
  Specifically, the term $(\Gamma-\rho\Lambda)T^{-1}$ in  Eq.~(\ref{eq:ent}) is 
  multiplied by
  \begin{equation}\label{coolxi} 
    \xi=\exp(-C|\nabla\zeta_\chi|^2),
  \end{equation} 
  where $\zeta_\chi$ is the shock diffusivity defined in Eq.~(\ref{shockdiff}). 
  Thus $\xi\approx1$ almost anywhere in the domain, but reduces towards zero in 
  strong shocks, where $|\nabla\zeta_\chi|^2$ is large.
  The value of the additional empirical parameter, $C\approx0.01$, was chosen
  to ensure numerical stability with minimum change to the basic physics.
  It has been verified that, acting together with other artificial diffusion
  terms, this does not prevent accurate modelling of individual SN remnants
  (see  Appendix~\ref{chap:EISNR} for details).
  
%-----------------------------------------------------------------------------
\section{\label{sect:bcs}Boundary conditions}

  Given the statistically homogeneous structure of the ISM in the horizontal
  directions at the scales of interest (neglecting arm-interarm variations), I
  apply periodic boundary conditions in the azimuthal ($y$) direction.
  Differential rotation is modelled using the shearing-sheet approximation with
  sliding periodic boundary conditions \citep{WT88} in $x$, the local analogue
  of cylindrical radius. 

%\subsection{Top and bottom boundaries}

  Unlike the horizontal boundaries of the computational domain, where
  periodic or sliding-periodic boundary conditions may be adequate 
  (within the constraints of the shearing box approximation), the boundary
  conditions at the top and bottom of the domain are more demanding.
  The vertical size of the galactic halo is of order of 10\kpc, and nontrivial
  physical processes occur even at that height, especially when galactic wind
  and cosmic ray escape are important. 
  Within a few kiloparsecs of the mid-plane, SN heating induces a flow of
  hot gas towards the halo. The results of e.g., \citet[][and references
  therein]{AB07} suggest that in excess of $\pm5\kpc$ is required for the hot
  gas to cool and return to the mid-plane.
  Their vertical extent ($z=\pm10\kpc$) excludes significant mass loss from the
  system, but at considerable numerical expense. 
  Because MHD runs may require simulation times exceeding 1\,Gyr, it is
  desirable to restrict resources to the region, which can be reliably modelled
  with the minimum vertical extent.
  \citet{Gressel08} applied double the height modelled here, but with half the 
  resolution, hence requiring a quarter of the resources. 
  In adopting $\pm1\kpc$ for the vertical boundary, there is no physical 
  mechanism for adequate cooling and return of the hot gas, so care must be 
  taken to address potential mass loss.
  For the HD runs, which require only a few Myr to acquire a statistically 
  steady turbulent state, this is not too problematic.
  For the MHD runs mass must be replaced, and this is addressed in
  Appendix~\ref{subsect:MOL}.

  It is also important to formulate boundary conditions at the top and
  bottom of the domain that admit the flow of matter and energy, while
  minimising any associated artefacts that might affect the interior.
  Stress-free, open vertical boundaries would seem to be the most appropriate, 
  requiring that the horizontal stresses vanish, while gas density, entropy and
  vertical velocity have constant first derivatives on the top and bottom
  boundaries. 
  These are implemented numerically using `ghost' zones; i.e., three outer grid
  planes that allow derivatives at the boundary to be calculated in the same
  way as at interior grid points.
  The interior values of the variables are used to specify their ghost zone
  values. 

  Shocks occur ubiquitously within the ISM and handling
  these is particularly problematic as they approach the boundary. 
  Within the computational domain and on the periodic boundaries, shocks are 
  absorbed into the ambient medium, placing a finite limit on the magnitude of
  the momentum or energy associated with the shock.
  When a sharp structure approaches the vertical boundary, however, the strong
  gradients may be extrapolated into the ghost zones.
  This artificially enhances the prominence of such a structure, and may cause
  the code to crash.
  Here I describe how these boundary conditions have been modified to ensure the
  numerical stability of the model.

\subsubsection{Mass  }
  To prevent artificial mass sources in the ghost zones, I impose a weak 
  negative gradient of gas density in the ghost zones.
  Thus, the density values are extrapolated to the ghost zones from the 
  boundary point as
  \[
  \rho(x,y,\pm Z\pm k\Delta)=(1-\Delta/0.1\kpc)\rho(x,y,\pm Z\pm(k-1)\Delta)
  \]
  for all values of the horizontal coordinates $x$ and $y$, where the boundary
  surfaces are at $z=\pm Z$, and the ghost zones are at $z=\pm Z\pm k\Delta$
  with $k=1,2,3$.
  The upper (lower) sign is used at the top (bottom) boundary. 
  This ensures that gas density gradually declines in the ghost zones. 
  
\subsubsection{Temperature }
  To prevent a similar artificial enhancement of temperature spikes in the
  ghost zones, gas temperature there is kept equal to its value at the boundary,
  \[
   T(x,y,\pm Z\pm k\Delta)=T(x,y,\pm Z)\,,
  \]
  so that temperature is still free to fluctuate in response to the interior
  processes. 
  This prescription is implemented in terms of entropy, given the density 
  variation described above.
  
\subsubsection{Velocity}
  Likewise, the vertical velocity in the ghost zones is kept equal to its 
  boundary value if the latter is directed outwards,
  \[
   u_z(x,y,\pm Z\pm k\Delta)=u_z(x,y,\pm Z)\,,
  \qquad 
   u_z(x,y,\pm Z) \gtrless 0\, .
  \]
  However, when gas cools rapidly near the boundary, pressure can decrease and
  gas would flow inwards away from the boundary. 
  To avoid suppressing inward flows, where 
  \[
    u_z(x,y,\pm Z)\lessgtr0
  \] 
  the following is applied:
  \[
  {\textrm {if}}\quad |u_z(x,y,\pm Z\mp \Delta)| < |u_z(x,y,\pm Z)|,
  \]
  set
  \[
  u_z(x,y,\pm Z\pm\Delta)=\tfrac{1}{2}\left[u_z(x,y,\pm Z)+u_z(x,y,\pm Z\mp\Delta)\right]\, ;
  \]
  otherwise, set
  \[
  u_z(x,y,\pm Z\pm\Delta)=2u_z(x,y,\pm Z)-u_z(x,y,\pm Z\mp \Delta)\,.
  \]
  In both cases, in the two outer ghost zones ($k=2,3$), set
  \[%begin{align*}
  u_z(x,y,\pm Z\pm k\Delta)=2u_z(x,y,\pm Z\pm (k-1)\Delta)
                            -u_z(x,y,\pm Z\pm (k-2)\Delta)\,,
  \]%end{align*}
  so that the inward velocity in the ghost zones is always smaller than its
  boundary value.
  This permits gas flow across the boundary in both directions, but ensures
  that the flow is dominated by the interior dynamics, rather than by anything 
  happening in the ghost zones.

  For the horizontal components of the velocity at the vertical boundaries, 
  symmetrical conditions are applied to exclude horizontal stresses, so
  \[
   u_x(x,y,\pm Z\pm k\Delta)=u_x(x,y,\pm Z \mp k\Delta)\,,
  \]
  and similarly for $u_y$.
  
\subsubsection{Magnetic field}
  For the MHD models, two alternative sets of vertical boundary conditions are
  considered: vertical field and open.
  The objective with the vertical field condition is to exclude external flux.
  Any flux across the horizontal periodic boundaries conserves the magnetic
  energy.
  To identify a genuine dynamo process, the addition of magnetic
  energy due to non-physical boundary effects can be excluded by ensuring zero
  vertical flux on the upper and lower surfaces.
  For the horizontal components of the vector potential
  \[
   A_x(x,y,\pm Z\pm k\Delta)=A_x(x,y,\pm Z \mp k\Delta)\,
  \]
  and similarly for $A_y$, and for the vertical component
  \[
   A_z(x,y,\pm Z\pm k\Delta)=-A_z(x,y,\pm Z \mp k\Delta)\,,
  \]
  ensuring on the boundary that 
  \[
    \frac{\partial A_x}{\partial z}=\frac{\partial A_y}{\partial z}=A_z=0=B_x=B_y\,.
  \]
  $B_z$ can differ from zero. 

  Such an approach creates a boundary layer, in which the non-zero horizontal 
  fields near the surfaces, which may be significant close to $z\pm=1\kpc$, must
  vanish nonphysically. 

  The alternative open boundary conditions do not exclude external flux, but 
  preserve the physical structure of the field up to the boundary. 
  \citet{Gressel08} applied these with a domain up to $\pm2\kpc$ in $z$, and 
  argue the external magnetic flux was negligible. 
  It is reasonable to expect, given the dominant flow of gas is outward in
  these simulations, that the outward magnetic flux might exceed the inward 
  flux, so that artificial growth of the magnetic field due to boundary effects
  may be discounted.
  The vertical component of the Poynting flux on the boundary is monitored with
  these boundary conditions and the total flux outwards is 2 orders of 
  magnitude greater than the total flux inwards. 
  For the vertical component 
  \[
   A_z(x,y,\pm Z\pm k\Delta)=A_z(x,y,\pm Z \mp k\Delta)\,,
  \]
  and for the horizontal components
  \[
   A_x(x,y,\pm Z\pm k\Delta)=2A_x(x,y,\pm Z \pm (k-1)\Delta)-A_x(x,y,\pm Z \pm (k-2)\Delta)\,
  \]
  and similar for $A_y$, such that 
  \[
    \frac{\partial^2 A_x}{\partial z^2}=\frac{\partial^2 A_y}{\partial z^2}=\frac{\partial A_z}{\partial z}=0=\frac{\partial B_x}{\partial z}=\frac{\partial B_y}{\partial z}\,.
  \]
  The constraint on ${\partial B_z}/{\partial z}$ is automatically satisfied
  since $\nabla\cdot\vect{B}=0$.

%--------------------------------------------------------------------------

%-----------------------------------------------------------------------------
\section{\label{sect:ICons}Initial conditions}

  An initial density distribution is adopted corresponding to
  \textit{isothermal\/} hydrostatic equilibrium in the gravity field of
  Eq.~(\ref{eq:grav}):
  \begin{equation}
  \label{eq:initrho}
  \rho(z)= \rho_0 \exp\left[a_1 \left(z_1- \sqrt{z_1^2+z^2}-
      \frac{a_2}{2a_1}\,\frac{z^2}{z_1}\right)\right].
  \end{equation}
  Since some models do not contain magnetic fields and none contain cosmic
  rays, which provide roughly one quarter each of the total pressure in the ISM
  (the remainder being thermal and turbulent pressures), the gas scale heights
  can 
  be expected to be smaller that those observed. Given the limited spatial
  resolution of the simulations, and the correspondingly weakened thermal 
  instability and neglected self-gravity, it is not quite clear in advance
  whether the gas density used in the model should include molecular hydrogen
  or, otherwise, include only diffuse gas.
  
  $\rho_0=3.5\times10^{-24}\g\cmcube$ is used for Models~RBN and \WSWa,
  corresponding to gas number density, $n_0=2.1\cmcube$ at the mid-plane.
  This is the total interstellar gas density, including the part confined to
  molecular clouds. These models, discussed in Section~\ref{subsect:COOL}, exhibit
  unrealistically strong cooling. Therefore, all other subsequent models have a
  smaller amount of matter in the computational domain (a 17\% reduction), with
  $\rho_0=3.0\times10^{-24}\g\cmcube$, or $n_0=1.8\cmcube$, accounting only
  for the atomic gas \citep[see also][]{Joung06}.
  
  As soon as the simulation starts, because of density-dependent heating and
  cooling the gas is no longer isothermal, so $\rho(z)$ given in
  Eq.~(\ref{eq:initrho}) is not a hydrostatic distribution. To avoid
  unnecessarily long initial transients, a non-uniform initial temperature 
  distribution is imposed so as to be near static equilibrium:
  \begin{equation}
  \label{eq:initT}
  T(z)=\frac{T_{0}}{z_1}\left(\sqrt{z_1^2+z^2}
                  +\frac{a_2}{2a_1}\frac{z^2}{z_2}\right),
  \end{equation}
  where $T_0$ is obtained from
  \[
  \Gamma(0)=\rho_{0}\Lambda(T_{0})\approx 0.0147\erg\g^{-1}\s^{-1}.
  \]
  The value of $T_0$ therefore depends on $\rho_0$ and the choice of the 
  cooling function.

  For the models that include a magnetic field, the initial seed
  field $\vect{B}=(0,B_I n(z),0)$.
  I use  $B_I=0.05\mkG\cm^3$ and $n_0=1.8\cmcube$, such that
  $\average{B\rrms}\simeq0.001\mkG$ with $\average{\cdot}$ indicating the
  average over the computational domain.

  Due to differential cooling and heating in the vertically stratified gas the
  initial setup is in an unstable equilibrium, resulting in collapse and 
  subsequent oscillations. 
  A new equilibrium requires the gradual build up of thermal and turbulent
  pressure, followed by an extended period while the oscillations of the disc 
  dissipate and the disc becomes statistically steady.
  \citet{Gressel08a} eliminate these transients by adjusting the vertical 
  heating profile to balance exactly the initial cooling. 
  While heating may be only marginal to the long term dynamics of the model, 
  this does add a long term unphysical ingredient.
  An alternative approach, for future reference, may be to begin without heating
  and cooling in isothermal, hydrostatic equilibrium.
  This would be a stable equilibrium, and the heating and cooling terms could be
  switched on after the thermal and turbulent pressures are sufficiently
  developed to support the gas against gravitational collapse.
  This may reduce the duration, and hence numerical expense, of the initial
  transients.

%-----------------------------------------------------------------------------
\section{\label{sect:time}Simulation vs. 'realised' time, and geometry}
  
  For this model, as with other models of this type, the unperturbed initial
  conditions, which cannot be expected to exist in nature, require an extended 
  period of simulation time to reach a quasi-steady turbulent state. 
  \citet{Joung06} allow 40\,Myr for the initial transients to be subsumed by
  turbulence, while using a similar model \citet{AB07} argue this should
  be more like 200\,Myr.
  The initial conditions are not in equilibrium due to 
  the heating and cooling terms, and in the absence of thermal stability
  there is a large scale collapse of the disc. The thickness of the disc is 
  gradually expanded by the build up of both thermal and ram pressure by the 
  injection of SNe.
  There remain large scale vertical oscillations from the initial collapse,
  which require longer to settle. 
  In my model the regular systematic oscillations dissipate over a
  timescale of at least 100\,Myr, but not more than 200\,Myr,
  although random fluctuations occur at later times.

  The quasi-steady turbulent state then needs to be tracked over a few hundred
  Myr to provide reliable statistics for a robust analysis.
  For models tracing the galactic dynamo, with the relatively low magnetic
  Reynolds numbers available in the model, it is necessary to run them for up
  to 2\,Gyr to see the magnetic field grow to saturation.
  Is it then reasonable to assume that the models have a time
  independent statistically steady state?
  \citet{Tutukov00} model the lifetime evolution of the Galaxy.
  They suggest significant variation of star formation and SN rates occur in
  timescales of a few  hundred Myr, with associated changes to the ISM mass and
  thickness of the disc. 
  On the scale of 1\,Gyr, qualitative as well as quantitative effects
  might be expected to occur.
  These scales are based on highly simplified premises and are only indicative
  of galactic trends.
  Nevertheless it is reasonable to question whether a steady hydrodynamical state
  should persist over such long timescales.

  Consider also the effect of time on the geometry of the numerical domain.
  Nominally the box is differentially rotating around the Galactic centre. 
  Due to the differential rotation in the azimuthal direction the boundary 
  nearest to the galactic centre moves ahead of the corresponding outer 
  boundary, and this is realised in the model by the sliding periodic boundary
  conditions. The numerical domain therefore is elongated in the azimuthal 
  direction over time as illustrated in Fig.~\ref{fig:shearbox}. The
  rectangular box in the bottom left follows the orbit about the galactic
  centre, being increasingly distorted as shown. If we regard the sliding  
  boundaries to be forever moving apart in this way, as the simulation box
  completes several orbits, the geometry would rapidly cease to resemble a 
  rectangle and be stretched much like an elastic band. 
   %-----------------------------------------------------------------------------
  \begin{figure}[tb]
  \begin{center}
  \includegraphics[width=0.7\linewidth]{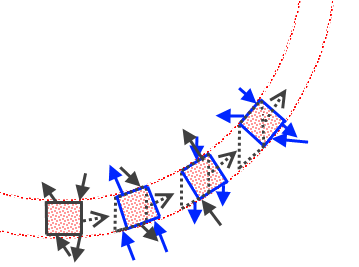}
    \caption[Geometry of shearing box.]{
  Illustration of the preservation of horizontal geometry of the simulation
  domain using the shearing box approximation. An arbitrary starting location 
  for the orbit of the numerical box about the galactic centre is indicated by
  the rectangular box bottom left. Differential rotation is represented by the
  inner boundary moving ahead of the outer ($2\nd$ position).
  The dotted arrows denote the angular velocity of the box.
  The rectangular geometry of the box is maintained by substituting
  statistically identical gas in the blue triangle ahead of the box for the
  gas in the trailing grey triangle.
  Arbitrary positions on the outer boundary subject to some velocity, are 
  indicated by the solid arrows, grey arrows for an original section and
  blue for a section of the substituted edge and its substitute.
  Corresponding arrows on the inner boundary indicate the offset postitions.
  The offset from the original edge increases, 
  while the offset from the substitute edge decreases ($3\rd$ position).
  By the time the sliding 
  boundaries have been offset by their full length ($4\uth$ position) the
  inner and outer edges are in exact correspondence. 
  Thus the blue rectangle in the $4\uth$ postion becomes a new arbitrary
  starting postion and the process repeats.  
    \label{fig:shearbox} }
  \end{center}
  \end{figure}
  %-----------------------------------------------------------------------------
   
  However to first approximation let us suppose the ISM in the neighbourhood of
  our box in the horizontal directions to be statistically homogeneous to that 
  within the box. 
  Consider the position of the $2\nd$ box in Fig.~\ref{fig:shearbox},   
  with the inner boundary in advance of the outer, illustrated by dotted grey
  lines. 
  Statistically there is little to distinguish gas within the triangular region
  bounded by the rear blue edge and the trailing corner from that bounded by 
  the leading and outer blue edges and leading dotted line.
  Preserving the geometry of the box (blue) as it orbits the galaxy, gas in the
  trailing triangle is replaced by equivalent gas occupying an 
  identical region of the ISM previously just ahead of the box.
  Some typical flow, denoted by the solid blue arrows, identifies the identical
  conditions on the trailing and substitute gas near the outer edge, and the
  corresponding position on the inner edge ($2\nd$ and $3\rd$ positions).
  The solid grey arrows denote such flow across original sections. 
  The offset between the black arrows increases with the shear, and the offset
  between the blue arrows reduces, until the offset for the original edges is
  the full length of the boundary ($4\uth$ position).
  At this point the inner and outer edges are no longer offset and the process
  is repeated from this new starting geometry (indistinguishable from the 
  $1\st$ position).
  
  With respect to the time, the total simulation time might be considered
  independent of the physical time to which it is scaled.
  Only the physical processes inside the calculation are actually tracking real
  time.
  Supernova remnants, the advection time of the hot gas from the disc to the
  boundary, the life times of superbubbles containing multiple SNe have 
  simulation times matching the corresponding physical times.
  The superbubbles have the longest duration of all the locally coherent
  physical features in the simulation domain.
  The premise of the periodic boundary conditions is that horizontally the 
  properties of the ISM in the neighbourhood are statistically identical to 
  first approximation.
  Thus as one superbubble dissipates and a new one evolves, rather than
  regarding the new event as occurring later in time it is reasonable to regard
  it as occurring at the same time in some random location in the neighbourhood.
  From this conception the entire simulation can be regarded as occurring 
  simultaneously in the quasi-steady state, whose physical time frame need 
  only match the simulation time of the longest lasting superbubble. 

  %-----------------------------------------------------------------------------
  \begin{figure}[tb]
  \begin{center}
  \includegraphics[width=0.7\linewidth]{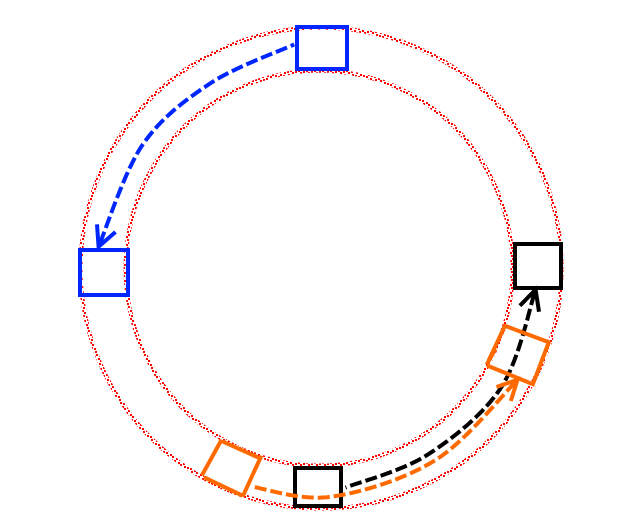}
    \caption[Statistical vs. real time diagram]{
  Illustration of the distinction between simulation time and realised time, as
  explained in the text.
%  An arbitrary starting time for the simulation in the steady state is 
%  indicated by the initial position of the box in black. Over about 60\Myr\ it
%  moves to it's second position.
%  The simulation continues, but the realised time is reset, but representing 
%  an arbitrary parcel of ISM in blue, which is considered to be concurrent, 
%  with the black. 
%  Such a concurrent parcel could even be tracing the close neighbourhood as 
%  indicated by the orange box.
    \label{fig:timeshear} }
  \end{center}
  \end{figure}
  %-----------------------------------------------------------------------------

  In Fig.~\ref{fig:timeshear} this principle is illustrated. Consider an 
  arbitrary starting time for the simulation in the statistically steady state
  to correspond to the initial position of the black box. 
  It orbits for 60--100\,Myr as shown.
  The simulation continues, but could be considered now to be modelling an 
  entirely different part of the orbit, shown in blue, but regarded in the 
  realised time frame to be concurrent with the part just modelled in black.
  Hence, 120--200\,Myr of simulation time has elapsed, but only 60--100\,Myr
  of realised time need be considered to have elapsed. 
  Given the statistical homogeneity of the ISM within the neighbourhood of the
  orbit, even a nearby parcel of ISM, indicated by the orange box, could be 
  regarded as statistically independent of the original sample.
  In this way each segment of 60--100\,Myr may be considered as repeated 
  sampling of the independent sections of the ISM in the same orbit over the 
  same time period, irrespective of how long the simulation lasts.
  Similarly, the extended simulation time may be considered in terms of 
  ensemble sampling from appropriate neighbourhoods within multiple galaxies, 
  where the common parameters apply.

  The only feature that is modelled here that has a structure larger than the
  local box, and evolves over time scales exceeding the longest lasting 
  superbubbles, is the mean magnetic field.
  However, the large scale structure  of the mean field, is in reality also a
  statistical property of the random magnetic field on the local scale.
  Rather than single field lines extending azimuthally over kpc scales, the
  field lines are highly scattered, but with a preferential orientation in the
  direction of shear.
  The large scale organization of the field in the model is therefore not due
  to any physical superstructures, but to alignment within the random small
  scale structures.  
  
  Finally, the large time scales required to follow the growth of the magnetic
  dynamo may contravene the assumption of a quasi-steady state.
  In this case the magnetic Reynolds numbers in the simulations are far smaller
  than in the real ISM, so it may be expected that the growth rates would be
  much slower in the simulations.
  Also here, however, it would be an error to regard the tracing of the dynamo
  as a physically continuous process. 
  Instead, the appropriate interpretation of time in the dynamo runs, is that
  each 100--300\,Myr of the simulation represents statistical sampling of
  hydrodynamically similar galaxies with magnetic fields at various stages of
  maturity. 

\end{chapter}

%-----------------------------------------------------------------------------
\begin{chapter}{Models explored\label{chap:Models}}
%-----------------------------------------------------------------------------
%-----------------------------------------------------------------------------
\section{Summary of models\label{sect:Mods}}
%-----------------------------------------------------------------------------

  I consider five numerical models which do not include magnetic fields
  (HD), and five which do (MHD), and these are named in 
  Table~\ref{table:models} Column~1, together with some of the more important
  input parameters. 
  Some significant statistical outcomes describing the results for each model
  are displayed in Table~\ref{table:results}.
  The HD models are labelled with prefix RBN or WSW to denote the cooling
  function used (as described in Section~\ref{sect:ssvrb}). 
  All the MHD models use the WSW cooling function, so WSW is dropped and labels
  identify the other key parameters applied.
%-----------------------------------------------------------------------------
  \begin{table*}[h]
    \centering
  \caption[List of models]{\label{table:models}
    Selected input parameters of 
    the numerical models explored in this paper, 
    named in Column~(1).
    Column: (2) numerical resolution;
    (3) galactic rotation rate $\Omega$; 
    (4) galactic shear rate $S$; 
    (5) galactic supernovae rate $\dot{\sigma}$; 
    (6) initial mid-plane gas number density $n_0$;
    (7) magnetic field $\vect{B}$;
    (8) vertical boundary condition on magnetic potential $\vect{A}$ 
    (Sect.~\ref{sect:eq});
    (9) Section where particular model is considered.
    The estimates for the solar neighbourhood of mean angular velocity
    $\Omega_0$ and rate of SN explosions $\dot{\sigma}_0$ (Sect.~\ref{sect:eq}
    and~\ref{sect:MSN}).
 }
  \begin{tabular}[htb]{lcccccccc}
  (1)      &(2)      &(3)          &(4)          &(5)          &(6)          &(7)             &(8)              &(9)     \\
  \hline\\                                                                                                       
   Model   &$\Delta$ &$\Omega$     &$S$          &$\dot{\sigma}$     &$n_0$        &$\vect{B}\neq\vect{0}$ & bc$z(\vect{A})$ &Section \\
           &[pc]     &$[\Omega_0]$ &$[\Omega_0]$ &$[\dot{\sigma}_0]$ &[$\!\cmcube$]&                &                 &        \\[3pt]
  \hline\\                                                                                                               
   \Op     &4        & 1           &      -1     &  1          &1.8          & no             & --              &\ref{chap:HDref},~\ref{chap:flow},~\ref{sect:ti}    \\
   \OpH    &2        & 1           &      -1     &  1          &1.8          & no             & --              &\ref{sect:TMPS}                       \\
   RBN     &4        & 1           &      -1     &  1          &2.1          & no             & --              &\ref{subsect:COOL},~\ref{subsect:MOL} \\
   \WSWa   &4        & 1           &      -1     &  1          &2.1          & no             & --              &\ref{subsect:COOL},~\ref{subsect:MOL} \\[9pt]
   $\Omph$ &4        & 1           &      -1     &  1          &1.8          & no             & --              &\ref{subsect:params},~\ref{chap:MHD}   \\[9pt] 
   $\Ompa$ &4        & 1           &      -1     &  1          &1.8          & yes            & vertical        &\ref{subsect:COOL},~\ref{chap:MHD}     \\
   $\Ompb$ &4        & 1           &      -1     &  1          &1.8          & yes            & open            &\ref{chap:MHD}                        \\
   $\Ompd$ &4        & 2           &      -2     &  1          &1.8          & yes            & vertical        &\ref{chap:meanB},~\ref{chap:MHD},~\ref{subsect:MOL} \\
   $\Ompc$ &4        & 1           &      -1     &0.8          &1.8          & yes            & vertical        &\ref{subsect:params},~\ref{chap:MHD}   \\
   $\Ompe$ &4        & 1           &      -1.6   &  1          &1.8          & yes            & vertical        &\ref{subsect:params},~\ref{chap:MHD}   \\ 
  \hline\hline
  \end{tabular}
   \end{table*}
%-----------------------------------------------------------------------------

  For the MHD models, $\Ompa$ and $\Ompb$ denote models with parameters
  matching the solar neighbourhood, but with alternative vertical boundary 
  conditions, as described in Section~\ref{sect:bcs};
  all models (HD also) have a rate and distribution of supernova explosions
  equivalent to the solar neighbourhood estimates $\dot{\sigma}_0$, as 
  explained in Section~\ref{sect:MSN}, except for $\Ompc$ at $0.8\dot{\sigma}_0$;
  $\Ompd$ has galactic rotation $\Omega$ and shearing $-S$ double the value
  $\Omega_0=25\kms\kpc^{-1}$ introduced in Section~\ref{sect:eq}; 
  and $\Ompe$ has a shearing rate $-1.6\Omega_0$. 
  The primary motivation for these variations is to investigate the
  sensitivity of the dynamo to the parameters.  

  Some adjustments to the model are implemented for the MHD runs, to stabilize
  the disc and to recycle mass, which is transported across the vertical 
  boundaries over the extended duration required for the dynamo.
  These are detailed in Appendix~\ref{subsect:MOL}.
  To assist direct comparison with the HD models, Model~\Op\ is repeated with
  these adjustments applied, and labelled in Table~\ref{table:models} as
  $\Omph$.

  Column~2 of Table~\ref{table:models} lists the resolution applied in each 
  model.
  The standard resolution is $(4\p)^3$, with only one model, \OpH, having
  $(2\p)^3$. 
  The latter model is applied to assess the sensitivity of the models to 
  resolution. 
  The starting state for this model is obtained by remapping a snapshot at
  $t=600$\,Myr from the standard-resolution Model~{\Op} (when the system has
  settled to a statistically steady state) onto a grid $512\times512\times1120$
  in size.
  Resolutions of $(1\p)^3$, $(2\p)^3$ and $(4\p)^3$ were tested for 
  separate analysis of single remnant models in uniform monotonic ambient ISM
  of various densities (see Appendix~\ref{sect:SNPL}), prior to conducting these
  comprehensive simulations.
   
  The initial conditions, described in Section~\ref{sect:ICons}, are intended 
  to be transient, with the statistically steady state into which the models
  evolve being independent of the initial state within 2--4\,Myr of the start.
  Subsequent evolution however is not independent of the initial density
  profile, and to facilitate comparison of my results with theory and 
  observations, it is important that the density profile in the evolved state
  should be close to the estimated values for the solar neighbourhood.  
  In the HD models, we would expect the combined thermal and RAM pressure 
  supporting the gas against gravitational acceleration to be weaker than 
  the observed ISM, due to the absence of magnetic and cosmic ray pressure.
  Cosmic ray pressure is absent from all models.
  Since the model resolution does not resolve the dense molecular  clouds, it
  is not clear in advance whether to include this contribution to the ISM
  density.
  Models RBN and \WSWa\, are initialised with the contribution from 
  molecular hydrogen included, while all other models exclude it, as listed in
  Column~6.

  The parameters of rotation, shear and supernova rate are listed in
  Columns~3, 4 and 5 of Table~\ref{table:models}. 
  Column~7 indicates whether a magnetic field is included, and Column~8
  indicates which boundary conditions apply to the vector potential $\vect{A}$
  on the upper and lower surfaces, which are described in 
  Section~\ref{sect:bcs}. Column~9 lists the Sections of this thesis in which
  the results for each model are described.

  These models are computationally demanding.
  The standard resolution models require parallel processing on around 280 
  nodes and consume around 100 000 cpu hours for 100\,Myr of simulation. 
  Therefore, to maximise the results from the available resources, only three 
  models have been run from the initial conditions at $t=0$\,yr. 
  These were Models~RBN, \WSWa\, and $\Ompa$, which were evolved until
  the systems were in the hydrodynamical steady turbulent state, at about 
  200\,Myr, and then continued to at least 600\,Myr.

  Model \Op\, was initialised from a turbulent snapshot of \WSWa\, at 400\,Myr,
  but had the gas density at every point reduced by about 17\%, so that
  the revised density profile corresponded to the estimated rates for the solar
  neighbourhood, excluding molecular hydrogen.
  It was then continued to 675\,Myr, sufficient for any transients arising from
  the adjustment to subside.
  Model~\OpH\, was initialised by remeshing a snapshot from Model~\Op\, at 
  about 600\,Myr, and was continued to 650\,Myr. 
  Due to the reduced grid spacing and the improved resolution of gradients
  in all variables, the iterative time step reduces substantially to satisfy 
  the Courant-Friedrichs-Lewy (CFL) condition. 
  The impact is an increase of 10--30 times in the computational demands to 
  advance the simulation each Myr.
  Consequently, only about 50\,Myr were completed for this model, which may 
  perhaps yield insufficient statistics for the steady state of this simulation.

  For the magnetic runs, $\Ompa$ was run to 400\,Myr, and then all of these 
  models were continued from this snapshot with the revised parameters as listed
  in Table~\ref{table:models}. 
  It becomes apparent from the rates of growth that if the dynamo is 
  successful, it would require in excess of 2\,Gyr for the magnetic field to 
  grow to saturation. 
  Each is run for another several hundred Myr, enough to identify the presence
  of a dynamo and provide a basis for comparisons between the models.
  To investigate the dynamo process to saturation, Model~$\Ompd$, with the 
  fastest growing dynamo, is continued to 1.7\,Gyr.
  The dynamo saturates at about 1.1\,Gyr and then continues with the system in
  a statistically steady state. 
  For comparisons in the saturated regime, models~$\Ompb$ and $\Ompa$ are 
  restarted from a snapshot of $\Ompd$ at 1.2\,Gyr and 1.4\,Gyr respectively, 
  and continued for a further 400\,Myr each.    
    
%-----------------------------------------------------------------------------
\section{Model data summary\label{sect:Stat}}
%-----------------------------------------------------------------------------
    
%-----------------------------------------------------------------------------
  \begin{table*}
  \centering
    \caption[Model results]{
      Column (1) lists the models from Table~\ref{table:models} with some
      indicative results: (10) average sound speed $\average{c\sound}$; 
      (11) average kinematic viscosity $\average{\nu}$;
      (12) average Reynolds number defined at the grid spacing 
       $\langle \Ry_\mesh\rangle$;
      (13) average r.m.s. perturbation velocity $\average{u\rrms}$;
      (14) turbulent velocity $\average{u\turb}$;
      (15) thermal energy density $\average{e_\mathrm{th}}$; 
      (16) kinetic energy density $\average{e_\mathrm{kin}}$; 
      (17) magnetic energy density $\average{e_\mathrm{B}}$;
  and (18) time span over which the models have been in steady state 
      ($\tau$ is the typical horizontal crossing time);
      Standard deviation over volume using eleven composite snapshots 
      (ten for \OpH) per model are given in brackets.
    \label{table:results}}
  \rotatebox{90}{
%      \hfill
 % {\small{
\begin{tabular}[htb]{lcccccccccccccccc}
  (1)     &(10)            &(11)                   &(12)                     &(13)            &(14)              &(15)              &(16)                  &(17)  &(18)                                \\
  \hline\\
  Model   &$\average{c\sound}$      &$\average{\nu}$ &$\average{\Ry_\mesh}$ &$\average{u\rrms}$       &$\average{u\turb}$ &$\average{e_{\rm th}}$      &$\average{e_{\rm kin}}$     &$\average{e_{\rm B}}$                &$\Delta t$      \\ 
          &$[\!\kms]$               &                &                       &$[\!\kms]$               &$[\!\kms]$      &[$E\SN\kpc^{-3}$] &[$E\SN\kpc^{-3}$] &[$E\SN\kpc^{-3}$] &[$\tau$]        \\[3pt]
  \hline\\ 
  \Op     &$108~(113)$              &$0.44$          &$0.88$                 &$\phantom{1}76~\phm(85)$ &$26~(27)$       &$30~\phm(87)$     &$13~(34)$         & --                         &3.9             \\
  \OpH    &$186~(207)$              &$0.77$          &$0.85$                 &          $103~(124)$    &$34~(37)$       &$19~\phm(59)$     &$10~(50)$         & --                         &0.5             \\
  RBN     &$\phantom{2}58~(118)$    &$0.24$          &$1.18$                 &$\phantom{1}37~\phm(42)$ &$18~(22)$       &$25~(115)$        &$\phm9~(25)$      & --                         &2.7             \\
  \WSWa   &$\phantom{2}65~\phm(75)$ &$0.27$          &$0.97$                 &$\phantom{1}45~\phm(49)$ &$20~(22)$       &$29~\phm(85)$     &$13~(46)$         & --                         &4.0             \\[9pt]
  $\Ompa$ &$\phm23~\phm(25)$        &$0.09$          &$0.64$                 &$\phm12~\phm(11)$        &$\phm5~\phm(6)$ &$25~\phm(93)$     &$\phm5~(13)$     &$5\phd~\phd(8)$             &1.5             \\
  $\Ompb$ &$\phm23~\phm(22)$        &$0.09$          &$0.65$                 &$\phm13~\phm(14)$        &$\phm6~\phm(8)$ &$25~\phm(96)$     &$\phm6~(15)$     &$3\phd~\phd(5)$             &1.9             \\
  $\Ompd$ &$\phm21~\phm(21)$        &$0.09$          &$0.59$                 &$\phm11~\phm\phm(9)$     &$\phm5~\phm(6)$ &$26~(102)$        &$\phm5~(15)$     &$7\phm~\phm(10)$            &2.0             \\[9pt]
  $\Omph$ &$\phm41~\phm(65)$        &$0.17$          &$0.95$                 &$\phm33~\phm(45)$        &$12~(16)$       &$26~\phm(74)$     &$\phm9~(19)$     &                            &4.8             \\[9pt] 
$\Ompa^*$ &$\phm58~\phm(75)$        &$0.24$          &$0.88$                 &$\phm38~\phm(47)$        &$15~(19)$       &$25~\phm(66)$     &$10~(26)$     &$0.1~(0.4)$                 &4.5             \\
  $\Ompc$ &$\phm21~\phm(23)$        &$0.09$          &$1.01$                 &$\phm18~\phm(13)$        &$\phm6~\phm(8)$ &$23~\phm(75)$     &$\phm7~(16)$     &$0.1~(0.4)$                 &2.5             \\
  $\Ompe$ &$\phm24~\phm(32)$        &$0.10$          &$0.96$                 &$\phm20~\phm(18)$        &$\phm7~(10)$    &$29~\phm(94)$     &$\phm7~(24)$     &$0.1~(0.4)$                 &3.6             \\ 
  \hline
  \hline
  \end{tabular}
}
    \end{table*}
%-----------------------------------------------------------------------------

  In Table~\ref{table:results} some indicative data from the various models
  identified in Table~\ref{table:models} are displayed. 
  Here angular brackets denote averages over the whole volume, taken
  from eleven snapshots (10 for \OpH) within the statistical steady state. 
  The standard deviations (in brackets) are spatial fluctuations for 
  the combined snapshots. 
  The time span, $\Delta t$, is given in Column~18, normalised by 
  $\tau=L_x/\average{u\turb}$, where $u\turb$ is the root-mean-square
  random velocity and $L_x\approx1\kpc$ is the horizontal size of the
  computational domain.
  For the HD models, this refers to the steady state.
  For the MHD models the system has been in a hydrodynamically steady state 
  much longer than the $\Delta t$ as listed.
  For $\Ompa$, $\Ompb$ and $\Ompd$, $\Delta t$ refers to the latter period
  when the magnetic field is also in a statistical steady state.
  For $\Ompc$ and $\Ompe$, the field is still growing and $\Delta t$ refers
  only to how long the system has been hydrodynamically steady. 
  Results are also taken from Model~$\Ompa$ during the kinematic
  stage, for comparison with the latter models; these are listed in the 
  table as $\Ompa^*$.
  As $\nu$ is set proportional to the speed of sound $c\sound$ (Column~10) it
  is variable, and the table presents its average value
  $\average{\nu}=\nu_1 \average{c\sound}$ as Column~11, where $\nu_1=0.004$ in 
  all models.
  
  The numerical resolution is sufficient when the mesh Reynolds number, 
  $\Ry\mesh = u\,\Mesh/\nu$, does not exceed a certain value (typically 
  between 1 and 10) anywhere in the domain.
  The indicative parameter values of the mesh Reynolds number in 
  Table~\ref{table:results} Column~12 are averages,
  $\langle\Ry\mesh\rangle=\Mesh\average{u\turb/c\sound}/\nu_1$, 
  where $\Mesh$
  is the grid spacing (4\,pc for all models, except for Model~{\OpH}, where
  $\Mesh=2\p$).
  Although $\Ry\mesh$ is typically close to 1\ in all models, for some locations
  this could be much greater. 
  However, $u\,\Mesh/\nu<5$, is ensured through the combination of temperature
  dependent bulk diffusion and shock enhanced diffusion.
  
  Other quantities shown in Table~\ref{table:results} have been calculated as
  follows. 
  In Column~13 $\average{u\rrms}$ is derived from the total perturbation
  velocity field $\vect{u}$, which excludes only the overall galactic rotation
  $\vect{U}$.
  In Column~14 $\average{u\turb}$ is obtained from the turbulent velocity only, 
  with the mean flows $\average{\vect{u}}_\ell$, defined in 
  Eq.~\eqref{eq:ugauss}, deducted from $\vect{u}$. 
  In Columns~15 and 16,
  $e_\mathrm{th}=\langle \rho e \rangle$ and 
  $e_\mathrm{kin}=\langle\tfrac12\rho u^2\rangle$ are the average
  thermal and kinetic energy densities respectively;
  the latter includes the perturbed velocity $\vect{u}$, and both are normalised
  by the SN energy $E\SN=10^{51}\erg$.
  In Column~17 $e_\mathrm{B}=\average{B^2/8\upi}$ is the average  
  magnetic energy density, also normalised by $E\SN$. 
  For the magnetic runs, the energies are given for the stage after the magnetic
  field has saturated, except for Models~$\Ompc$, $\Ompe$ and $\Ompa^*$, for
  which data 
  is only available for the kinematic stage.
  
\end{chapter}

%  \clearpage                            % End the current page making sure all
%  \thispagestyle{empty}                 % tables/figures are printed.
%  \cleardoublepage                      % Necessary for correct page numbering.
  \part{Multiphase description of the interstellar medium}\label{part:mphase}
%  \clearpage                            % End the current page making sure all
%  \thispagestyle{empty}                 % tables/figures are printed.
%  \cleardoublepage                      % Necessary for correct page numbering.
  %-----------------------------------------------------------------------------
\begin{chapter}{Is the ISM multi-phase?\label{chap:HDref}}
%-----------------------------------------------------------------------------

  As explained in Section~\ref{subsect:ism} the ISM has been described as 
  comprising dynamically distinct regions of hot, warm and cold gas 
  \citep{CS74,MO77}, which co-exist in relative pressure equilibrium.
  This was postulated theoretically; due to differential cooling rates of the
  gas at various states (temperature and density composition, which here I 
  shall refer to as {\textit{phases}}) compared to the rate of thermal
  conductivity in the ISM, it could be expected that the phases would settle to
  pressure equilibrium far faster than they diffuse into a thermal equilibrium.
  Since observations confirm gas exists at temperatures typical of the cold at
  $T\simeq10^2\K$, warm at $T\simeq10^4\K$ and hot at $T\simeq10^6\K$ this 
  picture of phases in pressure balance would appear to be inevitable.

  However the description becomes less convincing once the highly turbulent 
  nature of the ISM is considered.
  The continual shocks and heating produced by SNe create bubbles of 
  extremely hot gas surrounded by ballistically propelled cold dense shells.
  It
  cannot be excluded that in fact these phases are merely
  transitory features of a violently perturbed medium with a broad spectrum
  of randomly distributed pressures, where gases of different temperatures
  merely occupy part of the continuum of a qualitatively homogeneous gas
  \citep[][discussion on \emph{The controversy}]{V09}.
  
  In this chapter I report how my results address this issue using the model
  which most closely resembles the Milky Way, \Op. 
  The observational data and our understanding of the ISM is strongest for the
  solar neighbourhood
  and this also provides a good basis for assessing the accuracy of my model.
  If a distinction between the phases can be justified it may be useful in
  interpreting observations. 
  Trying to make sense of observations and statistical inferences from data,
  with such huge scale separation of 6 to 10 orders of magnitude, is highly
  unreliable. 
  If separation into phases is valid then it permits observers and 
  theoreticians to make inferences on more discrete data sets, with better
  defined characteristics in terms of for example mass, velocity, temperature
  and magnetic field distributions.

%-----------------------------------------------------------------------------
  \begin{figure}[h]\vspace{-0.5cm}
  \centering
\hspace{-2.8cm}  \includegraphics[width=.55\textwidth]{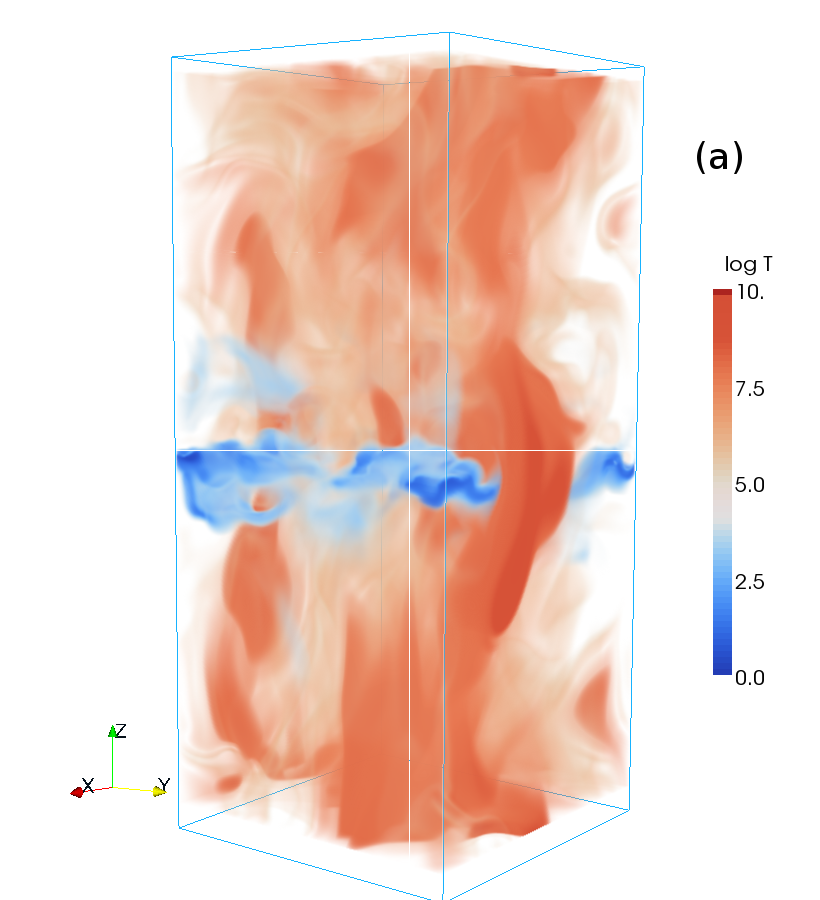}
 \hspace{-0.55cm} \includegraphics[width=.55\textwidth]{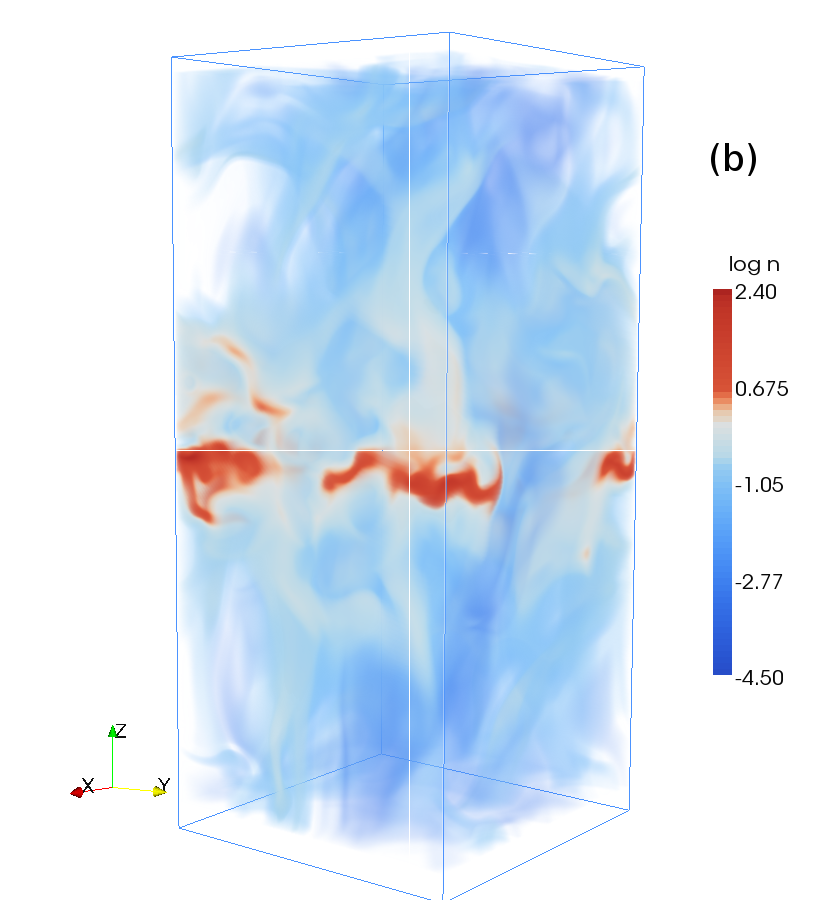}\hspace{-2.0cm}
  \caption[Volume plots of density and temperature for Model~\Op]{
  A three-dimensional rendering of \textbf{(a)} temperature and \textbf{(b)}
  density distributions in Model {\Op} at $t=551\Myr$.
  Cold, dense gas is mostly located near the mid-plane, whereas hot gas
  extends towards the upper and lower boundaries.
  To facilitate visualisation of the 3D structure, warm gas ($10^3<T<10^6\K$)
  and diffuse gas ($n<10^{-2}\cm^{-3}$) are plotted with high transparency,
  so that extreme temperatures, and dense
  structures are emphasized.
  \label{fig:tsnap}}
  \end{figure}
%------------------------------------------------------------------------

  Model~{\Op} is taken as a hydrodynamic (HD) reference model, because it has rotation
  corresponding to a flat rotation curve with the Solar angular velocity, and
  gas density comparable to the solar neighbourhood, but excluding the 
  molecular clouds 
  (See Appendix~\ref{subsect:MOL} for a discussion of the effect of this change of
  total mass). 

  Figure~\ref{fig:tsnap} shows typical temperature and density distributions 
  at $t=551\Myr$ (i.e., $151\Myr$ from the start of run \Op).  
  Supernova remnants appear as irregularly shaped regions of hot, dilute gas.
  A hot bubble breaking through the cold gas layer extends from the mid-plane
  towards the lower boundary, visible as a vertically stretched region in the 
  temperature snapshot near the $(x,z)$-face.
  Another, smaller bubble can be seen below the mid-plane near the $(y,z)$-face.
  Cold, dense structures are restricted to the mid-plane and occupy a small
  part of the volume.
  Very hot and cold regions exist in close proximity.

%-----------------------------------------------------------------------------
  \begin{figure}
 \hspace{-1.1cm}  \includegraphics[width=1.1\columnwidth]{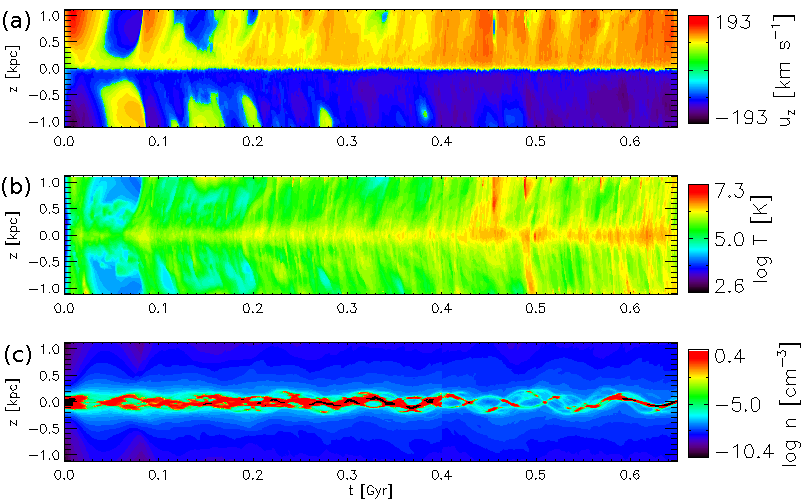}
  \caption[Horizontal averages $u_z$, $T$, and $\rho$ for Model~\Op(\WSWa)]{
  Horizontal averages of \textbf{(a)} the vertical velocity, \textbf{(b)}
  temperature and \textbf{(c)} gas density as functions of time for
  Model~{\Op} (Model {\WSWa} up to 0.4\,Gyr).
  \label{fig:pav}}
  \end{figure}
%---------------------------------------------------------------------------

  Horizontally averaged quantities, as functions of height and time, are plotted
  in Fig.~\ref{fig:pav} for Model \WSWa\ at $t<400\Myr$, and \Op~at later
  times, showing the effect of reducing the total mass of gas at the
  transition time.
  This figure shows the vertical velocity (panel a), temperature (b) and gas
  density (c).
  Average quantities may have limited physical significance, because the
  multi-phase gas structure encompasses an extremely wide range of conditions.
  For example panel (b) shows that the average temperature near the mid-plane,
  $|z|\la0.35\p$, is, perhaps unexpectedly, generally higher than that at
  the larger heights.
  This is due to SN~II remnants, which contain very hot gas with $T\ga10^8\K$
  and are concentrated near the mid-plane; even though their total volume is
  small, they significantly affect the average temperature.
   
  Nevertheless these average conditions help to illustrate some global 
  properties of the simulations.
  Before the system settles into a quasi-stationary state at about $t=250\Myr$,
  it undergoes a few large-scale transient oscillations involving quasi-periodic
  vertical motions. 
  The periodicity of approximately 100\Myr, is consistent with the breathing
  modes identified by \citet{WC01} in 1D models with vertical perturbations,
  attributable to the gravitational acceleration.
  Gas falling from altitude will pass through the mid-plane, where the net 
  vertical gravity is zero and so continues its trajectory until the increasing 
  gravity with height again reverses it. 
  Turbulent pressure and gas viscosity dampen these modes.
  At later times, a systematic outflow develops with an average
  speed of about $100\kms$; we note that the vertical velocity increases very
  rapidly near the mid-plane and varies much less at larger heights.
  The result of the reduction of gas density at $t\approx400\Myr$ is clearly
  visible, as it leads to higher mean temperatures and a stronger and more regular
  outflow, together with a less pronounced and more disturbed layer of cold gas.

%-----------------------------------------------------------------------------
  \section{Identification of a multi-phase structure}\label{sect:TMPS}
%-----------------------------------------------------------------------------

  All models described in this thesis, including the HD reference Model {\Op},
  have a well-developed multi-phase structure apparently similar to that
  observed in the ISM.
  Since the ISM phases are not genuine, thermodynamically distinct phases 
  \citep{V09},
  their definition is tentative, with the typical temperatures of the cold,
  warm and hot phases usually set at $T\simeq10^2\K$, $10^4$--$10^5\K$ and
  $10^6\K$, respectively.

%-----------------------------------------------------------------------------
  \begin{figure}[h]
  \centering\hspace{-2cm}
  \includegraphics[width=0.36\columnwidth,clip=true,trim=0 0 0 9mm]{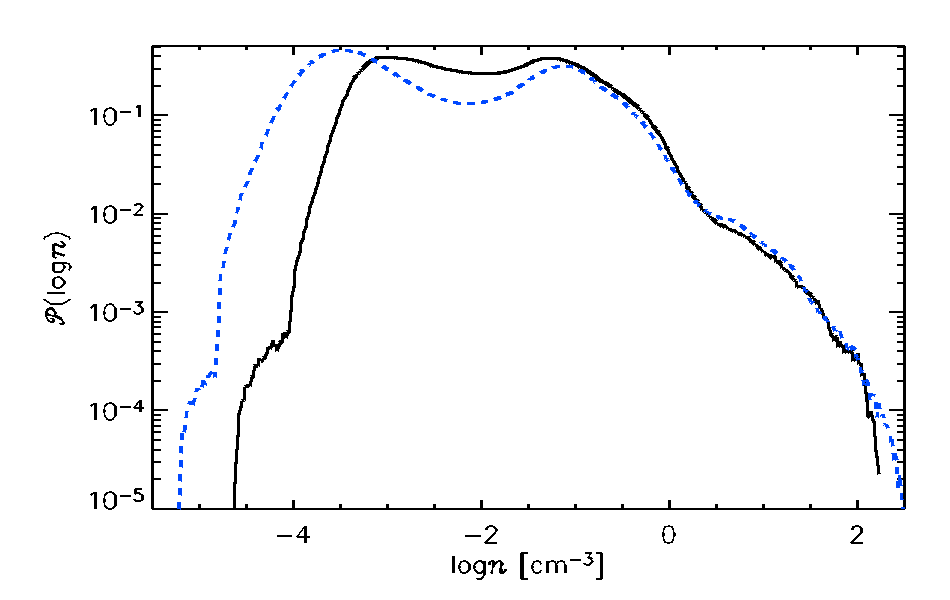}
  \includegraphics[width=0.36\columnwidth,clip=true,trim=0 0 0 9mm]{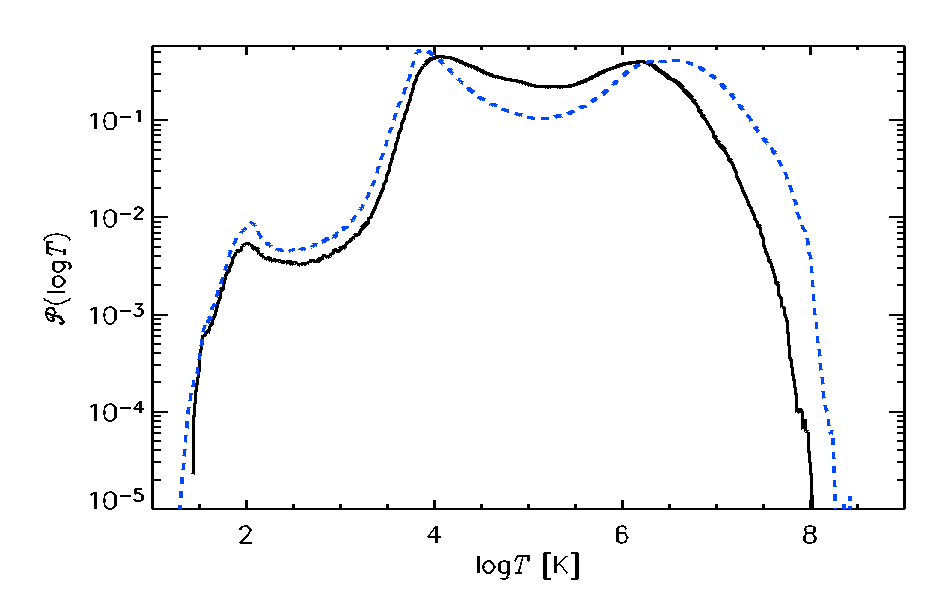}
  \includegraphics[width=0.36\columnwidth,clip=true,trim=0 0 0 9mm]{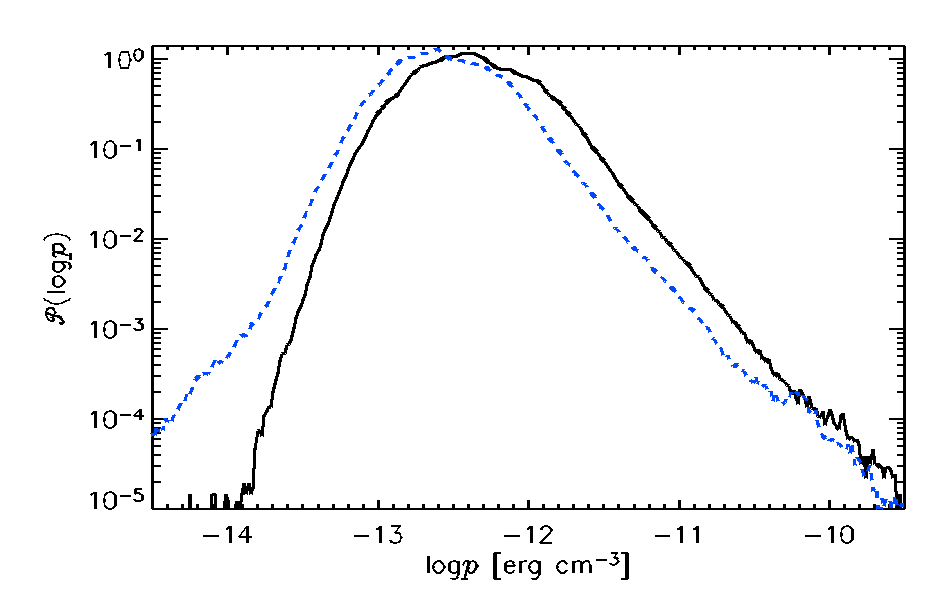}
  \hspace{-2cm}
    \caption[Total volume probability distributions for Model~\Op]{
  Volume weighted probability distributions of gas number density~{\textbf{(a)}},
  temperature~{\textbf{(b)}} and thermal pressure~{\textbf{(c)}} for  models {\Op} (black, solid) and
  {\OpH} (blue, dashed) for the total numerical domain $|z|\le1.12\kpc$.
    \label{fig:pdf2}}
  \end{figure}
%-----------------------------------------------------------------------------

  From inspection of the total volume probability density distributions shown
  in Fig.~\ref{fig:pdf2} it is not necessarily apparent that such a 
  description is justified.
  The black solid lines refer to Model~\Op\ and the blue dashed lines refer to
  Model~\OpH, which differs only in its doubly enhanced resolution and against 
  which the sensitivity to resolution may be measured.
  In panel~(b) there are clearly three distinct peaks in the temperature,
  corresponding roughly to the conventional definitions. 
  For the gas density and temperature probability distributions  
  (Fig.~\ref{fig:pdf2}a and b) the minima in the distributions appear 
  independent of resolution (at density $10^{-2}\,{\rm cm}^{-3}$, and at
  temperatures $10^2$ and $3 \times 10^{5}$ K).
  The distributions are most consistent in the thermally unstable 
  range 313 -- 6102\,K.
  The minimum about the unstable range above $10^5\K$ is more pronounced with
  the higher resolution, because the hotter gas has reduced losses to thermal
  conduction.
  The modal temperatures of the cold gas ($100$\,K) and warm gas ($10^4$\,K) are
  consistent.
  In Fig.~\ref{fig:pdf2} for Model~\OpH\ (blue, dashed lines) the bimodal
  structure of the density distribution (a) and the trimodal structure of the
  temperature distribution (b) are more pronounced.
  From Table~\ref{table:results} Column~15, note that the average thermal
  energy density for Model~\OpH\ is less than Model~\Op.
  This indicates that the higher temperatures are associated with even
  more diffuse gas, and the dense structures are even colder, reducing the 
  overall thermal energy in the system.  

  However the plots are not inconsistent with an alternative description. 
  \citet{AB05a} conclude, from very similar results, that the ISM gas density
  and temperature have a broad continuum of values, out of thermal pressure
  equilibrium. 
  In panel~(c), apart from a reduction of approximately one half for the higher
  resolution, the probability distributions of thermal pressure are almost 
  indistinguishable.
  They have a broad distribution, varying across about three orders of 
  magnitude for probability density above 0.001. 
  This is markedly narrow, however, when contrasted to the five to six 
  orders of magnitude across which density and temperature vary.

% -----------------------------------------------------------------------------
  \begin{figure}[h]
  \centering
  \includegraphics[width=1.0\columnwidth]{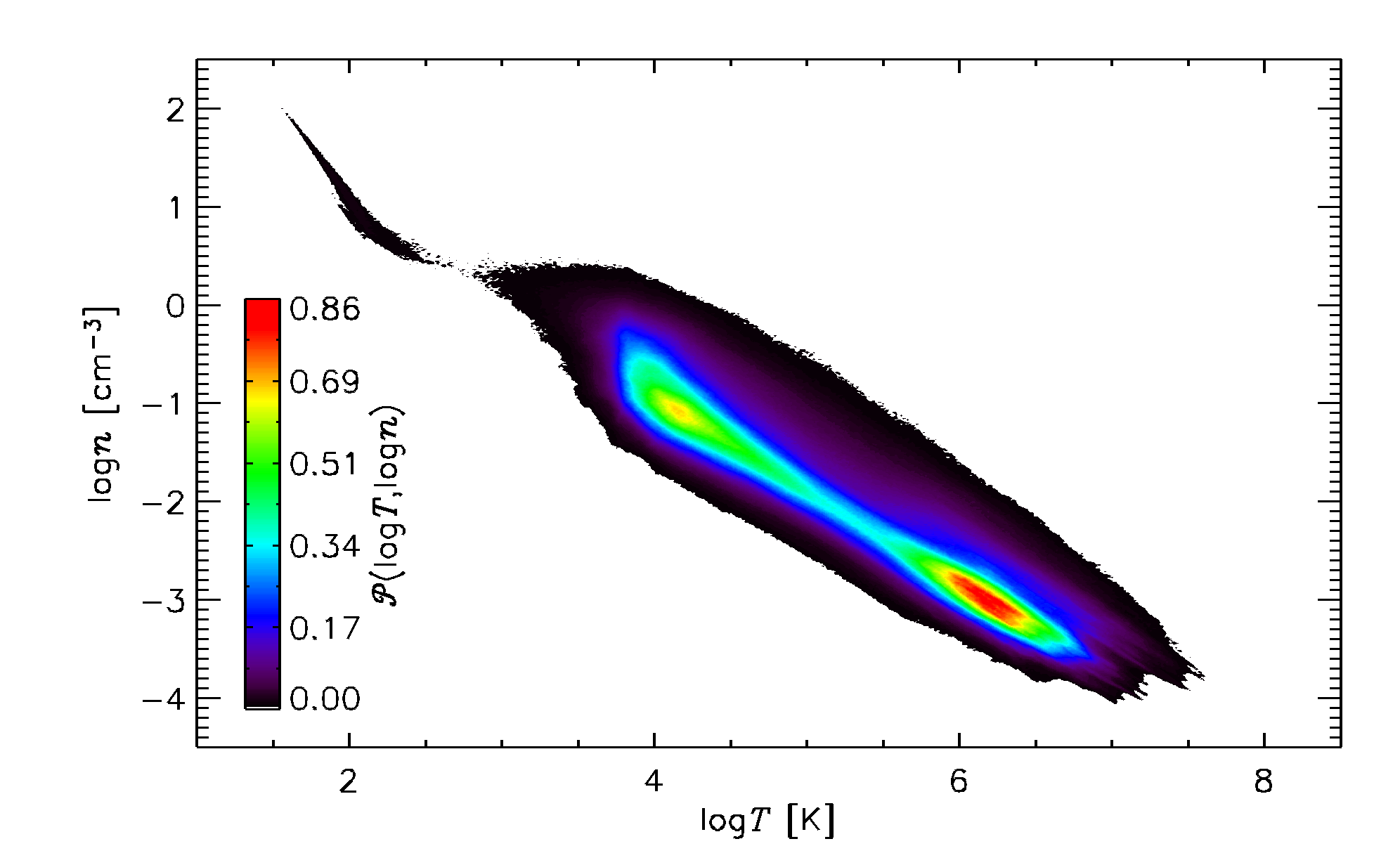}
  \caption[2D probability distribution of $n$ and $T$ for Model~\Op]{
  Total volume probability distributions of gas by temperature $T$ and density $n$.
  Eleven snapshots of Model~{\Op} in the interval $t=634$ to $644\Myr$ were used,
  with the system in a statistical steady state.
%  Contours of constant cooling time $10^5,~10^6$ and $10^8$yr, are overplotted
%  to indicate where the gas may be subject to rapid radiative cooling.
  \label{fig:pdf2d}}
  \end{figure}
%-----------------------------------------------------------------------------

  The probability distribution of gas number density and
  temperature is shown in Fig.~\ref{fig:pdf2d} (model~\Op).
  While it is evident that there are large regions of pressure disequilibrium,
  there is clear evidence of three well defined phases, in which the pressure
  is perturbed about a line of constant thermal pressure, described by,
  \[\log n = -\log T + {\rm constant}. \]
  There are three distinct peaks at $(T[\K],n[\cmcube])\simeq(10^2,10)$, 
  $(10^4,10^{-1})$ and $(10^6,10^{-3})$.
  The lines intersecting the maxima for the cold, warm and hot 
  distributions at $(\log T,\log n) = (2,1.2),(4,-0.8), (6,-2.8)$,
  respectively, correspond to $p=k_{\rm B}nT=10^{-12.7}\dyn\cm^{-2}$.
  A reasonable identification of the boundaries between these phases
  is via the minima at about $500\K$ and $5\times10^5\K$.

  Having divided the gas along these boundaries, the probability distributions
  within each phase of gas number density $n$, turbulent r.m.s. velocity
  $u\turb$, Mach number $\mathcal{M}$, thermal pressure $p$ and total pressure
  $P$ are displayed in Fig.~\ref{fig:pdf3ph}, from Model~{\Op}.

%-----------------------------------------------------------------------------
  \begin{figure}[h]
  \centering
  \hspace{25.0cm}
  \includegraphics[width=0.535\textwidth,clip=true,trim=0 0 0 9mm]{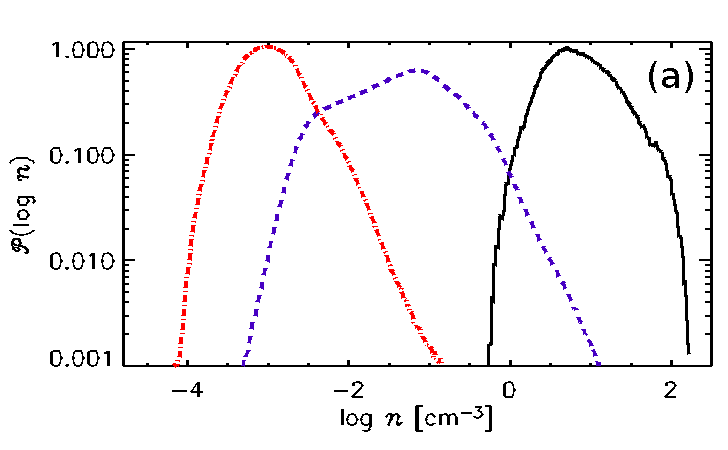}
  \hspace{25.0cm}\\
  \hspace{-1.5cm}
  \includegraphics[width=0.535\textwidth,clip=true,trim=0 0 0 9mm]{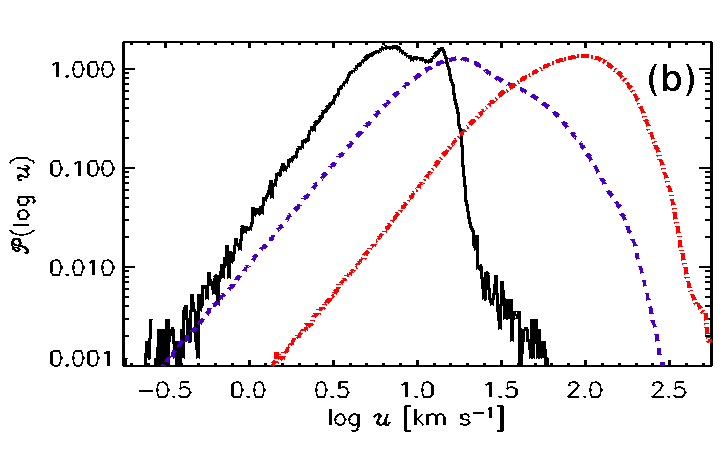}
  \includegraphics[width=0.535\textwidth,clip=true,trim=0 0 0 9mm]{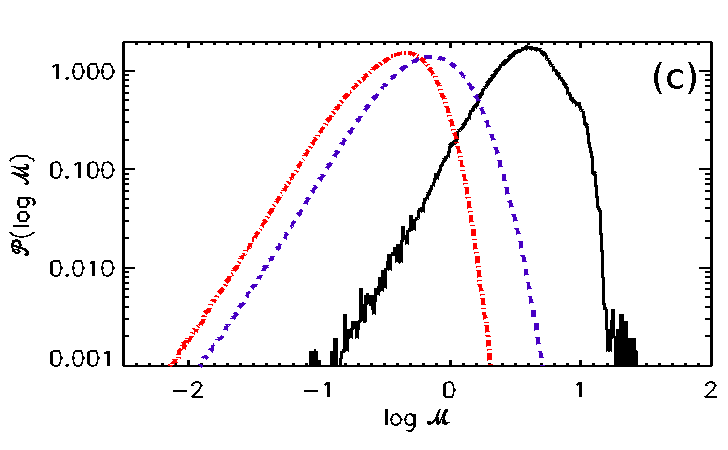}
  \hspace{-2.5cm}\\
  \hspace{-2.75cm}
  \includegraphics[width=0.535\textwidth,clip=true,trim=0 0 0 9mm]{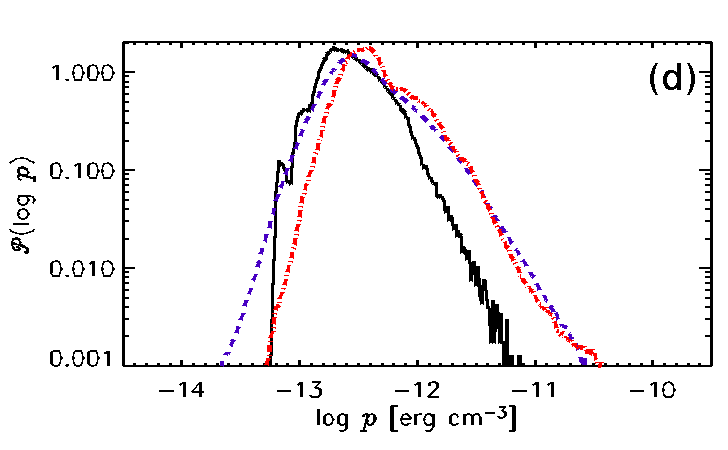}
  \includegraphics[width=0.535\textwidth,clip=true,trim=0 0 0 9mm]{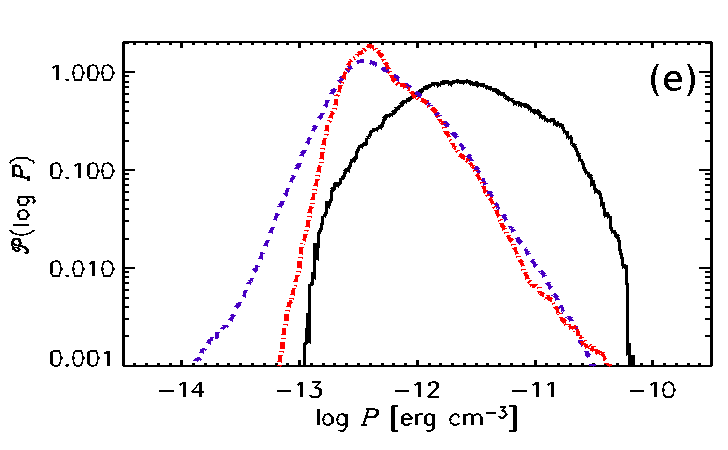}
  \hspace{-1.5cm}
\caption[Probability phase distributions Model~\Op]{Total volume probability distribution of
\textbf{(a)}~density,
\textbf{(b)}~turbulent velocity, $\vect{u}\turb$
\textbf{(c)}~Mach number (defined with respect to the local speed of sound 
and $\vect{u}\turb$),
\textbf{(d)}~thermal pressure, and
\textbf{(e)}~total pressure,
for each %the individual phases
phase of {Model~\Op}, using 
eleven snapshots spanning $t=634$ to $644\Myr$.
The phases are: cold $T<500\K$ (black, solid line), 
warm $500 \leq T < 5\times10^5\K$ (blue, dashed), and hot $T\geq5\times10^5\K$
 (red, dash-dotted).
\label{fig:pdf3ph}
\label{fig:apdf3ph}}
\end{figure}
%-----------------------------------------------------------------------------

  The overlap in the gas density distributions (Fig.~\ref{fig:pdf3ph}a) 
  is small.
  The modal densities that typify each of the hot, warm and cold gas are
  $10^{-3}$, $10^{-1}$ and $10\cmcube$, respectively.
  Cold, dense clouds are formed through radiative cooling facilitated by 
  compression; the latter, however, is truncated at the grid scale of 4\,pc,
  preventing compression to the higher densities in excess of about
  $10^2\cmcube$.
  From the comparison with the higher resolution Model~\OpH\ in 
  Fig.~\ref{fig:pdf2}, it is
  evident that the separation into phases is more distinct with increased
  numerical
  resolution and that the anti-correlation between temperature and density is
  stronger.
  Hence the distinction between the density distributions evident in 
  Fig.~\ref{fig:pdf3ph}a, is even more pronounced when applied to high
  resolution.

  The velocity probability distributions in Fig.~\ref{fig:pdf3ph}b reveal a clear
  connection between the magnitude of the turbulent velocity of gas and its
  temperature: the r.m.s.\ velocity in each phase scales with its speed of sound.
  This is confirmed by the Mach number distributions in Fig.~\ref{fig:pdf3ph}c: 
  both warm and hot phases are transonic with respect to their sound speeds. The 
  cold gas is mostly supersonic, having speeds typically under $10\kms$. 
  The double peak in velocity (Fig.~\ref{fig:pdf3ph}b) is a robust 
  feature, not dependent on the temperature boundary. 
  This likely includes ballistic gas in SN remnants, as well as bulk transport
  by ambient gas at subsonic or transonic speed with respect to the warm gas.
  
  Probability densities of thermal pressure, shown in Fig.~\ref{fig:pdf3ph}d, 
  are notable for their relatively narrow spread: one order of magnitude, 
  compared to a spread of six orders of magnitude in gas density.
  Moreover, the three phases have overlapping distributions, suggesting that
  the system is in statistical thermal pressure balance.
  However, thermal pressure is not the only part of the total pressure in the 
  gas.
  In the Galaxy, contributions to the pressure also arise from the turbulent
  pressure associated with the random motions and the cosmic ray and magnetic
  pressures. 
  The turbulent pressure, $p\turb=\frac{1}{3}\rho u^2\turb$
  is included in $P$ for Fig.~\ref{fig:pdf3ph}e. 
  The random motions $|\vect{u}\turb|^2$ are derived by subtracting the mean
  flows 
  $\average{|\vect{u}|^2}_\ell$ from the total $|\vect{u}|^2$
  (see Chapter~\ref{chap:meanB} for explanation with respect to $\vect{B}$).
  In Model~\Op\ the cosmic ray and magnetic pressures are absent.

  The total pressure distribution for the hot gas differs little
  from the thermal pressure.
  The modal warm pressure increases only by about 25\% at around
  $3\times10^{-13}\dyn\cm^{-2}$.
  For the cold gas the modal pressure increases about ten fold from 
  $2\times10^{-13}\dyn\cm^{-2}$ to $2\times10^{-12}\dyn\cm^{-2}$. 
  The cold gas appears over pressured as a result.
  It becomes apparent (cf. below Fig.~\ref{fig:pall4fits}) that this is due to
  the vertical pressure gradient.
  All the cold gas occupies the higher pressure mid-plane, while the warm and 
  hot gas distributions mainly include lower pressure regions away from the
  disc. 

  Theoretically, it might be expected that the densities of a gas subjected
  to multiple shocks may adopt a lognormal distribution rather than a 
  normal distribution \citep{KP09}.
  The probability distributions for density in Fig.~\ref{fig:pdf3ph}a can be 
  reasonably approximated by lognormal distributions, of the form
  \begin{equation}\label{ln}
  \mathcal{P}(n)=\Lambda(\mu_n,s_n)\equiv\frac{1}{n s_{n}\sqrt{2\upi}}
        \exp\left({\displaystyle-\frac{(\ln n-\mu_{n})^2}{2 s_{n}^2}}\right).
  \end{equation}
  
  The quality of the fits is good, but less so for the hot gas. 
  This is improved when the hot gas is subdivided into that near the 
  mid-plane ($|z|\le200\p$) and that at greater heights ($|z|>200\p$); the 
  former dominated by very hot gas in the interior of SN remnants, the latter
  predominantly more diffuse gas in the halo.
  These fits are illustrated in Fig.~\ref{fig:all4fits}, using 500 data bins in
  the range $10^{-4.8}<n<10^{2.5}\cm^{-3}$, where the warm gas has also been
  subdivided by height; the best-fit parameters are given in Table~\ref{phases}.

%-----------------------------------------------------------------------------
  \begin{figure}[h]\vspace{0.2cm}
  \centering
    \includegraphics[width=0.8\columnwidth,clip=true,trim=0 0 0 9mm]{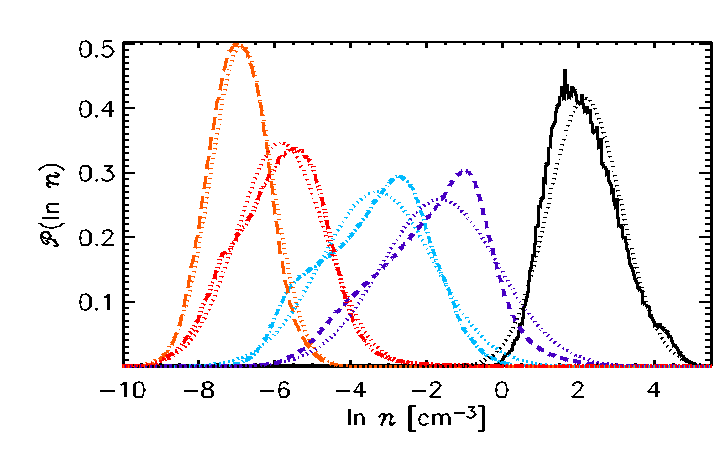} 
    \caption[Lognormal fit to $n$ probability distributions Model~\Op]{
  Density probability distributions for model~{\Op}, together with the best-fit
  lognormal distributions (dotted) for the cold (black, solid), warm and hot gas.
  The warm/hot gas have been divided into regions $|z|\le200\p$ 
  (blue/red, dashed/dash-3dotted) and $|z|>200\p$ 
  (light blue/orange, dash-dotted/long dashed) respectively.
    \label{fig:all4fits}
            }
  \end{figure}
%-----------------------------------------------------------------------------

  The two types of hot gas have rather different density distributions, and
  separating them in this way significantly improves the quality of the
  lognormal fits.
  The warm gas distributions remain very similar, with identical variance but  
  shifted to lower density above $|z|=200\p$.
  It therefore appears justified to consider the two hot gas components 
  separately, where it
  appears strongly affected by its proximity to the SN activity.
  The structure of the warm gas appears independent of the SN activity, and
  merely exhibits a dependence on the global density gradient, so it make sense
  to consider the warm gas as a single distribution.
  All distributions are fitted to lognormals at or above the 95\% level of
  significance applying the Kolmogorov-Smirnov test.
  Although the warm gas density is well described by a lognormal distribution, 
  there is the appearance of power law behaviour (Fig.~\ref{fig:all4fits}) in 
  its
  low density tail.

  Probability distributions of the pressure, displayed in Fig.~\ref{fig:pall4fits}, show 
  that although the thermal pressure of the cold gas near the mid-plane
  is out of equilibrium with the other phases the total pressures 
  are all much closer to equilibrium.
  The gas at $|z|>200\p$ (dotted lines) are in both thermal and total pressure
  balance.
  Thus by including the effect of the global pressure gradient, there is even 
  stronger evidence to support the concept of pressure equilibrium between the
  phases.

%-----------------------------------------------------------------------------
  \begin{figure}[h]
  \centering
  \hspace{-1.5cm}
  \includegraphics[width=0.535\columnwidth,clip=true,trim=0 0 0 9mm]{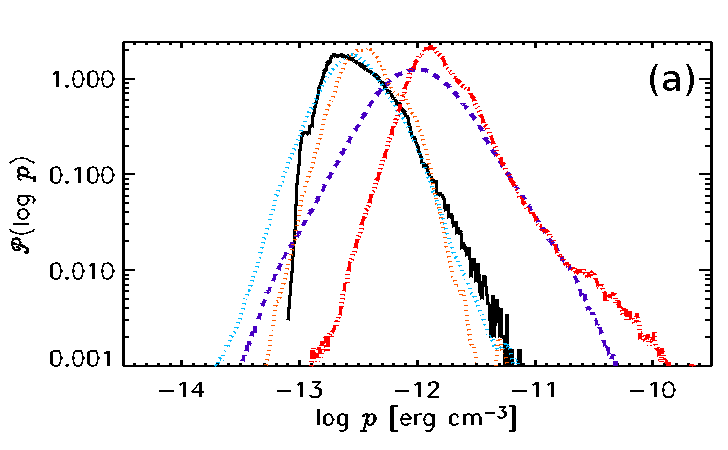}
  \hspace{-0.2cm}
  \includegraphics[width=0.535\columnwidth,clip=true,trim=0 0 0 9mm]{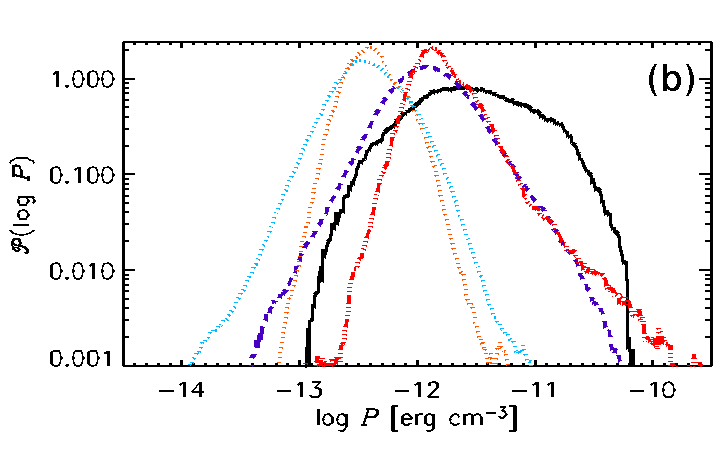}
  \hspace{-1.0cm}
    \caption[Probability distributions for pressure of Model~\Op]{
  Probability distributions for \textbf{(a)}~thermal pressure $p$ and
  \textbf{(b)}~total pressure $P$ in Model~{\Op}, separately for the cold
  (black, solid), warm and hot gas.
  The warm/hot gas have been divided into regions $|z|\le200\p$ 
  (blue/red, dashed/dash-3dotted) and $|z|>200\p$ 
  (light blue/orange, dash-dotted/long dashed) respectively.
    \label{fig:pall4fits}
           }
  \end{figure}
%-----------------------------------------------------------------------------

  The broad lognormal distribution of $P$ for the cold gas is consistent with 
  multiple rarefaction and compression due to shocks.
  The hot and warm gas pressure distributions however appear to follow a power
  law.
  Away from the mid-plane their pressure distributions have a similar shape, but
  slightly narrower.
  The density distribution for the hot gas is also narrower for $|z|>200\p$, 
  although not so the warm gas. 
  This indicates that the gas away from the SN active mid-plane is, 
  understandably, more homogeneous. 
  
  The time-averaged vertical density profiles obtained under the different
  numerical resolutions are shown in Fig.~\ref{fig:zrho_rho}.
  Although the density distribution in Fig.~\ref{fig:pdf2}a reveals higher 
  density contrasts with increased resolution, 
  there is little difference in the $z$-profiles of the models. 
  The mean gas number density at the mid-plane $n(0)$, which with the coarse
  grid resolution excludes the contribution from \HII, is about $2.2\cmcube$,
  double the observation estimates summarised in \citet{F01}.
  This might be expected in the absence of the magnetic and cosmic ray components
  of the ISM pressure, to support the gas against the gravitational force.

%-----------------------------------------------------------------------------
  \begin{figure}[h]
  \centering\hspace{-2cm}
  \includegraphics[width=0.535\columnwidth]{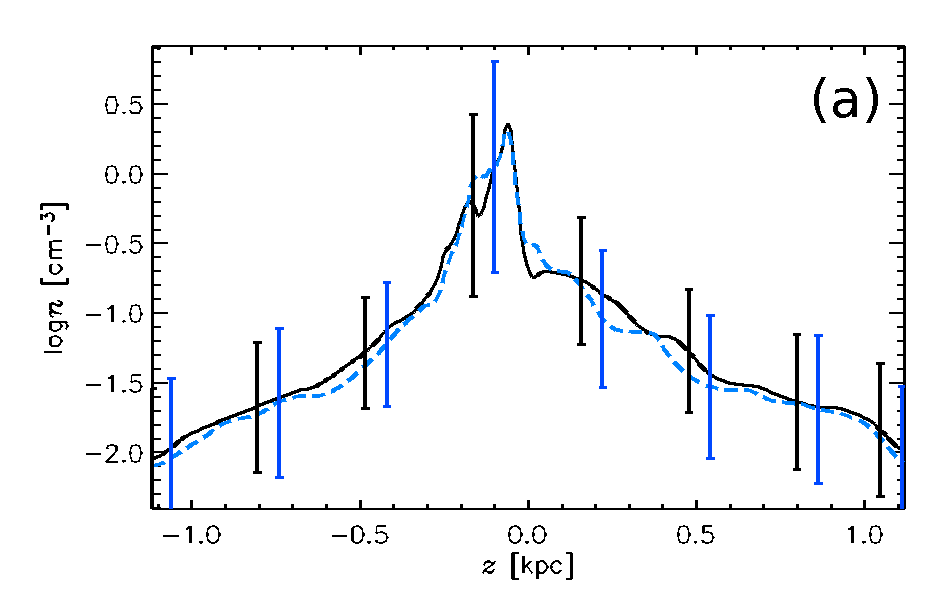}
  \includegraphics[width=0.535\columnwidth]{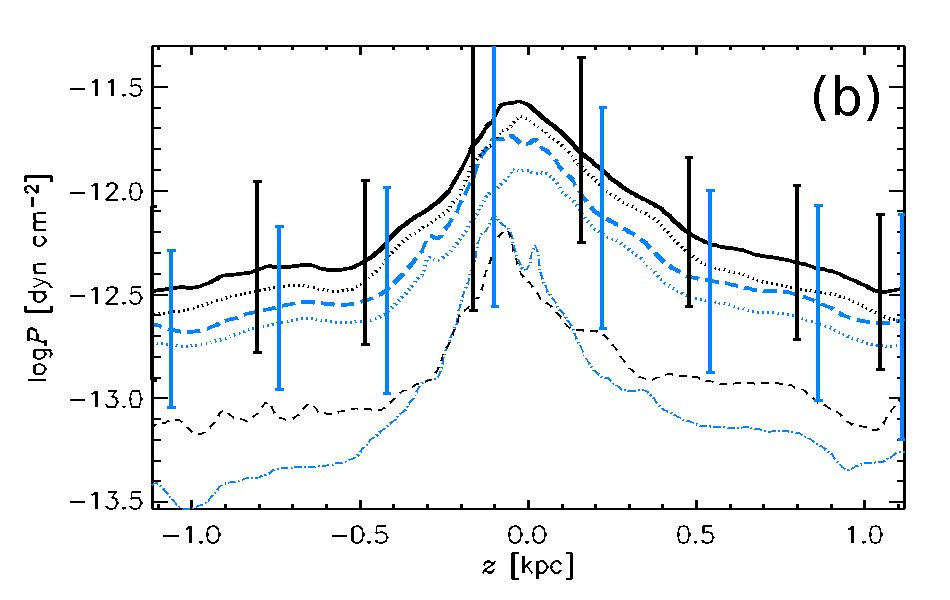}\hspace{-2cm}
    \caption[Horizontal averages of density and pressure of Model~\Op]{
  Horizontal averages of gas number density, $\mean{n}(z)$ {\textbf{(a)}}, and
  total pressure, $\mean{P}(z)$ {\textbf{(b)}}, for Model~{\Op} (solid, black),
  and Model~{\OpH} (dashed, blue). 
  Each are time-averaged using six and ten snapshots respectively, spanning 
  633 to 638\Myr.
  The vertical lines indicate standard deviation within each horizontal slice.
  The thermal $\mean{p}(z)$ (dotted) and ram $\mean{p}\turb(z)$ (fine dashed)
  pressures are also plotted {\textbf{(b)}}.
    \label{fig:zrho_rho}
            }
  \end{figure}

%-----------------------------------------------------------------------------

  However the vertical pressure distributions are consistent with the models of
  \citet[][their Fig.~1 and 2]{BC90}, which include the weight of the ISM up to
  $|z|=5\kpc$.
  The total pressure $P(0)\simeq2.5~(2.0)\times10^{-12}\dyn\cm^{-2}$ for the
  standard~(high) resolution model is slightly above their estimate of about 1.9 
  for hot, turbulent gas.  
  For the turbulent pressure alone 
  $p\turb(0)\simeq6.3~(7.9)\times10^{-13}\dyn\cm^{-2}$ falling to $1.0~(0.6)$ for
  $|z|=500\p$ and then reasonably level. 
  Together with the narrower pressure and density distributions for gas at
  $|z|>200\p$, discussed above, this indicates that the pressure gradient above 
  $500\p$ is weak and this gas remains relatively homogeneous in comparison 
  to gas near the mid-plane.
  The pressures are generally slightly reduced with increased resolution, except
  for $p\turb$ near the mid-plane. 
  Small scales are better resolved, so the turbulent structures are a stronger
  component of the SN active region.
  These pressures are consistent with \citet{BC90} even though our model does
  not explicitly include the pressure contributions from the ISM above $1\kpc$.

  Comparing the thermal pressure distribution (Fig.~\ref{fig:pdf2}c) with 
  \citet[][their Fig.~4a]{AB04} and \citet[][their Fig.~2]{Joung09} the peaks
  are at 3.16, 1.3 and $4.1\times10^{-13}\dyn\cm^{-2}$, respectively. 
  The model of \citet{AB04} include the gas up to $|z|=10\kpc$ and resolution
  up to 1.25\p, and has the lowest modal pressure.
  \citet{Joung09} model up to $|z|=10\kpc$, but include only gas within $125\p$
  of the mid-plane for the pressure distribution, and have the highest modal 
  value.
  Data in Fig.~\ref{fig:pdf2}c is for the total volume within $|z|=1\kpc$, 
  having a modal pressure between the other two models.

  It is reasonable to conclude that the main effects of the increased resolution
  are confined to the very hot interiors and to the thin shells of SN remnants;
  the interiors become hotter and the SN shell shocks become thinner with 
  increased resolution (see Appendix~\ref{chap:EISNR}). 
  Simultaneously, the higher density of the shocked gas enhances cooling, 
  producing more cold gas and reducing the total thermal energy.
  Otherwise, the overall structure of the diffuse gas is little affected: the
  probability distributions of thermal pressure are almost indistinguishable,
  with the standard resolution having a fractionally higher pressure 
  (Fig.~\ref{fig:pdf2}c).
  We can conclude that the numerical resolution of the reference model, $\Delta=4\p$, is sufficient
  to model the diffuse gas phases reliably.
  This choice of the working numerical resolution is further informed by tests
  described  in Appendix~\ref{chap:EISNR}.
   
  To summarise: the system is close to a state of statistical pressure
  equilibrium, with the total pressure having similar values and similar probability 
  distributions in each phase.
  \citet{Joung09} also conclude from their simulations that the gas is in both
  thermal and total pressure balance.
  This could be expected, since the only significant deviation in the
  statistical dynamic equilibrium of the system is the vertical outflow 
  of the hot gas and entrained warm clouds (see Section~\ref{subsect:GO}).

%-------------------------------------------------------------------------
  \section{The filling factor and fractional volume\label{sect:FF}}
%----------------------------------------------------------------------

%-----------------------------------------------------------------------------
\subsection{Filling factors: basic concepts\label{subsect:FFbc}}
%-----------------------------------------------------------------------------

  The fractional volume of the ISM occupied by the phase $i$ is given by
  \begin{equation}\label{fv}
    f_{V,i}=\frac{V_i}{V}
  \end{equation}
  where $V_i$ is the volume occupied by gas in the temperature range defining
  phase~$i$ and $V$ is the total volume. How the gas is distributed
  \emph{within} a particular phase is described by the 
  \emph{phase filling factor}
  \begin{equation}\label{phin}
    \phi_{i}=\frac{\mean{n_i}^2}{\mean{n_i^2}} \; ,
  \end{equation}
  where the over bar denotes a \emph{phase average\/}, i.e., an average only
  taken over the volume occupied by the phase $i$. $\phi_{i}$ describes whether
  the gas density of a phase is homogeneous $\phi_{i}=1$ or clumpy $\phi_{i}<1$.
  Both of these quantities are clearly important parameters of the ISM, allowing
  one to characterise, as a function of position, both the relative distribution
  of the phases and their internal structure.
  As discussed below, the phase filling factor is also directly related to the
  idea of an ensemble average, an important concept in the theory of random
  functions and so $\phi_{i}$ provides a useful connection between turbulence
  theory and the astrophysics of the ISM. Both $f_{V,i}$ and $\phi_{i}$ are easy
  to calculate in a simulated ISM by simply counting mesh-points.
  
  In the real ISM, however, neither $f_{V}$ nor $\phi_{i}$ can be directly
  measured.
  Instead the \emph{volume filling factor} can be derived 
  \citep[][]{Re77, Kulkarni88, Re91},
  \begin{equation}\label{ftilde}
    \tphi_{i}=\frac{\average{n_i}^2}{\average{n_i^2}},
  \end{equation}
  for a given phase $i$, where the angular brackets denote a 
  \emph{volume average\/}, i.e., taken over the total volume.
  Most observational work in this area to-date has concentrated on the diffuse
  ionized gas (or warm ionized medium) since the emission measure of the free
  electrons $\mathrm{EM}\propto n_e^2$ and the dispersion measure of pulsars
  $\mathrm{DM}\propto n_e$, allowing $\tphi$ to be estimated along many 
  lines-of-sight \citep[e.g.][]{Re77, Kulkarni88, Re91, BMM06, Hill08, Gaensler08}. 
  
  In terms of the volume $V_i$ occupied by phase $i$,
  \begin{equation}\label{phase_averaging}
    \mean{n_{i}}=\frac{1}{V_{i}} \int_{V_{i}} n_{i} \, dV ,
  \end{equation}
  whilst
  \begin{equation}\label{volume_averaging}
    \average{n_{i}}=\frac{1}{V} \int_{V} n_{i} \, dV =
    \frac{1}{V} \int_{V_i} n_{i} \, dV\,,
  \end{equation}
  the final equality holding because $n_i=0$ outside the volume $V_{i}$ by
  definition.
  Since the two types of averages differ only in the volume over which they are
  averaged, they are related by the fractional volume:
  \begin{equation}
    \average{n_i} = \frac{V_i}{V} \mean{n_i} = f_{V,i}\mean{n_i},
  \end{equation}
  and
  \begin{equation}
    \average{n_i^2} = \frac{V_i}{V} \mean{n_i^2} = f_{V,i}\mean{n_i^2}.
  \end{equation}
  Consequently, the \textit{volume filling factor\/} $\tphi_{n,i}$ and the 
  \textit{phase filling factor\/} $\phi_{n,i}$ are similarly related:
  \begin{equation}\label{FV}
    \tphi_{i}=\frac{\average{n_i}^2}{\average{n_i^2}}
          =f_{V,i}\frac{\mean{n_i}^2}{\mean{n_i^2}}=f_{V,i}\phi_{i}.
  \end{equation}
  
  Thus the parameters of most interest, $f_{V,i}$ and $\phi_{n,i}$,
  characterizing the fractional volume and the degree of homogeneity of a phase 
  respectively, are related to the observable quantity $\tphi_{n,i}$ by 
  Eq.~(\ref{FV}).
  This relation is only straightforward when the ISM phase can be assumed to be
  homogeneous or if one has additional statistical knowledge, such as the
  probability density function, of the phase.
  In next sub-section we discuss some simple examples to illustrate how these
  ideas can be applied to the real ISM and also how the properties of our
  simulated ISM compare to earlier observations. 
  
  As with the density filling factors introduced above,
  filling factors of temperature and other variables can be
  defined similarly to Eqs.~(\ref{phin}) and (\ref{ftilde}).
  Thus $\phi_{T,i}={\mean{T_i}^2}/{\mean{T_i^2}}$, etc.

\subsection{Homogeneous-phase and lognormal approximations\label{subsect:FFln}}
%%-----------------------------------------------------------------------

  To clarify the physical significance of the various quantities defined above,
  the relations between them are discussed in more detail.
  Consider Eqs.~(\ref{fv}), (\ref{phin}) and (\ref{ftilde}) for an idealised 
  two-phase system, where each phase is homogeneous.
  (The arguments can be generalised to an arbitrary number of homogeneous 
  phases.)
  This scenario is often visualised in terms of discrete clouds of one phase,
  of constant density and temperature, being embedded within the other phase, of
  different (but also constant) density and temperature.
  The two phases might be, for example, cold clouds in the warm gas or hot
  regions coexisting with the warm phase.
  Let one phase have (constant) gas number density $N_1$ and occupy volume $V_1$,
  and the other $N_2$ and $V_2$, respectively.
  The total volume of the system is $V=V_1+V_2$.
  
  The volume-averaged density of each phase, as required for Eq.~(\ref{ftilde}),
  is given by
  \begin{equation}\label{aven}
  \average{n_i}=\frac{N_i V_i}{V} = f_{V,i} N_i.
  \end{equation}
  where $i=1,2$.
  Similarly, the volume average of the squared density is
  \begin{equation}\label{avensq}
  \average{n_i^2}=\frac{N_i^2V_i}{V} = f_{V,i} N_i^2.
  \end{equation}
  The fractional volume of each phase can then be written as
  \begin{equation}\label{FVhomogen}
  f_{V,i} = \frac{\average{n_i}^2}{\average{n_i^2}}
  =\frac{\average{n_i}}{N_i}= \tphi_{n,i} ,
  \end{equation}
  with $f_{V,1}+f_{V,2}=1$,
  and $\tphi_{n,1}+\tphi_{n,2}=1$.
  The volume-averaged quantities satisfy
  $\average{n}=\average{n_1}+\average{n_2}=f_{V,1}N_1+f_{V,2}N_2$ and
  $\average{n^2}=\average{n_1^2}+\average{n_2^2}=f_{V,1}N_1^2+f_{V,2}N_2^2$,
  with the density variance
  $\sigma^2\equiv \average{n^2}-\average{n}^2 =f_{V,1} f_{V,2} (N_1-N_2)^2$.
  
  In contrast, the phase-averaged density of each phase, as required for 
  Eq.~(\ref{phin}), is simply $\mean{n_{i}}=N_{i}$, and the phase average of the
  squared density is $\mean{n_{i}^{2}}=N_{i}^{2}$, so that the phase filling
  factor is $\phi_{n,i}=1$.
  This ensures that Eq.~(\ref{FV}) is consistent with Eq.~(\ref{FVhomogen}).
  Thus, the phase filling factor is unity \textit{for each phase}
  of a homogeneous-phase medium,
  and these filling factors clearly do not sum to unity
  in the case of multiple phases.
  This filling factor can therefore be used as a measure
  of the homogeneity of the phase
  (with a value of unity corresponding to homogeneity). 
  On the contrary, the fractional volumes must always add up to unity,
  $\sum_i f_{V,i}=1$.
  
  Thus for homogeneous phases,
  the volume filling factor and the fractional volume
  of each phase are identical to each other,
  $\tphi_{n,i}=f_{V,i}$,
  and both sum to unity when considering all phases;
  in contrast,
  the phase-averaged filling factor is unity for each phase,
  $\phi_{n,i}=1$.
  If a given phase occupies the whole volume
  (i.e., we have a single-phase medium),
  then all three quantities are simply unity:
  $\phi_{n,i}=\tphi_{n,i}=f_{V,i}=1$.

  Apart from the limitations arising from the inhomogeneity of the ISM phases,
  an unfortunate feature of the above definition of the volume filling factor
  (\ref{ftilde}), which hampers comparison with theory, is that the averaging
  involved is inconsistent with that used in the theory of random functions.
  In the latter, the calculation of volume (or time) averages is usually
  complicated or impossible and, instead, ensemble averages (over the relevant
  probability distribution functions) are used;
  the ergodicity of the random functions is relied upon to ensure that the two
  averages are identical to each other \citep[Section~3.3 in][]{MY07,TL72}.
  But the volume filling factors above are not compatible with such a
  comparison, as they are based on averaging over the total volume, despite the
  fact that each phase occupies only a fraction of it.
  In contrast, the phase averaging \textit{is} performed only over the volume
  of each phase, and so should correspond better to results from the theory of
  random functions.

  To illustrate this distinction, first note that, for the lognormal
  distribution, $\mathcal{P}(n_i) \sim \Lambda(\mu_{n,i},s_{n,i})$,  
  Eq.~(\ref{ln}), the mean and mean-square densities are given by the following
  phase (`ensemble') averages:
  \begin{equation}\label{on}
  \mean{n_{i}}=e^{\mu_{n,i}+s_{n,i}^2/2},~~
  \sigma_i^2=\mean{(n_i-\mean{n_i})^2}=\mean{n_i}^2\left(e^{s_{n,i}^2}-1\right),
  \end{equation}
  where $\sigma_i^2$ is the density variance around the mean $\mean{n_i}$, so 
  that
  \begin{equation}\label{ophi}
  \phi_{n,i}=\frac{\mean{n_i}^2}{\mean{n_i^2}}=
  \frac{\mean{n_i}^2}{\sigma_i^2+\mean{n_i}^2}=\exp(-s_{n,i}^2).
  \end{equation}
  With the filling factor thus defined, $\phi_{n,i}=1$ \emph{only\/} for a
  homogeneous density distribution, $\sigma_i=0$ (or equivalently, $s_{n,i}=0$).
  This makes it clear that this filling factor, defined in terms of the phase
  average, is quite distinct from the fractional volume, but rather quantifies
  the degree of homogeneity of the gas distribution \emph{within\/} a given 
  phase.
  Both describe distinct characteristics of the multi-phase ISM, and, if 
  properly interpreted, can yield rich information about the structure of the
  ISM.

%-----------------------------------------------------------------
  \subsection{Application to simulations\label{subsect:TAB5}}
%-----------------------------------------------------------------

  The definitions of filling factors and fractional volumes from 
  Equations~\eqref{fv}, \eqref{ftilde} and \eqref{phin} are applied to
  the phases identified in Section~\ref{sect:TMPS} for the HD reference
  model (\Op).
  These are presented as horizontal averages in Fig.~\ref{fig:z3fill}.
  Volumes are considered as discrete $4\p$ thick slices spanning the full
  horizontal area. 
  The cold, warm and hot phases are $T<5\times10^2\K$, 
  $5\times10^2\K<T<5\times10^5\K$ and $T>5\times10^5\K$, respectively.

  % -----------------------------------------------------------------------------
  \begin{figure}[h]
  \centering
  \hspace{-1.75cm}
  \includegraphics[width=0.545\columnwidth]{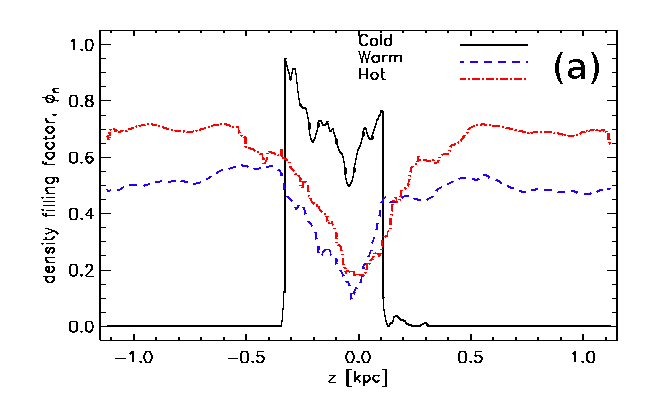}
  \hspace{-0.5cm}
  \includegraphics[width=0.545\columnwidth]{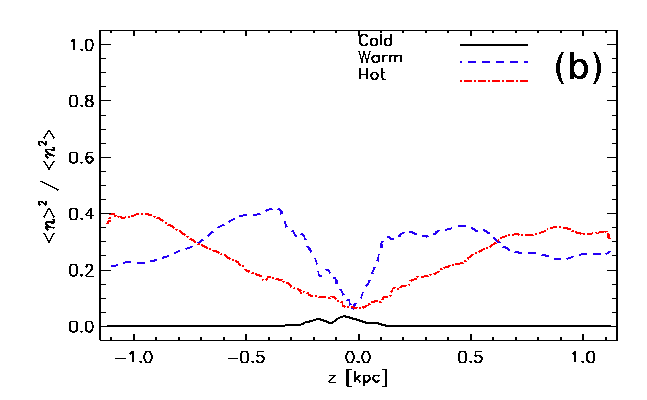}
  \hspace{-1.2cm}\\
  \includegraphics[width=0.545\columnwidth]{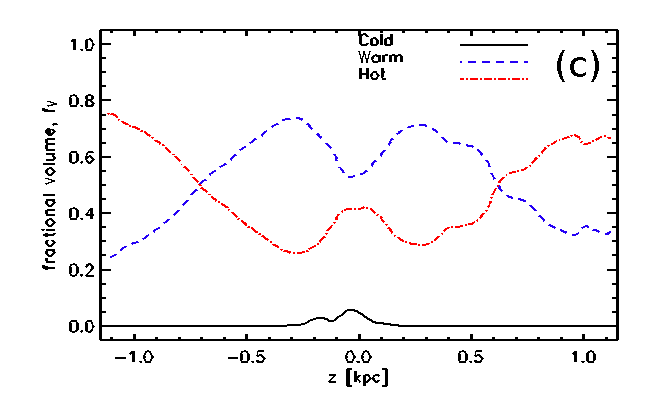}
    \caption[Filling factors for Model~\Op\ as function of $z$]{
  Horizontal averages for cold (black, solid line), warm (blue, dashed) and
  hot  (red, dash-dotted) 
  of the gas density
  \textbf{(a)}~phase-averaged filling factors $\phi_n$,
  \textbf{(b)}~volume-averaged filling factors $\tphi_n$ and
  \textbf{(c)}~fractional volumes $f_{V}$.
  Results plotted are from 21 snapshots in the interval $636$ to $646\Myr$ for
  Model~\Op.
  \label{fig:z3fill}
    \label{fig:3fill}
            }
  \end{figure}
  %-----------------------------------------------------------------------------
  
  The hot gas (Fig.~\ref{fig:3fill}c) accounts for about 60\% of the volume at
  $|z|\simeq1\kpc$ and about 50\% near the mid-plane.
  The local maximum of the fractional volume of the hot 
  gas at $|z|\la200\p$ is due to the highest concentration of SN remnants there.
  A conspicuous contribution to various diagnostics -- especially within $200\p$
  of the mid-plane, where most of the SNe are localised -- comes from the very 
  hot gas within SN remnants.
  Regarding its contribution to integrated gas parameters, it should perhaps be
  considered as a separate phase.
  
  Now how robust are results based on the assumption that each phase is
  homogeneous?
  How strongly does the inhomogeneity of the ISM phases affect the results?
  Note immediately that, unlike the fractional volumes, the filling factors do
  not add up to unity, $\sum_i \tphi_{n,i}\neq1$, if any phase $i$ is not 
  homogeneous.
  Figure~\ref{fig:3fill}b displays the vertical profiles of the density filling 
  factors $\tphi_{n,i}=\average{n_i}^2/\average{n_i^2}$ computed for each phase
  in the reference model, as in Eq.~(\ref{ftilde}).
  (As these are functions of $z$, the averaging here is in two dimensions, over
  horizontal planes; the distinction between the total plane and that part of 
  the plane occupied by the relevant phase remains relevant, however.)
  This should be compared with the fractional volumes shown in 
  Fig.~\ref{fig:3fill}c.
  For all three phases, the values of $\tphi_{n,i}$ are smaller than $f_{V,i}$
  and the peak near the mid-plane evident for the hot gas in $f_{V,i}$ is
  absent in $\tphi_{n,i}$.
  The volume density filling factors in Fig.~\ref{fig:3fill}a are closest to
  unity for the cold gas (near the mid-plane, where such gas is abundant),
  indicating that this phase is more homogeneous than the other phases.
   
%------------------------------------------------------------------------
\begin{table}
\caption[Filling factors for Model~\Op]{\label{phases}
{\footnotesize{
Statistical parameters of the distribution of gas number density $n$ in various phases
for Model~{\Op}, and their lognormal and homogeneous-phase approximations.
Figures in Part (A) have been calculated directly from a composite of 11 simulation snapshots,
those in (B) and (C) represent the best-fit lognormal and homogeneous-phase approximations
to the data in (A), respectively:
$\mu_{n}$ and $s_{n}$ are defined in Eq.~(\ref{ln});
$\mean{n}$ and $\sigma$ are the mean and standard deviation of $n$;
$\phi_n$, $f_V$ and $\tphi_n$ are the phase filling factor, the 
fractional volume and the volume filling factor, respectively,
as defined in Section~\ref{subsect:FFbc}.
The lower (${\rm Q}_1$) and upper (${\rm Q}_3$)
quartiles and the median of the density distributions are given in the last three columns.
Standard deviation for each mean between the simulation snapshots 
as \% of $\mean{n}$ is shown in brackets.
}}}
%\end{table}
%\begin{table}[h]
{\footnotesize{
\rotatebox{90}{%\noindent 
\begin{tabular}{@{}lccccccccccc@{}}
\vspace*{2.5cm}&&&&&&&&&&&\\
\hline
Phase                   &$\mu_n$           &$s_n$         &$\mean{n}$       &$\sigma$     &$\phi_n$&$f_V$   &$\tphi_n$ &$\displaystyle\frac{\phi_nf_V}{\tphi_n}$ &${\rm{Q}}_1$  & Median    & ${\rm{Q}}_3$\\
                        &$[\ln\cmcube]$    &$[\ln\cmcube]$&$[\!\cmcube]$    &$[\!\cmcube]$&        &        &          &                                         &$[\!\cmcube]$ &$[\!\cmcube]$ &$[\!\cmcube]$\\
\hline                                                                                                                                                                       
\multicolumn{11}{l}{\textbf{(A)} Gas density from                                                                                                                  the                                                                                                                    simulation}\\                                                                                                            
Cold                    &                  &              &$12.6\quad~(9\%)$&$15.90$      &$0.384$ &$0.004$ &$0.0016$    &$1.000$                                  &$3.9$         &$7.1$         &$14.2$\\
Warm                    &                  &              &$0.14\quad~(3\%)$&$0.48$       &$0.080$ &$0.608$ &$0.0486$    &$0.999$                                  &$0.019$       &$0.049$       &$0.16$\\
%Warm ($|z|\le200\p$)    &                  &              &$0.42\quad~(3\%)$&$1.01$       &$0.149$ &$0.118$ &$0.0982$    &$0.179$                                  &$0.059$       &$0.191$       &$0.46$\\
%Warm ($|z|>200\p$)      &                  &              &$0.079\quad(3\%)$&$0.13$       &$0.265$ &$0.490$ &$0.1580$    &$0.821$                                  &$0.012$       &$0.039$       &$0.095$\\
Hot\quad ($|z|\le200\p$)&                  &              &$0.0062~(13\%)$  &$0.019$      &$0.124$ &$0.057$ &$0.0039$    &$0.178$                                  &$0.0013$      &$0.0032$      &$0.0066$\\
Hot\quad ($|z|>200\p$)  &                  &              &$0.0013~(10\%)$  &$0.0016$     &$0.457$ &$0.331$ &$0.1850$    &$0.819$                                  &$0.00057$     &$0.00097$     &$0.0017$\\[3pt]
\multicolumn{11}{l}{\textbf{(B)} The
lognormal approximation
(\ref{ln}) to the gas 
density probability 
distribution in each phase}\\
Cold                    &$\phantom{-}2.02$ &$0.92$        &$11.5$           &$13.3$       &$0.43$  &$0.004$ &$0.0016^*$  &$1$                                      &$4.1$         &$7.5$         &$13.9$\\
Warm                    &$-3.03$           &$1.47$        &$0.14$           &$0.39$       &$0.115$ &$0.422$ &$0.0486^*$  &$1$                                      &$0.017$       &$0.048$       &$0.14$\\
%Warm ($|z|\le200\p$)    &$-1.64$           &$1.47$        &$0.57$           &$1.58$       &$0.115$ &$0.153$ &$0.0982^*$  &$0.179$                                  &$0.072$       &$0.048$       &$0.52$\\
%Warm ($|z|>200\p$)      &$-3.29$           &$1.47$        &$0.11$           &$0.30$       &$0.115$ &$1.128$ &$0.1580^*$  &$0.821$                                  &$0.014$       &$0.037$       &$0.100$\\
Hot\quad ($|z|\le200\p$)&$-5.78$           &$1.20$        &$0.0063$         &$0.011$      &$0.24$  &$0.003$ &$0.0039^*$  &$0.179$                                  &$0.0014$      &$0.0031$      &$0.0069$\\
Hot\quad ($|z|>200\p$)  &$-6.96$           &$0.77$        &$0.0013$         &$0.0011$     &$0.55$  &$0.276$ &$0.1850^*$  &$0.821$                                  &$0.00056$    &$0.00094$    &$0.0016$\\[3pt]
\multicolumn{11}{l}{\textbf{(C)} The homogeneous                                                                                                                           phase                                                                                                                   approximation}\\                                                                                                        
Cold                    &                  &              &$12.6$           &$0$          &$1$     &$0.002$ &$0.0016^*$  &$1$                                      &$12.6$        &$12.6$        &$12.6$\\
Warm                    &                  &              &$0.14$           &$0$          &$1$     &$0.049$ &$0.0486^*$  &$1$                                      &$0.14$        &$0.14$        &$0.14$\\
%Warm ($|z|\le200\p$)    &                  &              &$0.42$           &$0$          &$1$     &$0.017$ &$0.0982^*$  &$0.179$                                  &$0.42$        &$0.42$        &$0.42$\\
%Warm ($|z|>200\p$)      &                  &              &$0.079$          &$0$          &$1$     &$0.130$ &$0.1580^*$  &$0.821$                                  &$0.079$       &$0.079$        &$0.079$\\
Hot\quad ($|z|\le200\p$)&                  &              &$0.0062$         &$0$          &$1$     &$0.007$ &$0.0039^*$  &$0.179$                                  &$0.0062$      &$0.0062$      &$0.0062$\\
Hot\quad ($|z|>200\p$)  &                  &              &$0.0013$         &$0$          &$1$     &$0.152$ &$0.1850^*$  &$0.821$                                  &$0.0013$      &$0.0013$      &$0.0013$\\[3pt]
\hline\\
\multicolumn{11}{l} {$^*$~indicates a fixed value, taken from part (A)}
\end{tabular}
}}}
\end{table}

%-----------------------------------------------------------------------------

  Table~\ref{phases} illustrates the meaning and significance of the quantities
  introduced above.
  The first part, (A), presents results for the actual density distribution from
  Model~{\Op}.
  The volume filling factors $\Phi_n$ for the hot gas have been adjusted for
  the whole volume, since they are calculated over only 0.2 and 0.8 of the
  total volume, respectively.
  Thus, the hot gas at $|z|\leq0.2\kpc$ ($|z|>0.2\kpc$) occupies $0.275$
  ($0.410$) of that volume, but $0.055$ ($0.331$) of the total volume. 
  
  Part~(A) of the table allows confirmation, by direct calculation of the the
  quantities involved, that Eq.~(\ref{FV}) is satisfied to high accuracy, i.e.,
  $\phi_n f_V/\tphi_n=1$ for each phase.
  Because the statistical parameters involved are averaged over time, this
  relation is not exact, but is satisfied exactly for each snapshot.
  Similarly, $\phi_n=\mean{n}^2(\sigma^2+\mean{n}^2)$ only approximately, 
  whereas the equality is satisfied very accurately for each snapshot.
  
  Part~(B) of the table provides the parameters, $\mu_n$ and $s_n$, of the
  best-fit lognormal approximations to the probability distributions of the gas
  density shown in Fig.~\ref{fig:all4fits}.
  Using Eq.~\eqref{on}, $\mean{n}$ and $\sigma$ are derived and then 
  Eq.~\eqref{ophi} is used to determine each $\phi_n$.
  $f_V$ is taken directly from the simulation data, i.e. from Part~(A), and
  Eq.~\eqref{FV} is used to calculate $\tphi_n$. 
  
  The accuracy of the lognormal approximation to the density PDFs is 
  characterized by the values of the mean
  density $\mean{n}$, its standard deviation $\sigma$ and the two filling
  factors, as compared to the corresponding quantities in (A). 
  The approximation is quite accurate for the mean density and $\sigma$. 
  The median (Q$_2$) and the lower and upper quartiles (Q$_1$ and Q$_3$) of the
  density distribution, are derived
  \[
   {\rm Q}_2=e^{\mu_n},\quad
   {\rm Q}_{1,3}=e^{\mu_n\pm0.67s_n}
  \]
  and shown in the last three columns, are also reasonably consistent between
  Parts~(A) and (B).
  The most significant disparity is apparent in the cold gas, where the skew in
  the real distribution is stronger than evident in the lognormal 
  approximation.
  Even here the differences are modest, and for the warm and hot gas the
  agreement is excellent.
  
  Finally, Part~(C) allows one to assess the consequences of an assumption of 
  homogeneous-phases, where $\phi_n=1$ and $f_V=\tphi_n$ for each
  phase by definition.
  The values of $\tphi_n$ obtained under this approximation are very
  significantly in error (by a factor 3--10). 
  The last two columns of the table suggest the reason for that: perhaps
  unexpectedly, this approximation is strongly biased towards higher densities
  for all phases, except for the cold gas (which occupies negligible volume), 
  so that the gas density within each phase obtained with this approximation is 
  very close to the upper quartile of the probability distribution and thus
  misses significant amounts of a relatively rarefied gas in each phase.

  In conclusion the lognormal approximation to the gas density distribution
  provides much more accurate estimates of the fractional volume than the
  homogeneous-phase approximation, at least for the model ISM. 
  This should be borne in mind when observational data are interpreted in terms
  of filling factors.
  
  Comparing the gas density distribution shown in Fig.~\ref{fig:pdf2}a with
  the higher resolution run Model~\OpH, 
  the mean warm gas density ($0.14\,{\rm cm}^{-3}$) and the minimum in the 
  distribution ($10^{-2}\,{\rm cm}^{-3}$) appears to be independent of the 
  resolution. 
  However the natural log mean $\mu_n\simeq-8$ for the hot gas within and
  without $200\p$ of the mid-plane, but with larger standard deviation for the
  gas near the mid-plane.
  This compares with -6.96 and -5.78 in Table~\ref{phases} for $4\p$ resolution,
  i.e. about 1/3. 
  This reflects the improved resolution of low density in the remnant interior.

%-----------------------------------------------------------------------------
  \subsection{Observational implications\label{subsect:CWO}}
%-----------------------------------------------------------------------------

  Here I consider how these results might be used to interpret observational
  results.
  Observations can be used to estimate the volume-averaged filling factor
  $\tphi_{n,i}$, defined in Eq.~(\ref{ftilde}), for a given ISM phase.
  On its own, this quantity is of limited value in understanding how the phases
  of the ISM are distributed: of more use are the fractional volume occupied by
  the phase $f_{V,i}$, defined in Eq.~(\ref{fv}), and its degree of homogeneity
  which is quantified by $\phi_{n,i}$, defined by Eq.~(\ref{phin}). 
  Knowing $\tphi_{n,i}$ and $\phi_{n,i}$, $f_{V,i}$ follows via Eq.~(\ref{FV}):
  \begin{equation}\label{appl}
    f_{V,i}=\frac{\tphi_{n,i}}{\phi_{n,i}}.
  \end{equation}
  This formula is exact, but its applicability in practise is limited if
  $\phi_{n,i}$ is unknown.
  However $\phi_{n,i}$ can be deduced from the probability distribution of
  $n_i$: for example if the density probability distribution of the phase can
  be approximated by the lognormal, then $\phi_{n,i}$ can be estimated from
  Eq.~\eqref{ophi}. 
  
  \citet{BMM06} and \citet{BM08} estimated $\tphi_{n,\rm DIG}$ for the diffuse
  ionized gas (DIG) in the Milky Way using dispersion measures of pulsars and 
  emission measure maps.
  In particular, \citet{BMM06} obtain $\tphi_{n,\rm DIG}\simeq0.24$ towards
  $|z|=1\kpc$, and \citet{BM08} find the smaller value
  $\tphi_{n,\rm DIG}\simeq0.08$ for a selection of pulsars that are closer to
  the Sun than the sample of \citet{BMM06}. 
  On the other hand, \citet{BF08} and \citet{BF12} used similar data for
  pulsars with known distances to derive PDFs of the distribution of DIG volume
  densities which are well described by a lognormal distribution; the fitted
  lognormals have $s_{\rm DIG}\simeq0.22$ at $|b|>5\deg$
  \citep[Table~1 in ][]{BF12}.
  Using Eqs.~(\ref{ophi}) and (\ref{appl}), this implies that the fractional
  volume of DIG, with allowance for its inhomogeneity, is about
  \[
    f_{V,\rm DIG}\simeq 0.1\mbox{--}0.2.
  \]
  In other words these results imply that the DIG is approximately homogeneous.
  \citet{BF08} also fitted lognormal distributions to the volume densities of
  the warm \HI\ along lines-of-sight to 140 stars (although no filling factors
  could be calculated for this gas): they found $s_\mathrm{HI}\simeq 0.3$,
  again suggesting that this phase of the ISM is approximately homogeneous.
  Corrections for the inhomogeneity in the fractional volume only become
  significant when $s_i\ga0.5$; by a factor of 1.3 for $s_i=0.5$ 
  and a factor of 2 for $s_i=0.8$.

  %-----------------------------------------------------------------------------
\section{Three-phase structure defined using specific entropy\label{sect:entropy}}

%-----------------------------------------------------------------------------
  \begin{figure}
  \centering
  \includegraphics[width=\linewidth]{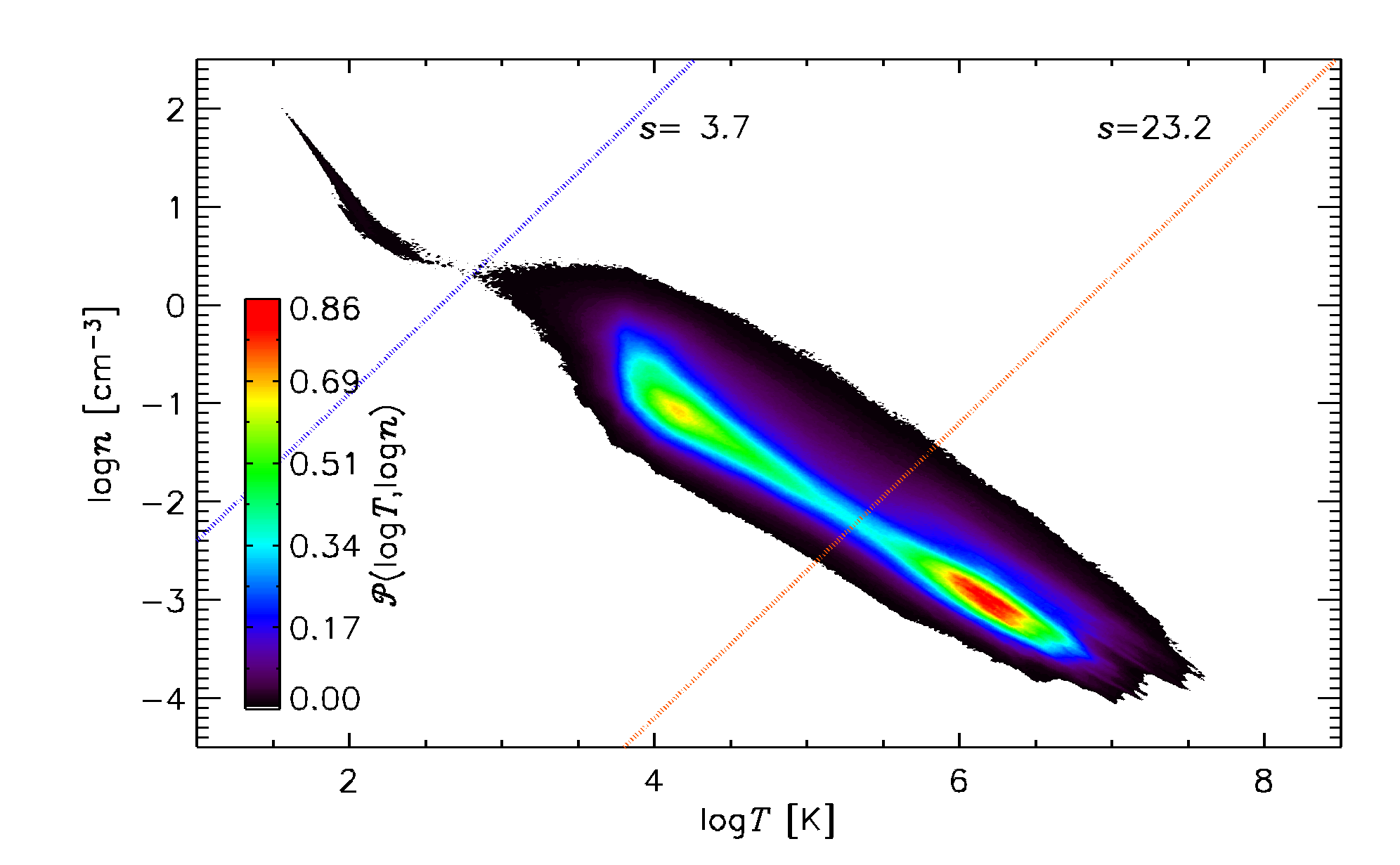}
  \caption[2D $n$ and $T$ distribution for Model~\Op\ over plotted with specific entropy]{
  PDF contour plot by volume of log$n$ vs log$T$ for Model~S$\Omp$, averaged
  over 200\,Myr. 
%  Contours of constant cooling time
%  $\tau_{\rm cool}=10^5,3\cdot10^6$ and $10^8\,{\rm yr}$ are included.
  The lines of constant specific entropy $s=3.7\cdot10^{8}$ 
  and $23.2\cdot10^{8}\erg\g^{-1}\K^{-1}$ indicate where the phases are defined
  as cold for $s\le3.7$ and as hot for $s>23.2$. 
  \label{fig:pdf2ds}}
  \end{figure}
%-----------------------------------------------------------------------------

  In Section~\ref{sect:TMPS} the three phase medium was defined according to
  temperature bands. 
  This definition unfortunately precludes any statistically meaningful 
  description of the temperature distribution of each phase, as these are
  effectively just the total temperature distribution split into three disjoint
  segments.
  It is evident that there is likely to be some overlap in the distribution of   
  temperatures between the phases; considering the similarities between the cold
  phase and colder gas in the warm phase, and between the hot phase and the
  hotter gas in the warm phase, as described in Section~\ref{subsect:multi}.
 
  With reference to Fig.~\ref{fig:pdf2d}, where the probability distribution 
  of $T,n$ is displayed as 2D contours, the gas is spread across regions 
  out of pressure equilibrium about a line of constant thermal pressure, 
  corresponding to $p\simeq k_{\rm B}nT=10^{-12.7}\dyn\cm^{-2}$.
  This is reproduced in Fig.~\ref{fig:pdf2ds}.
  The minima of the probability distribution may be considered to be aligned
  roughly perpendicular to the line of constant pressure, 
  intersecting at $\log T\simeq2.75\K$ and $\log T\simeq 5.25\K$.

  Specific entropy is related to temperature and density by
  \[
    s ={\cv}\left[\ln T-\ln T_0-(\gamma-1)(\ln \rho-\ln \rho_0)\right], 
  \]
  with $\cv$ denoting the specific heat capacity of the gas at constant
  volume and $\gamma=5/3$ the adiabatic index.
  Note that $s=0$ when both $T=T_0$ and $\rho=\rho_0$.
  Lines of constant specific entropy in Fig.~\ref{fig:pdf2ds} take the form 
  $\log n-1.5\log T=$\,constant.
  The dotted lines plotted identify the cold phase as occurring where
  $s<3.7\cdot10^{8}\erg \g^{-1}\K^{-1}$ (blue) and hot occurring where
  $s>23.2\cdot10^{8}\erg\g^{-1}\K^{-1}$ (orange). 
  Although these are not necessarily observable quantities, 
  they can be calculated directly for the simulated data, and the various
  probability distributions determined, including the temperatures.  

%-----------------------------------------------------------------------------
  \begin{figure}[h]
  \centering
  \hspace{-1.35cm}
  \includegraphics[width=0.535\linewidth]{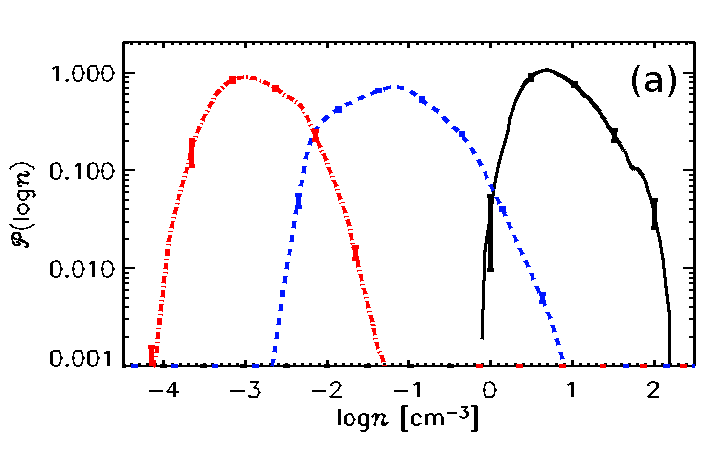}
  \hspace{-0.325cm}
  \includegraphics[width=0.535\linewidth]{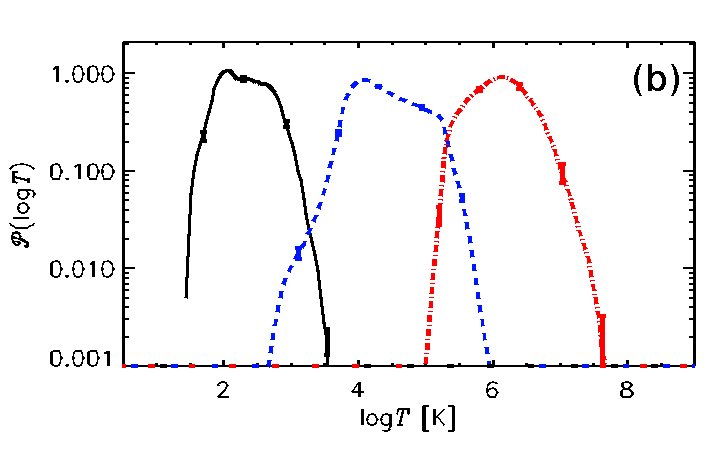}\hspace{-0.05cm}\\
  \hspace{-1.75cm}
  \includegraphics[width=0.535\linewidth]{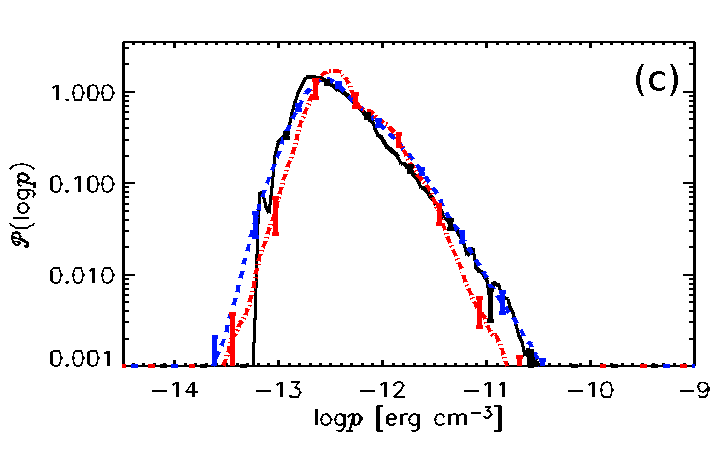}
  \hspace{-0.325cm}
  \includegraphics[width=0.535\linewidth]{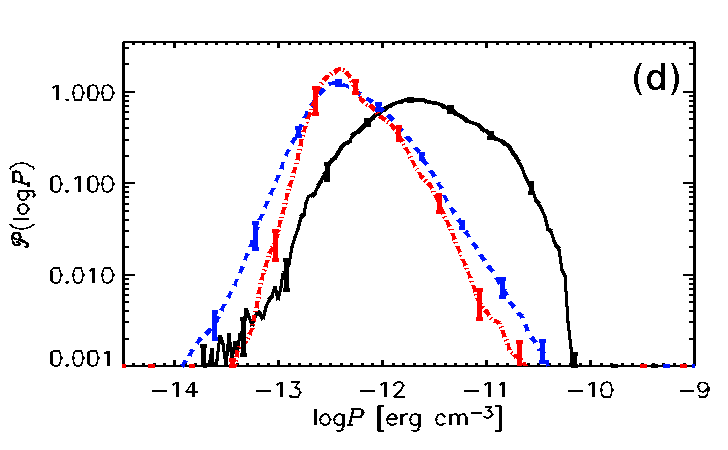}
  \hspace{-0.5cm}
    \caption[Probability distributions by phase for Model~\Op\ via specific entropy]{
  Distributions of entropy defined phases for Model~\Op: cold (black, solid), warm (blue, dashed)
  and hot (red, dash-dotted) for $\rho$, $T$, $p$, and $P$ (Panels a -- d,
  respectively).
  Eleven snapshots are used spanning $t=634$ to $644\Myr$.
  Error bars indicate the 95\% confidence interval accounting for deviations over
  time.   
  The cold phase is defined to exist where specific entropy
  $s<3.7\cdot10^{8}\erg \g^{-1}\K^{-1}$, hot where 
  $s>23.2\cdot10^{8}\erg \g^{-1}\K^{-1}$ and the warm phase in between. 
    \label{fig:epdf3}
            }
  \end{figure}
%-----------------------------------------------------------------------------

  In Fig.~\ref{fig:epdf3} the probability distributions for each phase defined
  by specific entropy from Model~\Op\ are displayed for $\rho$, $T$, $p$ and $P$ in
  panels (a -- d) respectively, where $p$ is thermal pressure and $P$ is total
  pressure, as described in Section~\ref{sect:TMPS}, so that it excludes
  cosmic ray and magnetic pressures, but include turbulent pressure.
  
  The distribution of $n$ for the hot gas is slightly narrower, but the modal
  density remains $10^{-3}\cmcube$.
  The warm gas distribution is however bi-modal with peaks at and more disjoint. 
  The peaks in the distributions are consistent with the definitions in 
  Section~\ref{sect:TMPS} at $10^{-3}, 10^{-1}$  and $10\cmcube$ for 
  hot, warm and cold respectively. 

  From panels (c) and (d) we can see that the peaks of the distributions are
  consistent with the line of constant pressure at $p=10^{-12.7}\erg\cmcube$
  referred to in Fig.~\ref{fig:pdf2ds}. 
  There is little change in the distributions of $p$, compared to 
  Fig.~\ref{fig:pdf3ph}. 
  The differences in $P$ between the cold gas and the rest can be accounted for
  by the vertical pressure gradient as explained with reference to 
  Fig.~\ref{fig:pall4fits}.
 
  With this approach we can obtain credible probability distributions for
  temperature $T$ shown in panel (b).
  These distributions may help to qualify the characteristics of a region of the
  ISM from measurements of temperature, as the means and variances can be well
  defined for each phase and the regions of intersection are small.

%\input{chapters/log2ln} 
  
%-----------------------------------------------------------------------------
  \section{Further results\label{subsect:multi}}
%-----------------------------------------------------------------------------
  
  In this section the distribution of the gas is explored more closely.
  The temperature of the ISM is more finely identified with a set of narrower 
  bands as detailed in Table~\ref{table:bands}. 
  As with the phases in Eq.~\eqref{fv} the fractional volume of each temperature
  range $i$ at a height $z$ can be given by
  \begin{equation}\label{fva}
    f_{V,i}(z)=\frac{V_i(z)}{V(z)}=\frac{N_i(z)}{N(z)},
  \end{equation}
  where $N_i(z)$ is the number of grid points in the temperature range
  $T_{i,\mathrm{min}}\leq T<T_{i,\mathrm{max}}$,
  with $T_{i,\mathrm{min}}$ and
  $T_{i,\mathrm{max}}$
  given in Table~\ref{table:bands}, and $N(z)$ is the total number of grid
  points at that height.
  The fractional mass is similarly calculated as
  \begin{equation}\label{fm}
  f_{M,i}(z)=\frac{M_i(z)}{M(z)},
  \end{equation}
  where $M_{i}(z)$ is the mass of gas within temperature range $i$ at a given
  $z$, and $M(z)$ is the total gas mass at that height.

%-----------------------------------------------------------------------------
  \begin{table}[h]
  \centering
    \caption[Fractional volumes for Model~RBN and \WSWa\ as function of $z$]{
  Key to Figs.~\ref{fig:zfill} and \ref{fig:zfill_RB_WSW},
  defining the gas temperature bands used there.
  \label{table:bands}}
  \begin{tabular}{ccl}
  \hline
  Temperature band   &Line style & Phase \\
  \hline
    \phantom{$5\times10^1\K<$}$T<5\times10^1\K$    &\rule[0.08cm]{1.05cm}{0.75pt}                                   & cold \\
    $5\times10^1\K\leq T<5\times10^2\K$            &{\Large\color{violet}{$\cdot$-$\cdot$-$\cdot$-$\cdot$-$\cdot$}} & cold \\
    $5\times10^2\K\leq T<5\times10^3\K$            &{\Large\color{blue}{- - - - -}}                                 & warm \\
    $5\times10^3\K\leq T<5\times10^4\K$            &{\Large\color{royalblue}{-- -- -- -}}                           & warm \\
    $5\times10^4\K\leq T<5\times10^5\K$            &{\Large\color{mygreen}{--}\color{black}{-}\hspace{-0.08cm} \color{mygreen}{$\cdot$}\hspace{-0.03cm}\color{black}{-}\hspace{-0.08cm} \color{mygreen}{$\cdot$}\hspace{-0.03cm}\color{black}{-}\hspace{-0.08cm}  \color{mygreen}{--}}                  & warm \\
    $5\times10^5\K\leq T<5\times10^6\K$            &{\Large\color{burntorange}{--$\cdots$--$\cdot$}}                & hot  \\
    \phantom{$5\times10^1\K<$}$T\geq5\times10^6\K$ &{\Large\color{red}{$\cdots\cdots$}}                             & hot  \\
  \hline
  \end{tabular}
  \end{table}
%-----------------------------------------------------------------------------
% -----------------------------------------------------------------------------
  \begin{figure}[h]
%  \vspace{-3.0cm}
 \centering
  \hspace{-1.0cm}
  \includegraphics[width=0.535\columnwidth]{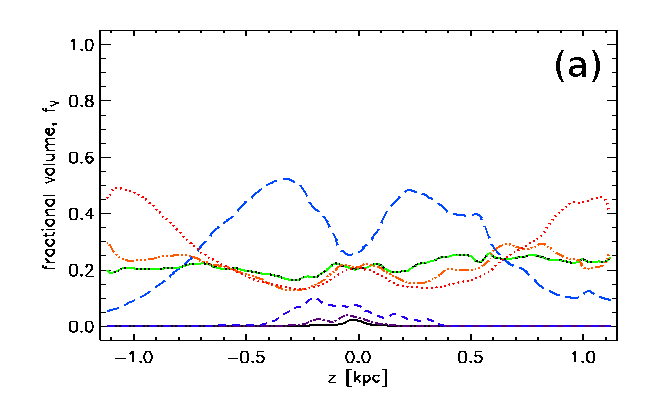}
  \hspace{-0.5cm}
  \includegraphics[width=0.535\columnwidth]{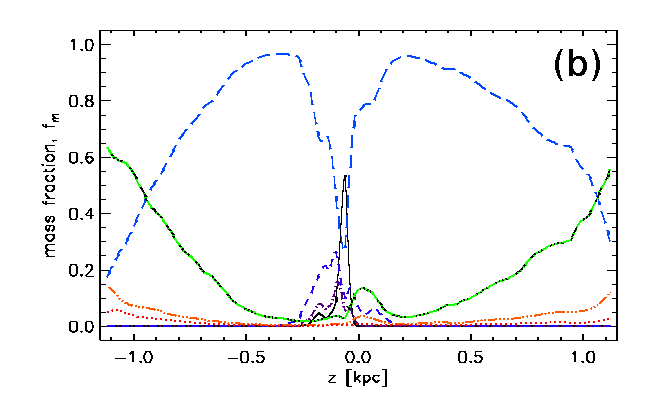}\hspace{-4.0cm}\\
  \hspace{-3.0cm}
  \includegraphics[width=0.535\columnwidth]{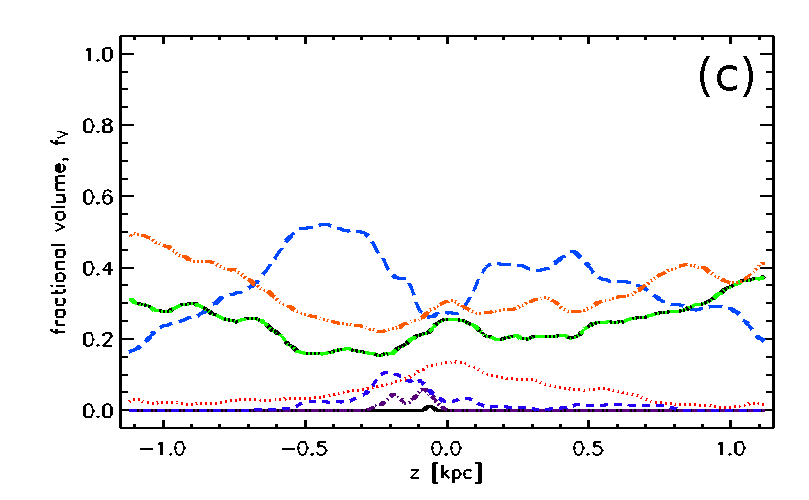}
  \hspace{-0.5cm}
  \includegraphics[width=0.535\columnwidth]{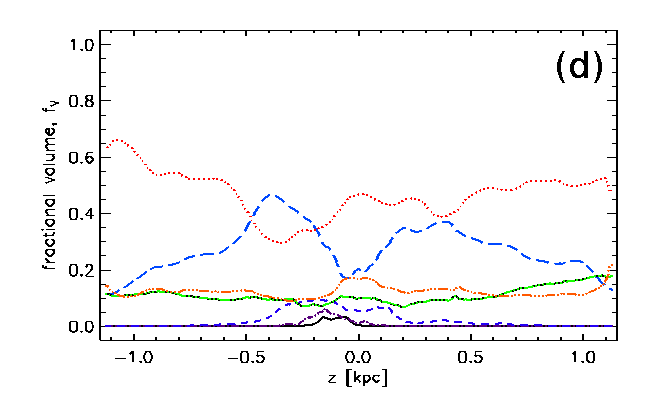}
  \hspace{-2.0cm}
%  \hspace{-3.0cm}\includegraphics[width=0.7\columnwidth]{fig/pqO1_av3fill.png}
  \caption[Volume and mass fractions for Model~\Op\ and \OpH\ -- function of $z$]{
  Vertical profiles of \textbf{(a)}~the fractional volume (Eq.~\ref{fva}) and
  \textbf{(b)}~the fractional mass (Eq.~\ref{fm}) for Model~\Op, calculated for  
  the temperature ranges given, along with the figure legend, in
  Table~\ref{table:bands}.
  Data from 21 snapshots were used spanning $636$ to $646\Myr$.
  Fractional volume \textbf{(c)}~for Model~\Op\ and \textbf{(d)}~for Model~\OpH.
  Direct comparison from 6~(10) snapshots for Model~{\Op}~({\OpH}) in the
  reduced interval $633$ to $638\Myr$.
    \label{fig:zfill_comp}
  \label{fig:zfill}}
  \end{figure}
%-----------------------------------------------------------------------------
  
  Note that the relative abundances of the various phases in these models
  might be affected by the unrealistically high thermal conductivity adopted.
  The coldest gas (black, solid), with $T<50\K$, is largely confined within
  about $200\p$ of the mid-plane.
  Its fractional volume (Fig.~\ref{fig:zfill}a) is small even at the mid-plane,
  but it provides more than half of the gas mass at $z=0$ (Fig.~\ref{fig:zfill}b).
  Gas in the next temperature range, $50<T<500\K$ (purple, dash-dotted), is
  similarly distributed in $z$.
  Models~{\Op} and {\OpH} differ only in their resolution, using 2 and 4\,pc,
  respectively. 
  Model~{\OpH} is a continuation of the state of {\Op} after 600\,Myr of
  evolution.
  With higher resolution the volume fraction of the coldest gas is significantly
  enhanced (Fig.~\ref{fig:zfill}d compared to b), but it is similarly distributed.

  Gas in the range $5\times10^2<T<5\times10^3\K$ (dark blue, dashed)
  has a similar profile to the cold gas for both the fractional mass and the
  fractional volume, and this is insensitive to the model resolution.
  This is identified with the warm phase, but exists in the thermally unstable
  temperature range.
  It accounts for about 10\% by volume and 20\% by mass of the gas near the 
  mid-plane.
  \citet{HT03} estimate at least 48\% of the WNM to be in the thermally
  unstable range $500\K<T<5000\K$, and that the WNM accounts for 50\% of the
  volume filling fraction at $|z|=0$. 
  From Fig.~\ref{fig:zfill}a, assuming the WMN corresponds to the two bands
  within $500\K<T<50\,000\K$, in the model this account for roughly 50\% of the
  total volume fraction for $|z|<400\p$, and about 40\% at $|z|\simeq0$, with
  only a third of this in the unstable range. 
  So the WMN is consistent with observation, although the model may slightly
  under represent the thermally unstable abundance. 
  %, which is consistent with observational evidence.
  It is negligible away from the supernova active regions. 
  
  The two bands with $T>5\times10^5\K$ (red, dotted and orange, dash-3dotted)
  behave similarly to each other, 
  occupying similar fractional volumes for $|z|\la0.75\kpc$, and with $f_{V,i}$ 
  increasing above this height (more rapidly for the hotter gas). 
  In contrast the fractional masses in these temperature bands are negligible
  for $|z|\la0.75\kpc$, and increase above this height (less rapidly for the
  hotter gas).
  The temperature band $5\times10^4<T<5\times10^5\K$ (green/black, dash-3dotted)
  is identified with the warm phase, based on the minimum in the combined 
  density and temperature distribution at $T\simeq5\times10^5\K$ in 
  Fig.~\ref{fig:pdf2d}.
  However as a function of $z$ it is similarly distributed to the hotter gas
  (orange) in all profiles.
  This indicates it is likely a transitional range of mainly hot gas cooling, 
  which accounts for a relatively small mass fraction of the warm gas.
  The dramatic effect of increased resolution (Fig.~\ref{fig:zfill}d compared 
  to c) is the
  significant increase in the very hot gas (red, dotted), particularly 
  displacing the hotter gases (orange and green) but also to some degree the
  warm gas (blue, dashed).
  This reflects the reduced cooling due to the better density contrasts resolved,
  associating the hottest temperatures to the most diffuse gas.   
 
  The middle temperature range $5\times10^3<T<5\times10^4\K$ has a distinctive
  profile in both fractional volume and fractional mass, with minima near the
  mid-plane and maxima at about $|z|\simeq400\p$, being replaced as the 
  dominant component by hotter gas above this height.
  The fractional volume and vertical distribution of this gas is quite 
  insensitive to the resolution.

%-----------------------------------------------------------------------------
  \begin{figure}
  \centering
  \hspace{-2.0cm}
  \includegraphics[width=0.425\linewidth]{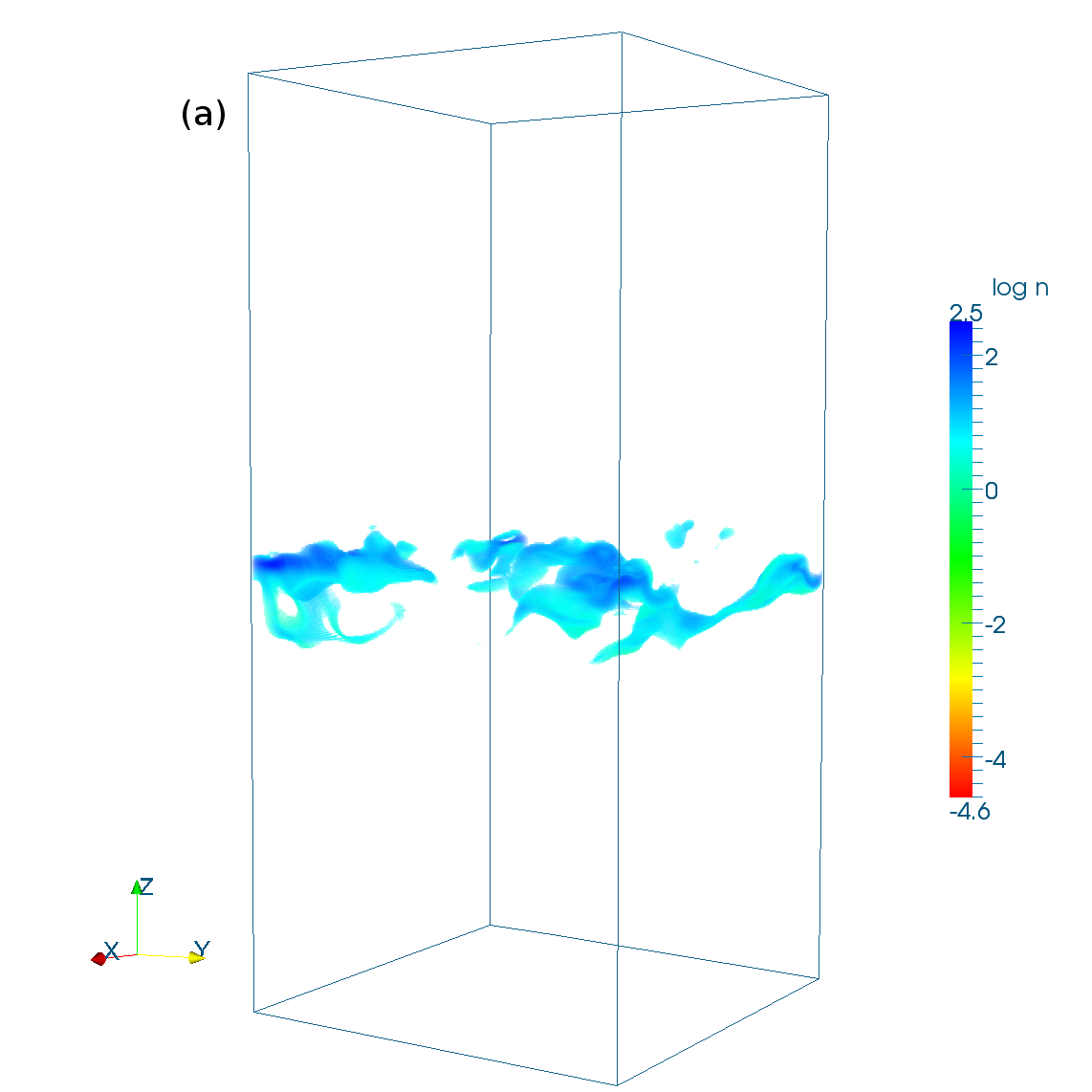}\hspace{-1.3cm}
  \includegraphics[width=0.425\linewidth]{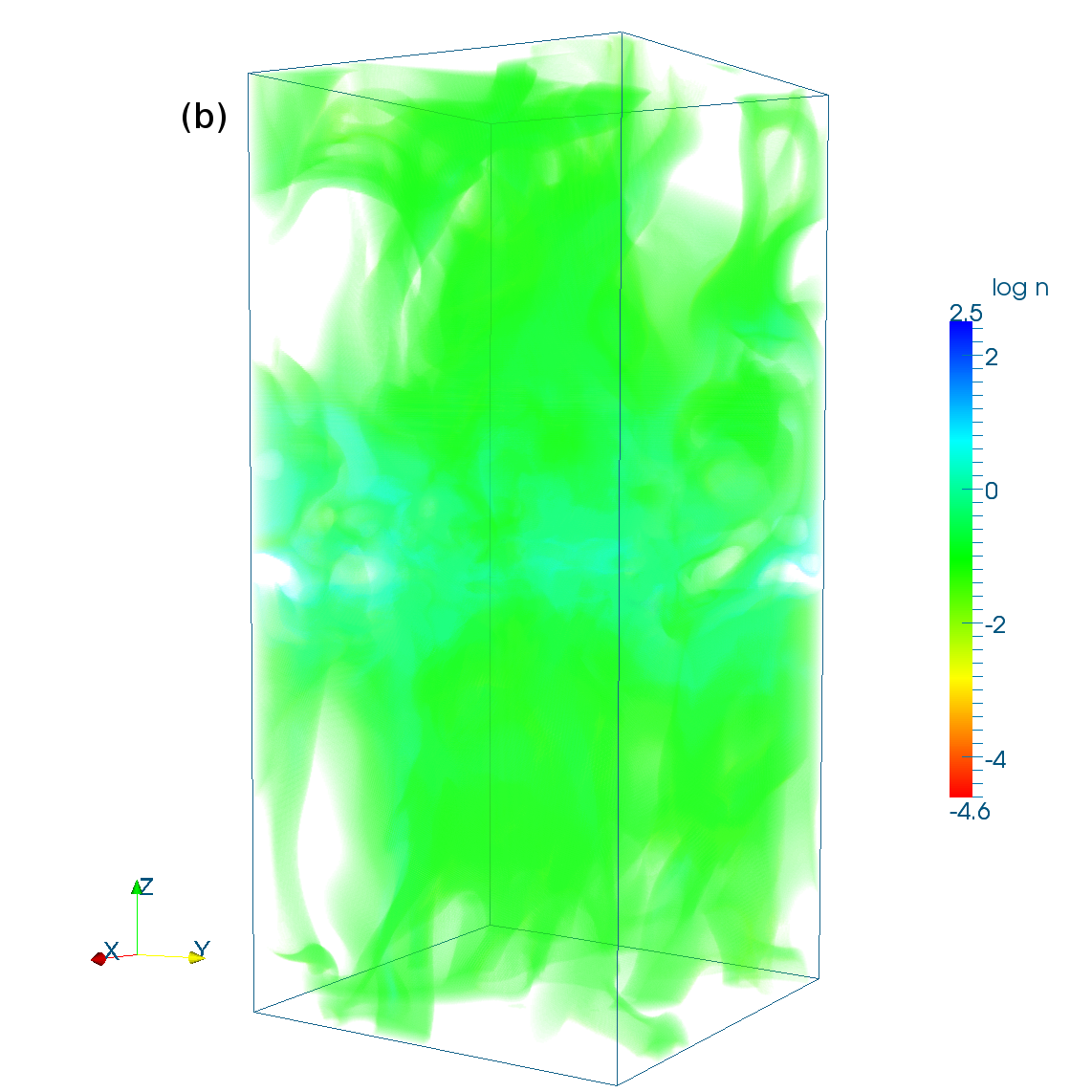}\hspace{-1.3cm}
  \includegraphics[width=0.425\linewidth]{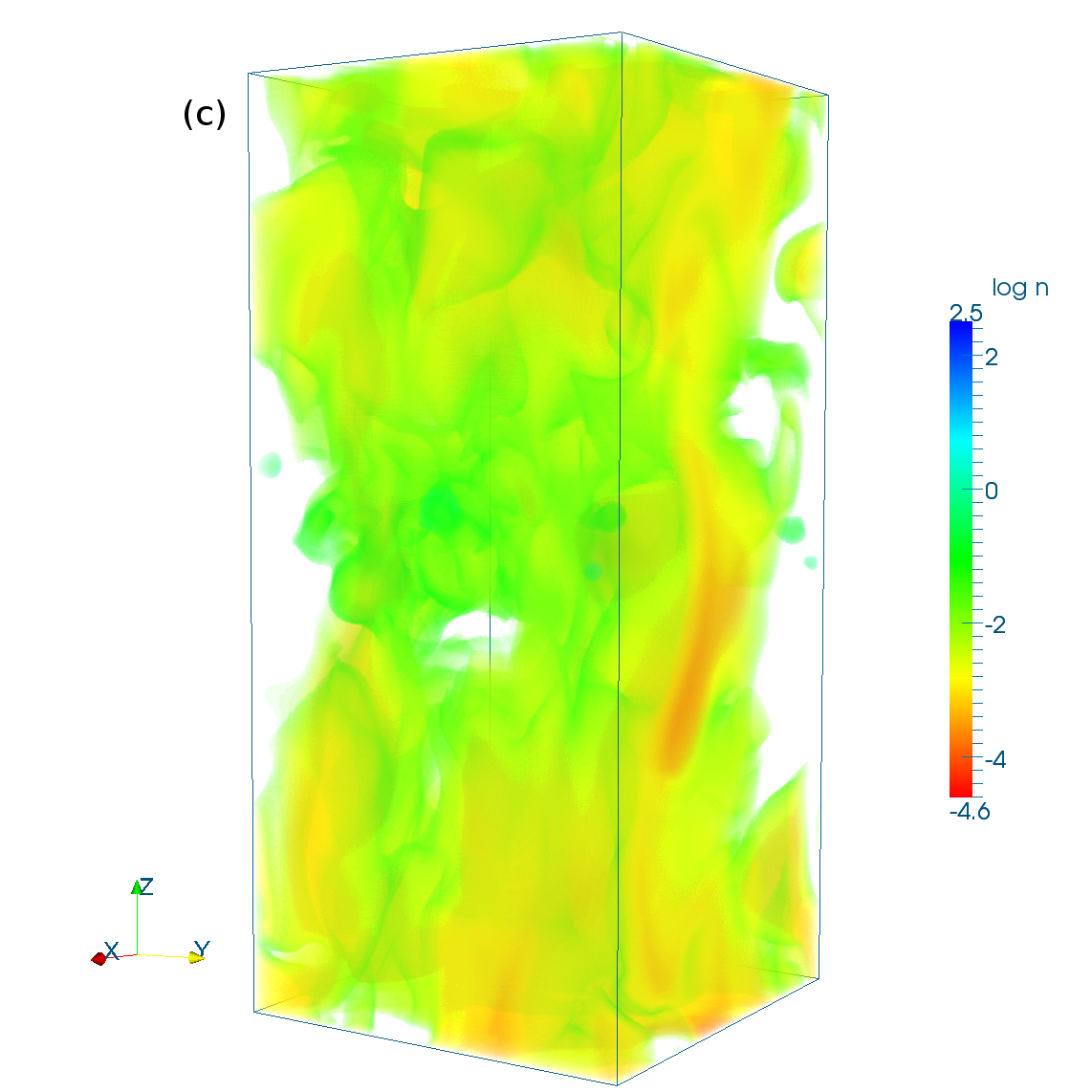}
  \hspace{-2.0cm}
    \caption[Volume snapshots of $\rho$ by phase for Model~\Op]{
  3D snapshots, from Model~{\Op}, of gas number density in \textbf{(a)} the
  cold gas, \textbf{(b)} the warm  gas, and \textbf{(c)} the hot gas.
  In each plot regions  that are clear (white space) contain gas belonging to 
  another phase.
  The phases are separated at temperatures $500\K$ and $5\times10^5\K$.
  The colour scale for log\,$n$ is common to all three plots.
    \label{fig:rho3d}
            }
  \end{figure}
%-----------------------------------------------------------------------------

  A three dimensional rendering of a snapshot of the density distribution from
  the reference model (\Op) is illustrated in Fig.~\ref{fig:rho3d}, showing the
  typical location and density composition of each phase separately.
  In panel~a the cold gas is located near the mid-plane, apparently in layers
  with some more isolated fragments. 
  The location and filling fraction of the warm (panel b) and hot (panel c) gas appear
  similar, although the hot gas has density typically 1/100 that
  of the warm gas density. 
  The density of the gas is more roughly correlated with $z$, with the stongest
  fluctuations in the structure of the hot gas, including very diffuse regions
  scattered about the mid-plane as well as near the vertical boundaries.

  The analysis of the warm gas density distribution within and outside the
  supernova active mid-plane in Section~\ref{sect:TMPS} indicated that its shape
  and standard deviation is fairly robust, and is merely shifted with respect to
  mean density, a predictable consequence of balancing the global pressure
  gradient.
  On the other hand the gas density distributions of the hot gas have been found
  to be quite different within and outside the mid-plane.
  In particular the distribution of hot gas near the mid-plane has a larger
  standard deviation, as well as increased mean.
  This is reflected in the pockets of very diffuse gas near the mid-plane, amongst
  the most dense hot gas.
  
  The significance of these features to observations, are that the effects of
  the warm gas are more predictable, with consistent standard deviation,
  independent of height and in principle a predictable trend in the mean with
  respect to the pressure balance to the gravitational potential. 
  For the hot gas observations along a line-of-sight above the supernova active
  region $|z|\simeq300\p$ will also have consistent standard deviation and 
  trend in the mean. 
  The structure of the hot gas near the mid-plane is significantly more
  complex and subject to strong local effects.

  \section{Summary}

  The ISM is reasonably described by separation into phases. 
  The boundaries between the phases can be identified by the minima in the
  volume weighted probability distributions of the gas temperature.
  An alternative representation in terms of specific entropy for the phase 
  boundaries also yields statistically meaningful phase temperature PDFs.
  The phases are well described by statistical thermal and total pressure 
  balance.
  The phase PDFs for the gas density are lognormal, and applying these fits
  provides  a better determination of the fractional volume from the
  phase filling factor than the assumption of homogeneous phases.

\end{chapter} 

  %-----------------------------------------------------------------------------
\begin{chapter}{Structure of the velocity field\label{chap:flow}}
%-----------------------------------------------------------------------------
  
%------------------------------------------------------------------------
  \begin{figure}[h]
  \centering
  \hspace{-2.8cm}
  \includegraphics[width=0.575\columnwidth]{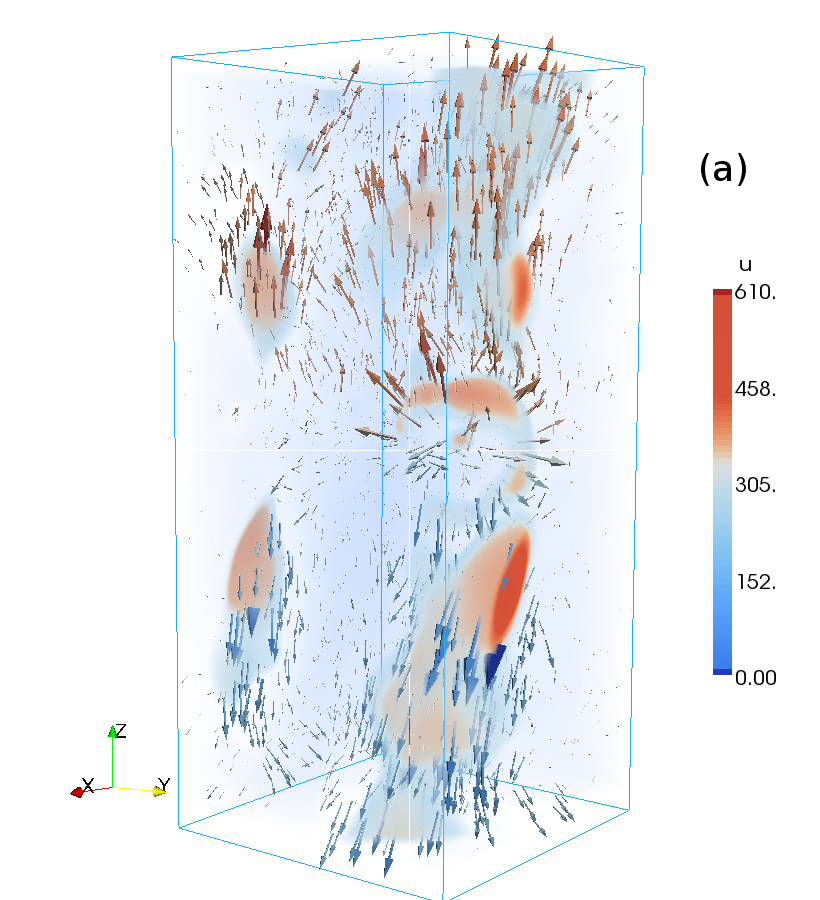}\hspace{-0.5cm}
  \includegraphics[width=0.575\columnwidth]{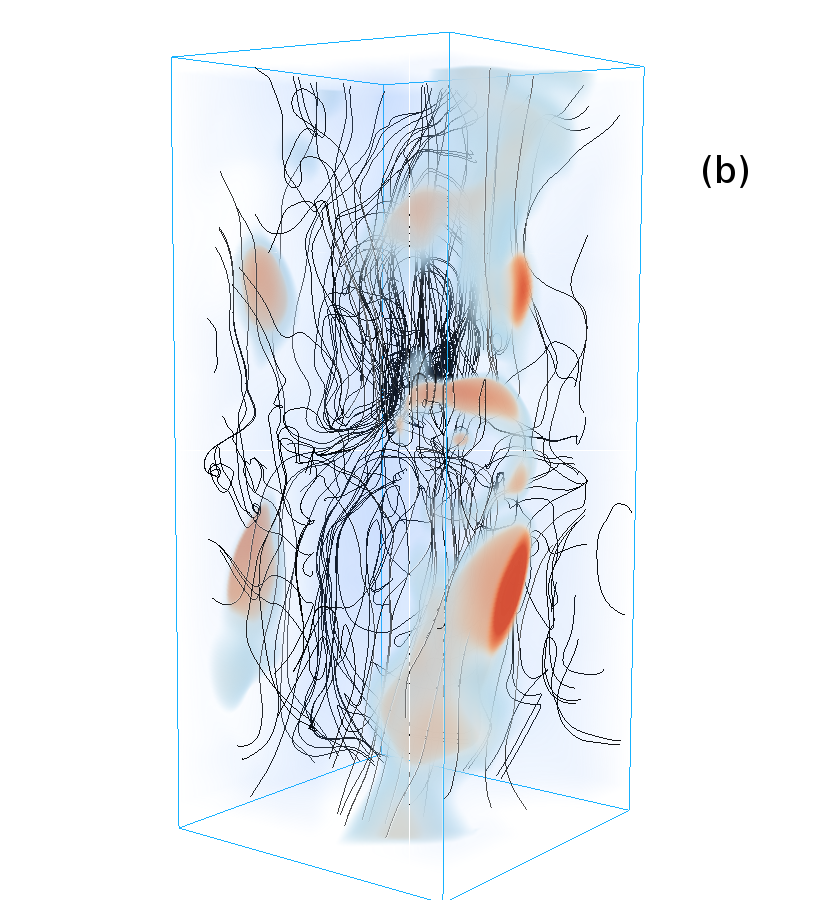}\hspace{-2.0cm}
    \caption[Volume snapshots of the velocity for Model~\Op]{
  The perturbation velocity field $\vect{u}$ in  Model~{\Op} at $t=550\Myr$.  
  The colour bar indicates the magnitude of the velocity depicted 
  in the volume shading,
  with rapidly moving regions highlighted with shades of red. 
  The low velocity regions, shaded blue, also have reduced opacity to 
  assist visualisation. 
  Arrow length of vectors \textbf{(a)} is proportional to the magnitude of 
  $\vect{u}$, with red (blue) arrows
  corresponding to $u_z>0$ ($u_z<0$) and independent of the colour bar.
  Trajectories of fluid elements \textbf{(b)} are also shown, indicating the 
  complexity of the flow and its pronounced vortical structure.
     \label{fig:vplot}
            }
   \end{figure}
%------------------------------------------------------------------------
  
  Understanding the nature of the flows in the ISM is important for appreciating
  many other inter-related properties of the ISM. 
  What are the systematic vertical flows? 
  Can they be characterized as a wind sufficient to escape the 
  gravitational potential of the galaxy and hence deplete the gas content or
  are they characteristic of a galactic fountain, in which case the gas is
  recycled?
  How does the velocity correlate with the density and hence contribute to 
  kinetic energy and the energy balance in the galaxy, against thermal,
  magnetic and cosmic ray contributions?
  How is the velocity field divided between a mean flow and fluctuations?
  To what extent do these properties vary by phase?
  Over what scales are the turbulent eddies of the ISM correlated, and how do they
  depend on phase or vary as a function of height?
  What is the vortical structure of the ISM and what contributions do various 
  features of the velocity field bring to the magnetic dynamo?
  In this section some of the general features of the velocity field are 
  described and some characteristics of the flow by phase are considered. 
  The correlation scales as a function of height are measured
  for the total ISM.

  In Fig.~\ref{fig:vplot} aspects of the velocity field in 3D for the
  reference Model~\Op\ are displayed.
  The volume rendering shows the regions of high speed.
  Speeds below about $300\kms$ are transparent to aid visualisation.
  Velocity vectors are illustrated (panel a) using arrows, size indicating
  speed and colour the $z$-component.
  These indicate that flows away from the mid-plane are dominant.
  Red patches are indicative of recent SNe and there is a strongly divergent
  flow close to middle on the $xz$-face.
  In addition stream lines (b) display the presence of many small scale 
  vortical flows near the mid-plane.

%--------------------------------------------------------------------------
  \section{Gas flow to and from the mid-plane}\label{subsect:GO}

%-----------------------------------------------------------------------------

%-----------------------------------------------------------------------------
  \begin{figure}[ht]
  \centering\hspace{-1.7cm}
\includegraphics[width=0.55\columnwidth]{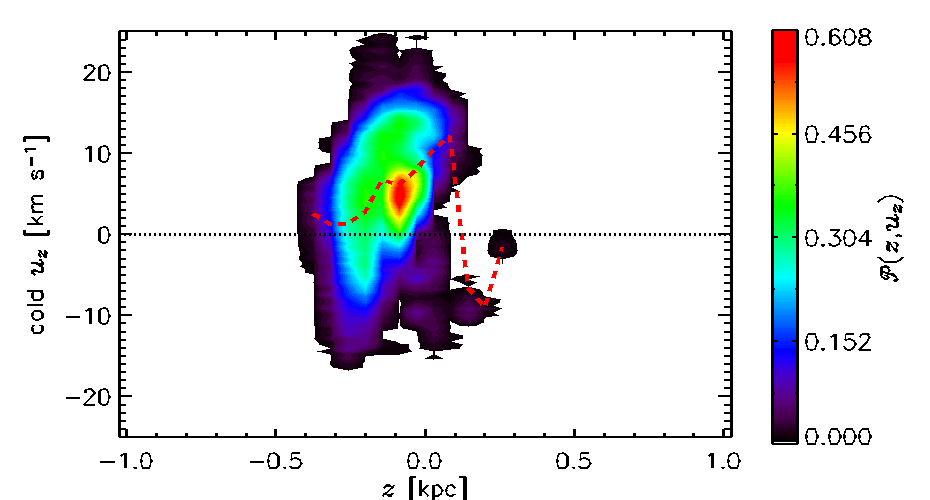}
\includegraphics[width=0.55\columnwidth]{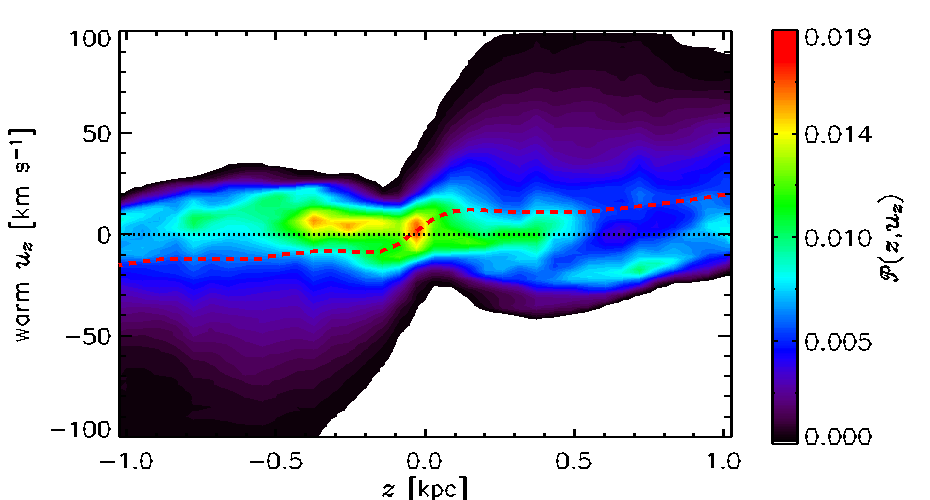}\hspace{-2.5cm}\\
\includegraphics[width=0.85\columnwidth]{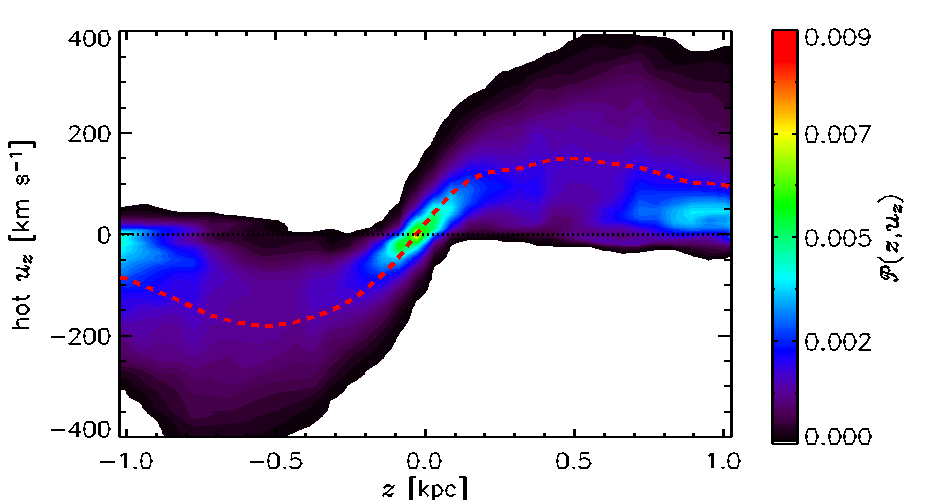}
%  \hspace{-1.5cm}
%  \includegraphics[width=0.55\columnwidth]{fig/phspqO1_uzzc.png}
%  \hspace{-0.5cm}
%  \includegraphics[width=0.55\columnwidth]{fig/phspqO1_uzzw.png}
%  \hspace{-1.5cm}\\
%  \hspace{-2.75cm}
%  \includegraphics[width=0.55\columnwidth]{fig/phspqO1_uzzh.png}
%  \hspace{-0.5cm}
%  \includegraphics[width=0.55\columnwidth]{fig/phspqO1_muzz.png}
%  \hspace{-1.5cm}
    \caption[Contour plots $u_z(z)$ for Model~\Op by phase]{
  Contour plots of the probability distributions $\mathcal{P}(z,u_z)$ for the
  vertical velocity $u_z$ as a function of $z$ in Model~{\Op} from eleven 
  snapshots 634--644\Myr.
  The cold $(T<500\K)$, warm $(500\K\leq T<5\times10^5\K)$ and hot 
  $(T\geq 5\times10^5\K)$ are shown in panels \textbf{(a)} to \textbf{(c)},
  respectively.
  The horizontal averages of the vertical velocity $u_z$ in each case are
  over plotted (red, dashed) as well as the mid-plane position (black, dotted).
    \label{fig:uzm}\label{fig:uzscat}
            }
  \end{figure}
%------------------------------------------------------------------------
  
  The mean vertical flow is dominated by the hot gas, so it is
  instructive to consider the velocity structure of each phase separately.
  Figure~\ref{fig:uzscat} shows plots of the probability 
  distributions $\mathcal{P}(z,u_z)$ as functions of $z$ and $u_z$ from eleven
  snapshots of Model~\Op, separately for the cold (a), warm (b) and hot gas (c).

  The cold gas is mainly restricted to $|z|<300\p$
  and its vertical velocity varies within $\pm20\kms$ for $|z|\lessgtr0$.
  On average, the cold gas moves towards the mid-plane, 
  presumably after cooling at larger heights. 
  The warm gas is involved in a weak net vertical outflow above $|z|=100\p$, 
  of order $\pm10\kms$.
  This might be an entrained flow within the hot gas.
  However, due to its skewed distribution, the modal flow is 
  typically towards the mid-plane.
  The hot gas has large net outflow speeds, accelerating to
  about $100\kms$ within $|z|\pm200\p$, but with small amounts of inward
  flowing gas at all heights.
  The mean hot gas outflow increases at an approximately constant rate
  to a speed of over $100\kms$ within $\pm100\p$ of the mid-plane,
  and then decreases with further distance from the mid-plane,
  at a rate that gradually decreases with height for $|z|\ga0.5\kpc$.
  This is below the escape velocity associated with a galactic wind.

  A significant effect of increased resolution is the increase in the magnitude
  of the perturbed velocity, from $\average{u_\rrms}=76\kms$ in Model~{\Op} to
  $103\kms$ in Model~{\OpH} (Table~\ref{table:results} Column~13) and
  temperatures ($\average{c\sound}$ increases from $150$ to $230\kms$);
  both $\average{u_\rrms}$ and the random velocity $\average{u_0}$ are increased
  by a similar factor of about 1.3. 
  However, the thermal energy $e_{\textrm{th}}$ is reduced by a factor of 0.6 
  with the higher resolution, while kinetic energy $e_{\textrm{K}}$ remains about
  the same. 
  This suggests that in the higher-resolution model, the higher velocities and 
  temperatures are associated with lower gas densities.

%--------------------------------------------------------------------------
  \section{The correlation scale of the random flows}\label{sect:CORR}
%--------------------------------------------------------------------------

  The correlation length of the random velocity $\vect{u}$ at a single time
  step of the Model~{\Op}, has been estimated by calculating the second-order
  structure functions $\str(l)$ of the velocity components $u_x$, $u_y$ and
  $u_z$, where
  \begin{equation}
    \str(l) = \langle \left[u(\vect{x}+\vect{l})-u(\vect{x})\right]^2\rangle,
    \label{eq:CORR:sf}
  \end{equation}
  with $\vect{x}$ the position in the $(x,y)$-plane and $\vect{l}$ a horizontal
  offset.
  The offsets in the $z$-direction have not been included because of the 
  systematic outflows and expected variation of the correlation length with 
  $z$.
  The aggregation of the squared differences by $|\vect{l}|$, presumes
  that the flow is statistically-isotropic horizontally.
  I will consider more detailed analysis of the three-dimensional properties
  of the random flows, including the degree of anisotropy and its dependence on
  height and phase in future work. 

%-----------------------------------------------------------------------------
  \begin{figure}[h]
  \centering
  \includegraphics[height=0.485\columnwidth,width=0.80\columnwidth]{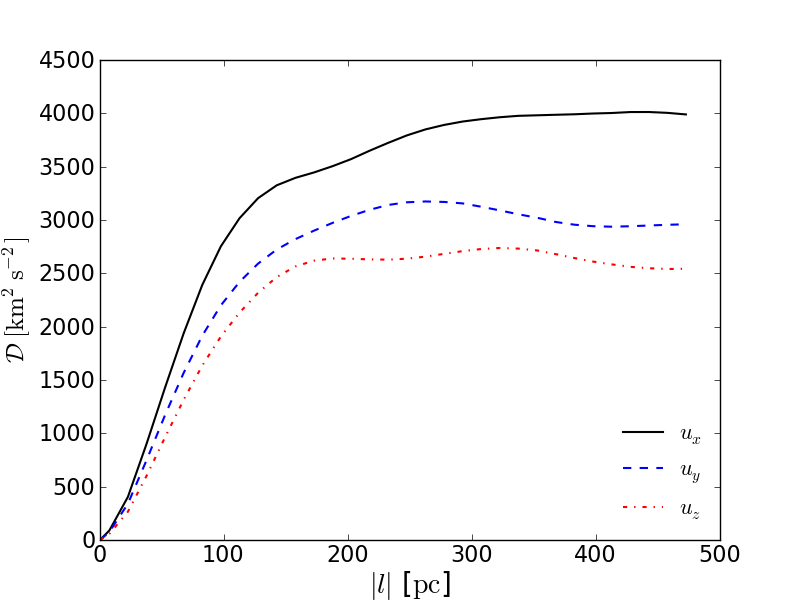}
    \caption[Structure function of velocity]{
  The second-order structure functions calculated using Eq.~(\ref{eq:CORR:sf}),
  for the layer $-10<z<10$\,pc, of the velocity components $u_x$ (black, solid
  line), $u_y$ (blue, dashed) and $u_z$ (red, dash-dot).
  The offset $l$ is confined to the $(x,y)$-plane only.
    \label{fig:CORR:sf}
            }
  \end{figure}
%------------------------------------------------------------------------
  
  $\str(l)$ was measured at five different heights, ($z=0, 100, -100$,
  $200$ and $800\p$).
  Averaging over six adjacent slices in the $(x,y)$-plane, each position   
  corresponds to a layer of thickness 20\,pc.
  The averaging took advantage of the periodic boundaries in $x$ and $y$;
  for simplicity choosing a simulation snapshot at a time for which the offset
  in the $y$-boundary, due to the shearing boundary condition, was zero.
  The structure function for the  mid-plane ($-10<z<10\p$) is shown in 
  Fig.~\ref{fig:CORR:sf}.

  The correlation scale can be estimated from the form of the structure function
  since velocities are uncorrelated if $l$ exceeds the correlation length
  $l\turb$, so that $\str$ becomes independent of $l$,
  $D(l)\approx 2u_\mathrm{rms}^2$ for $l\gg l\turb$.
  Precisely which value of $\str(l)$ should be chosen to estimate $l\turb$ in a
  finite domain is not always clear; for example, the structure function of 
  $u_y$ in Fig.~\ref{fig:CORR:sf} allows one to make a case for either the 
  value at which $\str(l)$ is maximum or the value at the greatest $l$.
  This uncertainty can give an estimate of the systematic uncertainty in the
  values of $l\turb$ obtained.
  Alternatively, and more conveniently, one can estimate $l\turb$ via the
  autocorrelation function $\corr(l)$,  related  to $\str(l)$ by
  \begin{equation}
    \corr(l) = 1-\frac{\str(l)}{2u^2_\mathrm{rms}}.
    \label{eq:CORR:ac}
  \end{equation}
  In terms of the autocorrelation function, the correlation scale $l\turb$ is
  defined as
  \begin{equation}
    l\turb=\int_0^\infty \corr(l)\,\dd l,
    \label{eq:CORR:l}
  \end{equation}
  and this provides a more robust method of deriving $l\turb$ in a finite domain.
  Of course, the domain must be large enough to make $\corr(l)$ negligible at
  scales of the order of the domain size; this is a nontrivial requirement,
  since even an exponentially small tail can make a finite contribution to
  $l\turb$. 
  In such estimates we are, of course, limited to the range of $\corr(l)$
  within the computational domain, so that the upper limit in the integral of
  Eq.~(\ref{eq:CORR:l}) is equal to $L_x=L_y$, the horizontal box size;
  this is another source of uncertainty in the estimates of $l\turb$.
  
%-----------------------------------------------------------------------------
  \begin{figure}[h]
  \centering
  \hspace{-1.5cm}
  \includegraphics[height=0.3\columnwidth,width=0.535\columnwidth,clip=true,trim=0 0 0 9mm]{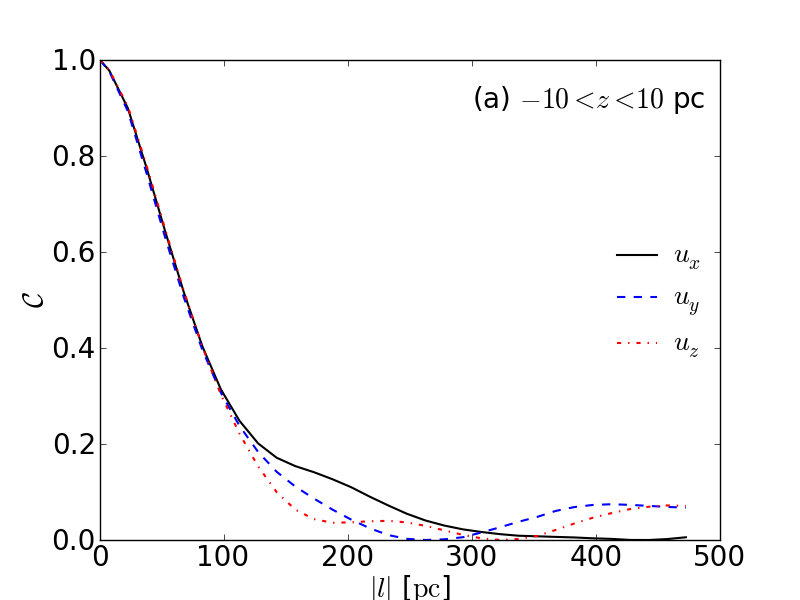}
  \hspace{-0.5cm}
  \includegraphics[height=0.3\columnwidth,width=0.535\columnwidth,clip=true,trim=0 0 0 9mm]{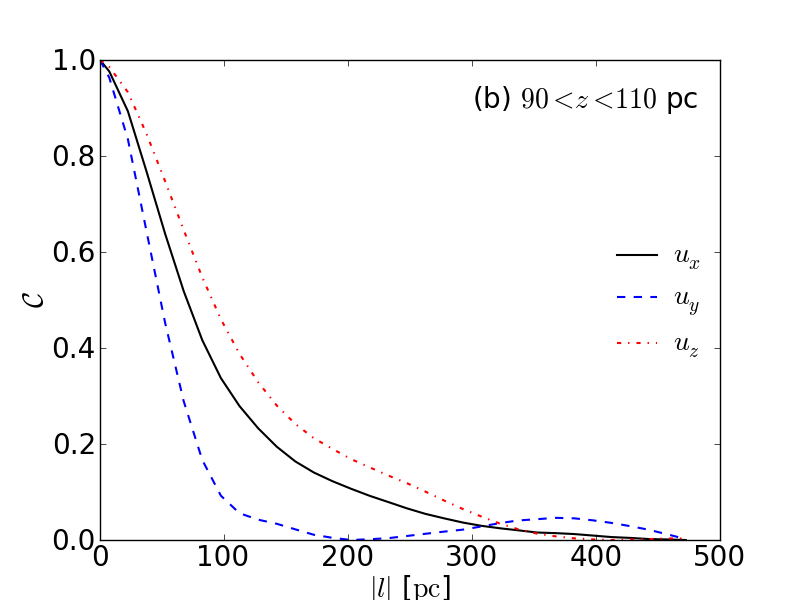}
  \hspace{-1.5cm}\\
  \hspace{-1.5cm}
  \includegraphics[height=0.3\columnwidth,width=0.535\columnwidth,clip=true,trim=0 0 0 9mm]{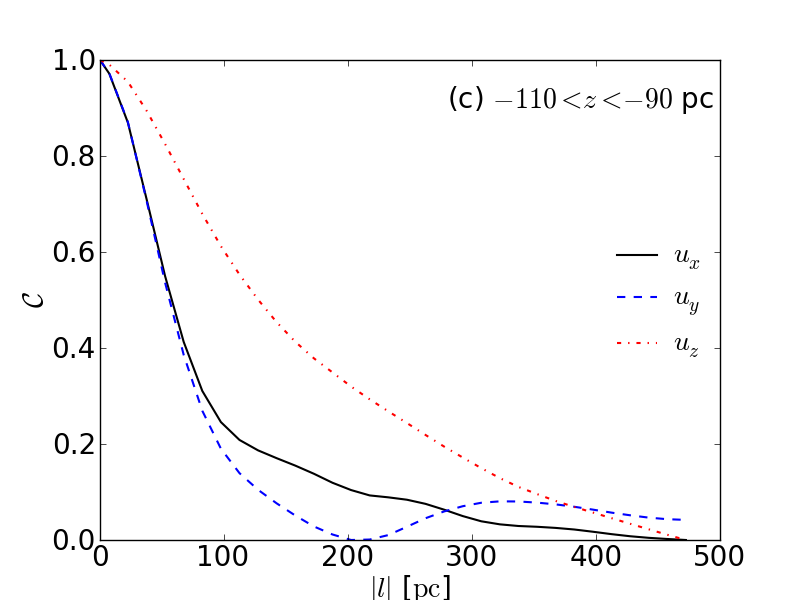}
  \hspace{-0.5cm}
  \includegraphics[height=0.3\columnwidth,width=0.535\columnwidth,clip=true,trim=0 0 0 9mm]{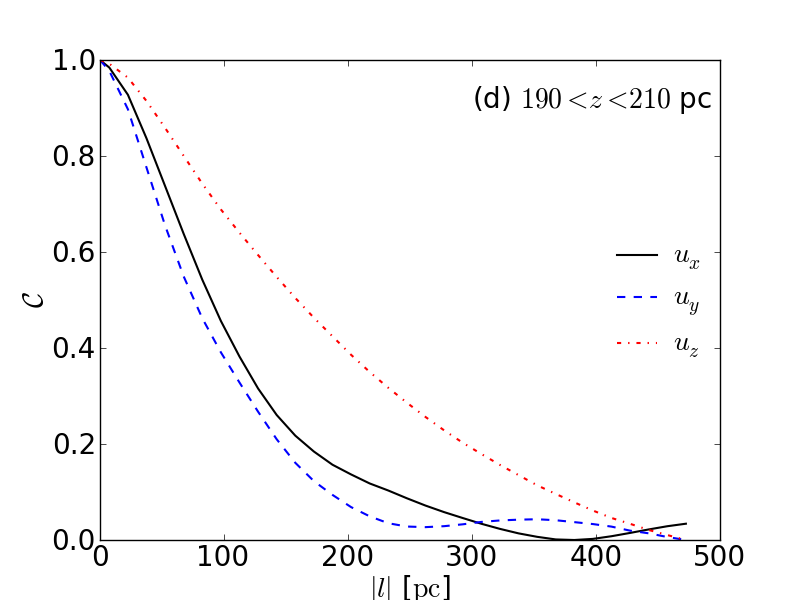}
  \hspace{-1.5cm}\\
  \hspace{-2.0cm}
  \includegraphics[height=0.3\columnwidth,width=0.535\columnwidth,clip=true,trim=0 0 0 9mm]{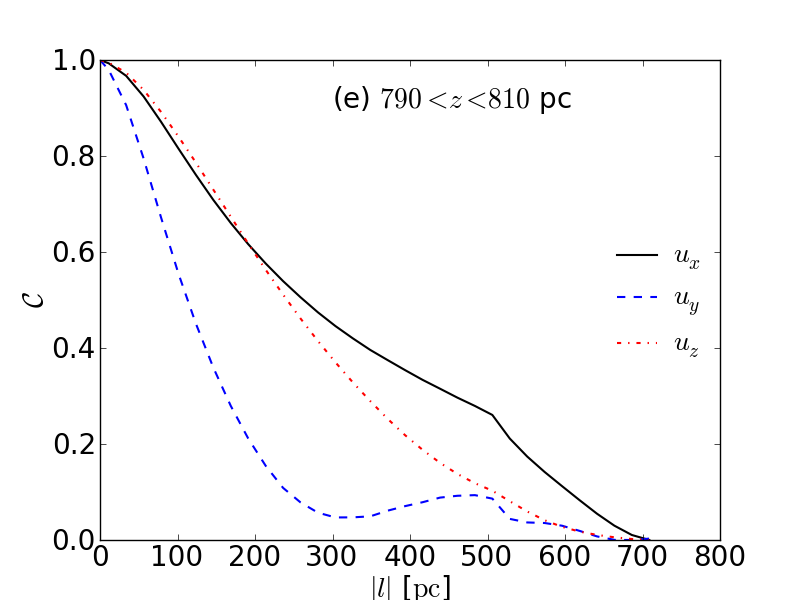}
  \hspace{-1.5cm}
    \caption[Autocorrelation function of velocity]{
  Autocorrelation functions for the velocity components $u_x$ (black, solid
  line), $u_y$ (blue, dashed) and $u_z$ (red, dash-dot) for $20\p$ thick layers
  centred on five different heights, from top to bottom: $-10<z<10\p$,
  $90<z<110\p$, $-110<z<-90\p$, $190<z<210\p$ and $790<z<810\p$.
    \label{fig:CORR:ac}
            }
  \end{figure}
%-----------------------------------------------------------------------------

  Figure~\ref{fig:CORR:ac} shows $\corr(l)$ for five different heights in the
  disc, where $u_\mathrm{rms}$ was taken to correspond to the absolute maximum
  of the structure
  function, $u^2_\mathrm{rms}=\max(\str)/2$, from Eq.~(\ref{eq:CORR:ac}).
  The autocorrelation function of the vertical velocity varies with $z$ more
  strongly than, and differently from, the autocorrelation functions of the 
  horizontal velocity components; it broadens as $|z|$ increases, meaning that
  the vertical velocity is correlated over progressively greater horizontal
  distances. 
  Already at $|z|\approx 200\p$, $u_z$ is coherent across a significant
  horizontal cross-section of the domain and at $|z|\approx 800\p$ so is $u_x$.
  An obvious explanation for this behaviour is the expansion of the hot gas
  streaming away from the mid-plane, which thus occupies a progressively larger
  part of the volume as it flows towards the halo.
  
  Table~\ref{table:CORR:l} shows the rms velocities derived from the structure
  functions for each component of the velocity at each height, and the
  correlation lengths obtained from the autocorrelation functions.
  Note that these are obtained without separation into phases.
  The uncertainties in $u_\mathrm{rms}$ due to the choices of local maxima in
  $\str(l)$ are less than $2\kms$. However, these can produce quite large
  systematic uncertainties in $l\turb$, as small changes in $u_\mathrm{rms}$ can 
  lead to $\corr(l)$ becoming negative in some range of $l$ (i.e.\ a weak
  anti-correlation), and this can significantly alter the value of the 
  integral in Eq.~(\ref{eq:CORR:l}).
  Such an anti-correlation at moderate values of $l$ is natural for
  incompressible flows; the choice of $u_\mathrm{rms}$ is thus not straight  
  forward.
  Other choices of $u_\mathrm{rms}$ in Fig.~\ref{fig:CORR:sf} can lead
  to a reduction in $l\turb$ by as much as $30\p$.
  Better statistics, derived from data cubes for a number of different 
  time-steps, will allow for a more thorough exploration of the uncertainties,
  but this analysis shall also be deferred to later work.

%-----------------------------------------------------------------------------
  \begin{table}[h]
  \centering
  \caption[Correlation scale of $l\turb$]{The correlation scale $l\turb$ and rms velocity $u_\mathrm{rms}$ at 
  various distances from the mid-plane.
  \label{table:CORR:l}}
  \begin{tabular}{cccccccc}
  \hline
   & \multicolumn{3}{c}{$u_\mathrm{rms}$ [$\!\kms$]}& &\multicolumn{3}{c}{$l\turb$ [$\!\p$]} \\
   \cline{2-4} \cline{6-8}
  $z$             &$u_x$ & $u_y$ & $u_z$ &&  $u_x$ & $u_y$ & $u_z$ \\
  \hline
  $\phantom{-10}0$& $45$ & $40$  & $37$  && $99$   & $98$  & $94$  \\
  $\phantom{-}100$& $36$ & $33$  & $43$  && $102$  & $69$  & $124$ \\
  $-100$          & $39$ & $50$  & $46$  && $95$   & $87$  & $171$ \\
  $\phantom{-}200$& $27$ & $20$  & $63$  && $119$  & $105$ & $186$ \\
  $\phantom{-}800$& $51$ & $21$  & $107$ && $320$  & $158$ & $277$ \\
  \hline
  \end{tabular}
  \end{table}
%-----------------------------------------------------------------------------
  
  The rms velocities given in Table~\ref{table:CORR:l} are compatible with the
  global values of $u_\mathrm{rms}$ and $u\turb$ for the reference run W$\Omega$
  shown in Table~\ref{table:models} (within their uncertainties).
  The increase in the root-mean-square value of $u_z$ with height, from about
  $40\kms$ at $z=0$ to about $60\kms$ at $z=200\p$, reflects the systematic
  outflow with a speed increasing with $|z|$.
  There is also an apparent tendency for the root-mean-square (rms) values of
  $u_x$ and $u_y$ to decrease with increasing distance from the mid-plane.
  This is understandable, as the rate of acceleration from SN forcing will 
  decline from the mid-plane. 
  However it is not excluded that these rms values might increase again at
  heights substantially away from the mid-plane, where the gas is hotter and
  gas speeds are generally higher in all directions. 
  
  The correlation scale of the random flow is very close to $100\p$ in the
  mid-plane, and this value has been adopted for $l\turb$ elsewhere in the thesis.
  This estimate is in good agreement with the hydrodynamic ISM simulations of
  \citet{Joung06}, who found that most kinetic energy is contained by
  fluctuations with a wavelength (i.e.\ $2l\turb$ in our notation) of $190\p$.
  In the MHD simulations of \citet{Korpi99}, $l\turb$ for the warm gas was $30\p$ 
  at all heights, but that of the hot gas increased from $20\p$ in the 
  mid-plane to $60\p$ at $|z|=150\p$.
  \citet{AB07} found $l\turb=73\p$ on average, with strong fluctuations in
  time.
  As in \citet{Korpi99}, there is a weak tendency for $l\turb$ of the horizontal
  velocity components to increase with $|z|$ in my simulations, but this 
  tendency remains tentative, and must be examined more carefully to confirm
  whether it is robust.
  The correlation scale of the vertical velocity, which has a systematic
  part due to the net outflow of hot gas, grows from about $100\p$ at the
  mid-plane to nearly $200\p$ at $z=200\p$.
  Due to the increase of the fractional volume of the hot gas with distance from
  the mid-plane, this scale might be expected also to increase further.

%-----------------------------------------------------------------------------
  \section{Summary}

  The heating and shocking induced by SNe produce a mean outflow of gas from the
  disc of up to $200\kms$ within $|z|\la500\p$, which slows to $100\kms$ 
  within $|z|\la1\kpc$, which is mainly comprised of hot gas.
  These velocities are below the escape velocity, that would be associated with
  a galactic wind, so represents only the outward part of the galactic fountain.
  The horizontal correlation scale of the turbulent flow is of order $100\p$ 
  near the mid-plane. 
  The correlation length increases with height, reaching $200$--$300\p$ for
  $|z|\simeq800\p$, and likely to continue to increase above this.

\end{chapter}

  \begin{chapter}{\label{chap:SMP}Sensitivity to model parameters}
%-----------------------------------------------------------------------

%-----------------------------------------------------------------------------
\section{\label{subsect:COOL}Sensitivity to the cooling function}
%-----------------------------------------------------------------------------

  An important factor in the persistence of a multi-phase ISM is the action of
  differential cooling. 
  Radiative cooling and UV-heating rates in the ISM
  are dependent on a variety of localised factors, such as gas density, the 
  presence and abundance of helium, heavier elements and metals, filtering 
  by dust grains, and the temperature and ionization of the gas
  \citep{Wolfire95}.
  In simulations of this kind, where the gas is usually 
  modelled as a single fluid with a mean molecular weight, the cooling and 
  heating processes are heavily parametrized.    
  Many authors adopt alternative cooling profiles, with little effort to 
  establish the utility of their choice.
  Some comparisons have been 
  recorded between cooling functions, but using alternative numerical 
  methods or modelling indirectly related physical problems. 
  \citet{ABnei10} use a time dependent cooling function, which recognises the 
  delay in recombination of previously ionized gas, and demonstrates that there
  is a substantial variation in the resulting cooling.
  There are inevitably overheads associated with this refinement. 
  So far, the extent to which this affects the primary global properties of the
  ISM (temperature, density, velocity distributions) has not been clarified.
 
  Here two models, RBN and \WSWa, with parameters given in 
  Table~\ref{table:models}, are compared directly to assess how the specific
  choice of the cooling functions affect the results.
  Apart from different parameterizations of the radiative cooling, the two
  models share identical parameters, except as follows: because of the
  sensitivity of the initial conditions to the cooling function
  (Section~\ref{sect:ICons}), the value of $T_0$ was slightly higher in
  Model~{RBN}.
  (The density, heating and gravity profiles were the same.)

% -----------------------------------------------------------------------------
  \begin{figure}[h]
  \centering
  \hspace{-1.5cm}
  \includegraphics[width=0.535\columnwidth]{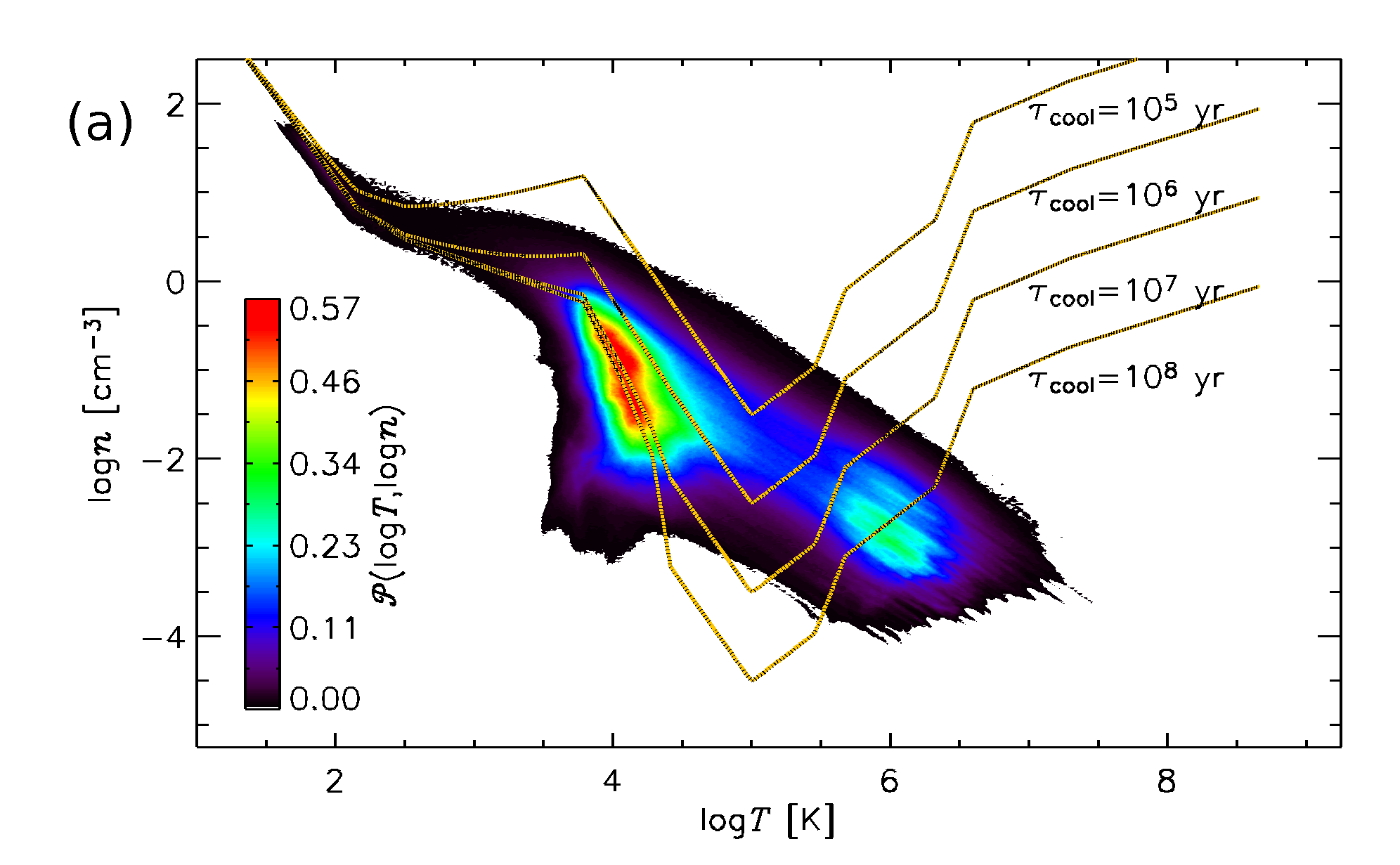}
  \includegraphics[width=0.535\columnwidth]{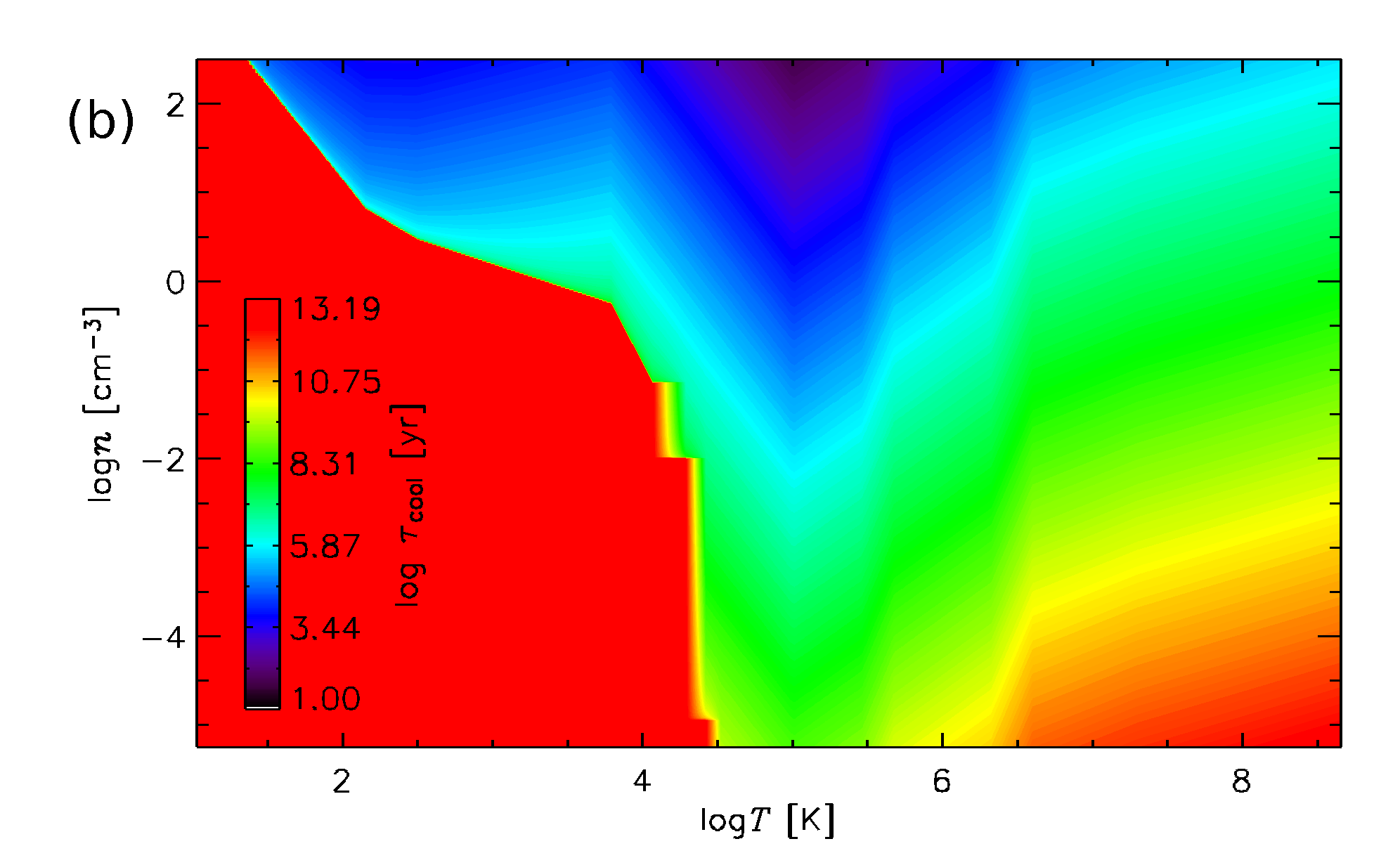}
  \hspace{-1.5cm}\\
  \hspace{-2.85cm}
  \includegraphics[width=0.535\columnwidth]{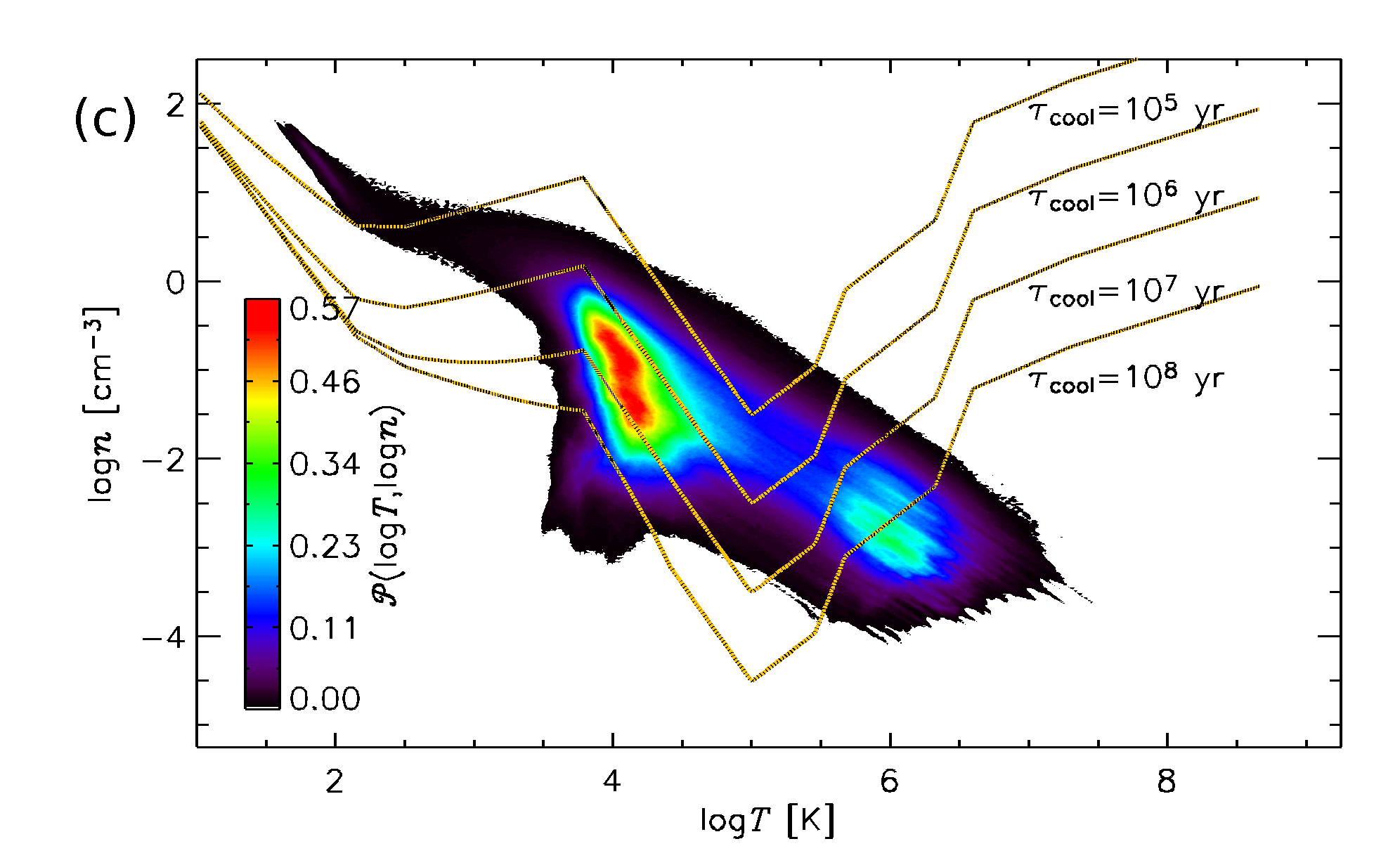}
  \includegraphics[width=0.535\columnwidth]{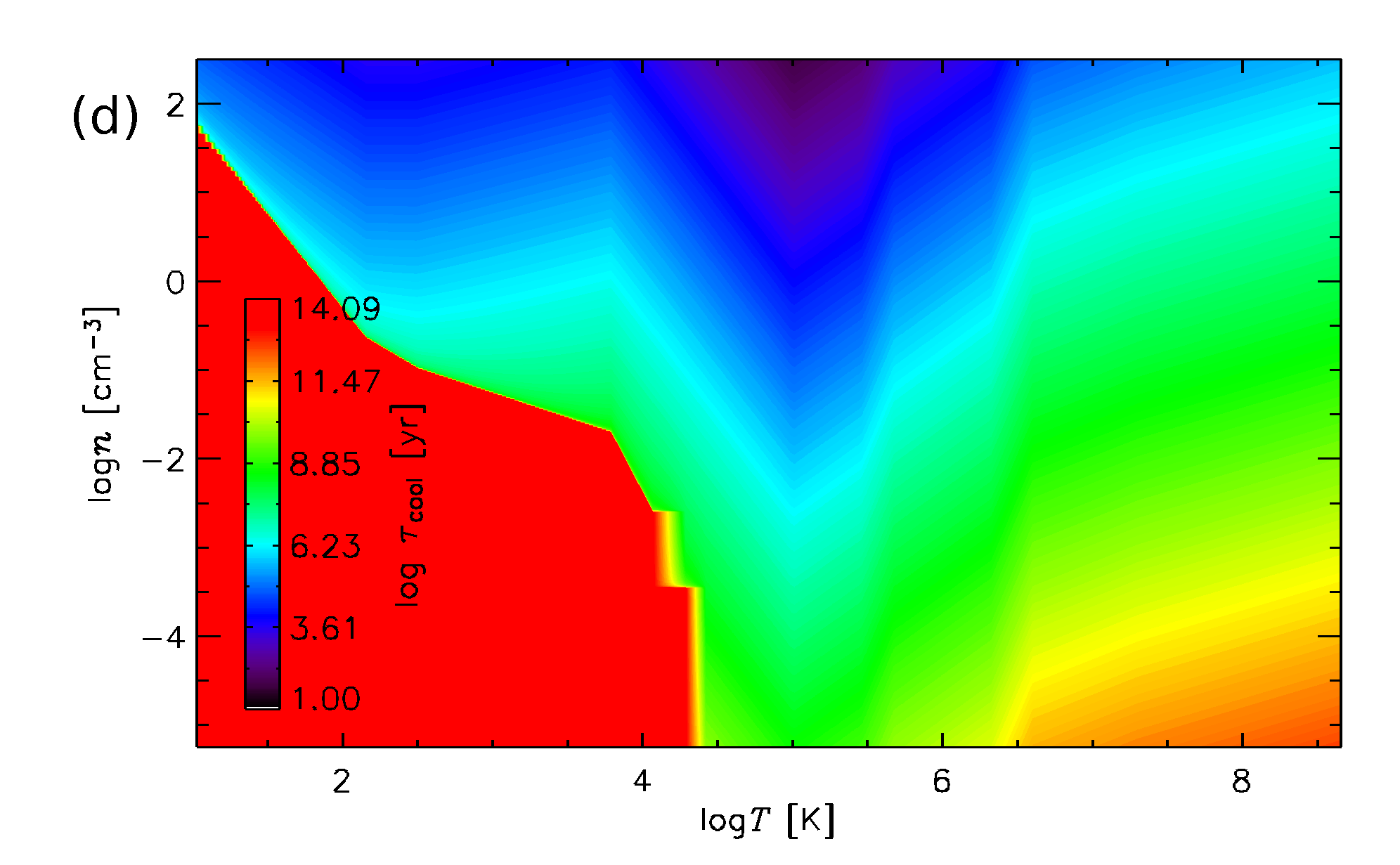}
  \hspace{-1.5cm}
  \caption[2D probability distribution of $n$ and $T$ for Model~{\WSWa}]{
  {\textbf{(a)}} The total volume probability distributions by gas number 
  density $n$ and temperature $T$.
  Eleven snapshots of Model~{\WSWa} in the interval $t=305$ to $325\Myr$ were
  used, with the system in a statistical steady state.
  Contours of constant cooling time ($10^5,~10^6,~10^7$ and $10^8$yr) are
  over plotted to indicate where the gas may be subject to rapid radiative
  cooling.
  These are drawn from the contour plot of cooling time $\tau\cool$
  {\textbf{(b)}}, given in Eq.~\eqref{eq:tau} with diffuse heating $\Gamma(0)$
  as applied at the mid-plane $z=0$. The solid red patch at $T\la10^4\K$ is a
  region where heating exceeds cooling so $\tau\cool$ is not defined.
  In {\textbf{(c)}} and {\textbf{(d)}} the contours again as {\textbf{(a)}} and
  {\textbf{(b)}} respectively, but for $\Gamma(z)$ at $|z|=1\kpc$.
  \label{fig:pdf2dop}}
  \end{figure}
%-----------------------------------------------------------------------------

  One consideration with the inclusion of a thermally unstable branch for
  the WSW cooling function, used in Model~\Op, is that the cooling times must be
  adequately resolved (see Section~\ref{sect:NS} and Appendix~\ref{sect:ti}).
  From Eq.~\eqref{eq:ent}, the net cooling rate due to the diffuse heating and
  cooling terms is 
  \[
    \frac{Ds}{Dt}=-\left(\frac{\rho\Gamma(z)-\rho^2\Lambda}{\rho T}\right);
      \quad[({\mathrm{in~cgs~units}})~~\erg\g^{-1}\K^{-1}\s^{-1}].
  \]
  The cooling time $\tau\cool$ is then given by
  \begin{equation}\label{eq:tau}
    \tau\cool=\frac{\cv T}{\rho\Lambda-\Gamma(z)}\quad[\s\,],
  \end{equation} 
  where $\Lambda$ and $\Gamma$ are the radiative cooling and ultra violet 
  heating respectively, as described in Section~\ref{sect:ssvrb}, and $\cv$ is
  the specific heat capacity of the gas at constant volume.

  In Fig.~\ref{fig:pdf2dop}a the combined probability distributions of the gas
  number density and temperature in Model~\WSWa\ are overplotted with
  contours of constant cooling time $\tau\cool$, from the WSW cooling function
  and with the heating as it is applied at the mid-plane.
  Alongside this (Fig.~\ref{fig:pdf2dop}b) are the contours of $\tau\cool$
  for the same range of temperature and density as in panel a. 
  Note that the distribution in panel a is considerably broader than that for 
  Model~\Op, displayed in Fig~\ref{fig:pdf2d}; i.e. there is greater pressure
  dispersion associated with the additional mass included in 
  Models~\WSWa\ and RBN.

  The heating $\Gamma(z)$ is a maximum at the mid-plane and
  reduces with $|z|$. 
  Ultra violet heating only applies at temperatures up to about $2\cdot10^4\K$,
  so $\tau\cool$ is independent of $z$ above these temperatures.
  Below $T\simeq10^4\K$ there are densities where heating exceeds cooling, so
  the cooling time is undefined, indicated by the solid red regions
  in the contour plot (b).
  From~(a) it is apparent that the distribution of the cold gas is bounded by this
  region, suggesting that gas cannot remain cold as it is reheated very quickly.
  The very narrow distribution of the cold gas is in part explained by the 
  equivalently narrow `corridor' between the rapid heating (red) and the rapid
  cooling (blue) bounding it.

  The effect of reduced heating with height is illustrated by the revised 
  contours of $\tau\cool$ (Fig.~\ref{fig:pdf2dop} c,d) for $|z|=1\kpc$.
  {\freply{The corridor between rapid heating and rapid cooling shifts well 
      below the 
  location of the cold gas in panel (c).
  It is also the case that the density of the gas reduces with height. 
  Therefore negligible quantities of gas in the thermally unstable range
  $313$--$6102\K$ will be subject to this faster cooling and longer cooling 
  times near the mid-plane will apply.}}
  
  Only warm gas $T\simeq10^5\K$ is significantly subject to cooling times less 
  than $10^5\yr$.
  There exists a substantial distribution of warm gas $T\simeq10^4\K$ inside
  the (red) region where $\Gamma(0)>\rho\Lambda$ (panel a), but not so for
  $\Gamma(1)$ (b).
  {\freply{From inspection of the distribution of gas near the mid-plane 
      (Fig.~\ref{fig:pdf2dcomp}a below) it is clear that very little gas 
      occupies the region of net heating (solid, red). Some gas
      below $10^4\K$ does spread into this region indicating that
  adiabatic cooling due to compression and expansion is dominant here.}}

% -----------------------------------------------------------------------------
  \begin{figure}[h]
  \centering
  \hspace{-1.5cm}
  \includegraphics[width=0.535\columnwidth]{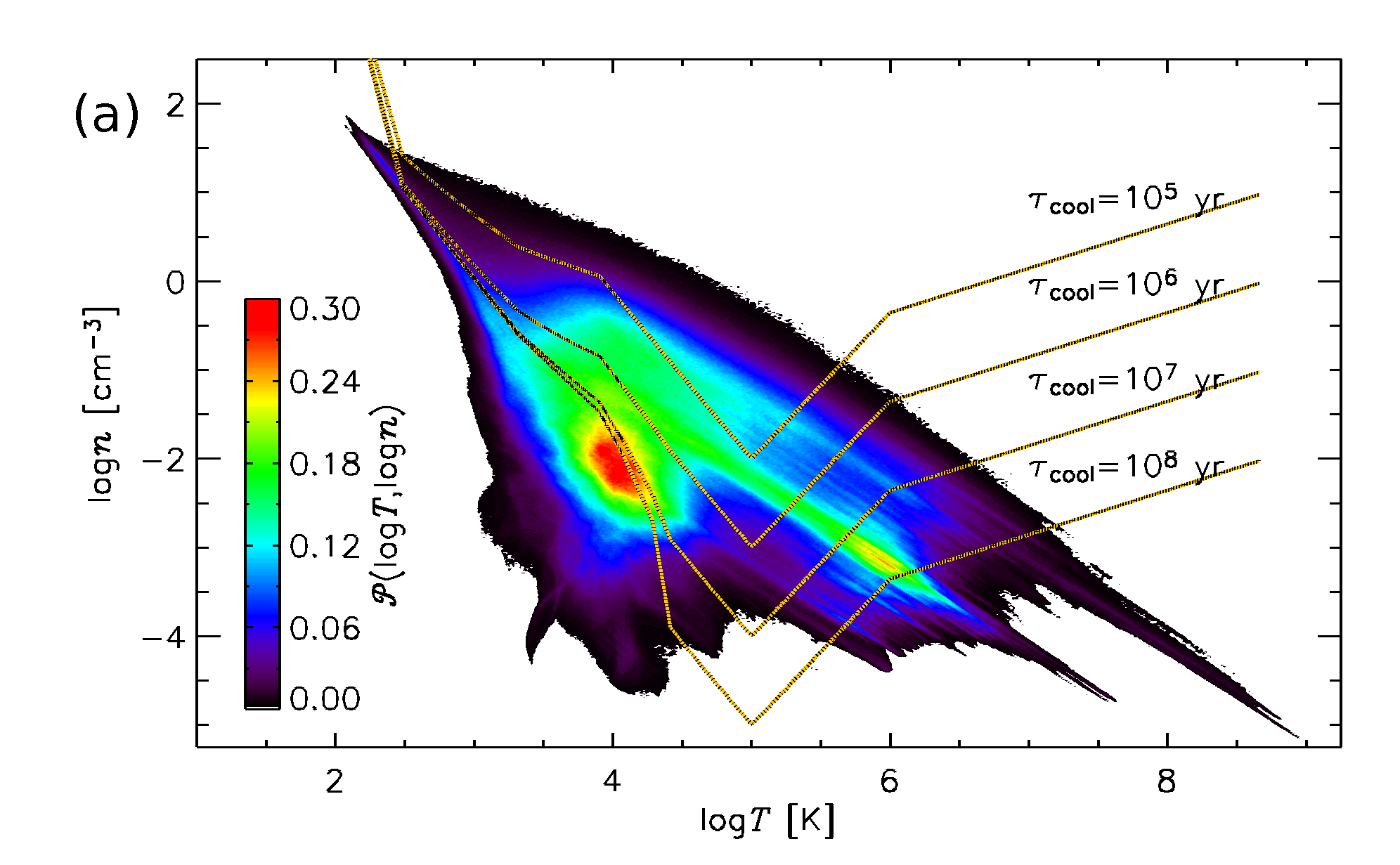}
  \includegraphics[width=0.535\columnwidth]{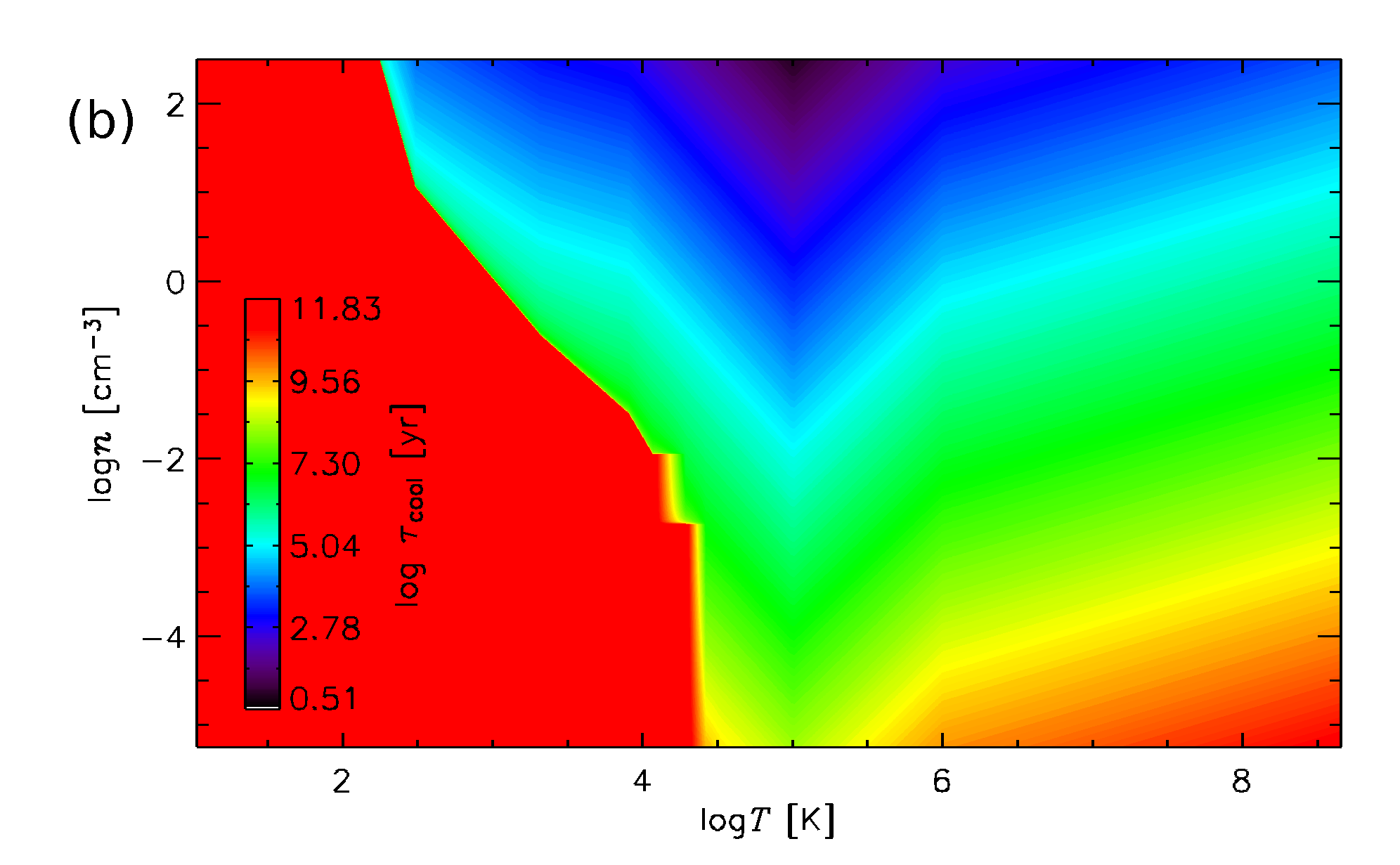}
  \hspace{-1.5cm}
  \caption[2D probability distribution of $n$ and $T$ for Model~{RBN}]{
  {\textbf{(a)}} The total volume probability distributions by gas number 
  density $n$ and temperature $T$ for Model~RBN.
  Contours of constant cooling time ($10^5,~10^6,~10^7$ and $10^8$yr) are
  over plotted to indicate where the gas may be subject to rapid radiative
  cooling.
  These are drawn from the contour plot of cooling time $\tau\cool$
  {\textbf{(b)}}, given in Eq.~\eqref{eq:tau} with diffuse heating $\Gamma(0)$
  as applied at the mid-plane $z=0$. The solid red patch at $T\la10^4\K$ is a
  region where heating exceeds cooling so $\tau\cool$ is not defined.
  \label{fig:rb2d}}
  \end{figure}
%-----------------------------------------------------------------------------

  In Fig.~\ref{fig:rb2d}, panels a and b of Fig.~\ref{fig:pdf2dop} are
  reproduced for Model~RBN. 
  There is far more gas in the temperature range around $10^3\K$, due to the
  absence of a thermally unstable branch between 313 and $6102\K$.
  The distribution of the cold gas near $\log n\simeq\log T\simeq2$ is very 
  similar to Model~\WSWa\ (Fig.~\ref{fig:pdf2dop}a,c), although in this case
  it is not aligned with the narrow corridor between rapid heating (red) and 
  rapid cooling (blue, panel b).
  This indicates that the composition of the cold gas is strongly dependent on
  the adiabatic shocks, rather than the balance between heating and cooling
  processes.
  The corresponding contours of $\tau\cool$ extend to much lower densities in
  the temperature range above $10^6\K$, so cooling times are much shorter for
  RBN cooling.
  Yet there is a greater abundance of very hot gas than with Model~\WSWa,
  which is explained by the much lower density of this gas, which mitigates 
  against the tendency for rapid cooling. 

  The distributions of gas near the mid-plane (within $100\p$) for both 
  models are displayed in Fig.~\ref{fig:pdf2dcomp}.
  From these it is evident that the low temperature -- low density boundary of
  the distributions aligns quite strongly with the region of net heating (red
  in Fig.~\ref{fig:pdf2dop}b and Fig.~\ref{fig:rb2d}b), although there is
  limited dispersion into this region. 
  Both models have distributions with extended peaks along this boundary 
  below $10^3\K$, corresponding to the most thermally stable location.
  This extends to higher temperature for RBN cooling than for WSW, through
  the range which is unstable for the latter.
  However the most significant alignment, common to both models, is along the
  line of constant thermal pressure.
  Note that the colour scale differs between the plots, such that these peaks 
  contain a similar proportion of the gas in both models.
  The cold dense tip of the RBN distribution extends into the region 
  ($\log n\simeq\log T\simeq2$) where net heating should make it unstable, yet
  lies on the line of constant pressure through the peaks of warm and hot gas.
  This is associated with the equilibrium pressure between the phases, which is
  slightly lower at the mid-plane for WSW cooling than for RBN.
  In contrast to the case for the total volumes, Model~\WSWa\ has more gas near 
  the 
  mid-plane at higher temperature and lower density than Model~RBN.
  Beyond this the cooling function has only a
  weak effect on the distribution for the gas with $T\ga10^3\K$.

% -----------------------------------------------------------------------------
  \begin{figure}[h]
  \centering
  \hspace{-1.5cm}
  \includegraphics[width=0.535\columnwidth]{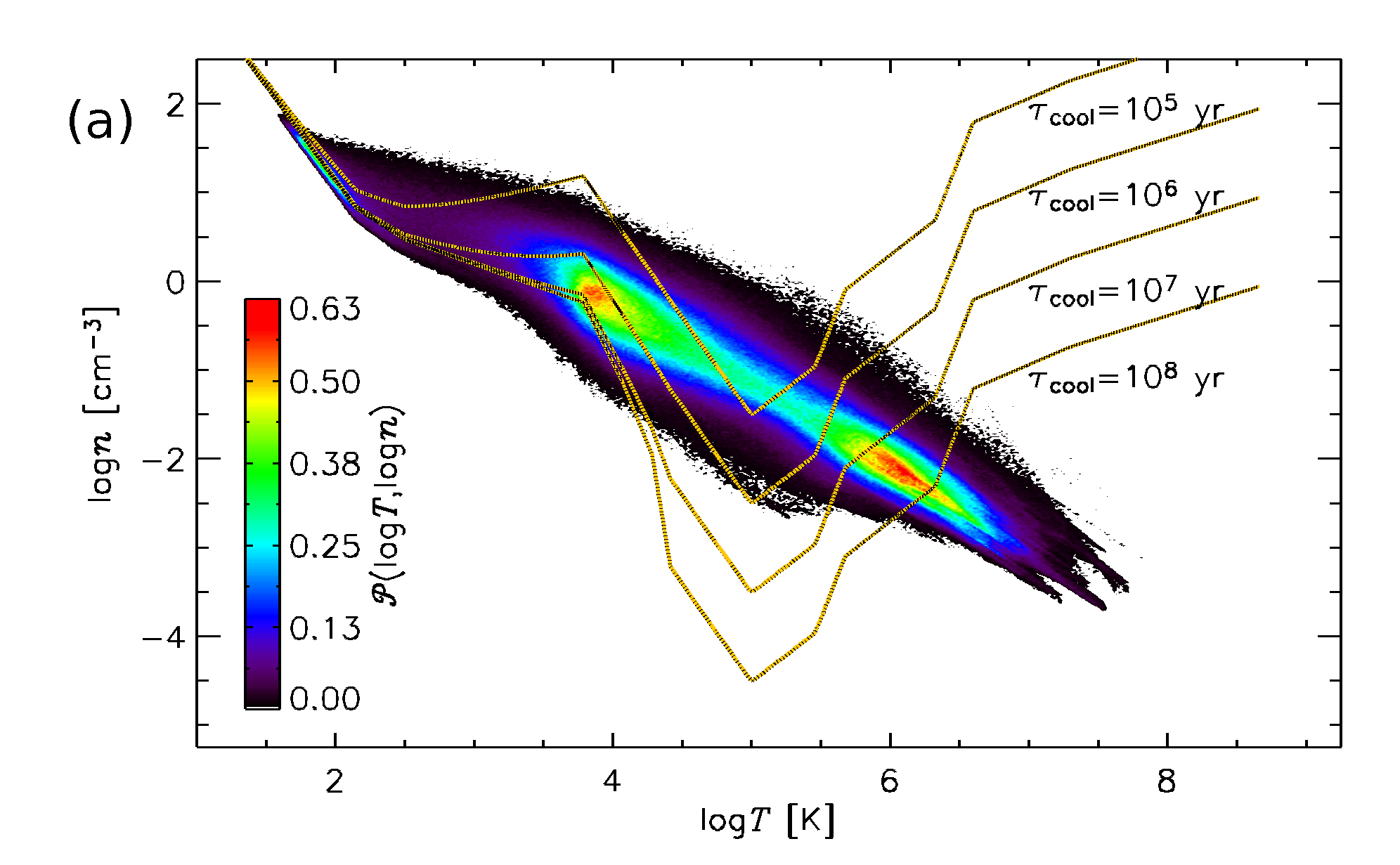}
  \includegraphics[width=0.535\columnwidth]{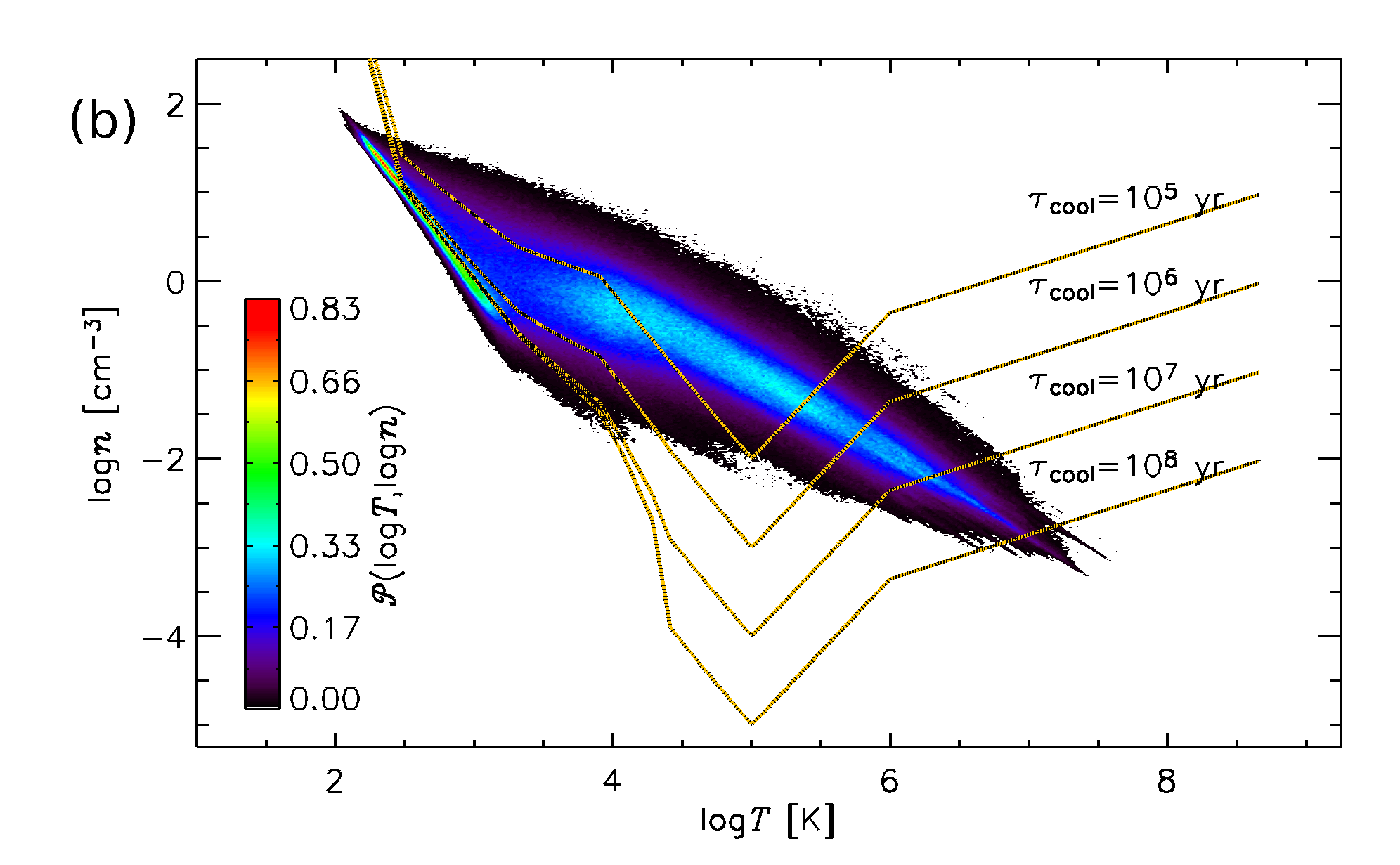}
  \hspace{-1.5cm}
  \caption[2D probability distribution of mid-plane $n$ and $T$]{
  The mid-plane probability distributions ($|z|<100\p$) by gas
  number density $n$ and temperature $T$ for {\textbf{(a)}} Model~\WSWa\ and
  {\textbf{(b)}} Model~RBN.
  Contours of constant cooling time ($10^5,~10^6,~10^7$ and $10^8$yr) are
  over plotted to indicate where the gas may be subject to rapid radiative
  cooling.
  \label{fig:pdf2dcomp}}
  \end{figure}
%-----------------------------------------------------------------------------
  
  The volume-averaged thermal and kinetic energy densities, the latter due to
  the perturbed motions alone, are shown in Fig.~\ref{fig:energetics} as
  functions of time.
  The averages for each are shown in Columns~15 and 16, respectively
  of Table~\ref{table:results}, using the appropriate steady state time 
  intervals given in Column~18.
  Models reach a statistically steady state, with mild fluctuations around a
  well defined mean value, very soon (within 60\,Myr of the start of the
  simulations).
  The effect of the cooling function is evident: both the thermal and kinetic
  energies in Model~RBN are about 60\% of those in Model~\WSWa.
  This is understandable as Model~RBN has a stronger cooling rate than 
  Model~\WSWa, only dropping below the WSW rate in the range $T<10^3\K$
  (see Fig.~\ref{fig:cool}).

%-----------------------------------------------------------------------------
  \begin{figure}[h]
  \centering
  \includegraphics[width=0.8\columnwidth]{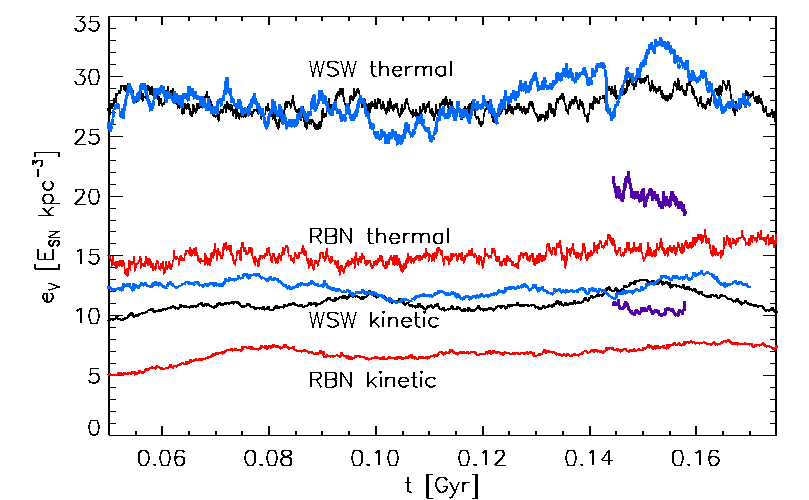}
    \caption[Comparison of energy density in HD models]{
  Evolution of volume-averaged thermal (black: Model~\WSWa,
  blue: Model~\Op, purple: Model~\OpH, red: Model~RBN) and kinetic energy
  density (lower lines) in the statistically steady regime, 
  normalised to the SN energy $E\SN \kpc^{-3}$.
  Models~{\WSWa}~(black) and RBN (red) essentially differ 
  only in the choice of the radiative cooling function.
    \label{fig:energetics}
            }
  \end{figure}
%-----------------------------------------------------------------------------

  Interestingly, both models are similar in that the thermal energy is about
  $2.5$ times the kinetic energy.
  These results are remarkably consistent with results by 
  \citet[][their Fig.~6]{Balsara04} and \citet[][Fig.\,3.1]{Gressel08b}.
  \citet{Gressel08b} applies WSW cooling and has a model very similar to 
  Model\,{\Op}, but with half the resolution and $|z|\leq2\kpc$. 
  He reports average energy densities of 24 and 10\,$E\SN\kpc^{-3}$ 
  (thermal and kinetic, respectively) 
  with SN rate  $\dot{\sigma}\SN$, comparable to 29 and 13\,$E\SN\kpc^{-3}$ 
  obtained here for Model\,{\Op}. 
  
  \citet{Balsara04} simulate an unstratified cubic region $200\p$ in size,
  driven at SN rates of 8, 12 and 40 times the Galactic rate, with resolution
  more than double that of Model\,{\Op}. 
  For SN rates $12\dot{\sigma}\SN$ and $8\dot{\sigma}\SN$, they obtain average
  thermal energy densities of about 225 and 160$\,E\SN\kpc^{-3}$,  and average
  kinetic energy densities of 95 and 60$\,E\SN\kpc^{-3}$, respectively 
  (derived from their energy totals divided by the $[200\p]^3$ volume). 

  To allow comparison with models here, where the SNe energy injection rate is
  $1\dot{\sigma}\SN$, we divide their energy densities by 12 and 8,
  respectively, to obtain energy densities of 19 and 20\,$E\SN\kpc^{-3}$ 
  (thermal), and 8 and 7.5\,$E\SN\kpc^{-3}$ (kinetic). 
  These are slightly smaller than our results with RBN cooling (25 and 
  9\,$E\SN\kpc^{-3}$) and those with WSW (29 and 
  13\,$E\SN\kpc^{-3}$ for WSWa, as given above).
  Since \citet{Balsara04} used an alternative cooling function \citep{Raymond77},
  leading to some additional uncertainty over the net radiative energy
  losses, the results appear remarkably consistent. 
  
  While cooling and resolution may marginally affect the magnitudes, it appears
  that the thermal energy density may consistently be expected to be about 
  $2.5$ times the kinetic energy density, independent of the model.
  It also appears, from comparing the stratified and unstratified models, that
  the ratio of thermal to kinetic energy is not strongly dependent on height,
  at least between the mid-plane and $\pm2\kpc$.

%-----------------------------------------------------------------------------
  \begin{figure}[h]
  \centering
  \hspace{-1.95cm}
  \includegraphics[width=0.55\columnwidth]{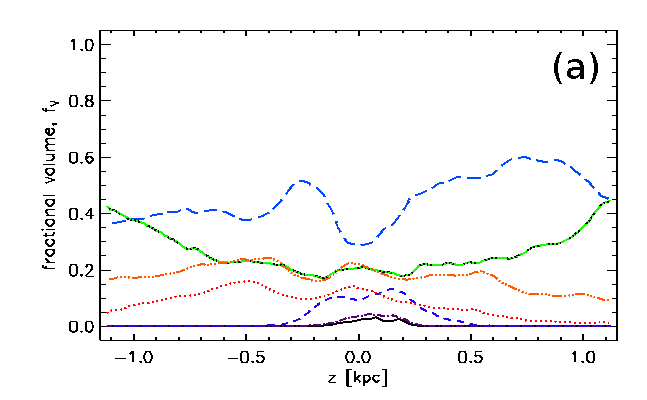}
  \hspace{-0.525cm}
  \includegraphics[width=0.55\columnwidth]{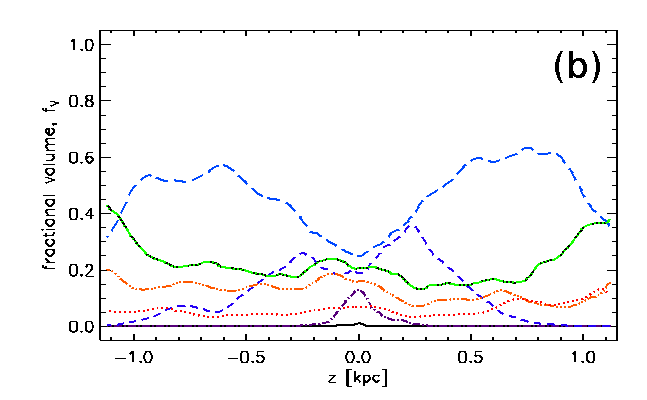}
  \hspace{-1.75cm}
    \caption[Fractional volumes by $z$ for Models~RBN and \WSWa]{
  Vertical profiles of the fractional volumes occupied by the
  various temperature ranges, with the key shown in Table~\ref{table:bands2}.
  \textbf{(a)}~Model~{\WSWa}, using 21 snapshots spanning 305 to 325\,Myr.
  \textbf{(b)}~Model~{RBN}, using 21  snapshots  spanning 266 to 286\,Myr.
    \label{fig:zfill_RB_WSW}
            }
  \end{figure}
%-----------------------------------------------------------------------------
  
  Figure~\ref{fig:zfill_RB_WSW}, which plots the horizontal fractional volumes
  $f_{V,i}(z)$ defined in Eq.~\eqref{fva},
  helps reveal how the thermal gas structure 
  depends on the cooling function. 
  Model~{\WSWa}, panel {{(a)}}, has significantly more very cold gas
  ($T<50\K$; black, solid) than RBN, panel {(b)}, but slightly warmer cold
  gas ($T<500\K$; purple, dash-dotted) is more abundant in RBN. 
  The warm and hot phases ($T>5\times10^3\K$) have roughly similar
  distributions in both models, although Model~RBN has less of both phases.
  Apart from relatively minor details, the effect of the form of the cooling
  function thus appears to be straightforward and predictable: 
  stronger cooling means more cold gas, and vice versa.
  What is less obvious, however, is that the very hot gas is more abundant near
  $\pm1\kpc$ in Model~RBN than in {\WSWa}, indicating that the typical
  densities must be much lower. 
  This, together with the greater abundance of cooler
  gas near the mid-plane, suggest that there is less stirring with RBN cooling.

%-----------------------------------------------------------------------------
  \begin{table}[tb]
  \centering
    \caption[Key to multiple temperature bands]{
  Key to Figs.~\ref{fig:zfill} and \ref{fig:zfill_RB_WSW},
  defining the gas temperature bands used there (same as
  Table~\ref{table:bands}).
  \label{table:bands2}}
  \begin{tabular}{ccl}
  \hline
  Temperature band   &Line style & Phase \\
  \hline
    \phantom{$5\times10^1\K<$}$T<5\times10^1\K$    &\rule[0.08cm]{1.05cm}{0.75pt}                                   & cold \\
    $5\times10^1\K\leq T<5\times10^2\K$            &{\Large\color{violet}{$\cdot$-$\cdot$-$\cdot$-$\cdot$-$\cdot$}} & cold \\
    $5\times10^2\K\leq T<5\times10^3\K$            &{\Large\color{blue}{- - - - -}}                                 & warm \\
    $5\times10^3\K\leq T<5\times10^4\K$            &{\Large\color{royalblue}{-- -- -- -}}                           & warm \\
    $5\times10^4\K\leq T<5\times10^5\K$            &{\Large\color{mygreen}{--}\color{black}{-}\hspace{-0.08cm} \color{mygreen}{$\cdot$}\hspace{-0.03cm}\color{black}{-}\hspace{-0.08cm} \color{mygreen}{$\cdot$}\hspace{-0.03cm}\color{black}{-}\hspace{-0.08cm}  \color{mygreen}{--}}                  & warm \\
    $5\times10^5\K\leq T<5\times10^6\K$            &{\Large\color{burntorange}{--$\cdots$--$\cdot$}}                & hot  \\
    \phantom{$5\times10^1\K<$}$T\geq5\times10^6\K$ &{\Large\color{red}{$\cdots\cdots$}}                             & hot  \\
  \hline
  \end{tabular}
  \end{table}
%-----------------------------------------------------------------------------

  The two models are further compared in Fig.~\ref{fig:pdfs_cool}, where
  separate 
  probability distributions are shown for the gas density, temperature and
  thermal pressure.
  With both cooling functions, the most probable gas number density is around
  $3 \times 10^{-2}\cm^{-3}$; the most probable temperatures are  also similar,
  at around $3 \times 10^{4}\K$.
  With the RBN (blue, dashed) cooling function, the density range extends to 
  smaller densities than with {\WSWa} (black, solid);
  on the other hand, the temperature range for {\WSWa}
  extends both lower and higher than for RBN. 
  It is evident that the isobarically unstable branch of WSW cooling 
  significantly reduces the amount of gas in the 313--6102\,K temperature range
  (dark blue, dashed in Fig.~\ref{fig:zfill_RB_WSW})
  and increases the amount below 100\,K. 
  However this is not associated with higher densities than RBN. 
  This may be indicative that multiple compressions, rather than thermal 
  instability, dominate the formation of dense clouds.
  The most probable thermal pressure is lower in Model~RBN than in {\WSWa}, 
  consistent with the lower thermal energy content of the former, and with
  the differences between the 2D plots in Fig.~\ref{fig:pdf2dop}a (WSW) and
  Fig.~\ref{fig:rb2d}a (RBN).
  However, in contrast, the pressure at the mid-plane (Fig.~\ref{fig:pdf2dcomp})
  is higher for Model~RBN than for Model~\WSWa. 
  
%-----------------------------------------------------------------------------
  \begin{figure*}[h]
  \centering
  \hspace{-2.0cm}
  \includegraphics[width=0.36\columnwidth,clip=true,trim=0 0 0 10mm]{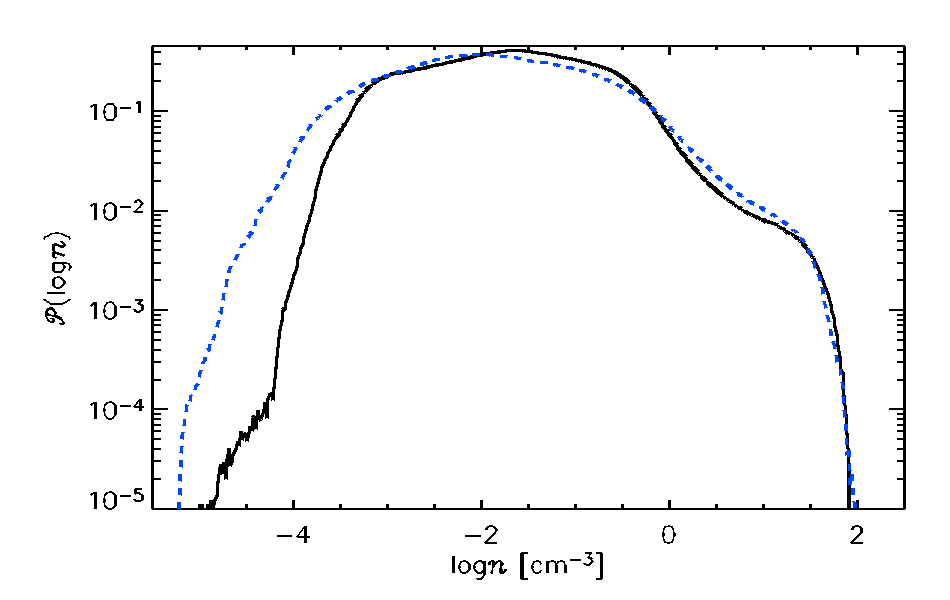}
  \includegraphics[width=0.36\columnwidth,clip=true,trim=0 0 0 10mm]{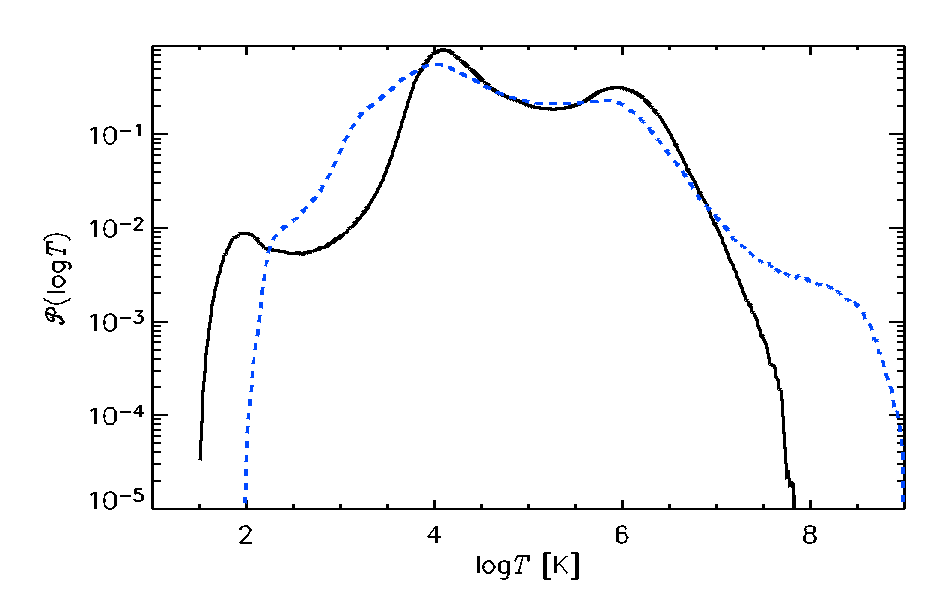}
  \includegraphics[width=0.36\columnwidth,clip=true,trim=0 0 0 10mm]{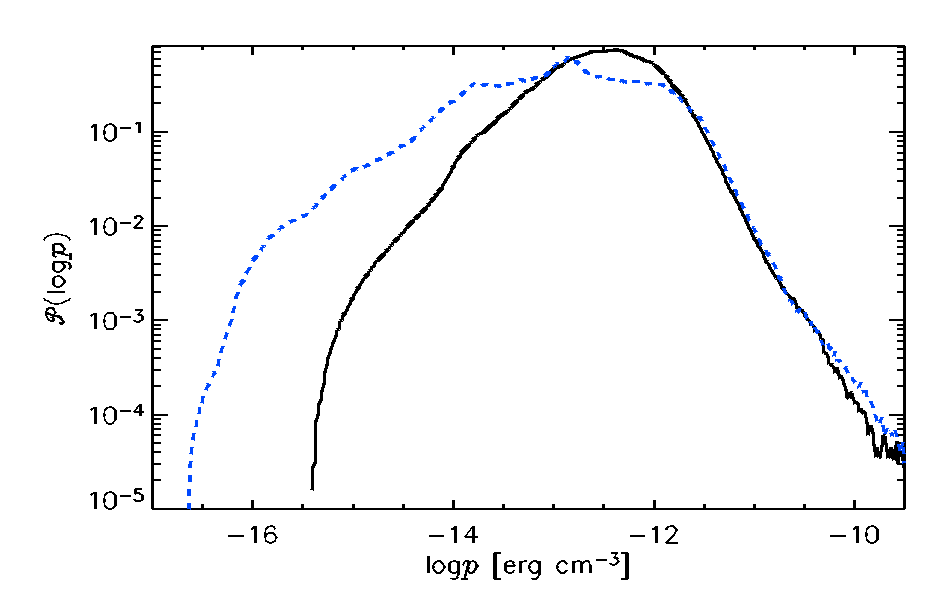}
  \hspace{-2.0cm}
    \caption[Total volume probability distributions for Models~RBN and \WSWa]{
  Volume weighted probability distributions for \textbf{(a)}~gas density,
  \textbf{(b)}~temperature and \textbf{(c)}~thermal pressure,
  for Model~{RBN} (blue, dashed) and Model~{\WSWa} (black, solid),
  in a statistical steady state, each averaged over 21 snapshots spanning 
  20\,Myr (RBN: 266 to $286\Myr$, and {\WSWa}: 305 to $325\Myr$) and the total
  simulation domain $|z|\le1.12\kpc$.
    \label{fig:pdfs_cool}
            }
  \end{figure*}
%-----------------------------------------------------------------------------
  
  The density and temperature probability distributions for {\WSWa} are similar
  to those obtained by \citet[][their Fig.~7]{Joung06}, who used a similar 
  cooling function, despite the difference in the numerical methods (adaptive
  mesh refinement down to $1.95\p$ in their case). 
  With slightly different implementation of the cooling and heating processes,
  again with adaptive mesh refinement down to $1.25\p$, \citet[][their 
  Fig.~3]{AB04} found significantly more cool, dense gas.
  It is noteworthy that the maximum densities and lowest temperatures 
  obtained in my study with a non-adaptive grid are of the same order of
  magnitude as those from AMR-models, where the local resolution is up to 
  three  times higher.
  At $4\p$ resolution, the mean minimum temperature is $34\K$, within the range 
  $15$--$80\K$ for 0.625--$2.5\p$ \citep[][their Fig.~9]{AB04}.  
  For mean maximum gas number, my value of $122\cmcube$ is within their range 
  $288$--$79\cmcube$.

%-----------------------------------------------------------------------------
  \begin{figure}[h]
  \centering
%  \hspace{-1.95cm}
  \includegraphics[width=0.55\columnwidth]{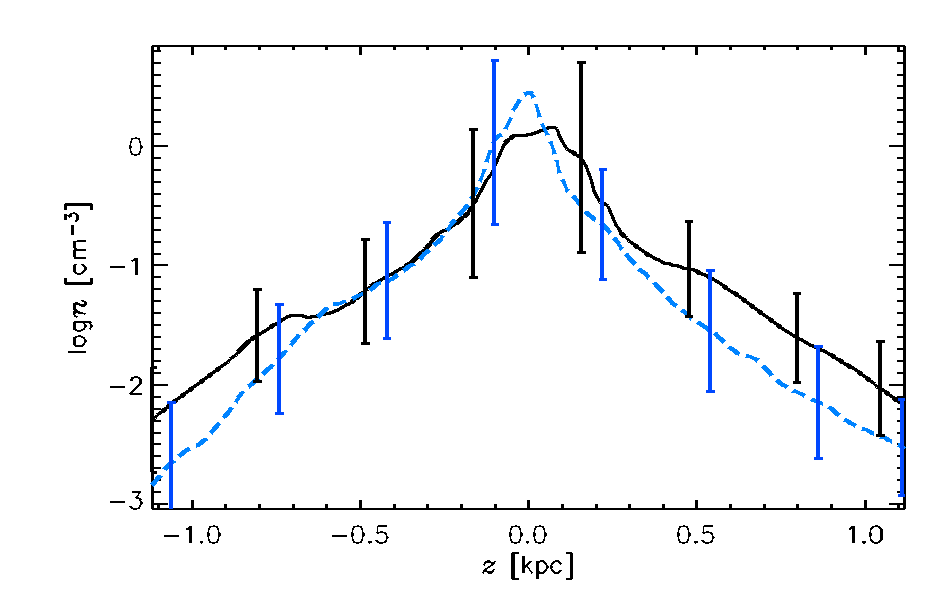}
%  \hspace{-0.525cm}
%  \includegraphics[width=0.55\columnwidth]{fig/zrprb.png}
%  \hspace{-1.75cm}
    \caption[Horizontal averages for $n$ from Models~RBN and \WSWa]{
  Horizontal averages of gas number density, $\mean{n}(z)$, % {\textbf{(a)}}, and
  %total pressure, $P(z)$ {\textbf{(b)}}, 
  for Model~\WSWa\ (solid, black) 
  and  Model~RBN (dashed, blue). 
  Each are time-averaged using eleven snapshots.
  The vertical lines indicate standard deviation within each horizontal slice.
%  The thermal $p(z)$ (dotted) and ram $p\turb(z)$ (fine dashed) pressures are
%  also plotted {\textbf{(b)}}.
    \label{fig:zrb}
            }
  \end{figure}
%-----------------------------------------------------------------------------
  
  In Fig~\ref{fig:zrb} the horizontal averages of the gas number density $n$
%  (panel a) and the thermal $p$, turbulent $p\turb$ and total $P$ pressures
%  (b) 
  are plotted for Models~\WSWa\ (black) and RBN (blue).
  The density scale height is lower for Model~RBN, 
  indicative that the stronger cooling is inhibiting the circulation of hot gas
  away from the mid-plane. 
    
%-----------------------------------------------------------------------------
  \begin{figure*}%[t]
  \centering
  \includegraphics[width=0.45\columnwidth,clip=true,trim=0 0 0 9mm]{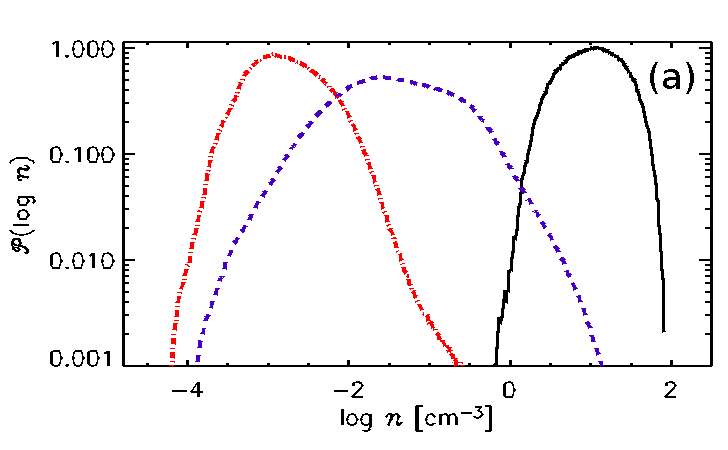}
  \includegraphics[width=0.45\columnwidth,clip=true,trim=0 0 0 9mm]{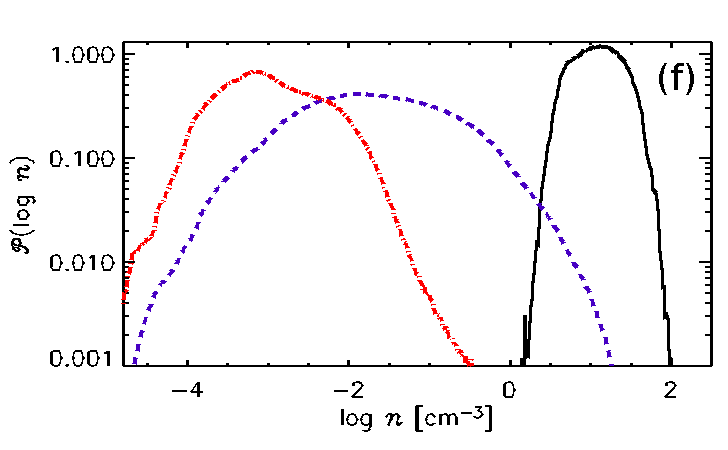}\\
%  \hspace{-1.5cm}
  \includegraphics[width=0.45\columnwidth,clip=true,trim=0 0 0 9mm]{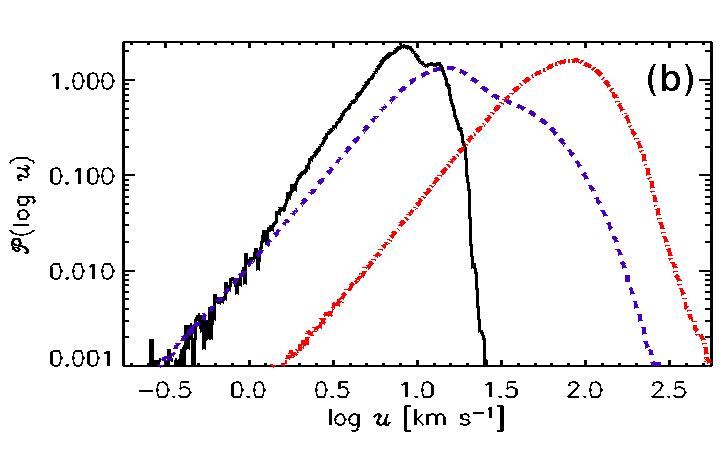}
  \includegraphics[width=0.45\columnwidth,clip=true,trim=0 0 0 9mm]{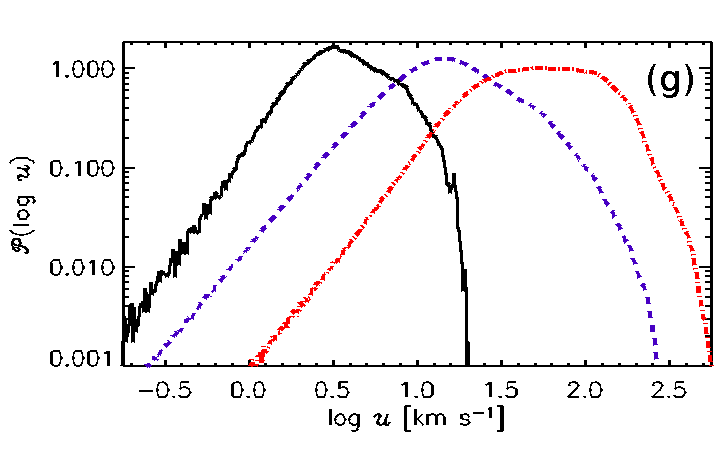}\\
%  \hspace{-1.5cm}
%  \hspace{-1.5cm}
  \includegraphics[width=0.45\columnwidth,clip=true,trim=0 0 0 9mm]{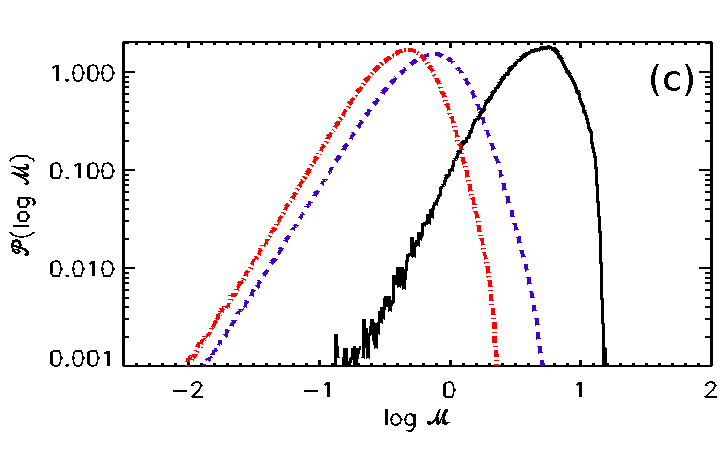}
  \includegraphics[width=0.45\columnwidth,clip=true,trim=0 0 0 9mm]{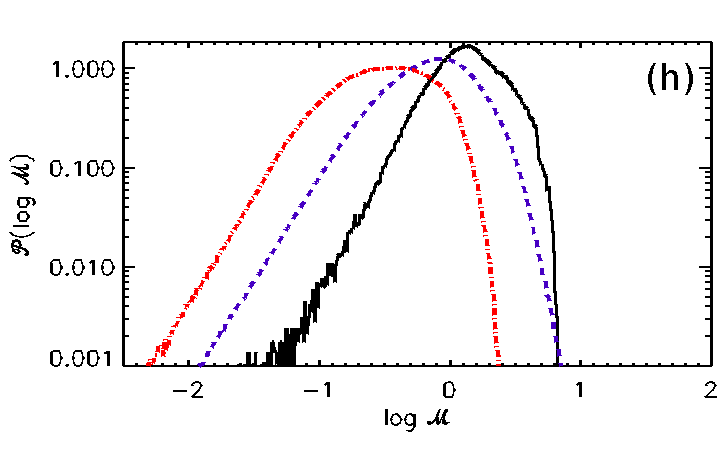}\\
%  \hspace{-1.5cm}
%  \hspace{-1.5cm}
  \includegraphics[width=0.45\columnwidth,clip=true,trim=0 0 0 9mm]{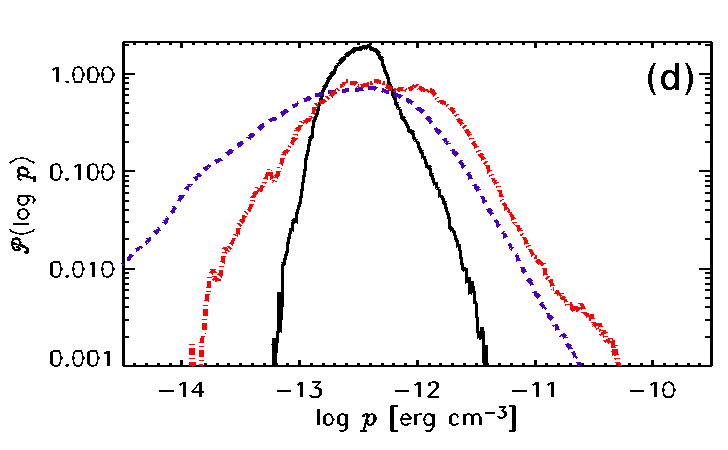}
  \includegraphics[width=0.45\columnwidth,clip=true,trim=0 0 0 9mm]{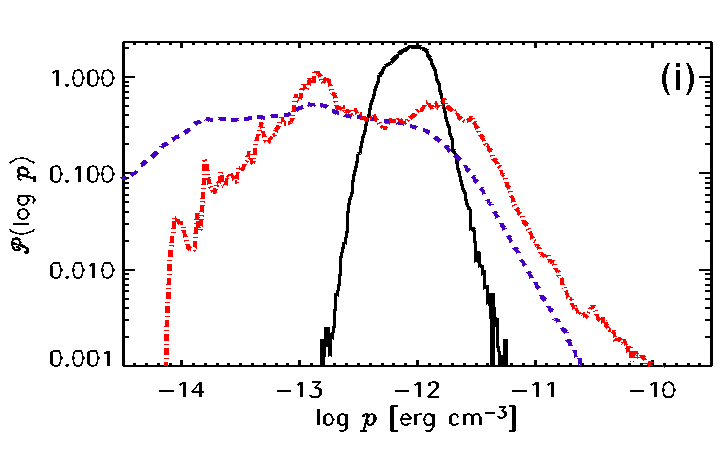}\\
%  \hspace{-1.5cm}
%  \hspace{-1.5cm}
  \includegraphics[width=0.45\columnwidth,clip=true,trim=0 0 0 9mm]{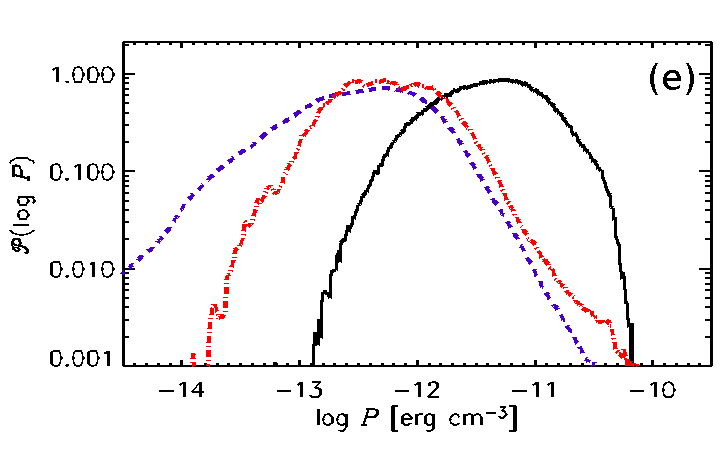}
  \includegraphics[width=0.45\columnwidth,clip=true,trim=0 0 0 9mm]{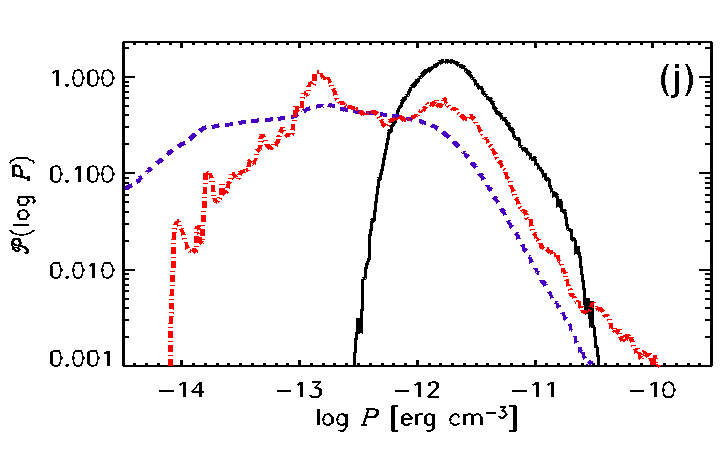}
%  \hspace{-2.5cm}
    \caption[Probability distributions by phase for Models~RBN and \WSWa]{
  Total volume phase probability distributions for Model~\WSWa\
  (left column) and Model~RBN (right column): 
  gas number density $n$ \textbf{(a)},\textbf{(f)}; 
  turbulent velocity $u\turb$ \textbf{(b)},\textbf{(g)};
  Mach number defined with respect to  
  local sound speed and $u\turb$ \textbf{(c)},\textbf{(h)};
  thermal pressure \textbf{(d)},\textbf{(i)}; and 
  total pressure \textbf{(e)},\textbf{(j)}.
  Cold phase $T<500\K$ (black, solid), warm $500 < T < 5\times 10^5\K$ (blue,
  dashed) and hot $T>5\times10^5\K$ (red, dash-dotted). 
  Eleven snapshots have been used for averaging, spanning $t=266$ to $286\Myr$
  for Model~RBN and  $t=305$ to $325\Myr$ for Model {\WSWa}.
    \label{fig:wsw_pdf3ph}
    \label{fig:wsw_pdf3pha}
            }
  \end{figure*}
%-----------------------------------------------------------------------------

  The probability distributions of Fig.~\ref{fig:pdfs_cool} do not show clear
  separations into phases (cf.\ the distributions shown by 
   \citet{Joung06,AB04}); thus division into three phases would
  arguably only be conventional, if based on these distributions alone. 
  Based on the analysis described in Section~\ref{sect:TMPS}, however the 
  assertion
  that the complicated thermal structure can indeed be reasonably described in
  terms of three phases remains justified.

  Distributions of density obtained for the individual phases from Models~\WSWa\
  and RBN, shown in Fig.~\ref{fig:wsw_pdf3pha}a and f respectively, 
  confirm the clear phase separation in both cases.
  This is also clear from the perturbation velocity in 
  Fig~\ref{fig:wsw_pdf3ph}b and g. 
  Here the same borderline temperatures are defined for individual phases as
  for Model~{\Op} (Fig.~\ref{fig:pdf3ph}).
  Despite minor differences between the corresponding panels 
  in Figs.~\ref{fig:pdf3ph} and \ref{fig:wsw_pdf3pha},
  the peaks in the gas density probability distributions are close to
  $10^1, 3 \times 10^{-2}$ and $10^{-3}\cm^{-3}$ in all models. 
  The similarity in the
  properties of the cold gas suggests that the radiative cooling (different in
  Models RBN and WSW) is less important than adiabatic cooling at these
  scales.
  Given the extra cooling of hot gas and reduced cooling of
  cold gas with the RBN cooling function, 
  more of the gas resides in the warm phase in Model~RBN. 
  The probability distribution for the Mach number in the warm gas extends to
  higher values with the RBN cooling function, perhaps because more shocks
  reside in the more widespread warm gas, at the expense of the cold phase. 
  It is useful to remember that, although each distribution in this figure is normalised to
  unit underlying area, the fractional volume of the warm gas is about a
  hundred times that of the cold.

  The thermal pressure distribution in the hot gas reveals the two `types'
  (see the end of section~\ref{sect:TMPS}),
  which are mostly found within $|z|\la200\p$ 
  (high pressure hot gas within SN remnants) 
  and outside this layer (diffuse, lower pressure hot gas). 
  The broader pressure distributions in both models, as compared to 
  Fig.~\ref{fig:pdf3ph}, indicate that the enhanced mass, particularly in the
  SN active regions, adds resistance to the release of hot gas from the 
  mid-plane through blow outs and convection.
  The pressure builds up at the mid-plane without release to the halo, leading
  to a stronger vertical pressure gradient.
  This is also evident in the MHD models (Sections~\ref{subsect:params} and
  \ref{chap:MHD}), where the circulation of hot gas out of the disc is 
  partially inhibited by the presence of a predominantly horizontal magnetic
  field. 
  Note that the location of the cold gas thermal pressure 
  (Fig.~\ref{fig:wsw_pdf3pha}i) is offset from 
  the hot pressure peak, but for total pressure (panel j) they are well aligned.

  In conclusion, the properties of the cold and warm phases 
  are not strongly affected by the choice of the cooling function. 
  The main effect is that the RBN cooling function produces slightly
  less hot gas, but extended to higher temperatures with significantly lower
  pressures.
  This can readily be understood, as this function provides significantly 
  stronger cooling at $T\ga10^3\K$.

%---------------------------------------------------------------------------- 
  \section{Sensitivity to rotation, shear and SN rate\label{subsect:params}}
%---------------------------------------------------------------------------- 
   
  More thorough analysis of the effects of the other model parameters included
  here will need to follow in future work.
  Direct comparison between Models~$\Ompd$ and $\Ompa$ in the saturated state
  may be used to consider the effect of rotation, but further studies with
  reduced rotation and a range of rotation rates would be more informative.
  The effects of shear and supernova rate can only be directly considered in
  the kinematic stage for the MHD runs with Models~$\Ompc$ and $\Ompe$, and
  it would also be helpful to consider a range of these parameters, to get a
  fuller appreciation of their effects.

  From Table~\ref{table:results}, it is evident that even in the kinematic stages there are significant 
  differences between the HD and MHD models, so
  that the presence of a weak magnetic field cannot be ignored. 
  For example in Column~10 of the table, the sound speeds $\average{c\sound}$
  of the MHD models are between one quarter and one half those of the HD models,
  indicating that the mean temperatures are much lower for MHD.
  Yet the thermal energies $\average{\eT}$ (15) are very similar. 
  The velocities for the MHD models, both perturbation $\average{u\rrms}$ (13) and turbulent 
  $\average{u\turb}$ (14), are around one fifth those of the
  HD models.
  The kinetic energy $\average{\eK}$ (16) is also reduced, even in the kinematic 
  stage, to about one half to one third of the HD values. 

  Two significant changes to the models were included for the MHD runs.
  The location algorithm for the SNe was corrected to eliminate unphysical oscillation 
  of the disc, and a numerical patch to replace physical mass loss was applied 
  (see Appendix~\ref{subsect:MOL} for details of both).
  The former may indicate that $\average{u\rrms}$ and $\average{\eK}$ have been 
  exaggerated in the
  HD runs.
  This also has the effect of stabilising the disc, so that the most dense gas 
  remains near the mid-plane and SN energy will be more readily absorbed. 
  This may also help to explain the reduced temperatures, without loss of
  thermal energy, and also the greatly reduced velocities, without a 
  commensurate loss of kinetic energy. 
  Similar models \citep{AB04,Joung06,Gressel08} have found clustering of SNe
  necessary to generate large enough bubbles of hot gas to blow out from the
  disc.
  It would appear to be necessary to include this, especially for the MHD models
  where the thickness of the disc increases.
  There may also be some unintended effect of replacing the mass.
  For a direct comparison, an HD model labelled $\Omph$ was run with the same parameters as Model~\Op, but
  with these two changes implemented 
  (see Section~\ref{sect:3BB}). 
  Without the magnetic field, temperatures are substantially higher with
  $\average{c\sound}\simeq44\kms$, but still much lower than before the
  numerical adjustments. 
  The thermal energy $\average{e_{\textrm{th}}}$ remains similar. 
  The velocities are larger than with MHD, but substantially reduced from the
  original HD runs.
  There is some reduction in $\average{e_{\textrm{kin}}}$ compared to the 
  original HD runs, but remains slightly higher than for MHD.

  In conclusion, the MHD models probably understate the temperature 
  differentials, with cold gas and hot gas under represented, because a 
  lack of clustering prevents adequate blowouts of hot gas. 
  Due to the improved stability of the disc, however, the velocity and in 
  particular the are
  kinetic energy is probably more accurately reproduced, and the higher values
  in the earlier HD models are unphysical manifestations of the vertical disc
  oscillations. 
  The vertical density distribution in Model~$\Omph$ is similar to that of 
  Model~\Op, suggesting that the numerical recycling of mass is not 
  strongly distorting the dynamics.

%-----------------------------------------------------------------------------  
  \subsubsection{Rotation}
%-----------------------------------------------------------------------------  
  
  Comparing the results in Table~\ref{table:results} Column~10, 
  $\average{c\sound}\simeq23\kms$ for Model~$\Ompa$ is slightly higher than 
  $21\kms$ for Model~$\Ompd$.
  However the thermal energy 
  $\average{\eT}\simeq25 E_{\mathrm{SN}}\kpc^{-3}$ 
  (Column~15) is lower than $26 E_{\mathrm{SN}}\kpc^{-3}$ for
  Model~$\Ompd$.
  The velocities (Columns~13 and 14) are very similar, as is the kinetic
  energy $\average{\eK}$ (16).
  The total energy is even higher for Model~$\Ompd$ with the doubled rotation,
  when the magnetic energy $\average{\eB}$ (17) is included.

  The probability distributions of the various quantities (density, temperature,
  pressure, etc.) do not differ greatly as a result of rotation, although the
  higher rotation inhibits very hot or cold gas.
  The horizontal averages of density, temperature and pressure are also very
  similar.
  So the dominant effect appears to be in the detailed structure of the 
  velocity field, particularly the extent of vorticity. 
  For 
  Model~$\Ompa$ $\average{|\nabla\times\vect{u}|}=194\kms\kpc^{-1}$ and
  $\average{|\nabla\cdot\vect{u}|}=185\kms\kpc^{-1}$, while these are 
  178 and $144\kms\kpc^{-1}$ respectively for Model~$\Ompd$. 
  So although there is more vorticity in Model~$\Ompa$, the ratio of the
  stream function to potential flow is greater for Model~$\Ompd$.
  %-----------------------------------------------------------------------------  
  \subsubsection{Shear}
%-----------------------------------------------------------------------------  
  
%-----------------------------------------------------------------------------
  \begin{figure}[h]
  \centering
  \hspace{-1.95cm}
  \includegraphics[width=0.55\columnwidth]{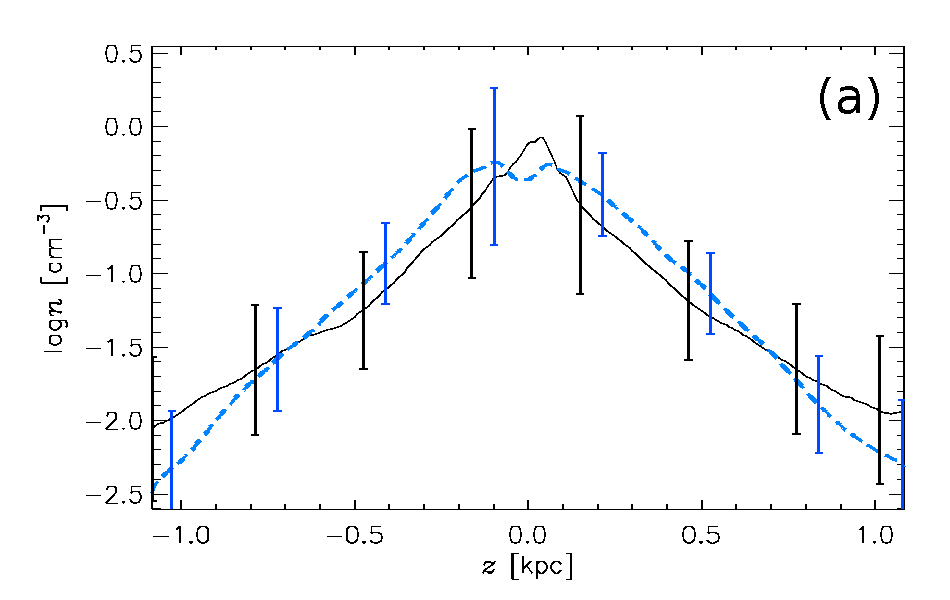}
  \hspace{-0.525cm}
  \includegraphics[width=0.55\columnwidth]{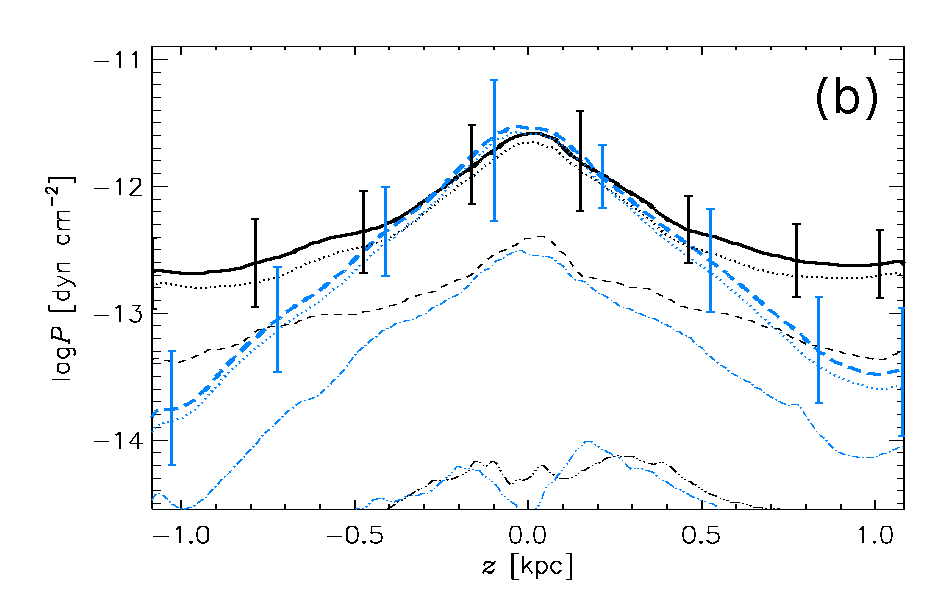}
  \hspace{-1.75cm}
    \caption[Horizontal averages of $n$ and $P$ for Models~{$\Ompa^*$} and {$\Ompe$}]{
  Horizontal averages of {\textbf{(a)}} gas number density, $\mean{n}(z)$, and
  {\textbf{(b)}} total pressure, $\mean{P}(z)$, for Model~{$\Ompa^*$} (solid, black), 
  and  Model~{$\Ompe$} (dashed, blue). 
  Each is time-averaged using eleven snapshots spanning 600 to 700\Myr.
  The vertical lines indicate standard deviation within each horizontal slice.
  The thermal $\mean{p}(z)$ (dotted) and ram $\mean{p}\turb(z)$ (fine dashed) pressures are
  also plotted {\textbf{(b)}}.
  The magnetic pressure $\mean{p}_B$ is also plotted (fine, dash-3dotted).
    \label{fig:zshear}
            }
  \end{figure}
%-----------------------------------------------------------------------------
  
  The effect of increasing the shear (Model~$\Ompe$) can be seen from the 
  comparisons with the results for Model~$\Ompa^*$, for the kinematic stage
  only.
  Temperatures are reduced, with
  $\average{c\sound}\simeq24\kms$ against $\average{c\sound}\simeq58\kms$,
  although in line with the other MHD models following the saturation of the
  dynamo, and also with $\Ompc$ for the kinematic stage only.
  However thermal 
  energy $\average{e_{\mathrm{th}}}\simeq29 E_{\mathrm{SN}}\kpc^{-3}$ is 
  enhanced compared to Model~$\Ompa^*$.
  The velocities are almost half that of Model~$\Ompa^*$, with kinetic energy
  only slightly diminished, and matching that of Model~$\Ompc$.
  
  Otherwise shear appears to make little difference to the hydrodynamics. 
  The probability distributions for the density, temperature, thermal and
  turbulent pressures are insensitive to the shear, except that the 
  hot gas extends to significantly lower densities than for the other models.
  However, the shear does affect the vertical structure of the model, unlike
  the effect of increased rotation.
  In Fig.~\ref{fig:zshear} the horizontal averages of density and pressure are
  plotted for Models~$\Ompa^*$ (black, taken from its kinematic 
  period) and $\Ompe$ (blue).
  The scale height of the density (panel a) is narrower for Model~$\Ompa^*$,
  with a higher density at the mid-plane. 
  Given the twin peaks in the density profile for Model~$\Ompe$, the higher
  scale height appears to be an indirect effect of the magnetic field, rather
  than of the shear itself.
  As discussed in Chapter~\ref{chap:MHD}, the regular part of the magnetic 
  field is strongest just outside of the SN active region, and can affect the
  location of the gas. 
  This is evident in the twin peaks of the magnetic pressure (panel b, blue,
  dash-triple-dotted).
  The total pressure near the mid-plane is about the same for both models, but
  falls more rapidly away from the mid-plane for Model~$\Ompe$.
  So increased shear enhances the vertical pressure gradient. 
  For 
  Model~$\Ompe$ $\average{|\nabla\times\vect{u}|}=342\kms\kpc^{-1}$ and
  $\average{|\nabla\cdot\vect{u}|}=188\kms\kpc^{-1}$.
  Although the potential flow is similar to Model~$\Ompa$, and stronger than 
  Model~$\Ompd$ after saturation of the dynamo, for Model~$\Ompa^*$ in the 
  kinematic stage $\average{|\nabla\times\vect{u}|}=737\kms\kpc^{-1}$ and
  $\average{|\nabla\cdot\vect{u}|}=295\kms\kpc^{-1}$.
  So the magnitude of the flows and the ratio of rotational to irrotational
  flow (1.8) are smaller than for Model~$\Ompa^*$ (2.5).
  
%-----------------------------------------------------------------------------  
  \subsubsection{Supernova Rate}
%-----------------------------------------------------------------------------  

  The reduction in SN rate in Model~$\Ompc$ predictably reduces the temperatures
  $\average{c\sound}$ and the thermal energy
  $\average{e_{\mathrm{th}}}$, but velocities and kinetic energy are similar 
  to Model~$\Ompe$, also in the kinematic stage (see Table~\ref{table:results}).
 
  There is more cold gas than for
  the other models, and there is less very diffuse gas.
  Otherwise the distributions of gas density, temperature and pressure, and the
  multi-phase structure are not strongly affected.
  
%-----------------------------------------------------------------------------
  \begin{figure}[h]
  \centering
  \hspace{-1.95cm}
  \includegraphics[width=0.55\columnwidth]{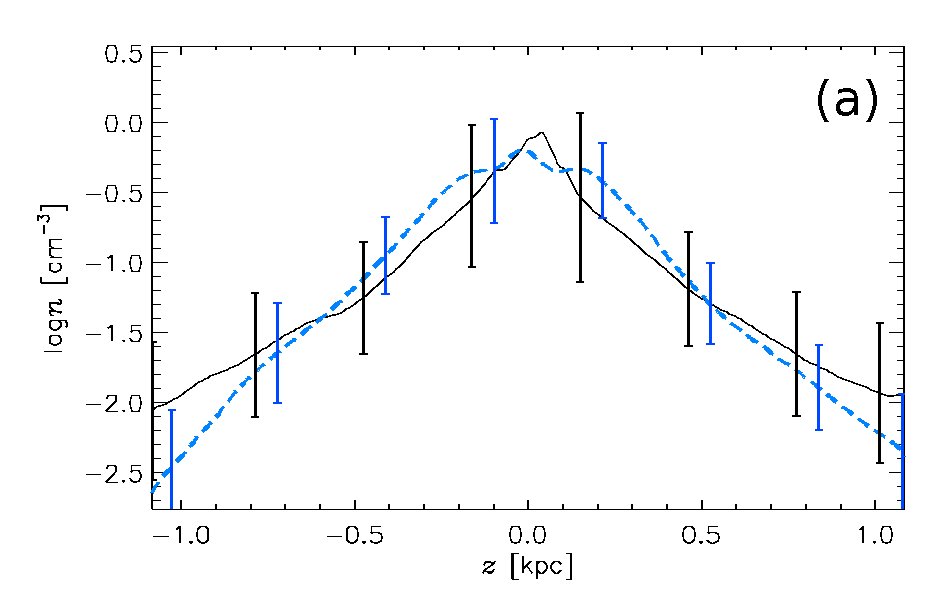}
  \hspace{-0.525cm}
  \includegraphics[width=0.55\columnwidth]{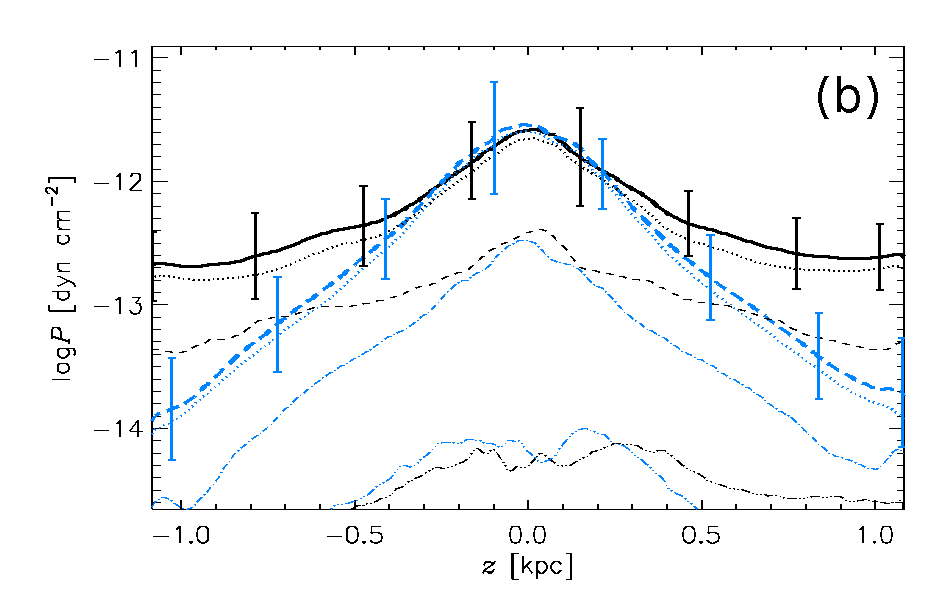}
  \hspace{-1.75cm}
    \caption[Horizontal averages of $n$ and $P$ for Models~{$\Ompa^*$} and      {$\Ompc$}]{
  Horizontal averages of {\textbf{(a)}} gas number density, $\mean{n}(z)$, and
  {\textbf{(b)}} total pressure, $\mean{P}(z)$, for Model~{$\Ompa^*$} (solid, black),    
  and Model~{$\Ompc$} (dashed, blue). 
  Each are time-averaged using eleven snapshots spanning 600 to 700\Myr.
  The vertical lines indicate standard deviation within each horizontal slice.
  The thermal $\mean{p}(z)$ (dotted) and ram $\mean{p}\turb(z)$ (fine dashed) pressures are
  also plotted {\textbf{(b)}}.
  The magnetic pressure $\mean{p}_B$ is also plotted (fine, dash-3dotted).
    \label{fig:zsn8}
            }
  \end{figure}
%-----------------------------------------------------------------------------
  
  However the vertical distribution of the gas is affected by the SN rate.
  Similar to the increased shear, the reduced SN rate in Model~$\Ompc$ 
  increases the scale height of the gas density (Fig.~\ref{fig:zsn8}a blue,
  dashed). 
  Unlike Model~$\Ompc$, there is a distinctive peak in density at the mid-plane.
  For 
  Model~$\Ompc$ $\average{|\nabla\times\vect{u}|}=251\kms\kpc^{-1}$ and
  $\average{|\nabla\cdot\vect{u}|}=153\kms\kpc^{-1}$.
  The potential flow is predictably reduced compared to Model~$\Ompa^*$, and
  also to Model~$\Ompe$.
  The ratio of rotational to irrotational flow is also reduced, at 1.6. 
  The scale height of the gas density may be related to this ratio, given the 
  similarities of Models~$\Ompc$ and $\Ompe$, with 1.6 and 1.8 respectively,
  and the similarities of Models~$\Ompa^*$ and $\Ompd$, with 1.0 and 1.2
  respectively.
  Alternatively there could be independent causes: the strong alignment of the magnetic field in the
  case of Model~$\Ompe$ and the reduced hot gas circulation induced by the SN 
  forcing in Model~$\Ompc$.

  \section{Summary}
  
  The cooling function affects the detail of the multi-phase structure, 
  particularly the abundance of gas in the thermally unstable range, and
  the overall energy, because of differences in the cumulative cooling effect.
  However the separation into phases is not strongly sensitive to the form 
  of the cooling function.
  Typical density, velocity and modal temperatures are quite consistent for 
  each phase, independent of the cooling function.
  Supernova rate affects the thermodynamic properties, so reduced SNe produces
  cooler, less turbulent gas, given the same gas surface density. 
  The perturbation and turbulent velocities, as well as the temperature are
  not sensitive to the rotation or shearing rate. 
  However the relative vorticity of the system is affected by changes to any
  of these parameters.

\end{chapter}

  \part{Galactic magnetism: mean field and fluctuations}\label{part:mvrb}
%  \clearpage                            % End the current page making sure all
%  \thispagestyle{empty}                 % tables/figures are printed.
%  \cleardoublepage                      % Necessary for correct page numbering.
  %----------------------------------------------------------------------------
\begin{chapter}{Mean field in anisotropic turbulence
\label{chap:meanB}}

  The ISM of spiral galaxies is extremely inhomogeneous, with SNe driving
  highly compressible, transonic turbulent motions,
  yet it supports magnetic fields at a global scale of a few kiloparsecs.
  Mean-field dynamo models have proven successful in modelling galactic 
  magnetic fields, and offer a useful framework to study them and to interpret
  their observations \citep[e.g.,][]{BBMSS96,S07}. 
  Turbulent dynamo action involves two distinct mechanisms.   
  The \textit{fluctuation dynamo\/} relies solely on the random nature of the
  fluid flow to produce \textit{random\/} magnetic fields at scales smaller
  than the integral scale of the random motions. 
  The \textit{mean-field\/} dynamo produces magnetic field at a scale 
  significantly larger than the integral scale, and requires rotation and
  stratification to do so.
  For any dynamo mechanism, it is important to distinguish the 
  \textit{kinematic\/} stage when magnetic field grows exponentially as it
  is too weak to affect fluid motions, and the \textit{nonlinear\/} stage when
  the growth is saturated, and the system settles to a statistically steady 
  state.  

  The scale of the mean field produced by the dynamo is controlled by the
  properties of the fluid flow. 
  For example, in the simplest $\alpha^2$-dynamo in a homogeneous, infinite
  domain, the most rapidly growing mode of the mean magnetic field has 
  scale of order $4\upi\eta\turb/\alpha$, where $\alpha$ can be understood as 
  the helical part of the random velocity and $\eta\turb$ is the turbulent
  magnetic diffusivity \citep{SSR83}. 
  For any given $\alpha$ and $\eta\turb$, 
  this scale is finite and the mean field produced is, of course, not uniform. 

  In galaxies, the helical turbulent motions and differential rotation drive
  the so-called $\alpha\omega$-dynamo, where the mean field has a radial scale
  of order $\Delta r\simeq3|{\cal{D}}|^{-1/3}(hR)^{1/2}$ at the kinematic stage
  \citep{SS89}, with ${\cal{D}}$ the dynamo number, 
  $h\simeq0.5\kpc$ the half-thickness of the dynamo-active layer and
  $R\simeq3\kpc$ the scale of the radial variation of the local dynamo number. 
  For $|{\cal{D}}|=20$, this yields $\Delta r\simeq1.3\kpc$.

  These estimates refer to the most rapidly growing mode of the mean magnetic 
  field in the kinematic dynamo; it can be accompanied by higher modes that
  have a more complicated structure.
  Magnetic field in the saturated state can be even more inhomogeneous due to
  the local nature of dynamo saturation.
  The mean magnetic field can have a nontrivial, three-dimensional
  spatial structure, and any analysis of global magnetic structures must 
  start with the separation of the mean and random (fluctuating) parts.
  However, many numerical studies of mean-field dynamos define the mean
  magnetic field as a \textit{uniform\/} field obtained by averaging over the
  whole volume available; or in the case of fields showing non-trivial
  variations in a certain direction, as planar averages, e.g., over horizontal
  planes for systems that show $z$-dependent fields \citep[e.g.,][]{BS05}.

  The mean and random magnetic fields are assumed to be separated by a scale,
  $\lambda$, of order the integral scale of the random motions, $l\turb$;
  $\lambda$ is not necessarily precisely equal to $l\turb$, however, and must be
  determined separately for specific dynamo systems.
  The leading large-scale dynamo eigenmodes themselves have extended Fourier
  spectra, both at the kinematic stage and after distortions by the dynamo
  nonlinearity. 
  Thus, both the mean and random magnetic fields are expected to have a broad
  range of scales, and their spectra can overlap in wavenumber space.
  It is therefore important to develop a procedure to isolate a mean magnetic field
  without unphysical constraints on its spectral content.
  This problem is especially demanding in the multi-phase ISM, where the
  extreme inhomogeneity of the system can complicate the spatial
  structure of the mean magnetic field.

  The definition of the mean field as a horizontal average may be appropriate 
  in some simplified numerical models, where the vertical component of the mean 
  magnetic field, $\mean{B_z}$, vanishes because of periodic boundary
  conditions applied in $x$ and $y$; otherwise,
  $\nabla\cdot\mean{\vect{B}}=0$ cannot be ensured.
  An alternative averaging procedure that retains three-dimensional spatial
  structure within the averaged quantities is volume averaging with a kernel
  $G_\ell(\vect{r}-\vect{r'})$,  where $\ell$ is the averaging length:
$\average{f(\vect{r})}_\ell=\int_V  f(\vect{r}')
  G_\ell(\vect{r}-\vect{r'})\,d^3\vect{r}'$, for a scalar random field $f$. 
  
  \subsubsection{Reynolds rules}
  A difficulty with kernel volume averaging, appreciated early in the
  development of
  turbulence theory, is that it does not obey the Reynolds rules of the mean
  (unless $\ell\to\infty$); i.e.
  \begin{equation}
    \label{eq:RR}
    \average{\average{f}_\ell g}_\ell\neq\average{f}_\ell\average{g}_\ell \qquad
    {\textrm{and}} \qquad
    \average{\average{f}_\ell}_\ell\neq\average{f}_\ell
  \end{equation}
  \citep[Sect.~3.1 in][]{MY07}.
  If instead $\ell\to\infty$ and the mean is simply computed over the whole
  volume, then the total field can be decomposed into the mean and
  random parts, $\vect{B}=\vect{B}_0+\vect{b}$, where 
  $\vect{B}_0=\mean{\vect{B}}$ and 
  $\mean{\vect{b}}=\vect{0}$.
  The decomposition for the magnetic energies then follows conveniently as
  \begin{equation}\label{eq:bar}
    \mean{\vect{B}^2}=\mean{(\vect{B}_0+\vect{b})^2}=
                 \mean{\vect{B}_0^2+2{{\vect{B}_0\cdot\vect{b}}}+
          \vect{b}^2}
                   =\mean{\vect{B}}_0^2+\mean{\vect{b}^2}
  \end{equation}
   
  Horizontal averaging represents a special case with $\ell\to\infty$ in two
  dimensions, or alternatively can be considered an average over the total
  `volume' of each horizontal slice, and thus satisfies the Reynolds rules; 
  however, the associated loss of a large part of the spatial structure 
  of the mean field limits its usefulness.
  Kernel volume averaging retains the spatial structure, but critically 
  $\average{\vect{b}}_\ell$ in general does not equal $\vect{0}$.
  The decomposition of the total magnetic energy $\vect{B}^2$ into mean and fluctuating 
  energies cannot so conveniently be recovered from the 
  quadratic terms $\average{\vect{B}}_\ell^2$ and $\vect{b}^2$.
  An alternative formulation must be devised.

  \subsubsection{Germano approach to turbulence}
  \citet{G92} suggested a consistent approach to volume averaging which does
  not rely on the Reynolds rules.
  A clear, systematic discussion of these ideas is provided by
  \citet[][Chapter~2]{E12}, and an example of their application can be found in
  \citet{E05}.
  The averaged Navier--Stokes and induction equations remain unaltered,
  independent of the averaging used, if the mean Reynolds stresses and the mean
  electromotive force are defined in an appropriate, generalised way.
  The equations for the fluctuations naturally change, and care must be taken
  for their correct formulation.
  An important advantage of averaging with a Gaussian kernel (Gaussian
  smoothing) is its similarity to astronomical observations, where such
  smoothing arises from the finite width of a Gaussian beam, or is applied
  during data reduction.

  In this chapter, I consider the magnetic field $\vect{B}$ produced by the rotational shear
  and random motions in Model~$\Ompd$ with $\Omega=-S=2\Omega_0$. 
  $\Omega_0=25\kms\kpc^{-1}$ is the angular velocity in the Solar vicinity.
  This model has the fastest dynamo of all the magnetic runs in 
  Table~\ref{table:models}, so was adopted pragmatically to allow me to follow the dynamo 
  process from early in the kinematic stage through to saturation and beyond.
  All of the magnetic models included in this thesis, however, appear to have 
  produced a dynamo, so this analysis would be equally valid for all cases, 
  were there sufficient resources to track them fully to the non-linear stage.

  Using Gaussian smoothing, an approach is suggested, below, to determine the
  appropriate length $\ell$, and then obtain the mean magnetic field
  $\vect{B}_\ell$.
  The procedure ensures that $\nabla\cdot\vect{B}_\ell=0$. 
  The random magnetic field $\vect{b}$ is then obtained as
  $\vect{b}=\vect{B}-\vect{B}_\ell$.
  The Fourier spectra of the mean and random magnetic fields overlap in 
  wavenumber space, but their maxima are well separated.

%-----------------------------------------------------------------------------
  \section{The mean magnetic field}\label{sect:dyn}\label{subsect:mb}
%-----------------------------------------------------------------------------

%-----------------------------------------------------------------------------
  \begin{figure}
  \centering
  \includegraphics[width=0.8\columnwidth]{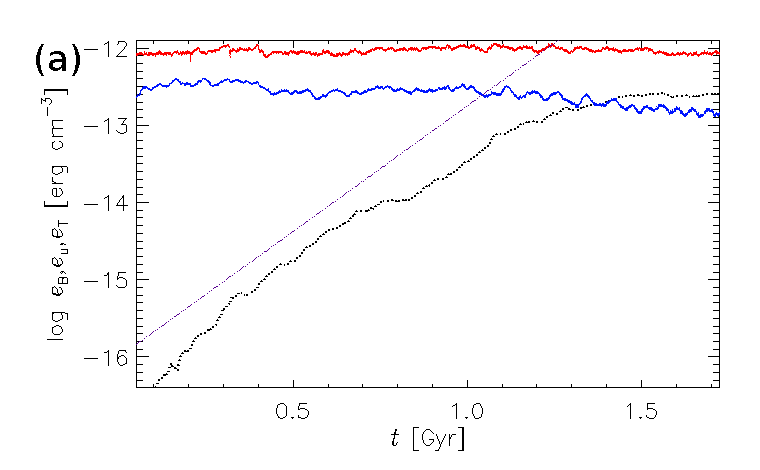}
    \caption[Growth of energy densities for Model~$\Ompd$]{
  Evolution of energy densities averaged over the whole computational domain;
  thermal $\averaged{e_{\mathrm{T}}}$ (red), kinetic
  $\averaged{e_{\mathrm{u}}}$ (blue) and magnetic $\averaged{e_{\mathrm{B}}}$
  (black).
  Guide line (purple, dotted) indicates $\exp(7.5t\Gyr^{-1})$.
  \label{fig:bbelog}}
  \end{figure}
%-----------------------------------------------------------------------------

  Figure~\ref{fig:bbelog} demonstrates the approximately exponential growth of
  magnetic energy density against the (relatively) stationary background of
  turbulent motions and thermal structure.
  As the average magnetic energy density 
  $\averaged{e_{\mathrm B}}=\averaged{|\vect{B}|^2}/8\upi$ becomes comparable to
  the average kinetic energy density $\averaged{e_{\mathrm{u}}}$ at $t>1\Gyr$, 
  the latter shows a modest reduction, as expected for the conversion of
  the kinetic energy of random motions to magnetic energy.
  The use of $\average{}$ without subscript indicates averaging over the 
  total numerical domain, analogous to the use of overbar in Eq.~\eqref{eq:bar}.
  For $t>1.4\Gyr$, the magnetic field settles to a statistically steady state
  with $\averaged{e_{\mathrm B}}\approx2.5\times10^{-13}\erg\cm^{-3}$, somewhat
  larger than $\averaged{e_{\mathrm{u}}}\approx1.6\times10^{-13}\erg\cm^{-3}$,
  while the thermal energy density
  $\averaged{e_\mathrm{T}}\approx10^{-12}\erg\cm^{-3}$ 
% is not affected by magnetic field.
  {\frcorr{appears only weakly affected. Due to changes to the flow and 
      thermodynamic composition, Ohmic heating is offset by reduced viscous 
      heating or more efficient radiative cooling.}}

  Magnetic field $\vect{B}$ can be decomposed into $\vect{B}_\ell$, the part
  averaged over the length scale $\ell$,  and the complementary fluctuations 
  $\vect{b}$,
  \begin{equation} \label{eq:meanB}
    \vect{B}=\vect{B}_\ell+\vect{b},
    \quad 
    \vect{B}_\ell=\average{\vect{B}}_\ell,
    \quad 
    \vect{b}=\vect{B}-\vect{B}_\ell,
  \end{equation}
  using volume averaging with a Gaussian kernel:
  \begin{align}\label{eq:Bxgauss}
  \average{ \vect{B}}_\ell(\vect{x})
	&=\int_{V}\vect{B}(\vect{x}')G_\ell(\vect{x}-\vect{x}')\,\dd^3\vect{x}',\\
    G_\ell(\vect{x})&=\left(2\upi \ell^2\right)^{-{3}/{2}}
	\exp\left[-{\vect{x}^2}/({2\,\ell^2})\right],\nonumber
  \end{align}
  where $V$ is the volume of the computational domain.
  This operation preserves the solenoidality of both $\vect{B}_\ell$ and 
  $\vect{b}$, and retains their three-dimensional structure.
  For computational efficiency, the averaging was performed in the Fourier
  space where the convolution of Eq.~\eqref{eq:Bxgauss} reduces to the product
  of Fourier transforms.
  \[
    \vect{\hat{B}}_\ell(\vect{k},\ell)=\vect{\hat{B}}(\vect{k})
        \exp\left\{-2( \ell \upi)^2\vect{k}\cdot
        \vect{k}\right\}\,, 
  \]
  where $\vect{\hat{B}}(\vect{k})={\cal{F}}\left[\vect{B}(\vect{x})\right]$
  and $\vect{\hat{B}}_\ell(\vect{k},\ell)={\cal{F}}\left[\vect{{B}}_\ell(\vect{x},\ell)\right]$.

  Since the averaging \eqref{eq:Bxgauss} does not obey the Reynolds rules for 
  the mean, the definitions of various averaged quantities should be 
  generalised as suggested by \citet{G92}.
  In particular, the local energy density of the fluctuation field is
  given by
  \begin{equation}\label{energyb}
  e_b(\vect{x})=\frac{1}{8\upi}\int_{V}
	|\vect{B}(\vect{x}')-\vect{B}_\ell(\vect{x})|^2
	G_\ell(\vect{x}-\vect{x}')\,\dd^3\vect{x}',
  \end{equation}
  while
 \begin{align*}
\average{e_B(\vect{x})}_\ell=&\frac{1}{8\upi}\int_{V}|\vect{B}(\vect{x}^\prime)|^2
        G_\ell(\vect{x}-\vect{x}^\prime)\dd^3\vect{x}'\\
e_{B\ell}(\vect{x})=&\frac{1}{8\upi}
        {|\vect{B}_\ell(\vect{x})|^2}
     \end{align*}
  This ensures that $\average{e_B}_\ell=e_{B_\ell}+e_{b}$.

  Expanding $\vect{B}(\vect{x}')$ in a Taylor series around $\vect{x}$, and 
  using
  \[\int_{V}G_\ell(\vect{x}-\vect{x}')\,\dd^3\vect{x}'=1
  \]
  (normalisation),
  and
  \[
  \int_{V} (\vect{x}-\vect{x}')G_\ell(\vect{x}-\vect{x}')\,\dd^3\vect{x}'=0
  \]
 (symmetry of the kernel), it can be shown that
  $e_b=|\vect{b}|^2/(8\upi)+O(\ell/L)^2$, where $L$ is the scale of the averaged
  magnetic field, defined as $L^2=|\vect{B}_\ell|/|\nabla^2\vect{B}_\ell|$ in
  terms of the characteristic magnitude of $\vect{B}_\ell$ and its second
  derivatives.
  Thus the difference between the ensemble and volume averages rapidly 
  decreases as the averaging length decreases, $\ell/L\to0$, 
  or if the averaging is performed over the whole space, $L\to\infty$;
  this quantifies the deviations from the Reynolds rules for finite $\ell/L$.

%-----------------------------------------------------------------------------
  \begin{figure}[h]
  \centering
  \hspace{-1.5cm}
  \includegraphics[width=0.535\linewidth]{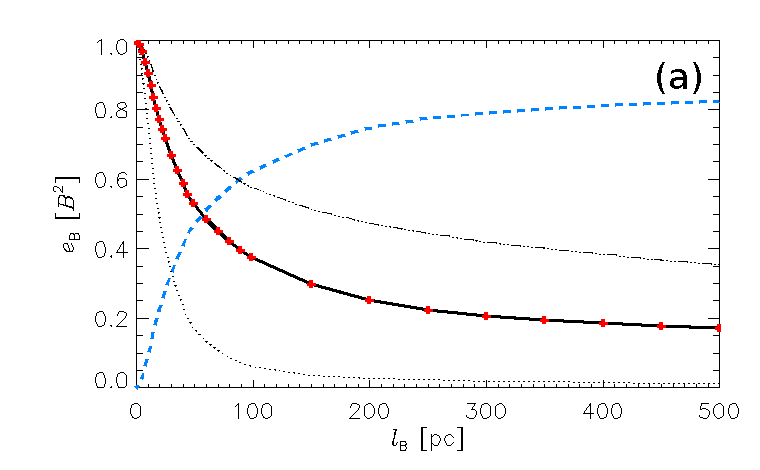}
  \includegraphics[width=0.535\linewidth]{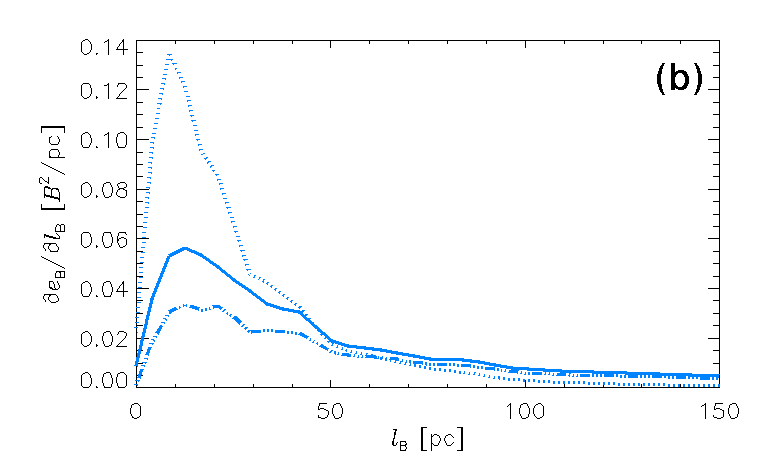}
  \hspace{-1.5cm}
  \caption[Mean and fluctuating $\eB$ for Model~$\Ompd$]{
  \textbf{(a)} Energy densities at $t=1.2\Gyr$ of mean,   
  $\averaged{e_{B_\ell}}$ (black, solid), and fluctuating, 
  $\averaged{e_b}$ (blue, dashed), magnetic fields 
  as functions of the averaging length $\ell$, normalised by 
  $\averaged{\averaged{e_{\mathrm B}}_\ell}$ and averaged over the region $|z|<0.5\kpc$; also
  $\averaged{e_{B_\ell}}$ at $t=0.8\Gyr$ (dotted) and $t=1.6\Gyr$ 
  (dash-{\frcorr{triple}}-dotted). 
  Values of $\ell$ sampled are indicated by red crosses.
  \textbf{(b)}~Derivative of $\averaged{e_b}$ from Panel (a) with respect to
  $\ell$, 
with
the same line types.
  \label{fig:blb}}
  \end{figure}
%-----------------------------------------------------------------------------

  The appropriate choice of $\ell$ is not obvious. 
  The decomposition into $\vect{B}_\ell$ and $\vect{b}$ was considered, through
  applying the averaging over a range $0<\ell<500\p$ to 37 snapshots of the
  magnetic field between $t=0.8$ and 1.7\,Gyr.
  The results for $t=0.8$, 1.2 and $1.6\Gyr$ are shown in Fig.~\ref{fig:blb}. 
  The smaller is $\ell$, the closer the correspondence between the averaged
  field and the original field (since the average is effectively sampling
  a smaller local volume), and hence the smaller the part of the total field 
  considered as the fluctuation.
  Hence $\averaged{B_\ell^2}$ is a
  monotonically decreasing function of $\ell$, whereas $\averaged{b^2}$
  monotonically increases (Fig.~\ref{fig:blb}a). 
  The curves for $t=0.8$ and 1.6\,Gyr merely demonstrate that as the field 
  grows over time, the mean field becomes more dominant.
  There is a corresponding shift downward with time of the curve for 
  $\average{e_b}$,
  so that the length $\ell_B$, where they are in equipartition, increases. A 
  length scale defining the mean would not be expected to vary, at least in 
  the kinematic stage. 
  
  It may be helpful to identify that value of $\ell$ where the variation of
  $\vect{B}_\ell$ and $\vect{b}$ with $\ell$ becomes weak enough.
  To facilitate this, consider the rate of change of the relevant quantities 
  with $\ell$, shown in Fig.~\ref{fig:blb}b (note the different scale of the
  horizontal axis in this panel). 
  The length $\ell\approx50\p$ is clearly distinguished:
  all the curves in Panel~(b) are rather featureless for $\ell>50\p$. 
  The values in Fig.~\ref{fig:blb} have been obtained from the part of the
  domain with $|z|<0.5\kpc$, where most of the gas resides and where dynamo 
  action is expected to be most intense.
  The results, however, remain quite similar if the whole computational
  domain $|z|\leq1\kpc$ is used.
  While this value of $\ell$ has been estimated in a rather heuristic 
  and approximate manner,
  the analysis of the magnetic power spectra in 
  Section~\ref{subsect:scales} confirms
  that $\ell=50\p$ is close to the optimal choice.

  To put the estimate $\ell\approx50\p$ into context, note that it
  is about half the integral scale of the random motions, $l\turb$, 
  estimated in Section~\ref{sect:CORR}.
  Simulations of a single SN remnant in an ambient density similar to that at
  $z=0$ (Appendix~\ref{sect:SNPL}) also show that the expansion speed of the
  remnant reduces to the ambient speed of sound when its radius is 50--70\,pc. 

%-----------------------------------------------------------------------------
  \begin{figure*}[h]
  \centering
  \includegraphics[width=0.3765\linewidth]{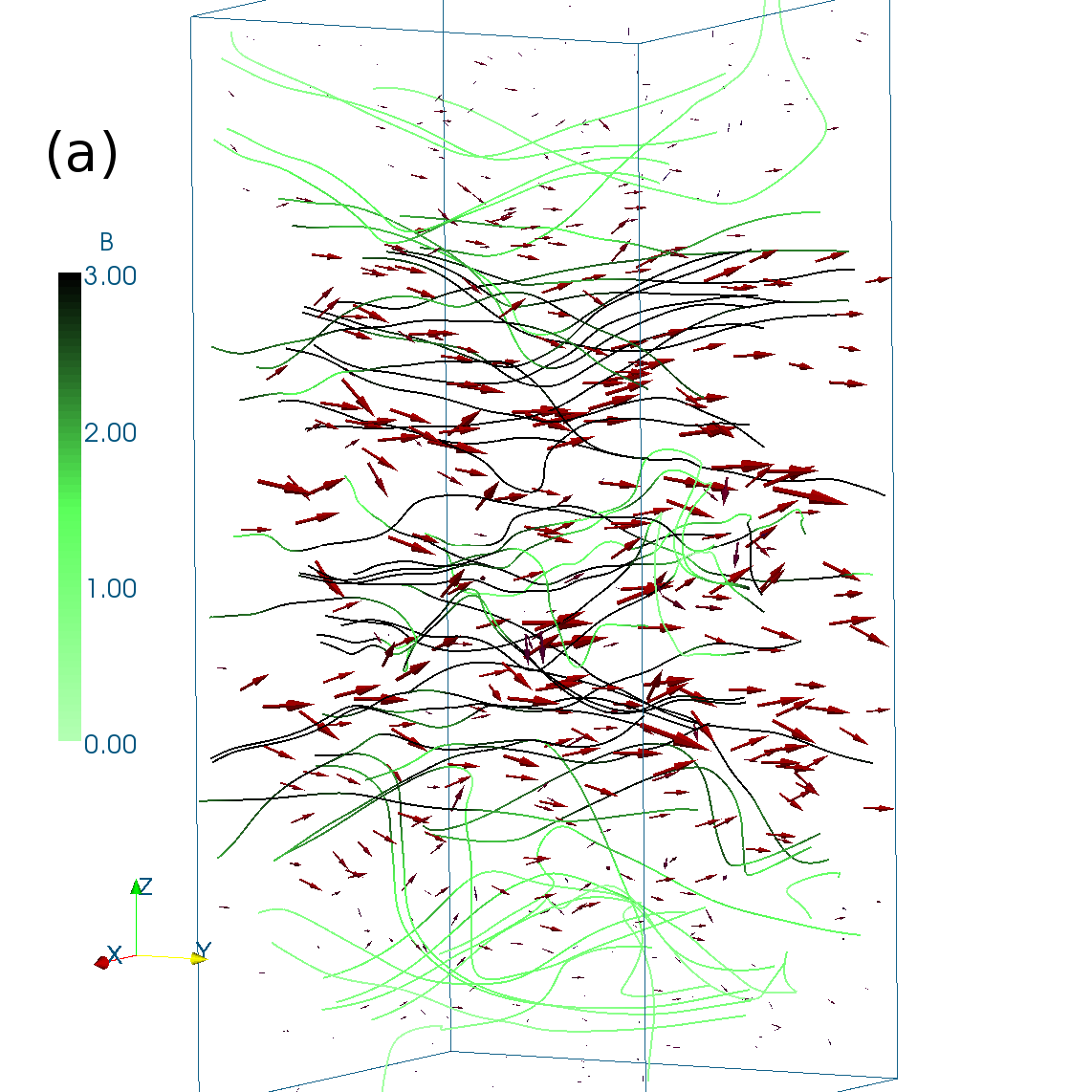}\hspace{-1.25cm}
  \includegraphics[width=0.3765\linewidth]{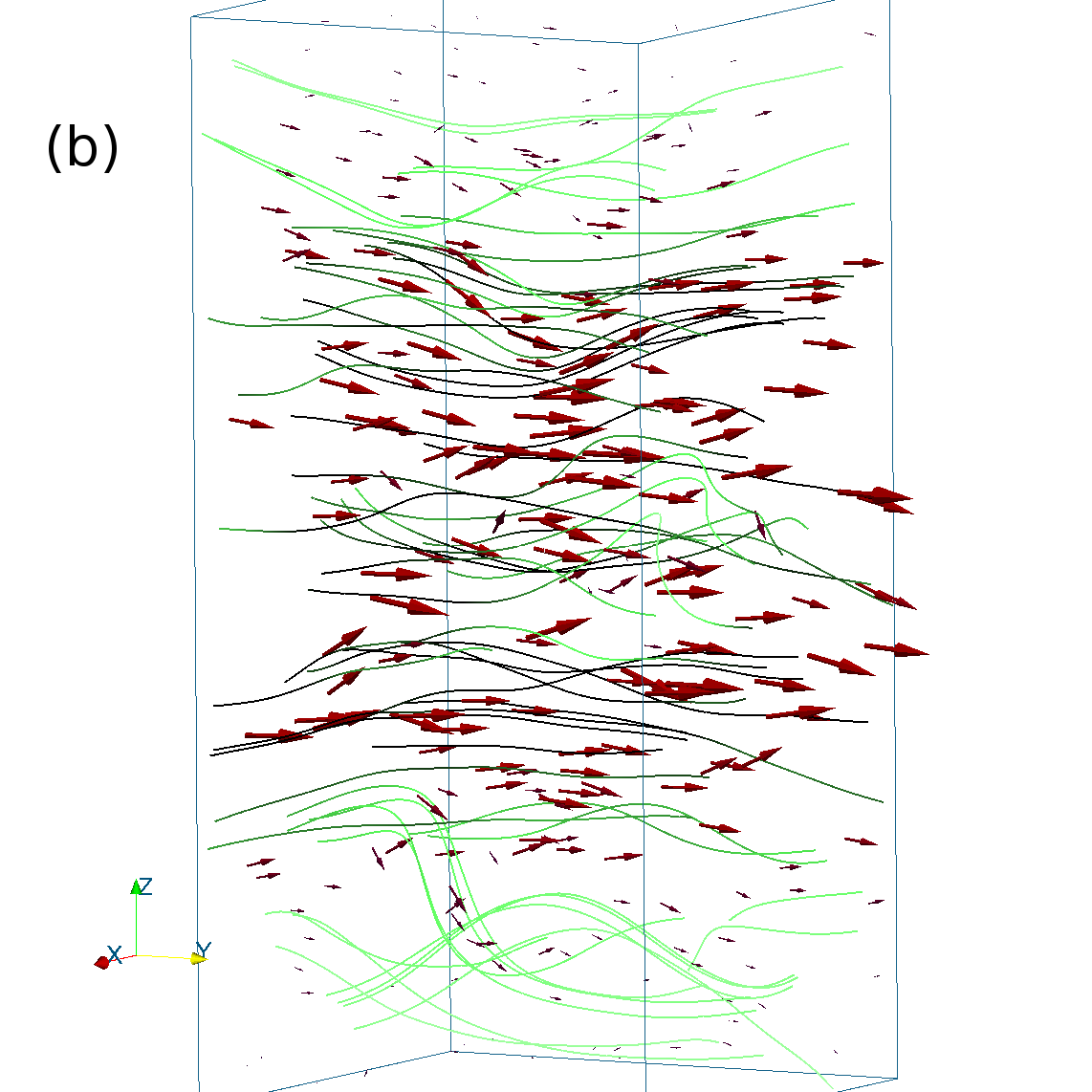}\hspace{-1.25cm}
  \includegraphics[width=0.3765\linewidth]{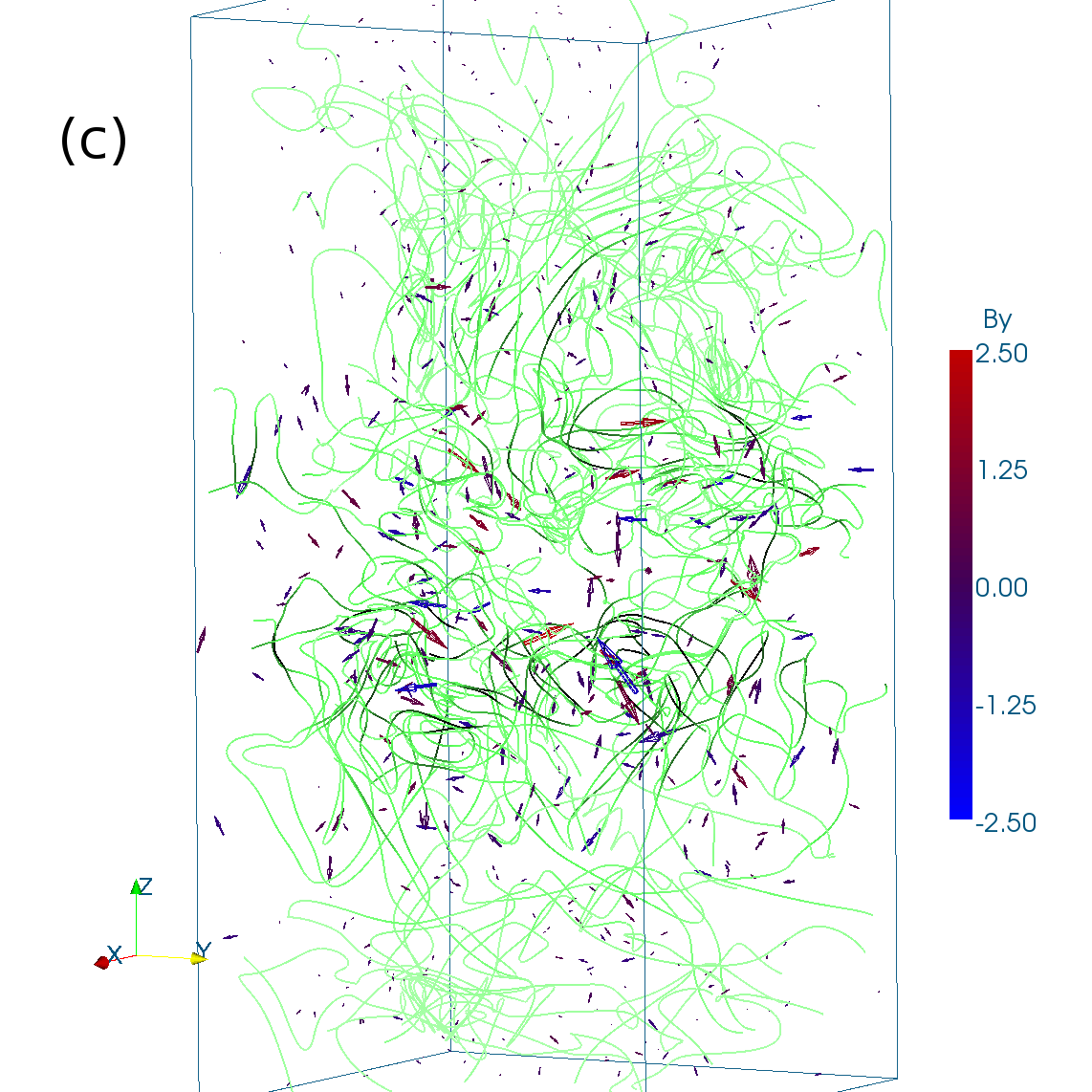}
  \caption[Volume snapshot of mean and fluctuating $\vect{B}$]{
  Field lines of (a) the total magnetic field $\vect{B}$, 
  (b) its averaged part $\vect{B}_\ell$, (c) the fluctuations $\vect{b}$,
  obtained by averaging with $\ell=50\p$, for $t=1.625\Gyr$.
  Field directions are indicated by arrows.
  The colour of the field lines indicates the field strength 
  (colour bar on the left), 
  whereas the vectors are coloured according to the strength of the
  azimuthal ($y$) component (colour bar on the right).
  \label{fig:box}}
  \end{figure*}
%-----------------------------------------------------------------------------

  Figure~\ref{fig:box} illustrates the total, mean and random magnetic fields
  thus obtained.
  For this saturated state, the field has a very strong uniform
  azimuthal component and a weaker radial component.
  The orientation of the field is the same  above and below the mid-plane
  ($B_y>0$ and $B_x<0$), with maxima located at $|z|\approx0.2\kpc$;
  the results will be reported in greater detail in Part~\ref{part:dynamo}.

%-----------------------------------------------------------------------------
  \section{Scale separation}{\label{subsect:scales}}
%-----------------------------------------------------------------------------

  Scale separation between the mean and random magnetic fields in natural 
  and simulated turbulent dynamos remains a controversial topic. 
  The signature of scale separation sought for is a pronounced minimum in
  the magnetic power spectrum at an intermediate scale, larger than the 
  energy-range scale of the random flow and smaller than the size of the 
  computational domain. 
  {\frcorr{The power spectrum for $\vect{B}$ is}} 
%  \begin{equation*}
  $ {\frcorr{M(k)=k^{-2}{\average{|\mathcal{F}({k})|}_k},}}$ 
%    M(k)=\frac{|\mathcal{F}(\vect{k})|}{|\vect{k}|^2}
%  \end{equation*}
  {\frcorr{for spherical shells of 
thickness $\delta k$ at radius $k=|\vect{k}|$, from}}
      $ {\frcorr{\mathcal{F}(\vect{k})
    ={\int_V \vect{B}(\vect{x})\exp({-2\pi i\vect{k}\cdot\vect{x}})\dd^3\vect{x}}.}}$
  The spectra of the mean and random magnetic fields obtained by Gaussian
  smoothing, shown in Fig.~\ref{fig:ffta}, have maxima at significantly 
  different wavenumbers.
  Note however that the spectrum of the total magnetic field does not have any
  noticeable local minima and, with the standard approach (e.g., based on
  horizontal averages), the system would be considered to lack scale 
  separation. 
     
%-----------------------------------------------------------------------------
  \begin{figure}[h]
  \centering
  \includegraphics[width=0.6535\linewidth]{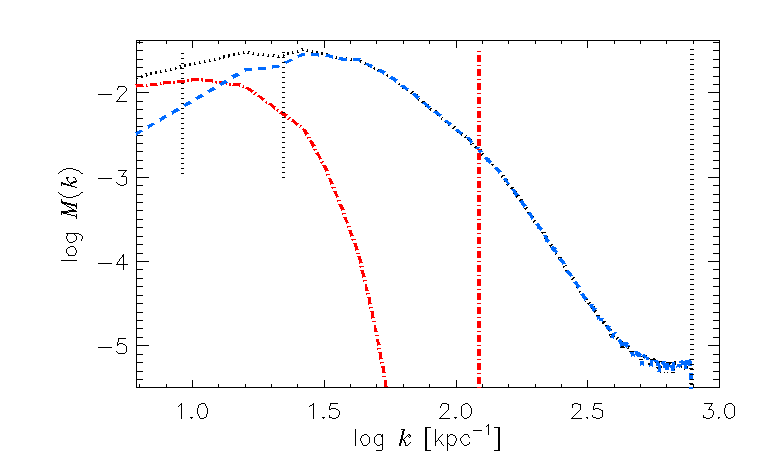}
  \caption[Power spectra of mean and fluctuating $\vect{B}$, $\ell=50\p$]{
  Power spectra of total (black, dotted), mean (red, dash-dotted) 
  and fluctuating (blue, dashed) magnetic fields, for 
Gaussian smoothing  
  with $\ell=50\p$, at $t=1.05\Gyr$.  
  Vertical red (dash-dotted) line marks the averaging wavenumber 
  $2\upi\ell^{-1}$ and vertical black (dotted) line indicates the Nyquist wave
  number $\upi\Delta^{-1}$.
  Short vertical segments mark the energy-range scales 
  $2\upi\lcorr_\ell^{-1}$ and 
  $2\upi\lcorr_b^{-1}$ of the mean and random magnetic fields, respectively,
obtained from Eq.~\eqref{Bcorr}.
  \label{fig:ffta}}
  \end{figure}
%-----------------------------------------------------------------------------

  figure~\ref{fig:ffta} shows the magnetic power spectra of $\vect{B}$, 
  $\vect{B}_\ell$ and $\vect{b}$.
  Note that $\ell$ is not located between the maxima in the power
  spectra of $\vect{B}_\ell$ and $\vect{b}$; 
  in fact, the spectral density of the mean field is negligible for 
  $k\simeq2\upi\ell^{-1}$. 
  This can be understood from the transform of the kernel $G_\ell(\vect{x})$,
  i.e.\ $\widehat{G}_{\ell}(\vect{k})=\exp(-\ell^2 \vect{k}^2/2)$;
  this kernel would divide a purely sinusoidal field equally into the mean and
  random parts at the wavelength $\lambda_{\rm eq}=\sqrt{2/\ln 2}\upi\ell$.
  For $\ell=50\pc$, $\lambda_{\rm eq}=0.27\kpc$,
  and the latter figure is a better guide to the expected separation scale.
  The separation of scales is immediately apparent in the spectra of the mean 
  and random fields, $M_\ell(k)$ and $M_b(k)$, with the former having a broad 
  absolute maximum at
  about $0.56\kpc$, and the latter a broad
  maximum near $0.2\kpc$. 
  The effective separation scale $\lambda\approx0.48\kpc$ 
  can be identified where $M_\ell(k)=M_b(k)$, i.e. where
  the curves cross at $\log k\simeq1.1$.     

%-----------------------------------------------------------------------------
  \begin{figure}[t]
  \centering
  \hspace{-2.95cm}
  \includegraphics[width=0.58\linewidth]{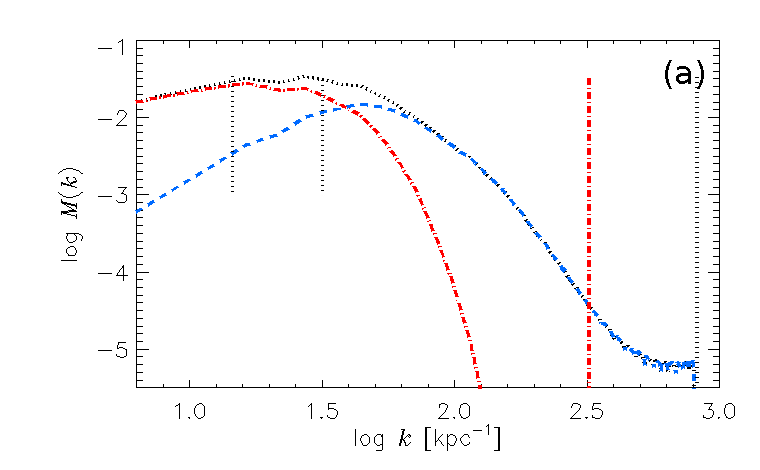}
  \hspace{-0.5cm}
  \includegraphics[width=0.58\linewidth]{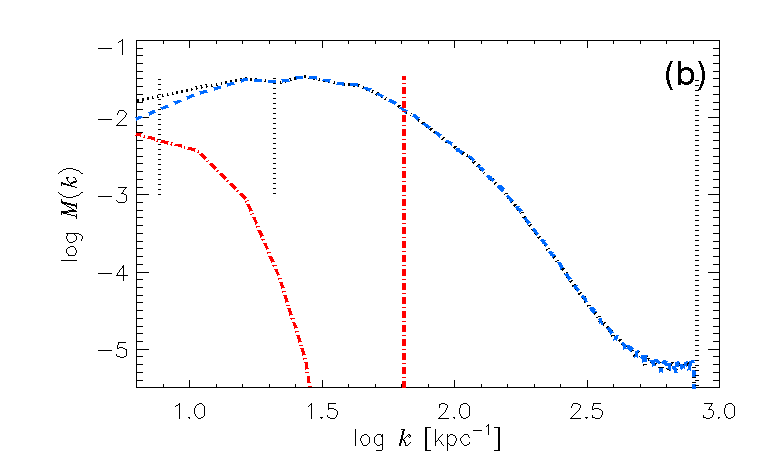}
  \hspace{-1.5cm}
  \caption[Power spectra of mean and fluctuating $\vect{B}$, $\ell\neq50\p$]{
  As Fig.~\ref{fig:ffta}, but for \textbf{(a)} $\ell=20\p$ and \textbf{(b)} 
  $\ell=100\p$.
  \label{fig:fftb}}
  \end{figure}
%-----------------------------------------------------------------------------
  
  The integral scales of $\vect{B}_\ell$ and $\vect{b}$ can be obtained from
  their spectral densities as
  \begin{equation}
  \lcorr_\ell=\frac{\upi}{2}\,
  {\int_{{2\upi}/{D}}^{{\upi}/{\Delta}} k^{-1}M(k)\,\dd k} \left[
  {\int_{{2\upi}/{D}}^{{\upi}/{\Delta}} M(k)\,\dd k}\right]^{-1},
  \label{Bcorr}
  \end{equation}
  where $\Delta$ is the numerical grid separation and $D$ is the size of the 
  domain \citep[Section 12.1,][]{MY07}. 
  This yields $\lcorr_\ell\simeq0.67\kpc$ and $\lcorr_b\simeq0.28\kpc$ for
  $\vect{B}_\ell$ and $\vect{b}$, respectively (Fig.~\ref{fig:ffta}).

  As the magnetic field strength grows, the magnitudes of the spectral
  densities change but the characteristic wavenumbers vary rather weakly. 
  $\lcorr_\ell$ remains close to $0.7\kpc$ throughout the
  kinematic phase, increasing only beyond 1.1\,Gyr to about $0.9\kpc$. 
  $\lcorr_b$ increases from $0.23$ to $0.28\kpc$ during the kinematic phase,
  but rises to $0.4\kpc$ after the system saturates.
  The stability of $\lcorr_\ell$ is consistent with the eigenmode  
  amplification of the 
  mean magnetic field as expected for a kinematic dynamo, and supports 
  $\ell\approx50\p$ as a reasonable choice of the averaging scale.

  Figure~\ref{fig:fftb} presents magnetic energy spectra obtained  
  using  $\ell=20\p$   and $100\p$.
  In the 
  former, the effective separation scale $\lambda\approx0.16\kpc$ is less
  than $\lcorr_b\simeq0.21\kpc$.  
  This is inconsistent, implying that energy at scales about $\lcorr_b$ lies
  predominantly within $\vect{B}_\ell$. 
  For $\ell=100\p$, 
  $\lambda\geq1\kpc$ is greater than $\lcorr_\ell\simeq0.81\kpc$,
  which is also inconsistent.
  Significantly, applying $\ell=50\p$ satisfies $\lcorr_b<\lambda<\lcorr_\ell$.
  Hence the scale $\lambda$, at which the dominant energy contribution
  switches between the mean and fluctuating parts, 
  is here consistent only with $\ell\simeq50\pc$.

  As well as different spatial scales, the mean and random magnetic field
  energies have different exponential growth rates (Fig.~\ref{fig:bvt}), given
  in Table~\ref{tab:fit}.
  The growth for the mean field component of the total magnetic energy is 
  denoted by $\Gamma_e$ and the growth of the random field energy by $\gamma_e$.
  Results from horizontal averaging are also given; they differ 
from those obtained with Gaussian smoothing,
  especially for the mean field.
  The growth rate $\Gamma_e$ of the mean-field energy is controlled by the shear
  rate, mean helicity of the random flow, and the turbulent magnetic 
  diffusivity. 
  Any mean magnetic field is accompanied by a random field, which is part of
  the mean-field dynamo mechanism (as any large-scale magnetic
  field is tangled by the random flow); 
  the energy of this part of the small-scale field should grow at the same rate
  $\Gamma_e$ as the mean energy.
  The fluctuation dynamo produces another part of the random field whose energy
  growth rate depends on the turbulent kinematic time scale $l/u$, and the
  magnetic Reynolds and Mach numbers.
  The difference between $\Gamma_e$ and $\gamma_e$ 
obtained suggests
  that both 
  mean-field and fluctuation dynamos are present
in our model.
  The growth rate of the mean magnetic energy is roughly double that of 
  the random field for both types of averaging.
  This is opposite to what is usually expected, plausibly because of the
inhibition of the fluctuation dynamo by the  
  strongly compressible nature of the flow 
  {\frcorr{and low Reynolds numbers available at this resolution.
  We would expect the fluctuation dynamo to be stronger with more realistic
  Reynolds numbers}}.

%----------------------------------------------------------------
  \begin{table}[t]
  \caption[Growth rates of $\vect{B}$ for Model~$\Ompd$]{\label{tab:fit}
  Exponential growth rates of energy in the mean and random 
  magnetic fields, $\Gamma_e$ and $\gamma_e$, respectively, 
  with associated values of reduced $\chi^2$. 
  From exponential fits to the corresponding 
  curves in Fig.~\ref{fig:bvt}, for $0.8<t<1.05\Gyr$.
  (Growth for $t<0.8\Gyr$ is similar to this interval
  -- see Fig.~\ref{fig:bbelog}.)
  }
  \centering
    \begin{tabular}{lcccc}
  \hline
                        &$\Gamma_e$      &$\chi^2$   &$\gamma_e$         &$\chi^2$ \\
                        &[Gyr$^{-1}$]  &           &[Gyr$^{-1}$]     &         \\
  \hline 
Gaussian smoothing   	&$10.9$        &$1.00$     &5.5              &$1.15$   \\
  Horizontal averaging	&$13.6$        &$0.25$     &$6.2$            &$0.25$   \\
  \hline
    \end{tabular}
  \end{table}
%------------------------------------------------------------

%-----------------------------------------------------------------------------
  \begin{figure}[h]
  \centering
  \includegraphics[width=0.8\linewidth]{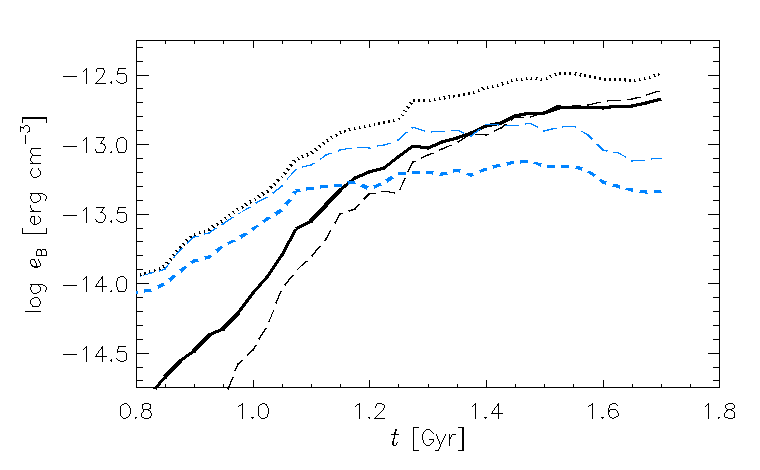}
  \caption[Growth of $\eB$ for Model~$\Ompd$]{
  Evolution of magnetic energy densities, averaged for full domain:
  total magnetic field (black, dotted), mean (black, solid), and random 
  (blue, dashed) obtained from 
Gaussian smoothing
   with $\ell=50\p$.
  Energy densities of mean and random magnetic fields
  obtained from horizontal averaging are shown long-dashed (black and blue,
  respectively).
  \label{fig:bvt}}
  \end{figure}
%-----------------------------------------------------------------------------

%---------------------------------------------------------------
  \section{Evaluation of the method} \label{sect:eval}
%---------------------------------------------------------------

  The approach used here to identify the appropriate averaging length $\ell$
  (and thus the effective separation scale $\lambda$) is simple (but 
  not oversimplified); $\ell$ can in fact depend on position, and it may remain 
  constant in time only at the kinematic stage of all the dynamo effects 
  involved.
  Wavelet filtering may prove to be more efficient than Gaussian smoothing in
  assessing the 
  variations of $\ell$.

  At the kinematic stage, the spectral maximum of the mean field is
  already close to the size of the computational domain, and it cannot be
  excluded that the latter is too small to accommodate the most
  rapidly growing dynamo mode.  These results should therefore be
  considered as preliminary with respect to the mean field;
  simulations in a bigger domain are clearly needed.
  
  The physically motivated averaging procedure used here, producing a mean
  field with three-dimensional structure, may facilitate fruitful 
  comparison of numerical simulations 
  with theory and observations.
  Although 
Gaussian smoothing
  does not obey all the Reynolds rules, 
  it is possible that a consistent mean-field theory can be developed,
  e.g.\ in the framework of the
  $\tau$-approximation \citep[see e.g.][and references therein]{BS05}. 
  This approach does not rely upon solving the equations for the
  fluctuating fields, and hence only requires the linearity
  of the average and its commutation with the derivatives.
  The properly isolated mean field is likely to exhibit different spatial and
  temporal behaviour than the lower-dimensional magnetic field obtained by 
  two-dimensional averaging. 

  \citet[][Ch.11.5]{BS05} include estimates for the strength of the mean field
  dynamo in rapidly rotating galaxies. Here I apply their method to 
  Model~${\Ompd}$, to compare their estimates with the simulation results.
  The turnover time is given by $\tau=l\turb/u\turb$.
  From Section~\ref{sect:CORR}, $l\turb\simeq0.1\kpc$.
  The estimates assume incompressibility, so I shall use the upper bound of the 
  subsonic turbulence. The mean field dynamo grows primarily in the warm 
  gas, as is detailed in Part~\ref{part:dynamo}. 
  From Fig.~\ref{fig:epdf3}, the modal temperature of the warm gas is $10^4\K$,
  which equates to $c\sound\simeq15\kms$, the upper bound for $u\turb$.
  Hence $\tau\simeq1/150\Gyr$.

  The turbulent resistivity is 
  $\eta\turb=\sfrac{1}{3}u\turb l\turb\simeq0.5\kms\kpc$.
  From Fig.~\ref{fig:pavB} in Part~\ref{part:dynamo}, it is apparent that the 
  scale height of the dynamo active region may be as high as $h\simeq0.6\kpc$.
  So $\alpha=\tau^2\Omega(u\turb^2/h)\simeq5/6\kms$.
  The $\alpha\Omega$-dynamo has two control parameters: 
  $C_\Omega=S h^2/\eta\turb\simeq-36$ and $C_\alpha=\alpha h/\eta\turb\simeq1$,
  so the dynamo number $D=C_\Omega C_\alpha\simeq-36$.

  If $|D|>D_{\mathrm{crit}}$, a critical dynamo number which the authors expect
  to be in the range $6$--$10$, then exponential growth of the magnetic field is
  possible during the kinematic stage.
  The growth rate exponent $\Gamma$ can be estimated by
  \begin{equation}
    \label{eq:dynamo}
    \Gamma\approx\frac{\eta\turb}{h^2}\left(\sqrt{|D|}-
        \sqrt{D_{\textrm crit}}\right)
  \end{equation}
  So for Model~$\Ompd$, the expected growth rate of the magnetic field
  $\Gamma$ is of order 4--5$\Gyr^{-1}$.

  Note the growth rates $\Gamma_e$ shown in Table~\ref{tab:fit} refer to the 
  magnetic energy, so $\Gamma_e=2\Gamma$ as estimated here. 
  For Gaussian smoothing $\Gamma_e=10.9\Gyr^{-1}$, while for horizontal 
  averaging it is $13.6\Gyr^{-1}$. 
  Although the analytic estimates apply to incompressible flow and there is
  considerable uncertainty over the choices of $l\turb,~u\turb,~h$ and 
  $D_{\textrm{crit}}$, a reasonable upper bound for this model is 
  $\Gamma=6\Gyr^{-1}$. 
  In addition, given the complexity of the system, other mean field dynamo
  processes could be present in addition to the shear dynamo to which the analytics refer.
  Nevertheless, the agreement of theory and experiment is very encouraging.

  \section{Summary}

  Using kernel smoothing to identify the mean field permits separation of the
  fluctuations from the systematic part of the magnetic field, while 
  preserving its basic structure.
  The departure from the Reynolds rules is addressed by redefining how the 
  magnetic energy terms are derived. 
  A heuristic approach has been adopted to determine the appropriate smoothing
  scale, which appears to correctly identify the separation between the mean
  and random parts of the field.
  This allows us to examine the growth of the mean and random field, and
  a mean field dynamo has been modelled, with growth consistent with the 
  analytic estimates for the galactic shear dynamo. 
  In the remainder of this thesis, the method described in this chapter
  is applied to determine the mean $\vect{B}$ field and the mean flow
  $\vect{u}$ whenever they are required. 
  
\end{chapter}

  \part{The galactic dynamo and magnetic structure}\label{part:dynamo}
%  \clearpage                            % End the current page making sure all
%  \thispagestyle{empty}                 % tables/figures are printed.
%  \cleardoublepage                      % Necessary for correct page numbering.
%  \include{chapters/hdruns}
%
%%  \part{Magnetohydrodynamical simulations}\label{part:mhd}
  %-----------------------------------------------------------------------------
\begin{chapter}{\label{chap:MHD}The magnetic field}

%------------------------------------------------------------------------
  In this chapter the magnetic field generated dynamically in the various MHD
  models is investigated in some detail. 
  All the models in this section show sustained amplification of the magnetic
  field, and evident organisation into a mean field.
  Of interest to theorists is what is driving the dynamo, and where the dynamo 
  or dynamos are acting. 
  Of general interest is the general shape and structure of the magnetic 
  field, to what extent is it ordered or fluctuating, and how strongly 
  correlated is it to the density or temperature in the ISM.
  The vertical magnetic structure of the ISM is particularly difficult to
  observe, so how the simulated field varies with distance from the mid-plane 
  is of great interest.
  How does the magnetic field interact with the velocity field, and to what
  extent are the properties of the ISM, such as distributions of temperature, 
  pressure and density, or various filling factors, affected by the magnetic
  field?
  These issues will be considered by comparing the MHD models with each 
  other and with the non-magnetic models.

%-----------------------------------------------------------------------------
  \section{\label{sect:dyn}The magnetic dynamo}
%-----------------------------------------------------------------------------

%-----------------------------------------------------------------------------
  \begin{figure}[h]
  \centering
  \hspace{-1.5cm}
  \includegraphics[width=0.535\linewidth]{fig/bbelog.png}
  \includegraphics[width=0.535\linewidth]{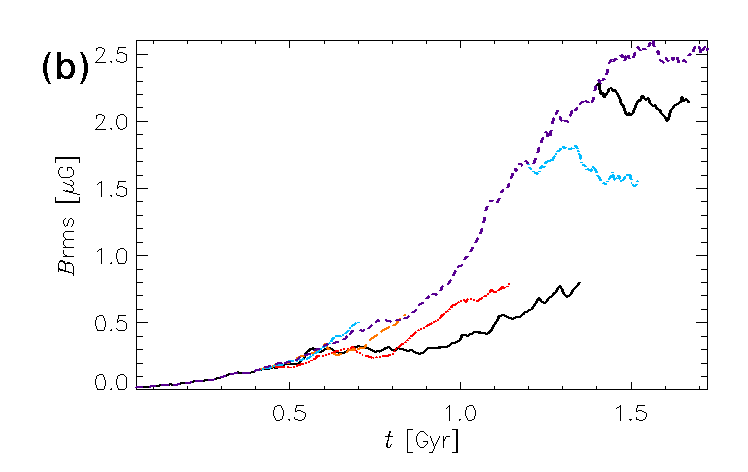}
  \hspace{-1.5cm}
    \caption[Magnetic energy for MHD models]{
  {\textbf{(a)}} Time evolution for the log of thermal (red, top jagged), 
  kinetic  (blue, $2\nd$ top jagged) and magnetic (black, dotted) energy
  densities from 
  Model~$\Ompd$, averaged over the total volume.
  The dotted line indicates $6.6\Gyr^{-1}$.
  {\textbf{(b)}} $\average{B\rrms}$ as a function of time for
  Models~$\Ompa$ (solid, black), 
  $\Ompd$ (dashed, purple),
  $\Ompb$ (dash-dotted, blue),
  $\Ompe$ (long dashed, orange), 
  and $\Ompc$ (dash-3dotted, red).
  The later continuations of Models~$\Ompa$ from 1.4\,Gyr and $\Ompb$ from
  1.2\,Gyr are Model~$\Ompd$ continued with parameters amended. 
    \label{fig:bbe}}
  \end{figure}
%-----------------------------------------------------------------------------

  In Fig.~\ref{fig:bbe}a the time evolution of magnetic (purple), kinetic 
  (blue) and thermal (red) log energy density is plotted for Model~$\Ompd$.
  The growth in kinetic energy density from rest, and thermal energy due to
  SN heating, is near instantaneous.
  The growth in $\average{B\rrms}$ is much slower and irregular, 
  intermittently exponential, linear and even occasionally exhibiting decay. 
  Dynamo theory anticipates that, as magnetic energy density
  ($\eB=\frac{1}{2\mu_0}|\vect{B}|^2$) approaches that of the kinetic energy
  density ($\eK=\frac{1}{2}\rho|\vect{u}|^2$), 
  the magnetic field saturates to a steady state, with kinetic and magnetic 
  energy near equipartition.
  The Lorentz Force feeds back onto the velocity field suppressing the 
  kinematic dynamo.
  In Fig.~\ref{fig:bbe}a $\average{\eB}$ and $\average{\eK}$ 
  intersect at about 1.3\,Gyr corresponding to $\average{B\rrms}\simeq2.5\mkG$.
  This is remarkably consistent with the observed estimates of the
  field strength in the solar neighbourhood.

  The thermal energy density ($\eT=\rho\cv T$), with $\cv$ the specific 
  heat capacity at constant volume) is quite insensitive to $\average\eB$, as 
  is $\average\eK$ up to 1\,Gyr, with $\average{\eT}\simeq9\times10^{-13}$ and
  $\average{\eK}\simeq3\times10^{-13}\erg\cmcube$.
  However around 1.1\Gyr, when $\average{\eB}$ grows to within half of 
  $\average{\eK}$, kinetic energy is transferred to magnetic energy.
  There is also a slight reduction in $\average\eT$ after 1.4\Gyr, such that
  total energy density is roughly conserved, with 
  $\average{\eB}\simeq2.5\times10^{-13}$,
  $\average{\eK}\simeq1.5\times10^{-13}$ and
  $\average{\eT}\simeq8  \times10^{-13}\erg\cmcube$.

  In Fig.~\ref{fig:bbe}b the time evolution of $\average{B\rrms}$ is 
  plotted for the MHD models.
  It is evident that all the models exhibit periods of exponential growth.
  At 400\,Myr $\Ompa$ is in quasi-steady hydrodynamic turbulence with 
  negligible magnetic energy and the other models are initialised using this
  snapshot.
  Subsequently the lowest growth rate appears to apply for $\Ompa$. 
  However $\Ompb$, plotted in blue dash-dotted in Fig.~\ref{fig:bbe}a, 
  differs only in the open vertical boundary condition, and appears to attain
  stronger growth.
  For this model the Poynting flux is monitored on the boundary and the net
  outflow of magnetic energy throughout this phase is consistently of order
  $10^2$ times the inflow or greater.
  The growth would appear therefore to be driven by the internal dynamics, 
  and potentially assisted by the flux of magnetic helicity outward.

  In their models matching these parameters, \cite{Gressel08} found no dynamo.
  Only with differential galactic rotation $\geq4\Omega_0$ did they observe a
  mean field dynamo.   
  They resolved a grid of $\Delta=8\p$, while here $\Delta=4\p$, and I also use
  temperature dependent viscosity and thermal conductivity, which permit much
  larger Reynolds numbers.
  Consistent with their results, however, I do find the strongest dynamo
  corresponds to the higher rotation rate (plotted in purple, dashed in 
  Fig.~\ref{fig:bbe}a).  

  Increasing the shear alone in Model~$\Ompe$ (plotted in orange, long dashes)
  also enhances magnetic field growth relative to Model~$\Ompa$. 
  For spiral galaxies, which often have nearly flat rotation curves with
  respect to the galactocentric radius $r$, with angular velocity of the form
  $\Omega \propto r^{-1}$, the shear parameter $S=-\Omega$.
  It is convenient to introduce the ratio of the shear rate to the rotation
  rate:
  \begin{equation}\label{eq:q}
    q = -\frac{S}{\Omega}=-\frac{\dd\ln\Omega}{\dd\ln r}.
  \end{equation}
  Flat rotation curves correspond to $q=1$.
  Such rotation laws are known to be hydrodynamically stable, as the Rayleigh
  instability only sets in for $q>2$.
  To exclude additional Rayleigh instability effects $-S/\Omega<2$ is required, 
  so $S=-40\kms\kpc^{-1}=-1.6\Omega_0$ is adopted, yielding $q=1.6$.
  There may however remain a contribution of magnetorotational instability
  (MRI), which has an instability condition of $q>0$. 
  There is evidence, however, that this effect is suppressed in the presence of 
  SN turbulence \citep[][]{KKV10,Piontek07a,BH91}.
  Future work, comparing a model in which the sign of $S$ or $\Omega$ is
  reversed (so that $q<0$) to a standard model, may help to isolate the
  influence of MRI from the shear dynamo.
  Such a run, with the outer disc modelled to rotate faster than the
  inner disc, would not be relevent for any known galaxies, but other than a 
  sign change in the mean horizontal components of the flow and the mean 
  magnetic field, the ISM temperature, density and pressure distributions 
  should be retained.
  The turbulent structure should be unaffected, at least in the kinematic 
  stage. 

%-----------------------------------------------------------------------------
  \begin{figure}[h]
  \centering
  \includegraphics[width=\linewidth]{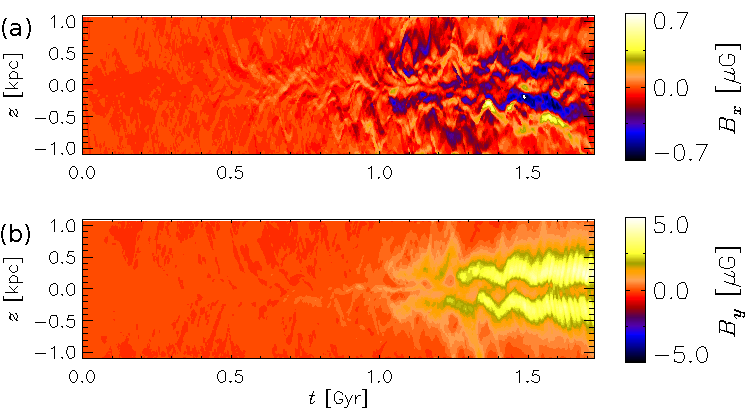}
    \caption[Horizontal averages for $B_x$ and $B_y$ in Model~$\Ompd$]{
  Time evolution of the horizontal averages for {\textbf{(a)}} $B_x$ and 
  {\textbf{(b)}} $B_y$ for Model~$\Ompd$, with galactic rotation twice 
  $\Omega_0$ and parameters otherwise typical of the Galaxy in the solar 
  neighbourhood.
  The maximum strength of the seed field is $B_y<0.01\mkG$.
    \label{fig:pavB}}
  \end{figure}
%-----------------------------------------------------------------------------
%
  In Fig.~\ref{fig:pavB} the time evolution of $B_x$ and 
  $B_y$, averaged horizontally, are plotted against $z$ for Model~$\Ompd$.
  Horizontally averaged $B_z\simeq0$ for all $z$, although the random
  fluctuations are typically stronger than for $B_x$ and $B_y$.
  It is evident that a strong mean field is generated, that it is 
  dominated by the azimuthal ($y$) component, and that the magnitudes of both 
  $B_y$ and $B_x$ vary with $z$ and are roughly symmetric about the mid-plane.

  That $\average{B_z}\simeq0$ is not a physical result, but imposed by the 
  periodic and sliding periodic boundary conditions applied horizontally to all 
  three components of the magnetic potential.
  Given the divergence free condition on the magnetic field in the galaxy, it
  is reasonable to expect that the mean radial field, which is oriented inwards 
  on both sides of the mid-plane throughout the simulation, cannot meet at the
  centre of the galaxy.
  Towards the centre of the galaxy, there must be a significant mean vertical 
  component.
  Of course for small galactic radius, the shearing box approximation breaks
  down anyway.
  Nevertheless in future work horizontal boundary conditions, which permit the
  free movement of the field lines, may reveal non trivial vertical field
  structure. 
   
  Reducing the supernova rate by 20\% with Model~$\Ompc$ (plotted in red, 
  dash-dotted) also drives a stronger dynamo than $\Ompa$.
  As discussed in Section~\ref{subsect:COOL}, the investigation by
  \cite{Balsara04} with SN rates of 8$\sigma_0$ to 40$\sigma_0$ indicates that
  as SN rates exceed some critical level, the amplification of the small scale
  field is quenched.
  The critical rate will be at least above $12\sigma_0$, based on their 
  periodic box calculations, and will likely be higher, given the release of 
  hot gas away from the mid-plane in the stratified ISM.
  It may also be reasonable to expect a minimum SN rate, below which the 
  dynamo cannot operate. 
  \citet{Gressel08b} also found that the strength of the magnetic field 
  increased for lower SN rates over a range from $0.25\sigma_0$ to $\sigma_0$.

  The strongest ordered field corresponds to the layers between 200 and 600\,pc
  away from the mid-plane in Fig.~\ref{fig:pavB}.
  Although the ordering of the field is stronger in Model~$\Ompd$, all the
  MHD models share this structure. 
  The weaker mean field near the mid-plane may be a result of the scrambling of 
  the field lines by SN turbulence.
  The strongest mean field occurs where the filling factors of the warm gas are
  highest, and outside the most active SN layer. 
  This is consistent with the loss of field amplification as SN rates increase.

%-----------------------------------------------------------------------------
  \begin{figure}[h]
  \centering
%  \hspace{-1.25cm}
  \includegraphics[width=\linewidth]{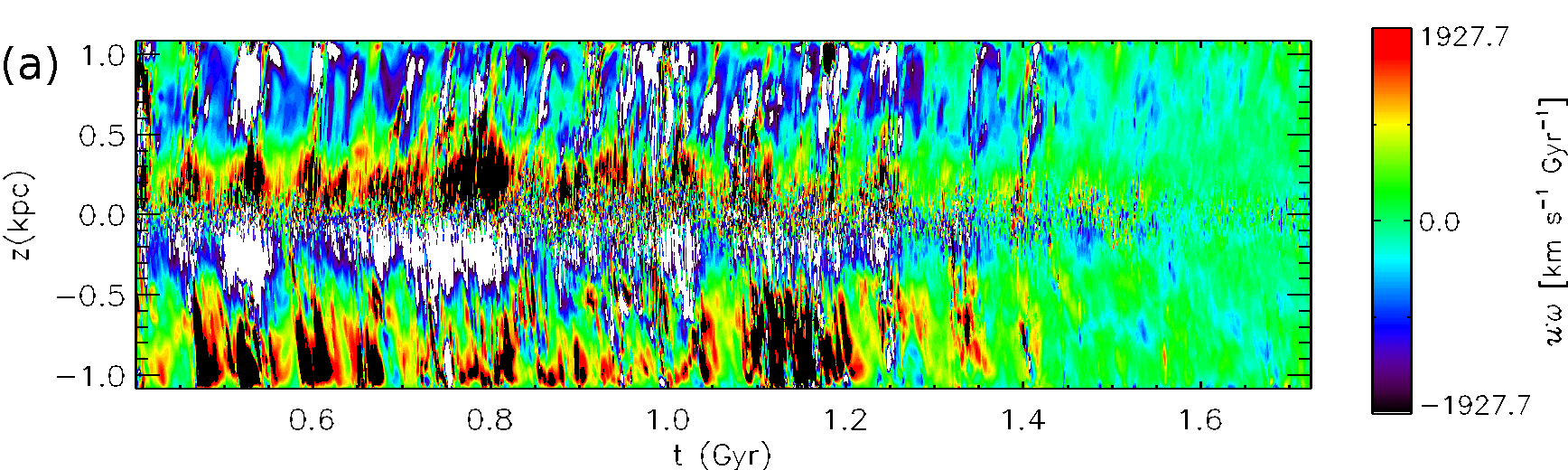}
  \includegraphics[width=\linewidth]{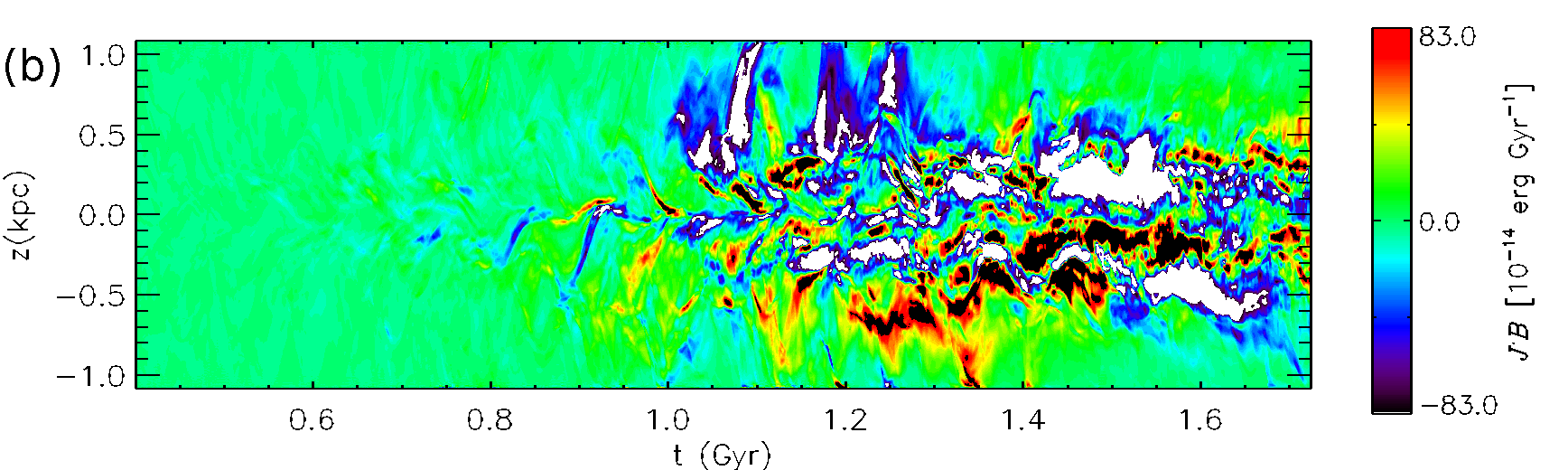}
  \includegraphics[width=\linewidth]{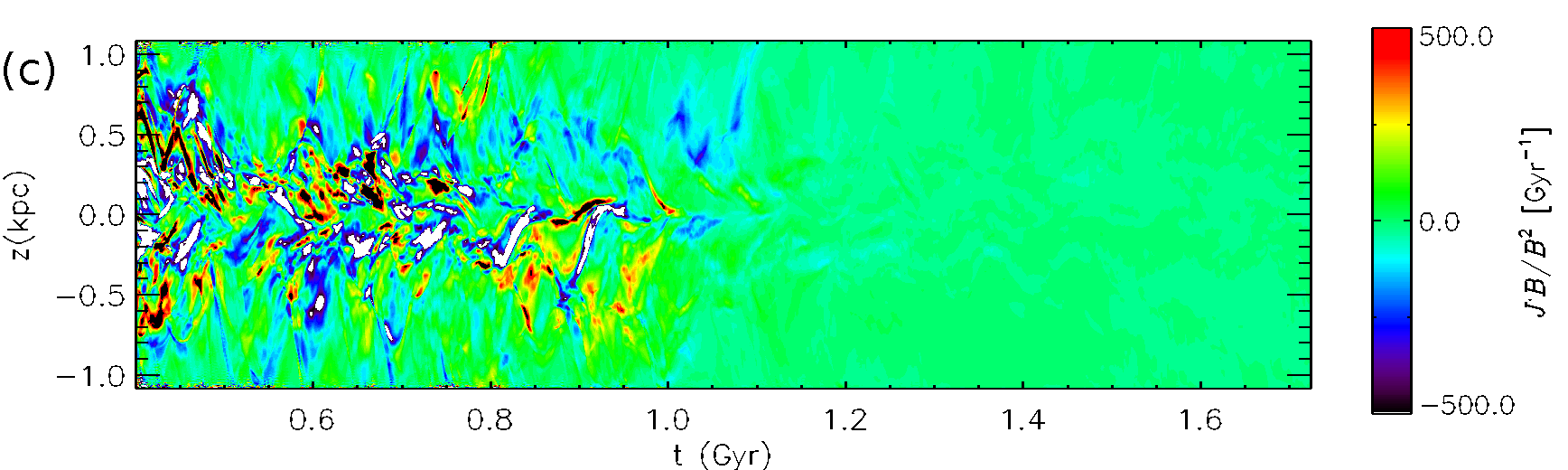}
%  \hspace{-1.5cm}
    \caption[Horizontal averages for magnetic helicity in Model~$\Ompd$]{
  Time evolution from Model~$\Ompd$ of horizontal averages for {\textbf{(a)}}
  kinetic helicity ($\vect{u}\cdot\vect{\omega}$), and {\textbf{(b)}} the 
  current helicity
  $\vect{J}\cdot\vect{B}$, a gauge invariant proxy for magnetic helicity. 
  In {\textbf{(c)}} the relative current helicity $\vect{J}\cdot\vect{B}$ 
  normalised by $\average{B\rrms^2}(t)$ is displayed.
  Regions of black (white) indicate values above (below) the colour bar range.
    \label{fig:hel}}
  \end{figure}
%-----------------------------------------------------------------------------

  To understand what is driving the dynamo, it may help to look at the kinetic
  and magnetic helicity.
  The growth of magnetic helicity may suppress the dynamo. 
  In Fig.~\ref{fig:hel}, horizontal averages from Model~$\Ompd$ as a function 
  of $z$ and evolving in time are displayed for (panel a) kinetic helicity
  ($\vect{u}\cdot\vect{\omega}$) and (panel b) current helicity 
  $\vect{J}\cdot\vect{B}$, a proxy for the magnetic helicity 
  ($\vect{A}\cdot\vect{B}$), but gauge invariant.
  $\vect{J}=\frac{c}{4\upi}\nabla\times\vect{B}$ in the time independent electric field.
  Hence, $\vect{J}\sim\lcorr^{-2}\vect{A}$, where $\lcorr$ is some integral
  scale of the magnetic potential, can be used to indicate the evolving
  structure without the requirement to deduct the contribution of the arbitrary 
  gauge $\nabla\Phi$ from $\vect{A}$. 
  Fig.~\ref{fig:hel}c shows $\vect{J}\cdot\vect{B}$, normalised by the time 
  dependent $\average{B\rrms^2}$, the relative current helicity, of use for
  comparing the kinematic stage with the saturated state.
  
  From the first two plots it appears that the saturation of the dynamo beyond
  $1\Gyr$ coincides with the transfer of mean helicity from kinetic to 
  magnetic form. 
  This transfer is present in the saturated stages for Models~$\Ompa$ and 
  $\Ompb$ also.
  In (a), the kinetic helicity either side of the mid-plane is of opposite sign,
  negative in the south and positive in the north, as would be expected.
  What is surprising is that there is another sign change at 
  $|z|\simeq0.3\sim0.5\kpc$.
  The pattern is most clearly revealed for Model~$\Ompd$, but is common to all
  of the MHD models and is also a consistent feature of the HD models. 
  In the saturated stage the magnetic helicity appears to exhibit strong mean
  structures, and there is some tendency for sign asymmetry about the 
  mid-plane, although this is fragmented.
  The sign of $\average{B_y}$ is quite consistently positive in the latter 
  stages, so the intermittent reversals in helicity are primarily due to 
  the fluctuations in $\average{B_x}$.

  The stronger current helicity at later times, however, is primarily only a 
  weak effect of a much stronger field.
  In Fig.~\ref{fig:hel}c, where the normalised $\vect{J}\cdot\vect{B}$ is
  plotted, it is evident that the peak relative helicity occurs prior to field
  saturation up to $1.1\Gyr$. 
  Throughout the simulation the net $\vect{J}\cdot\vect{B}$ is negative.
  The normalised 
  $\mean{\average{\vect{J}\cdot\vect{B}/\average{B^2}}}=-6.0\Gyr^{-1}$ before
  saturation, where $\average{}$ indicates averaging over the whole volume and
  the over-bar here refers to averaging over time, $0.4<t<1.1\Gyr$, and 
  $-0.6\Gyr^{-1}$ post-saturation. 
  Further details of this quantity are given in Table~\ref{table:hel} for all 
  the MHD models.

%-----------------------------------------------------------------------------
  \begin{table}[h]
  \caption[Magnetic helicity in MHD models]{\label{table:hel}
  For each MHD model the time and total volume averaged $\vect{J}\cdot\vect{B}$ 
  normalised by the time dependent $|\vect{B}|^2$ are listed $[\Gyr^{-1}]$.
  $\average{}_{S(N)}$ indicates averaging over the total volume below (above)
  the mid-plane.
  $\Delta t$ is the period to which the averages relate $[\Gyr]$.
  Part~({\textbf{a}}) applies to the models prior to the field saturating
  and ({\textbf{b}}) to the models afterwards.
  Standard deviations are given in brackets.
          }
  \begin{tabular}{lrrrrr}
  \hline
          & 
   \multicolumn{1}{c}{$\Ompa$} & 
   \multicolumn{1}{c}{$\Ompb$} & 
   \multicolumn{1}{c}{$\Ompc$} & 
   \multicolumn{1}{c}{$\Ompd$} & 
   \multicolumn{1}{c}{$\Ompe$}   \\ 
  \hline
  \multicolumn{6}{l}{({\textbf{a}}) The kinematic phase} \\
  ${\average{\vect{J}\cdot\vect{B}}  }$
  &$\phm-7.3\,(484)$                               
  &$-48.0\,(404)$                               
  &$\phm-8.1\,(183)$                               
  &$\phm-6.0\,(170)$                               
  &$\phm-6.0\,(210)$  \\                           
  ${\average{\vect{J}\cdot\vect{B}}_S}$
  &$\phm+9.3\,\phantom{(484)}$                              
  &$-17.0\,\phantom{(484)}$                               
  &$-13.5\,\phantom{(484)}$                             
  &$\phm+7.0\,\phantom{(484)}$                              
  &$-20.7\,\phantom{(484)}$  \\                         
  ${\average{\vect{J}\cdot\vect{B}}_N}$
  &$-23.8\,\phantom{(484)}$
  &$-78.0\,\phantom{(484)}$
  &$\phm-2.7\,\phantom{(484)}$
  &$-19.1\,\phantom{(484)}$
  &$\phm+8.6\,\phantom{(484)}$  \\ [3pt]
  $\phantom{(4)}\Delta t$                          
  &$0.0$--$0.6\phm\quad$
  &$0.4$--$0.7\phm\quad$
  &$0.4$--$1.1\phm\quad$
  &$0.4$--$1.1\phm\quad$
  &$0.4$--$0.9\phm\quad$ \\ [3pt]
  \multicolumn{6}{l}{({\textbf{b}}) The non-linear phase}\\
  ${\average{\vect{J}\cdot\vect{B}}  }$
  &$\phm+0.2\,(\,4.2\,)$
  &$\phm-1.2\,(\,9.8\,)$
  &                         
  &$\phm-0.6\,(6.4\,)$
  & \\
  ${\average{\vect{J}\cdot\vect{B}}_S}$ 
  &$\phm-0.04\phantom{(~84)}$
  &$\phm-0.6~\phantom{(484)}$
  &      
  &$\phm+1.4\,\phantom{(484)}$
  &  \\
  ${\average{\vect{J}\cdot\vect{B}}_N}$ 
  &$\phm+0.46\phantom{(~84)}$
  &$\phm-1.8~\phantom{(484)}$
  &      
  &$\phm-2.7\,\phantom{(484)}$
  &  \\[3pt]
  $\phantom{(4)}\Delta t$                           
  &$1.4$--$1.7\phm\quad$
  &$1.2$--$1.7\phm\quad$
  &        
  &$1.2$--$1.7\phm\quad$
  & \\
  \hline
  \end{tabular}
  \end{table}
  
%-----------------------------------------------------------------------------
 
  The initial condition has a very weak purely azimuthal magnetic field, which 
  has zero helicity. 
  For the models with a vertical field boundary condition there can be no
  net flux of helicity into or out of the domain as $\vect{A}\cdot\vect{B}=0$ 
  on the upper and lower surfaces, and flux across the periodic boundaries sums
  to zero. 
  Therefore the generation of net magnetic helicity (Table~\ref{table:hel}a)
  in these models is only possible through the electrical resistivity present
  in the induction equation, Eq.~\eqref{eq:ind}.  
  For Model~$\Ompb$ helicity is free to be transported across the vertical 
  boundary, and although net helicity can be also generated through resistive 
  transfer, it is clear that significant vertical transport does occur, with
  $|{\average{\vect{J}\cdot\vect{B}}|=48}\Gyr^{-1}$ being much
  larger than $6-8\Gyr^{-1}$ for the other models.
  Given that $\average{\vect{J}\cdot\vect{B}}$ is negative, it appears that 
    positive helicity is being 
  preferentially transported out of the domain, and this enhances the dynamo.

  {\freply In all cases, the fluctuations (standard deviation) in helicity, shown in 
  brackets in Table~\ref{table:hel}, are 1--2 orders of magnitude larger than
  the net values. 
  With such large fluctuations, the case could be made that the net helicity 
  does not significantly differ from zero. 
  However even the small net changes in helicity may be sufficient to 
  influence the dynamo, given the efficiency of Model~$\Ompb$ compared to
  Model~$\Ompa$. 
  The standard deviation is about the same, but the signature of the faster
dynamo is stronger.}

  After saturation, when the field has grown by an order of $10^2$, the
  relative helicity is reduced to order unity in all models. 
  From Fig~\ref{fig:hel}b, for the unnormalised magnetic helicity, it is 
  evident that the net helicity at this stage can switch sign, so in fact
  over a sufficiently long period it would appear that the net relative 
  helicity fluctuates around zero. 
  For the dynamo in the rotating galaxy, the removal of positive helicity is
  required. 
  Where the net helicity is of opposite sign in the North and South, 
  (as in $\Ompa$, $\Ompd$ and $\Ompc$ Table~\ref{table:hel}a), the positive
  side can be in either half.
  There is evidence from Fig~\ref{fig:hel}c, for the normalised magnetic
  helicity, that the sign alternates in both hemispheres during the kinematic
  phase also for Model~$\Ompd$. 
  Given that helicity cannot be transported away, this may be the positive 
  helicity, being transported back and forth within the domain, whereas
  for Model~$\Ompb$ both hemispheres are strongly negative as the positive
  helicity can be transported completely out of the domain.
  {\freply Of course, substantial positive helicity persists in both 
  hemispheres as a result of the powerful fluctuations.}
%-----------------------------------------------------------------------------
  \section{Mean and fluctuating field composition}
%-----------------------------------------------------------------------------

  Applying the approach in Section~\ref{subsect:mb} to Model~$\Ompa$ yields a
  similar outcome to the plot in Fig.~\ref{fig:blb} for Model~$\Ompd$.
  As displayed in Fig.~\ref{fig:blb1}, the dependence of the separation of the
  total field into mean and fluctuating parts is very similar to Model~$\Ompd$.
  The appropriate choice of $\ell\simeq50\p$ for the smoothing length appears
  independent of model, and instead closely related to the typical SN remnant
  scale.  
  This choice is applied to all the MHD models and analysis of the resulting
  mean and fluctuating fields are presented in Table~\ref{table:dynamo} and 
  Fig.~\ref{fig:bfits}.
  
%-----------------------------------------------------------------------------
  \begin{figure}[h]
  \centering
  \hspace{-1.5cm}
  \includegraphics[width=0.535\linewidth]{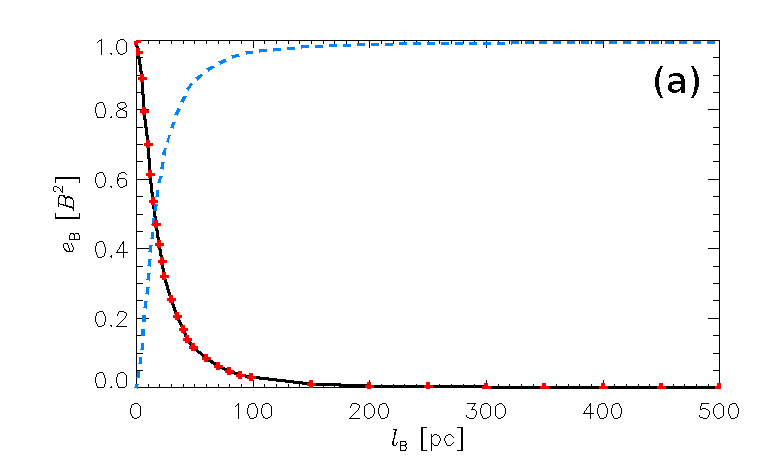}
  \includegraphics[width=0.535\linewidth]{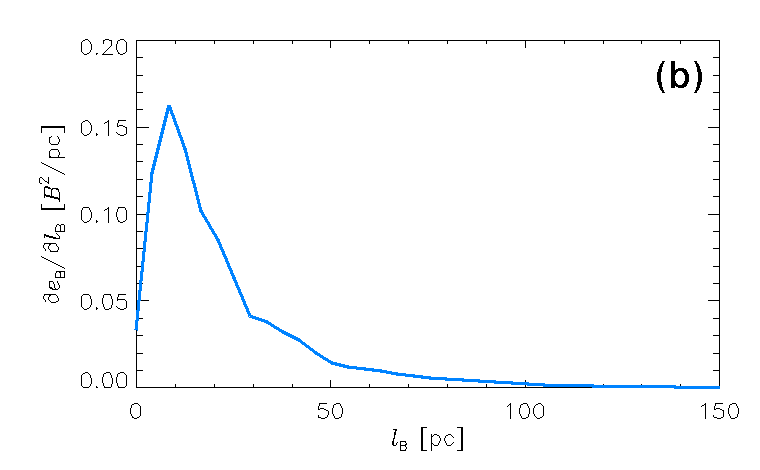}
  \hspace{-1.5cm}
    \caption[Separation into mean and fluctuating magnetic energy for
    Model~$\Ompa$]{
  For Model~$\Ompa$ at $t=0.49\Gyr$ {\textbf{(a)}} the proportion of 
  $\average{\eB}$ 
  contribution from ${B}_\ell$ and ${b}$ depending on $\ell_B$ and
  {\textbf{(b)}} $\upartial\average{\eB}/\upartial\ell_B$. 
  Compare with Fig.~\ref{fig:blb} for Model~$\Ompd$.
    \label{fig:blb1}
            }
  \end{figure}
%-----------------------------------------------------------------------------

%-----------------------------------------------------------------------------
  
  \begin{table}[b]
    \caption[Magnetic field growth rates]{\label{table:dynamo} Growth rate for the mean ($\Gamma_{e}$) and
  fluctuating ($\gamma_{e}$) magnetic energy during a portion
  of the kinematic stage for each model. 
  $\hat{\Gamma}$ is the estimated growth rate of the mean magnetic field using
  the approximation described in Section~\ref{sect:eval}.
  $\Delta t$ is the period analysed, and $N$ is the number of snapshots used.
  The fit of the sum of both exponentials to the total measured field energy
  is indicated by the reduced $\chi^2$.
          }
    \centering
    \begin{tabular}{l|cccccc}
\hline
Model   &$2\hat\Gamma$          &$\Gamma_{e}$ &$\gamma_e$   &$\Delta t$ & $\chi^2$ &$N$  \\
        &$[Gyr^{-1}]$           &$[Gyr^{-1}]$ &$[Gyr^{-1}]$ &$[Gyr]$    &          &     \\
\hline                                                                       
$\Ompa$ & \phd0 -- 1.4          & 2.1         &2.6          &0.4 -- 0.83  &0.42      &31   \\
$\Ompb$ & \phd0 -- 1.4          &10.1         &9.0          &0.4 -- 0.70  &0.01      &16   \\
$\Ompc$ & \phd0 -- 1.4          & 4.2         &2.4          &0.4 -- 0.82  &0.43      &36   \\
$\Ompd$ &\phd8 -- \phantom{.}10 & 5.5         &5.1          &0.4 -- 0.84  &0.04      &37   \\
$\Ompe$ & 1.8 -- 3.8            & 5.8         &5.3          &0.4 -- 0.82  &0.10      &36   \\
    \end{tabular}
  \end{table}

%-----------------------------------------------------------------------------
    
  The growth rates of the mean and fluctuating parts of the magnetic energy 
  from the MHD models are listed in Table~\ref{table:dynamo}. 
  The plots of the actual growth together with the fitted exponentials are 
  displayed in Fig.~\ref{fig:bfits}. 
  These can be compared to the growth rates of the total field energy shown in 
  Fig.~\ref{fig:bbe}b, which includes the whole simulation period.
  The table data and the plots in Fig.~\ref{fig:bfits} are derived over a time
  interval of about $400\Myr$, and strong temporal fluctuations are evident,
  even over so long a time frame. 
  This time frame is chosen because it is well within the kinematic regime
  for analysis of the dynamo,
  and because there is snapshot data for all models during this period.
  From the steady growth in Models~$\Ompb$ and $\Ompd$ (panels a and b)
  it is clear that they can be well approximated as exponential.
  This is less clear from the other models (in panels c and d), which have
  more pronounced dips and plateaux. 
  However from Fig.~\ref{fig:bbe} it is evident that further exponential 
  growth follows.
  There would be greater certainty over the parameters derived for $\Gamma_e$
  and $\gamma_e$ if resources permitted the extension of these models for at
  least another $400\Myr$, and preferably up to saturation. 
%-----------------------------------------------------------------------------
  \begin{figure}[h]
  \centering
  \hspace{-1.25cm}
  \includegraphics[width=0.535\linewidth]{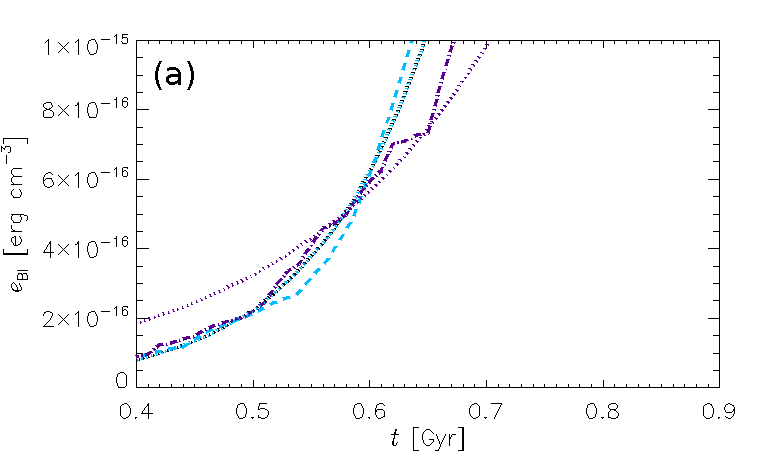}
  \includegraphics[width=0.535\linewidth]{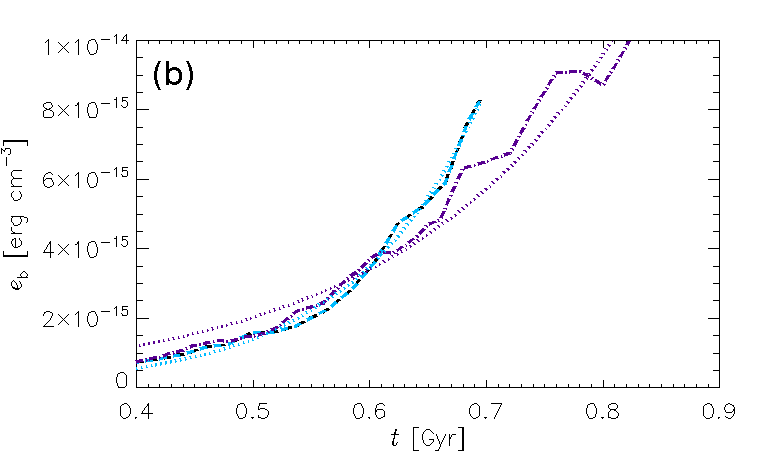}
  \hspace{-1.5cm}\\
  \hspace{-2.5cm}
  \includegraphics[width=0.535\linewidth]{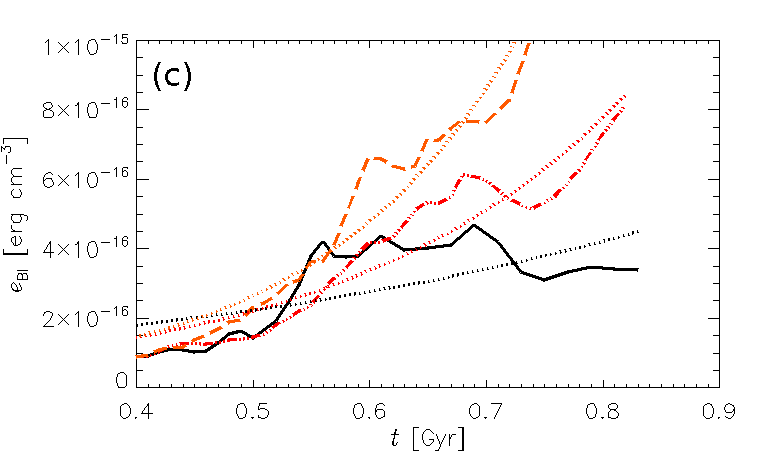}
  \includegraphics[width=0.535\linewidth]{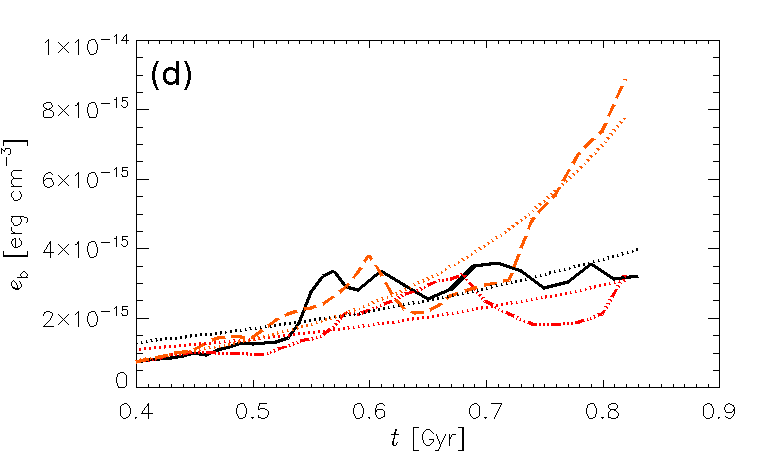}
  \hspace{-1.5cm}
    \caption[Exponential fits for magnetic growth rates]{
  Growth of the mean ({\textbf{(a)}} and {\textbf{(c)}}) and fluctuating 
  ({\textbf{(b)}} and {\textbf{(d)}}) field.
  Models~$\Ompb$ (light blue, dashed) and $\Ompd$ (purple, dash-dotted) are 
  shown in 
  panels (a) and (b);
  Models~$\Ompa$ (black, solid),
  $\Ompc$ (red, dash-3dotted) and
  $\Ompe$ (orange, long-dashed) in panels (c) and (d).
  Plotted with the actual growth rates are the exponential fits for each 
  model (matching color, dotted).
  The parameters of the fits are detailed in Table~\ref{table:dynamo}.
    \label{fig:bfits}}
  \end{figure}
%-----------------------------------------------------------------------------

  The estimate $\hat\Gamma$ is derived using Eq.\eqref{eq:dynamo} from 
  Section~\ref{sect:eval}.
  The estimates included are $l\turb=0.1\kpc$, $u\turb=15\kms$, $h=0.6\kpc$ and
  $D_{\textrm{crit}}=6\sim10$.
  From the location of the strongest field in Fig.~\ref{fig:pavB} it may be
  that for the mean field dynamo, the active region is $0.2\la|z|\la0.6$ with
  the mid-plane more active for the fluctuation dynamo.
  The height of the strongest mean field in Model~$\Ompc$ appears slightly
  lower, $h\la0.4\kpc$, from its equivalent plot to Fig.~\ref{fig:pavB}.
  This might be expected as the reduced SN activity will not inflate the disc
  as much.
  Otherwise $h$ appears quite consistent across the models.
  $l\turb$ would not appear to be dependent upon the parameters, but on the size
  of the remnants.
  Again, it might be expected to be somewhat smaller for Model~$\Ompc$, because
  increased density in the most SN active region could restrict typical remnant
  size. 
  Since the gas in all models is transonic, the characteristic turbulent 
  velocity is likely to be around the speed of sound for the gas. 
  $u\turb$ may therefore vary modestly and it will be worth obtaining improved 
  approximations. 
 
  Comparing the model fitted growth rates $\Gamma_e$ to the estimates for the
  growth rates from the shear dynamo $\hat\Gamma$ all of the actual growth rates
  exceed the estimates, except for Model~$\Ompd$. 
  In Section~\ref{sect:eval} this model was analysed for a later period in its
  evolution and $\Gamma_e=10.9\Gyr^{-1}$ obtained is comparable to 
  $2\hat\Gamma=8\sim10\Gyr^{-1}$. 
  Perhaps there remains significant transients from the Model~$\Ompa$ snapshot
  at $400\Myr$ from which this model is initialised. 
  Otherwise the model growth rates are only slightly above the upper estimates,
  except for Model~$\Ompb$ with $\Gamma_e=10.1\Gyr^{-1}$ much larger than the 
  upper estimate of 1.4.
  This model has the open vertical boundary condition for the magnetic field
  and the efficiency of the dynamo may benefit from the advection of helicity.

  Also of interest, the model growth rates $\gamma_e$ for the fluctuating field
  energy are generally smaller than for the mean field energy.
  This is not the case for Model~$\Ompa$ where $\Gamma_e=2.1\Gyr^{-1}$ is 
  less than $\gamma_e=2.6\Gyr^{-1}$.
  As can be seen from Fig.~\ref{fig:bfits}c and d, the growth of
  this model had stalled temporarily, so this might not be significant over the 
  longer times scales.
  For Model~$\Ompc$ the relative difference between $\gamma_e=2.4\Gyr^{-1}$ and
  $\Gamma_e=4.2\Gyr^{-1}$ is somewhat larger than in the other models.
  This could also be due to the short term dip in the total magnetic energy,
  but may indicate a higher SN rate contributes to the fluctuation dynamo.
  In both cases further extended analysis in the kinematic phase is required.
\section{Three-phase structure of the Field}\label{sect:3BB}
%-----------------------------------------------------------------------------

  The total volume probability distributions for gas number density $n$, 
  temperature $T$ and thermal pressure $p$ are displayed in Fig.~\ref{fig:pdfm}
  from Models~$\Ompa$ (black, solid) and $\Omph$ (blue, dashed). 
  Note that Model~$\Omph$ includes the correction to the SN distribution,
  which stabilises the disc against unphysical cyclic oscillations, although
  it is still subject to natural random vertical fluctuations.
  Hence the mean density in the SN active region remains consistently higher
  than in Model~\Op.
  
%-----------------------------------------------------------------------------
  \begin{figure}[h]
  \centering\hspace{-2cm}
  \includegraphics[width=0.36\columnwidth,clip=true,trim=0 0 0 9mm]{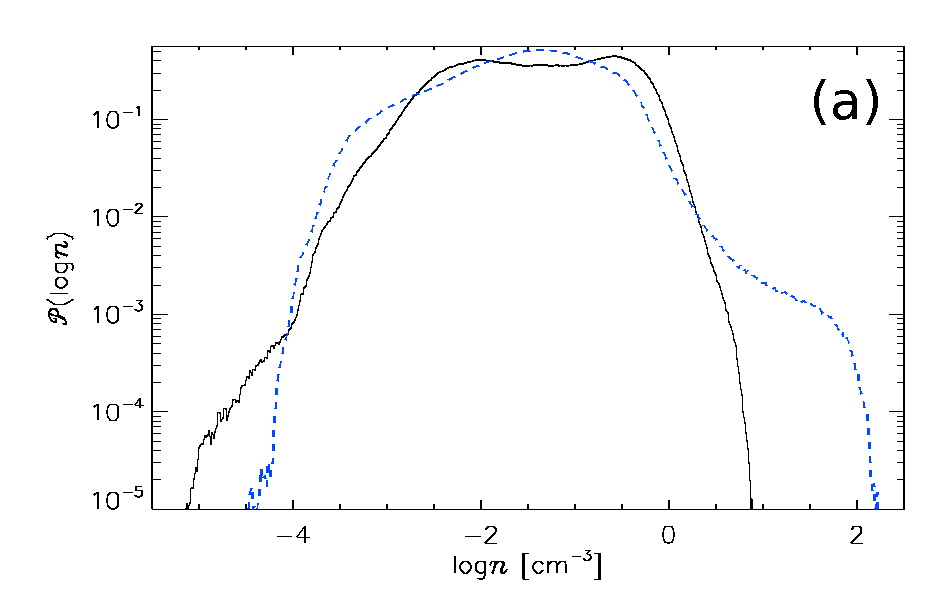}
  \includegraphics[width=0.36\columnwidth,clip=true,trim=0 0 0 9mm]{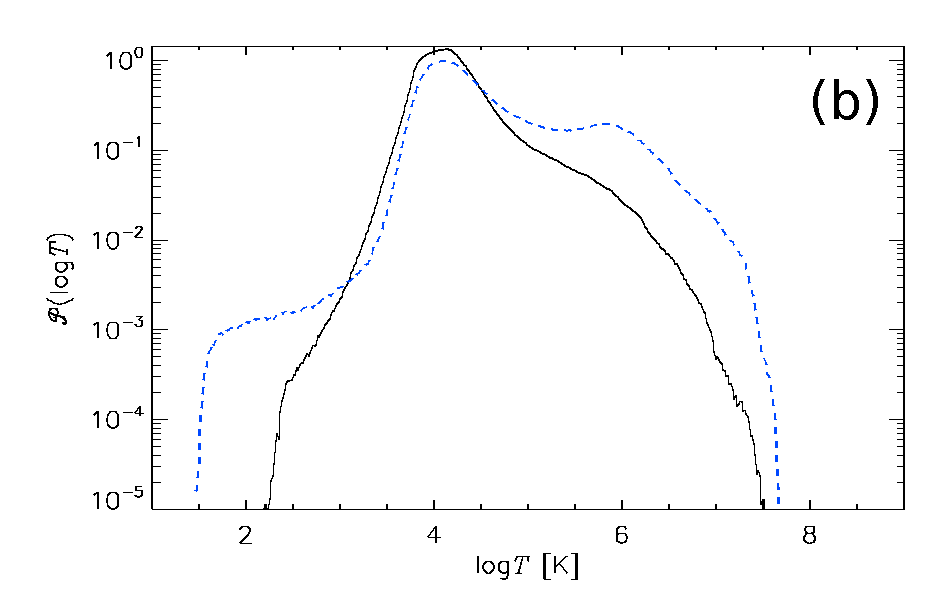}
  \includegraphics[width=0.36\columnwidth,clip=true,trim=0 0 0 9mm]{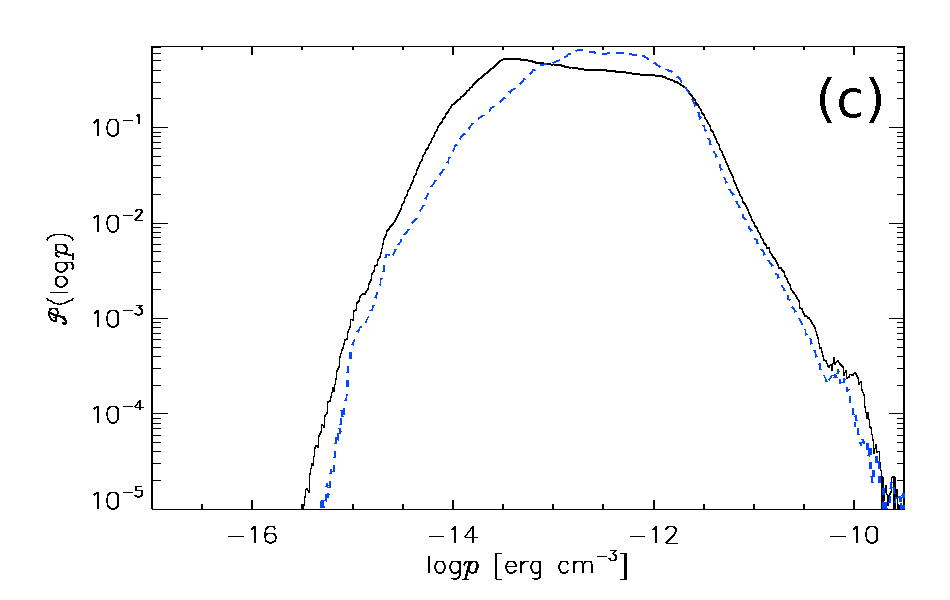}
  \hspace{-2cm}
    \caption[Total volume probability distributions for Model~$\Ompa$ and $\Omph$]{
  Volume weighted probability distributions of gas number
  density~{\textbf{(a)}}, temperature~{\textbf{(b)}} and thermal
  pressure~{\textbf{(c)}} for  models {$\Omph$} (black, solid) and
  {$\Ompa$} (blue, dashed) for the total numerical domain $|z|\le1.12\kpc$.
    \label{fig:pdfm}}
  \end{figure}
%-----------------------------------------------------------------------------

  Although the three phase temperature distribution is still visible for this
  model in Fig.~\ref{fig:pdfm}b, it is less pronounced than the results from
  Model~\Op\ shown in Fig.~\ref{fig:pdf2}b.
  However the three phase structure for the MHD Model~$\Ompa$ is not at all
  apparent in panel b, with a significantly narrower range of temperatures.
  The bulk of the density distributions (a) are quite similar between the 
  HD and MHD models, except that the high densities are not as well resolved in
  the MHD models.
  The thermal pressure distributions are very similar, but with the HD modal
  pressure approximately one third the MHD modal pressure. 

%-----------------------------------------------------------------------------
  \begin{figure}[h]
  \centering\hspace{-2cm}
  \includegraphics[width=0.535\columnwidth]{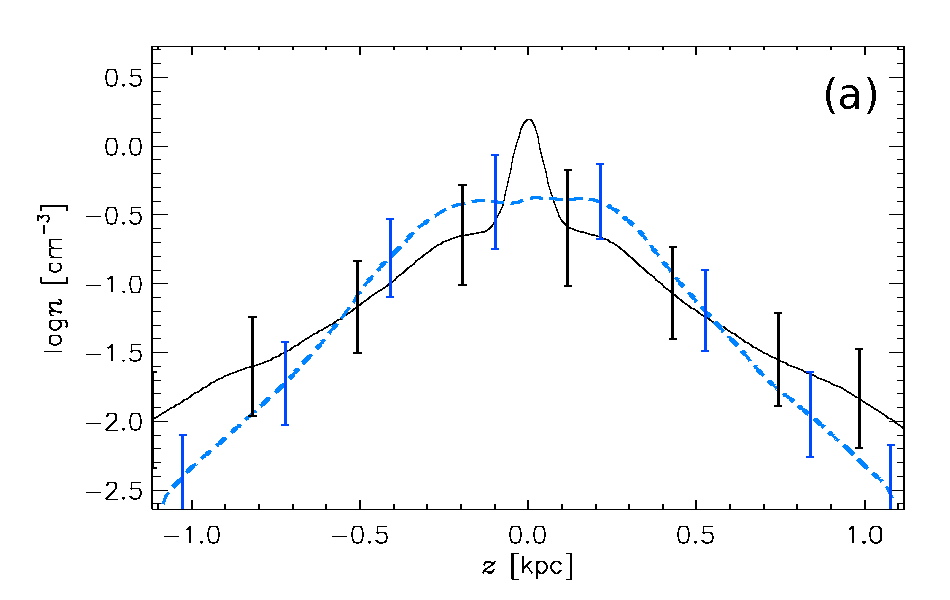}
  \includegraphics[width=0.535\columnwidth]{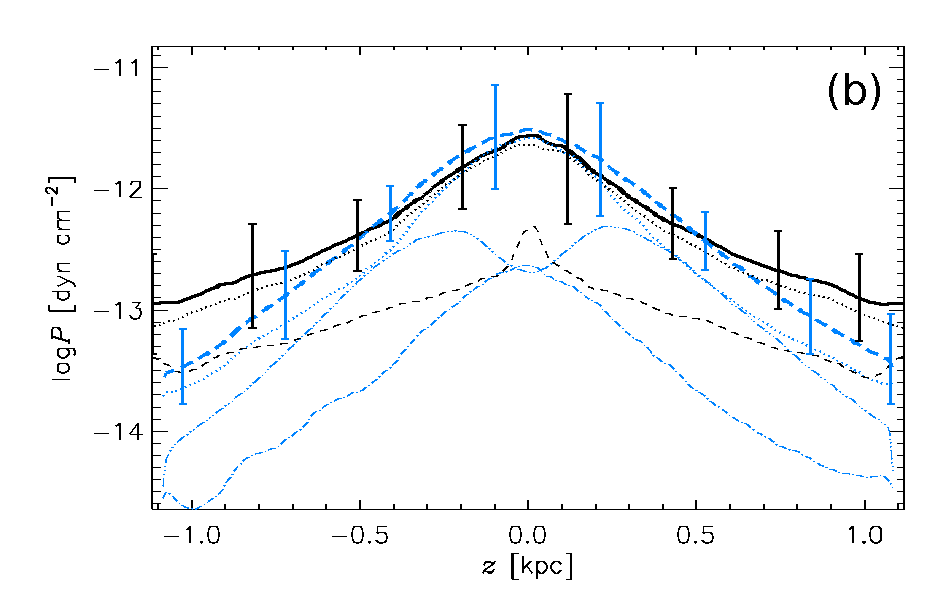}\hspace{-2cm}
    \caption[Horizontal averages of $n$ and $P$ for Model~$\Ompa$ and $\Omph$]{
  Horizontal averages of gas number density, $\mean{n}(z)$ {\textbf{(a)}}, and
  total pressure, $\mean{P}(z)$ {\textbf{(b)}}, for Model~{$\Ompa$} (solid,
  black), and Model~{$\Omph$} (dashed, blue). 
  Each are time-averaged using eleven snapshots respectively, spanning 
  100\Myr.
  The vertical lines indicate standard deviation within each horizontal slice.
  The thermal $\mean{p}(z)$ (dotted) and ram $\mean{p}\turb(z)$ (fine dashed)
  pressures are also plotted {\textbf{(b)}}.
  For Model~{\OpH} the magnetic pressure $\mean{p}_B$ is also plotted (fine, 
  dash-3dotted).
    \label{fig:zrhom}
            }
  \end{figure}
%-----------------------------------------------------------------------------

  As discussed in Section~\ref{subsect:params} the stability of the disc
  reduces the effectiveness of the SNe to generate and circulate hot gas, in
  the absence of SN clustering. 
  Hence Model~$\Omph$ has a slightly thicker disc, with less hot gas than 
  Model~\Op.
  On top of this the effect of the magnetic pressure in Model~$\Ompa$ is to 
  expand the thick disc even further and this is illustrated in 
  Fig.~\ref{fig:zrhom}a, where the horizontal averages of gas number density
  $n(z)$ are plotted against $z$ for both models. 
  The strong peak in the density at the mid-plane is evident for Model~$\Omph$
  (black, solid), while for Model~$\Ompa$ (blue, dashed) there is a broad
  plateau in the density, extending to $|z|\simeq300\p$, where the mean 
  magnetic field is strongest.
  
%-----------------------------------------------------------------------------
  \begin{figure}[h]
  \centering
  \includegraphics[width=0.45\linewidth]{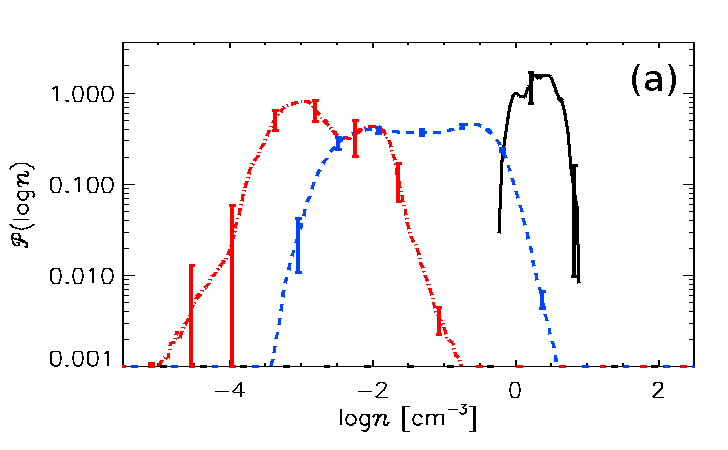}  
  \includegraphics[width=0.45\linewidth]{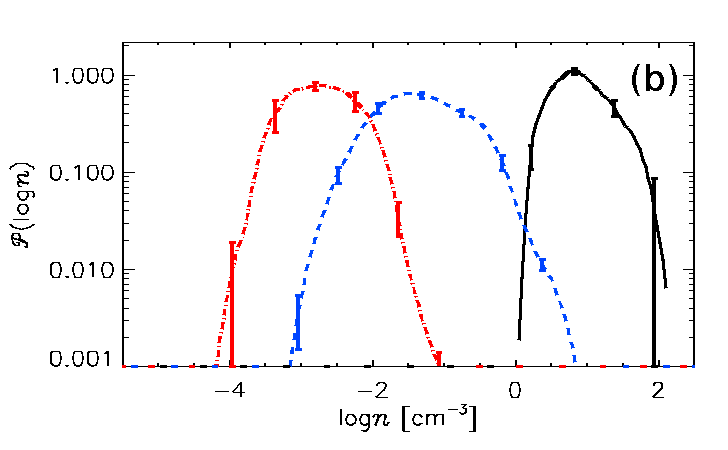}\\  
  \includegraphics[width=0.45\linewidth]{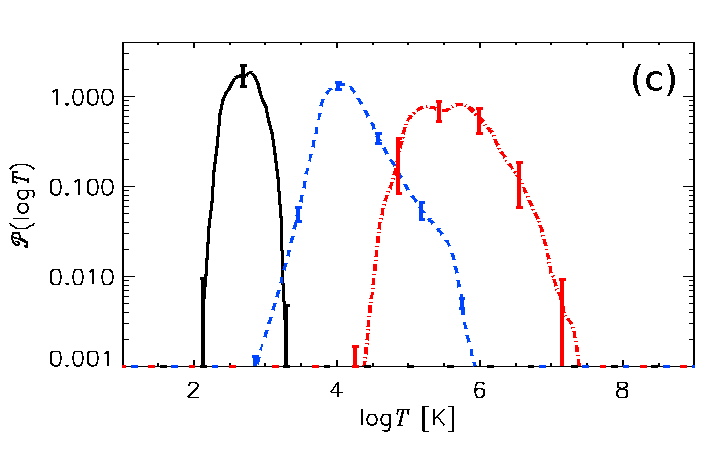}
  \includegraphics[width=0.45\linewidth]{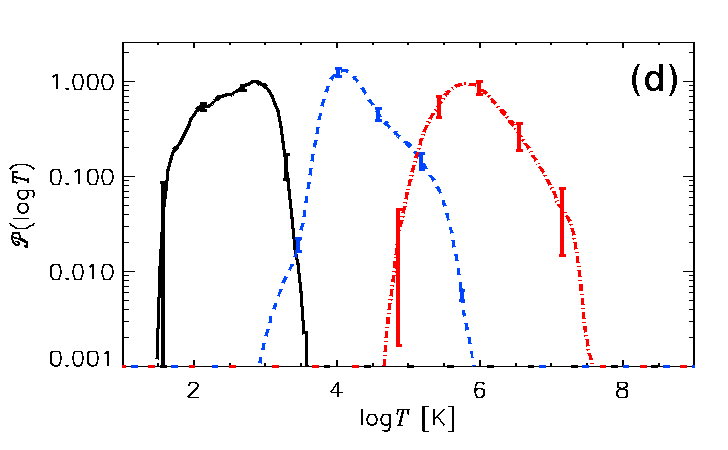}  
    \caption[Probability distributions by phase for $n$ and $T$]{
  Probability distributions by phase: cold (blue, dashed), warm (black, solid)
  and hot (red, dash-dotted) for gas number density ($n$ {\textbf{(a), (b)}})
  and temperature ($T$ {\textbf{(c), (d)}}) 
  for Model~$\Ompa$ ({\textbf{(a), (c)}}) and Model~$\Omph$
  ({\textbf{(b), (d)}}).
%  Data are averaged over 100\,Myr using eleven snapshots. 
  95\% confidence intervals for temporal deviation are shown as error bars.   
  \label{fig:npdf3s}
    }
  \end{figure}
%-----------------------------------------------------------------------------

  The horizontal averages of the pressure are plotted in Fig.~\ref{fig:zrhom}b
  for both models. 
  There is a strong peak at the mid-plane in the turbulent pressure for the HD
  model (black, dashed), but a much weaker profile for the MHD model (blue,
  dash-dotted). 
  There are two peaks in the magnetic pressure (blue, dash-3dotted) near 
  $|z|\simeq200\p$, which supports the extended density profile.
  Another possible effect, which might constrain the circulation of the hot
  gas, and hence enhance the pressure at the mid-plane, is the strong 
  horizontal orientation of the field.
  As mentioned in Section~\ref{sect:dyn} the periodic boundary conditions
  exclude a non-zero vertical component to the mean field, so the magnetic
  tension predominantly acts against the vertical flows.
  Some understanding of the multi-phase structure of the magnetised ISM is
  still possible from these models, but the extreme temperatures and densities
  are significantly under represented.
  To improve this in future work it will be desirable to allow unrestricted 
  evolution of vertical field and to apply realistic clustering of the 
  SNe to generate more supperbubbles (composite multiple SN remnants forming a
  single superstructure) or chimneys (plumes venting hot gas from the disc 
  towards the halo).

  Results for separation of the ISM into three phases using the method 
  detailed in Section~\ref{sect:entropy} are shown for Models~$\Ompa$ and
  $\Omph$ in Fig~\ref{fig:npdf3s} with total volume probability distributions.
  The phases are defined using entropy $s$ such that for cold 
  $s<4.4\cdot10^{8}\erg \g^{-1}\K^{-1}$ and hot 
  $s>23.2\cdot10^{8}\erg \g^{-1}\K^{-1}$ with warm in between.
  Apart from the higher densities for the cold phase with the HD model (panel 
  b), 
  anticipated by the volume distributions (Fig.~\ref{fig:pdfm}), the warm and
  hot distributions for the MHD density (panel a) are broad, with a bimodal 
  structure to the hot gas. 
  The distributions for the warm gas (panels c and d) are very similar and for the
  cold (hot) distributions the MHD model does not extend to as low (high)
  temperatures. 
  
%-----------------------------------------------------------------------------
  \begin{figure}[h]
  \centering
  \includegraphics[width=0.475\linewidth]{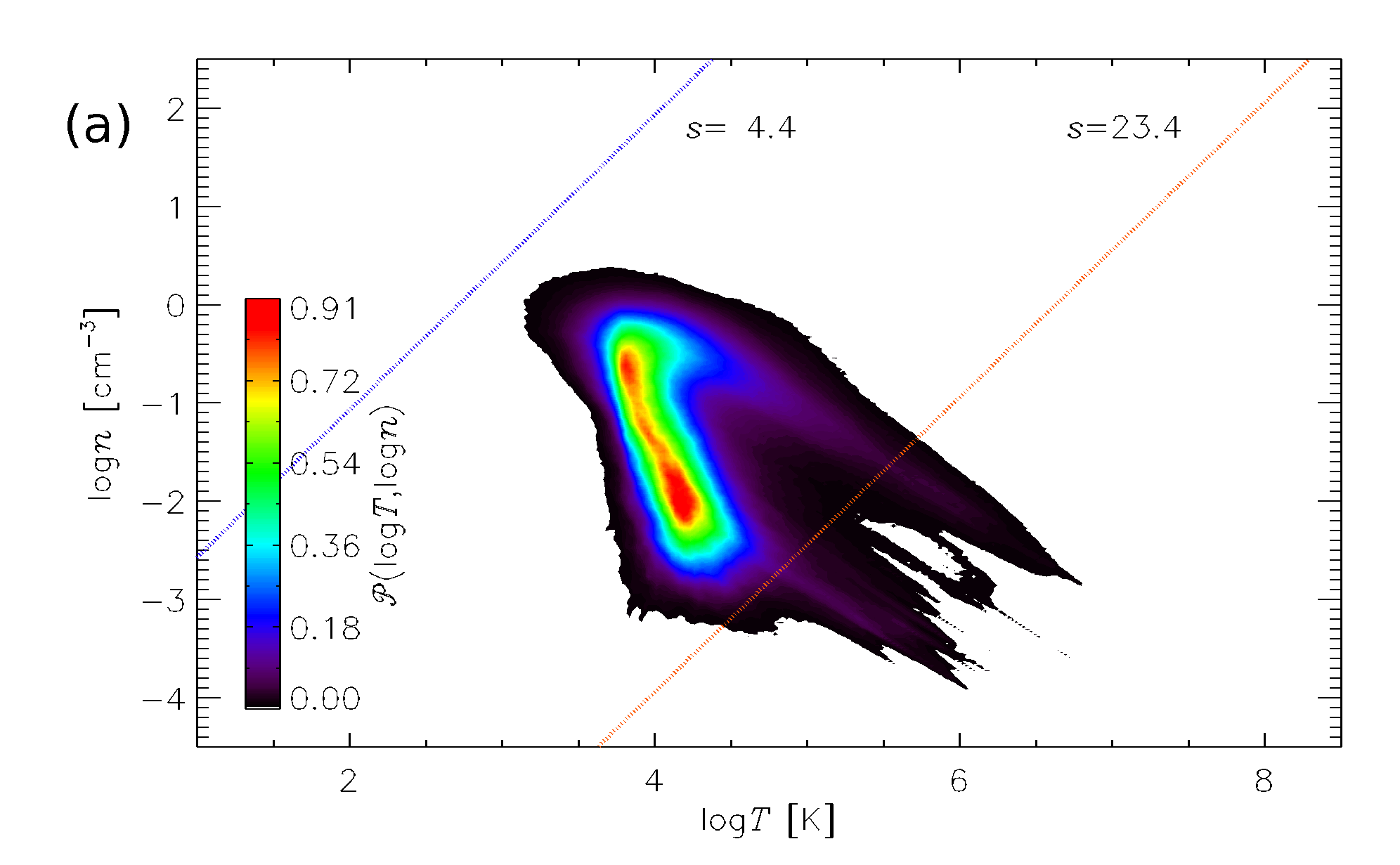}  
  \includegraphics[width=0.475\linewidth]{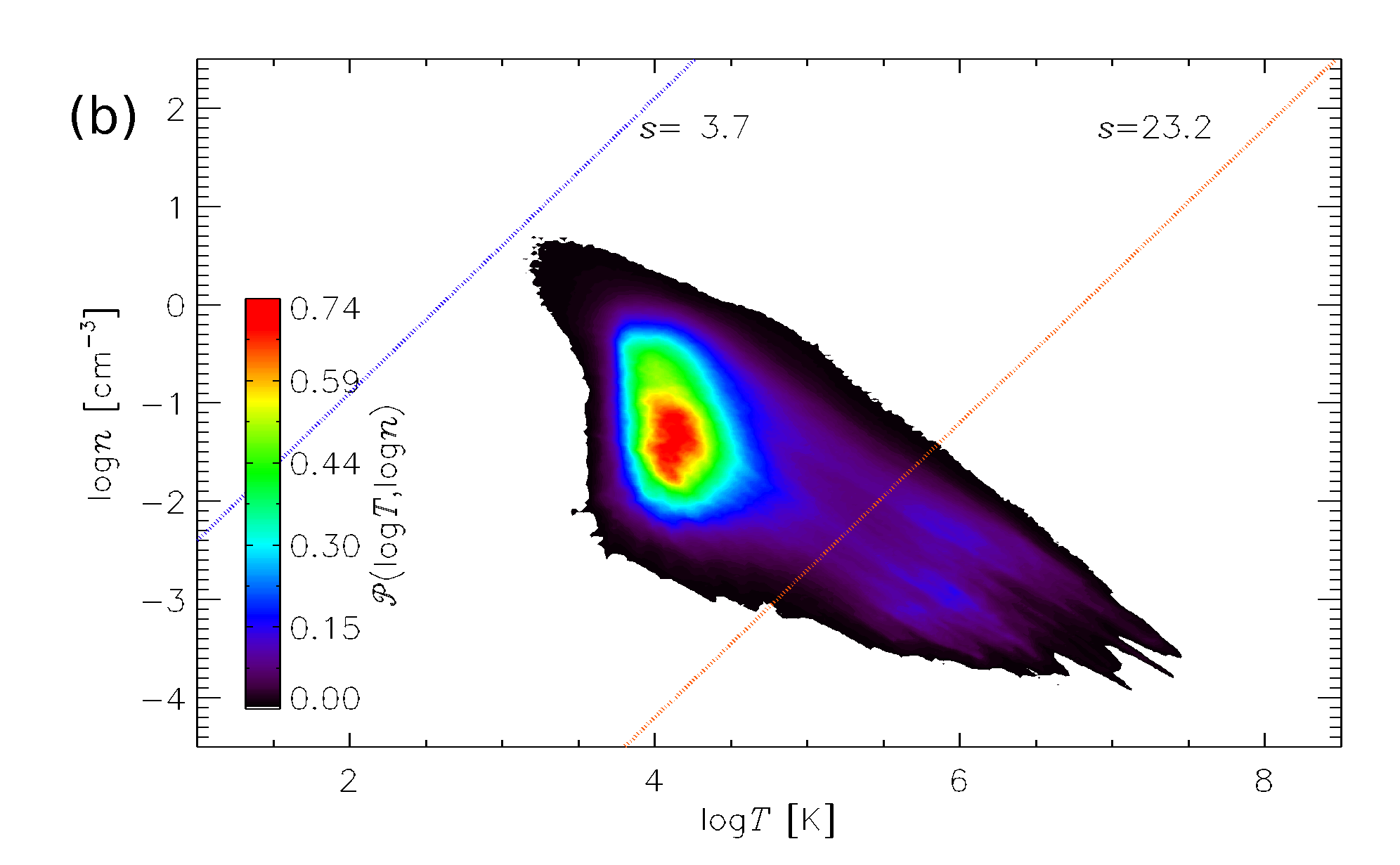}
    \caption[2D probability distribution of $n$ and $T$ for Models~$\Ompa$ and $\Omph$]{
  Probability contour plot by volume of log$n$ vs log$T$ for Model~$\Ompa$
  {\textbf{(a)}}
  and Model~$\Omph$ {\textbf{(b)}}.
  The lines of constant entropy $s=4.4\cdot10^{8}$ 
  and $23.2\cdot10^{8}\erg\g^{-1}\K^{-1}$ indicate where the phases are defined
  as cold for $s\le4.4$ and as hot for $s>23.2$. 
  \label{fig:b2dv}
    }
  \end{figure}
%-----------------------------------------------------------------------------

  Comparing the combined probability distribution of density and temperature
  for both of these models in Fig.~\ref{fig:b2dv} with those of Model~\Op\
  in Fig.~\ref{fig:pdf2d} the spread is more broad and not obviously 
  aligned along a line of constant pressure.
  The HD distribution here is less compact than with MHD. 
  However when considering only the mid-plane distributions, as displayed in
  Fig.~\ref{fig:b2dh} the distributions match better with Model~\Op\ and the 
  pressure alignment is evident.
  So the broad distributions for the total volumes are explained by the 
  stronger gradient in the pressure distribution, due to the reduced 
  stirring of the hot gas.
  The thermal pressure at the mid-plane is also very similar in both models,
  reflected also in the agreement of the total and thermal pressure near the
  mid-plane in the plot of horizontal averages (Fig~\ref{fig:zrhom}b).
  The magnetic and turbulent pressure in Model~$\Ompa$ combine to match the
  mid-plane turbulent pressure alone of Model~$\Omph$.
  For the temperature in Model~$\Ompa$ the hot gas has two modes, evident in 
  Fig.~\ref{fig:b2dv}a at $10^5\K$ and $10^6\K$, but at the mid-plane there 
  only the single $10^3\K$ mode. 
  The structure of the ISM at the mid-plane is therefore common to both models
  with modes at $10^6\K,~10^{-2}\cmcube$ and $10^4\K,~1\cmcube$.
  The cold gas is insufficiently resolved in Model~$\Ompa$ for comparison. 
  
%-----------------------------------------------------------------------------
  \begin{figure}[h]
  \centering
  \includegraphics[width=0.475\linewidth]{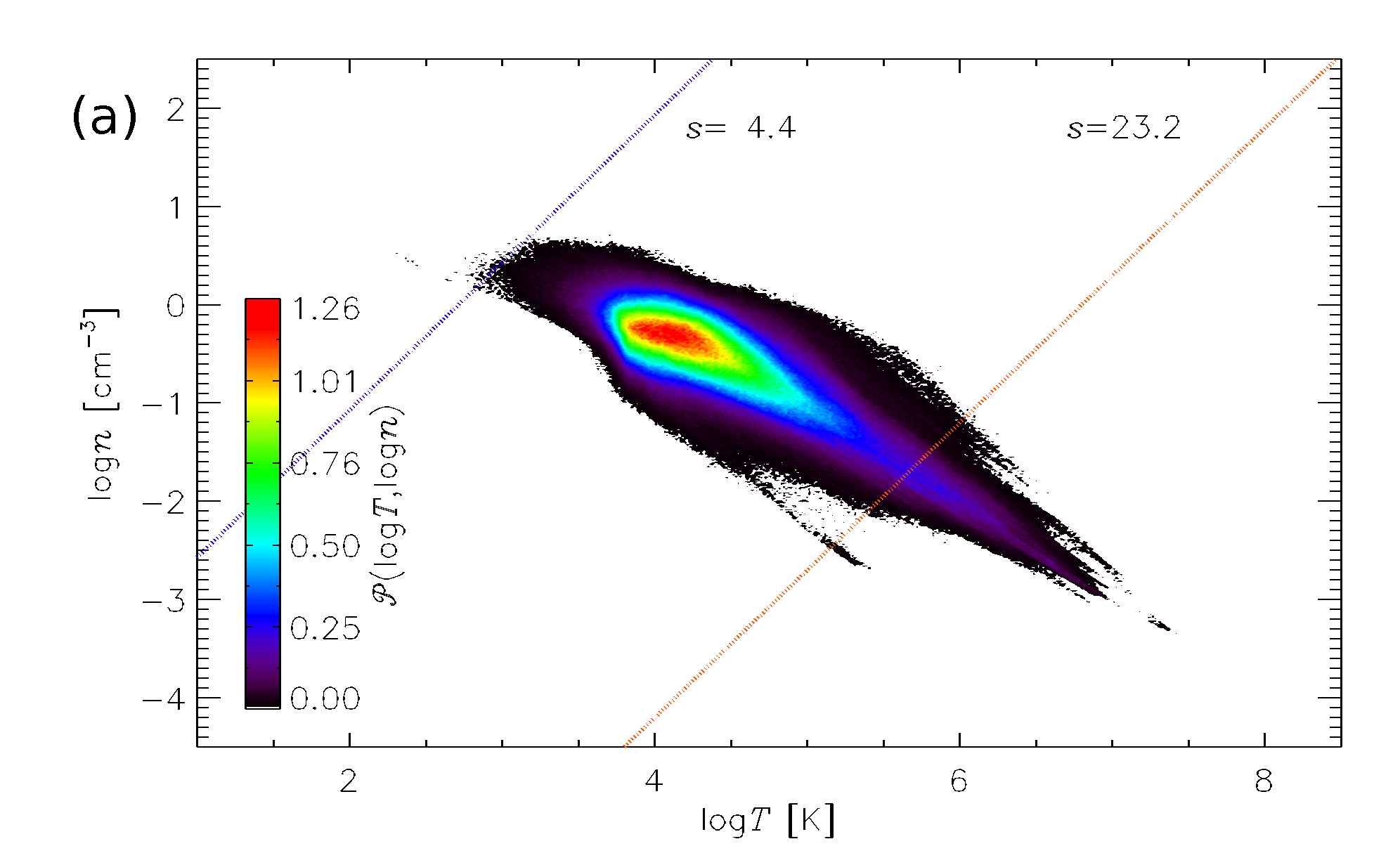}  
  \includegraphics[width=0.475\linewidth]{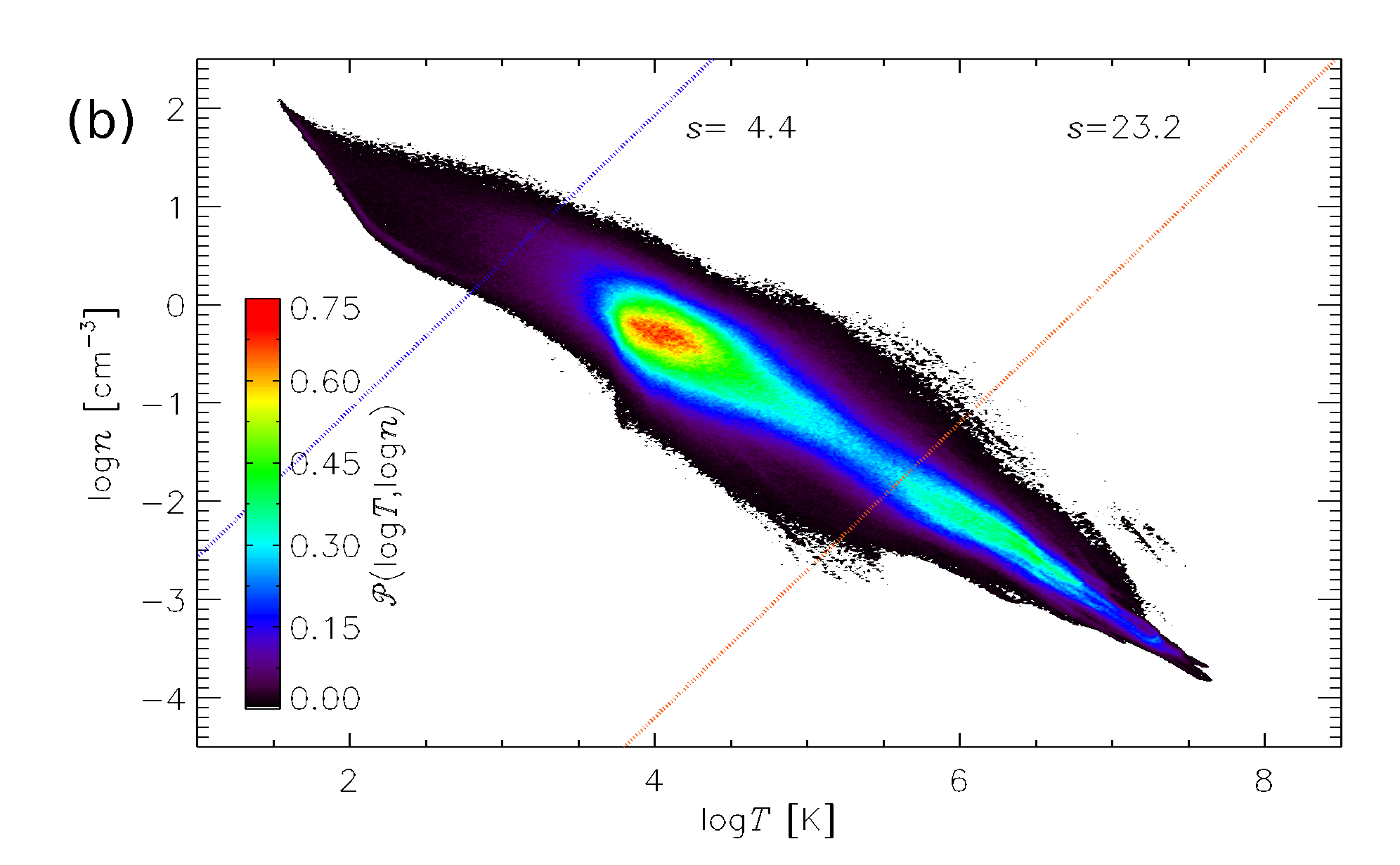}
    \caption[2D mid-plane probability distribution for Models~$\Ompa$ and  $\Omph$]{
  The mid-plane probability distributions ($|z|<100\p$) by gas number density 
  $\log n$ and temperature $\log T$ for {\textbf{(a)}} Model~$\Ompa$ and
  {\textbf{(b)}} Model~$\Omph$.
  \label{fig:b2dh}
    }
  \end{figure}
%-----------------------------------------------------------------------------

  There is no evident dependence in the probability distributions between the 
  MHD models differing in rotation, shear or SN rate.
  The models in the kinematic stage extend to lower densities and higher 
  temperatures than either the HD Model~$\Omph$ or the MHD models in the 
  dynamo saturated state, and also extend to lower pressures.
  
%-----------------------------------------------------------------------------
  \begin{figure}[h]
  \centering
  \includegraphics[width=0.475\linewidth]{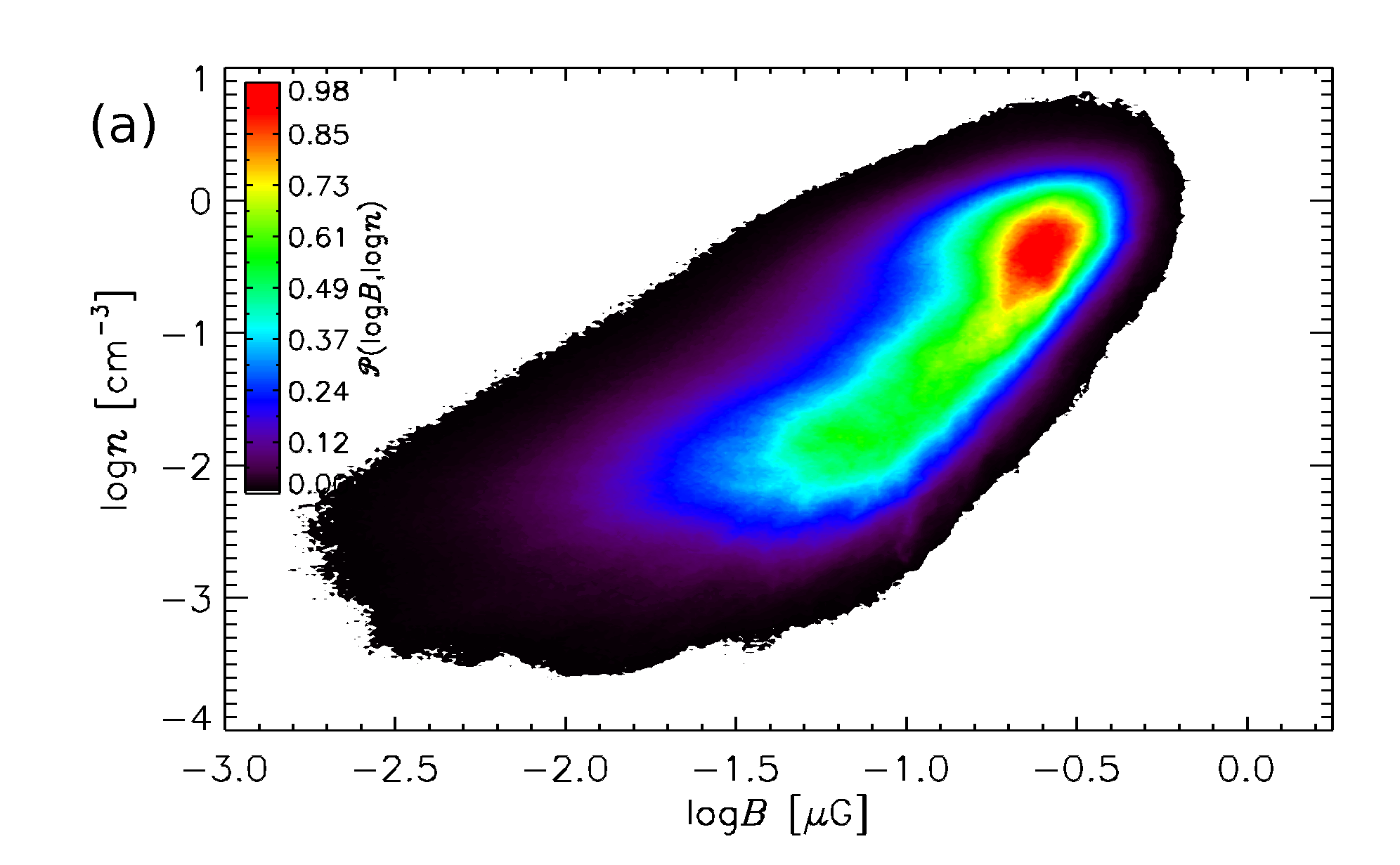}  
  \includegraphics[width=0.475\linewidth]{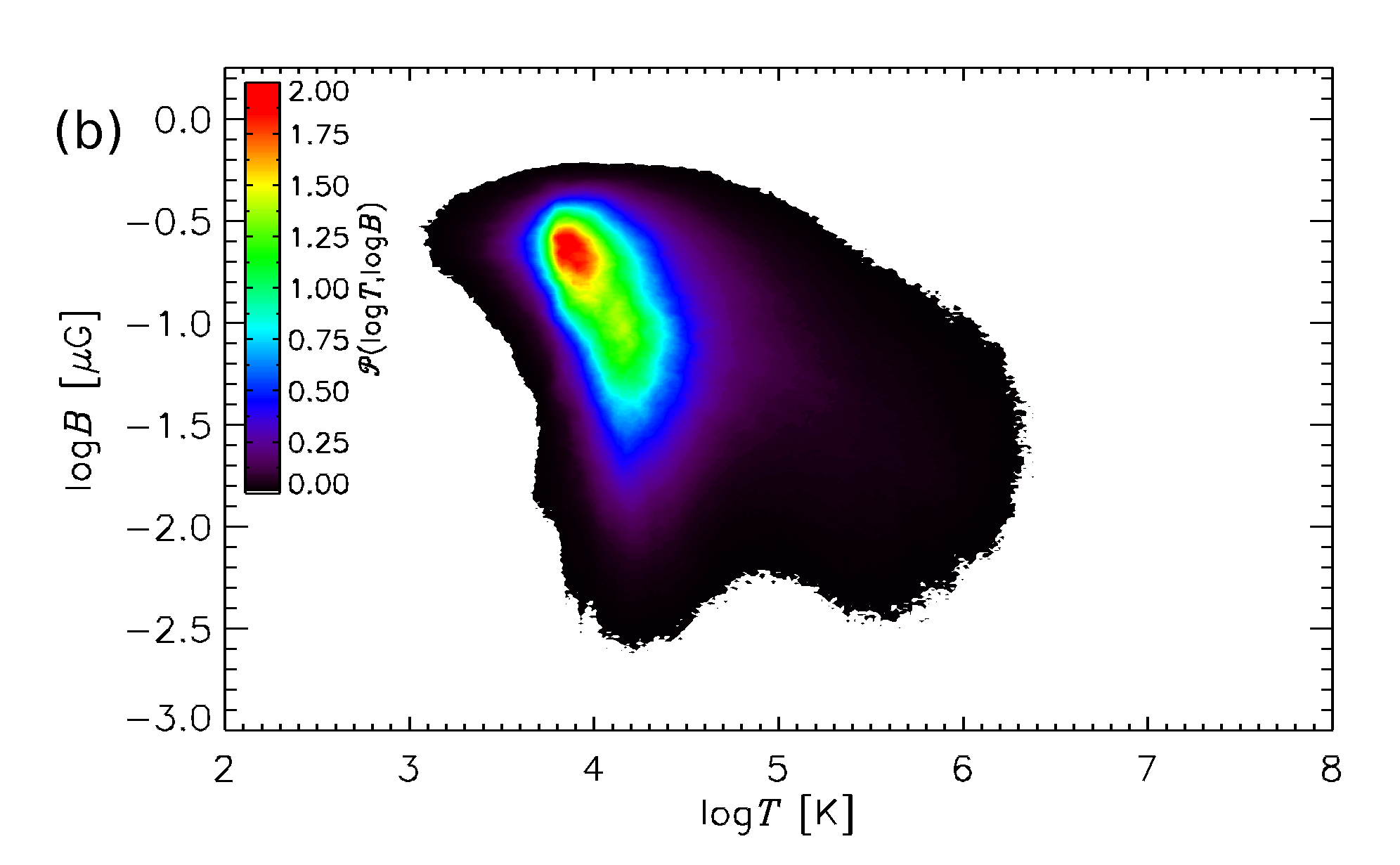}
    \caption[2D probability distribution of $n,B$ and $B,T$ for Model~$\Ompa$]{
  Total volume probability distributions ($|z|<100\p$) by gas number density 
  $\log n$ and magnetic field strength $\log |B|$ {\textbf{(a)}} and
  temperature $\log |T|$ and magnetic field strength $\log |B|$ 
  {\textbf{(b)}} for Model~$\Ompa$.
  \label{fig:b2dnt}
    }
  \end{figure}
%-----------------------------------------------------------------------------

  In Fig.~\ref{fig:b2dnt}a the joint probability distributions of gas number 
  density with magnetic field strength is shown and in Fig.~\ref{fig:b2dnt}b
  of temperature with magnetic field strength for Model~$\Ompa$.
  From (a) it is clear there is a strong positive correlation between magnetic
  field strength and density and from (b) a weak negative correlation between
  temperature and field strength.
  The $T,B$ distribution has very strong peak at $T=10^4\K$.

%-----------------------------------------------------------------------------
  \begin{figure*}[h]
  \centering
  \hspace{-1.25cm}
  \includegraphics[width=0.3965\linewidth]{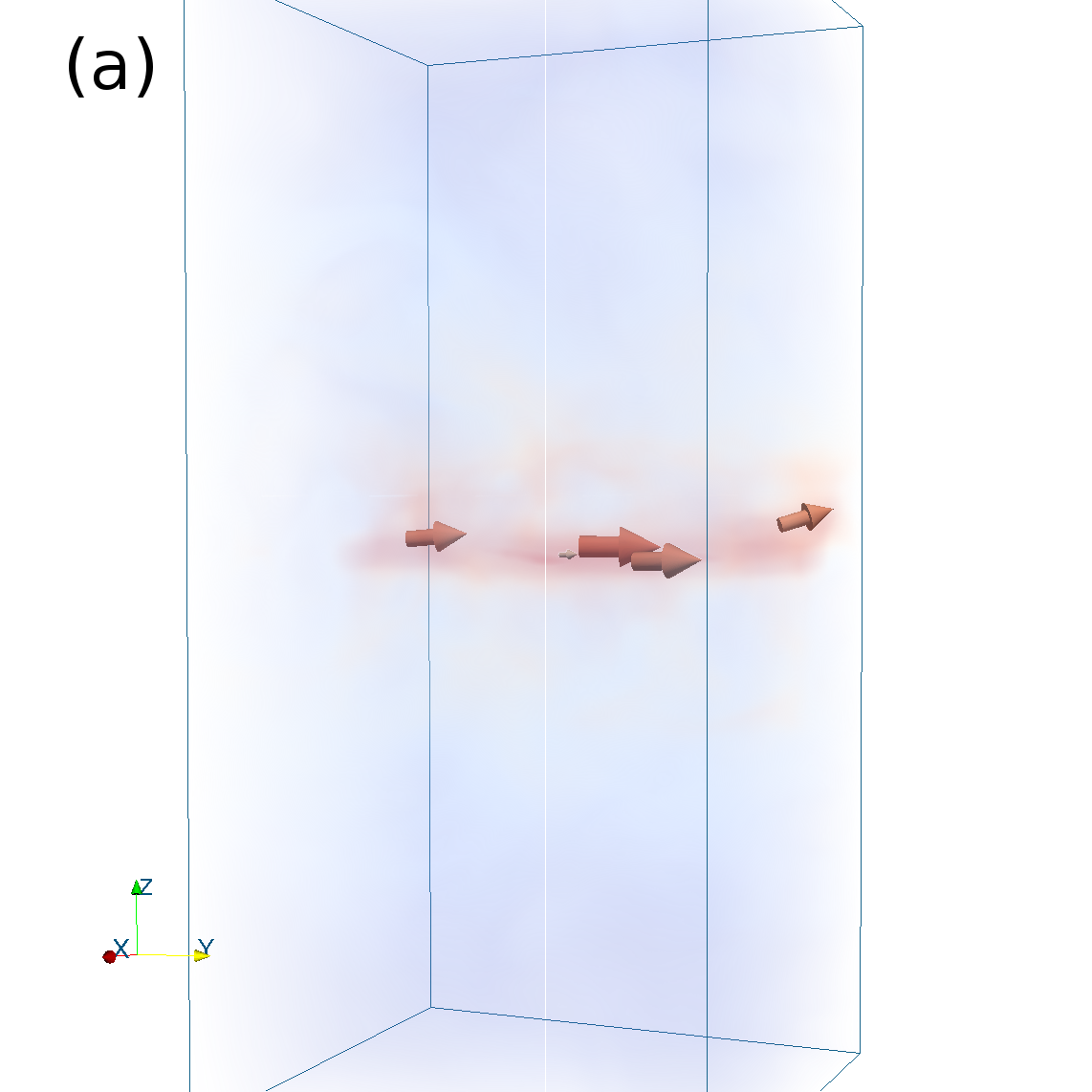}\hspace{-1.15cm}
  \includegraphics[width=0.3965\linewidth]{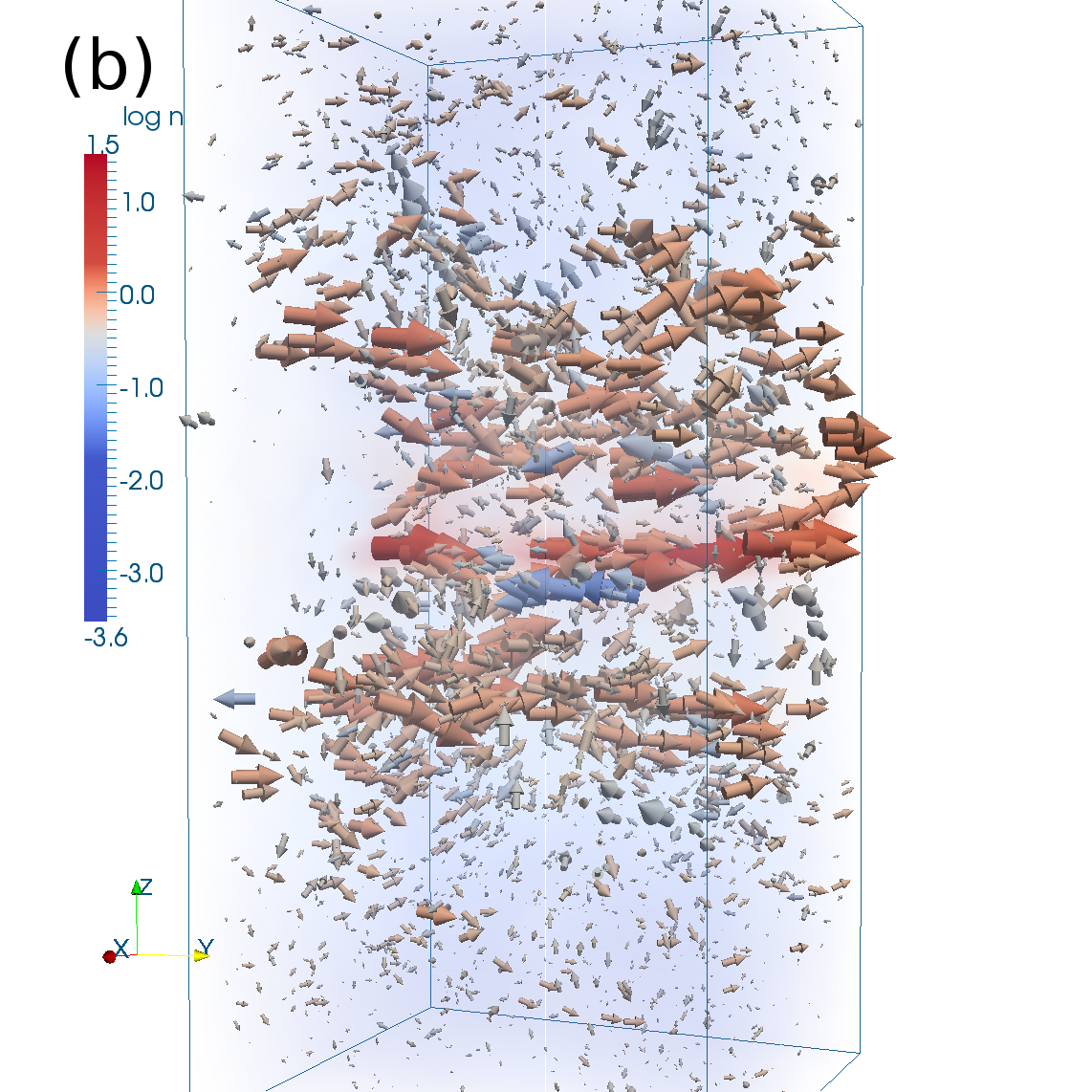}\hspace{-1.15cm}
  \includegraphics[width=0.3965\linewidth]{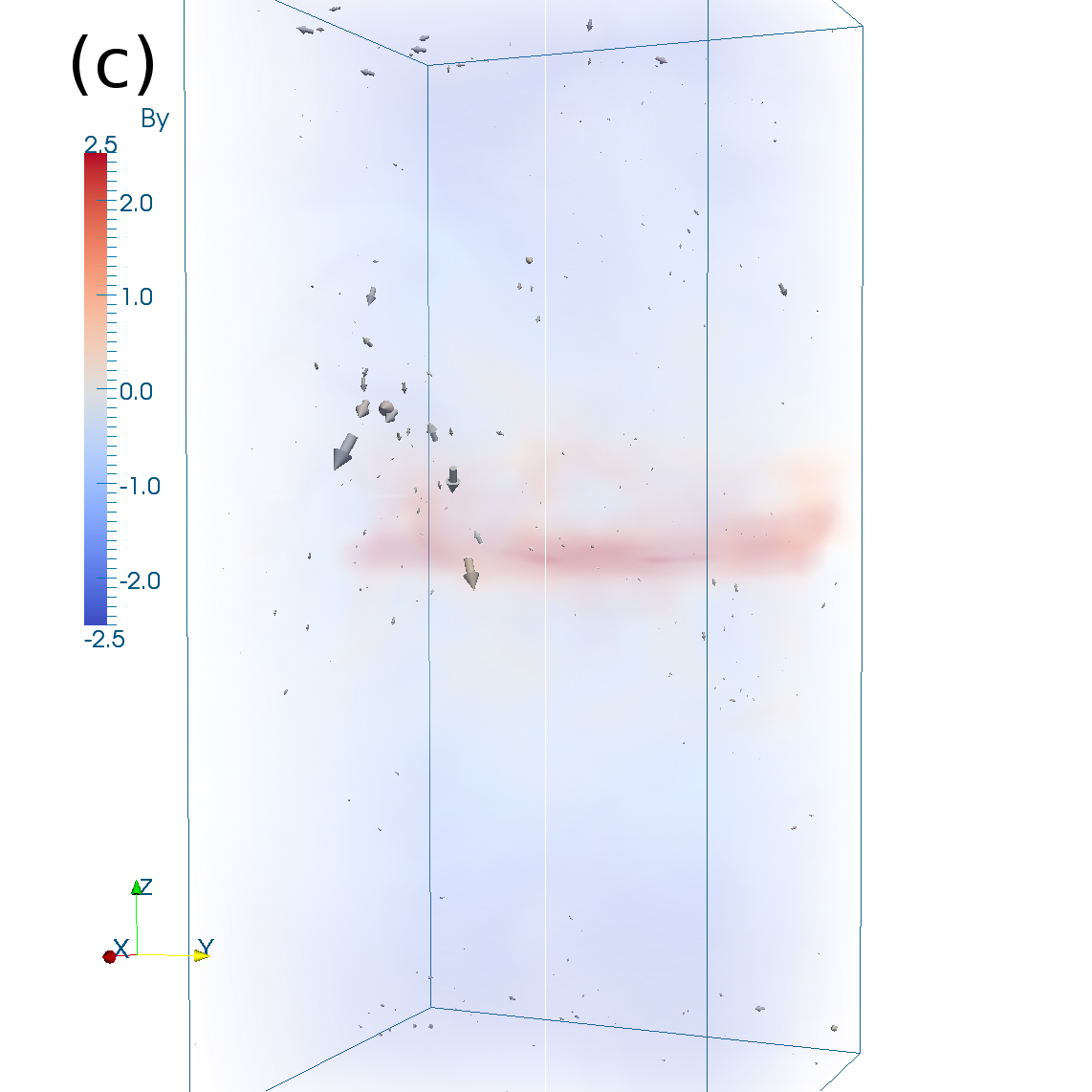}
  \hspace{-1.25cm}
  \caption[Volume snapshots of $\vect{B}$ by phase for Model~$\Ompa$]{
  Vector plots of the magnetic field $\vect{B}$ {\text{(a)}} in the cold phase
  {\text{(b)}} the warm phase and {\text{(c)}} the hot phase.
  Field directions are indicated by arrows and strength by their thickness.
  The colour of the arrows indicates the strength of the
  azimuthal ($y$) component (colour bar on the right).
  The background shading illustrates the density of the ISM.
  \label{fig:b3box}}
  \end{figure*}
%-----------------------------------------------------------------------------

  For Model~$\Ompa$ the ISM for a single snapshot is decomposed into the 
  three phases and the magnetic field for each plotted separately in 
  Fig.~\ref{fig:b3box}.
  In panel a the cold gas occupies only a limited volume near the mid-plane, 
  but the magnetic field is very strong and organised in alignment with the
  mean field surrounding it in the warm gas. 
  This is represented by the length and thickness of the vector arrows.  
  The colour of the arrows emphasises that the alignment has a strong 
  azimuthal component. 
  No arrows are present away from the mid-plane, because the cold gas is absent 
  there.
  In panel b the warm gas is present throughout the numerical volume.
  Field vectors are present almost throughout and the field is highly aligned,
  mainly in the azimuthal direction. 
  The strength of the field increases towards the mid-plane.
  The presence of some vectors in blue or grey indicates that there are 
  significant perturbations where the field includes reversals, some of these
  strong.
  Some of the field exhibits significant vertical orientation, but it is 
  mainly horizontal.
  In panel (c) the hot gas is also present throughout the volume, although 
  in smaller amounts near the mid-plane. 
  Despite this there is very little magnetic field.
  What field there is generally weak and lacks much systematic alignment, 
  although any orientation tends to be vertical, consistent with the
  field lines being stretched by the gas flowing away from the mid-plane. 
  Effectively the hot gas has a very weak field, which is highly disordered.
  Most of the magnetic field, and particularly the mean field, occupies the 
  warm gas.
  Detailed quantitative analysis of the structure of the gas will be deferred to
  future work.

%-----------------------------------------------------------------------------
\section{Structure of the Field}\label{sect:pitch}
%-----------------------------------------------------------------------------
   
%-----------------------------------------------------------------------------
  \begin{figure}[h]
  \centering
  \includegraphics[width=0.75\linewidth]{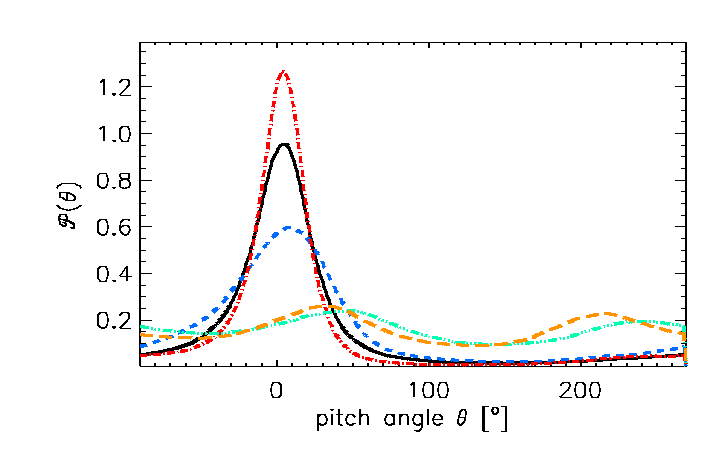}  
    \caption[Pitch angles for MHD models]{
  Total volume weighted probability distributions of magnetic field pitch 
  angles in models
  $\Ompa$~(black, solid),
  $\Ompb$~(blue, dashed),
  $\Ompd$~(red, dash-dotted),
  $\Ompc$~(green, dash-3dotted) and 
  $\Ompe$~(orange, long-dashed).
  Composite data for eleven snapshots each spanning $100\Myr$ from the 
  saturated dynamo (kinematic stage) for the first three (latter two). 
  \label{fig:pitch}
    }
  \end{figure}
%-----------------------------------------------------------------------------

  In future work the structure of the magnetic field shall be analysed in more
  depth for the total field as well as each phase individually.
  The correlation of the field in the kinematic stage and once the dynamo has
  saturated shall be measured, with attention to how it varies with height.
  
  As a preliminary inspection the pitch angles, $\arctan(B_x/B_y)$,
  for the total volume have been
  calculated for each model using eleven snapshots spanning $100\Myr$ and the
  probability distributions plotted in Fig.~\ref{fig:pitch}.
  As the dynamo evolves the magnetic field becomes increasingly ordered, with 
  the orientation close to the azimuthal. 
  Model~$\Ompd$ (red,dash-dotted), with the double rotation rate, predictably
  has the most regular field, with the smallest standard deviation and smallest
  modal pitch angle of $\theta\simeq3^\circ$, close to the azimuthal direction.
  Increased rotation will increase the laminar tendency of the flow, whereas
  Models~$\Ompa$ (black, solid) and $\Ompb$ (blue, dashed) have higher
  perturbation velocities relative to their rotation.
  So although the modal pitch angles are also close to azimuthal, $4^\circ$ and 
  $9^\circ$ respectively, their standard deviations are higher. 
  With open boundaries Model~$\Ompb$ the field is more irregular than 
  Model~$\Ompa$.
 
  In the kinematic stage for Models~$\Ompc$ (green, dash-3dotted) and $\Ompe$
  (orange, long-dashed) the fields are clearly less organised and the modal
  pitch angles are further from the azimuthal with $\theta\simeq50^\circ$ and
  $32^\circ$ respectively. 
  There are also significant modes in almost the reverse directions, at
  $\theta\simeq240^\circ$ and $216^\circ$ respectively.
  This is consistent with the higher magnitude of the relative helicity during
  the dynamo growth stage, as there is likely to be reduced tangling of the
  field lines if they have the same horizontal orientation.
  It may be, however, that the highly ordered field in these simulations is
  enhanced by the reduced motions of the hot gas.
  The lack of clustering in the SNe, now that the mass in the disc is more 
  stable, and the constraint for the mean magnetic field to have zero vertical
  component, due to the periodic boundary conditions 
  (Sections~\ref{subsect:params} and \ref{sect:dyn}), mitigate against the
  transport of hot gas away from the disc.
  However the pitch angles may be expected to align within the range $9^\circ$ 
  to $32^\circ$, assuming more scattering than in the saturated stage with 
  Model~$\Ompb$ for the former and more ordering than in the kinematic stage 
  for Model~$\Ompe$ for the latter angle.

  \section{Summary}
  All of the models investigated have produced a galactic mean field dynamo.
  In general the simulation dynamo grows more rapidly than the analytic
  predictions, so there would appear to be an additional elements in the 
  theoretical model required to explain the dynamo.
  The efficiency of the dynamo is improved if positive magnetic helicity can 
  be removed, either by vertical transport or resistive diffusion.
  The magnitude of the magnetic field is strongly aligned to the density of the
  ISM and indirectly the warm and cold phases.
  More particularly the mean field is stronger in the warm and cold gas, with 
  the hot gas containing a more random field. 
  The mean magnetic field and the magnetic energy is strongest at 
  $|z|\simeq300\p$, just outside the SNe active region. 
  The fluctuating dynamo is likely to be strongest in this SNe active region,
  but due to the low magnetic Reynolds numbers in the simulations, it is likely
  that the field and energy is significantly weaker in the simulations than 
  might be expected.

%%-----------------------------------------------------------------------------
%  \begin{figure}
%  \centering
%  \includegraphics[width=0.8\linewidth]{fig/o1pr_pdf2d.png}
%  \includegraphics[width=0.8\linewidth]{fig/o1pr_pdf2db.png}
%  \includegraphics[width=0.8\linewidth]{fig/o1pr_pdf2tb.png}
%  \caption{PDF contour plot by volume of log$T$ vs log$B$ for Model~S$\Omp$, 
%  averaged over 200\,Myr. 
%  \label{fig:2dpdfb}}
%  \end{figure}
%%-----------------------------------------------------------------------------

%   In Fig~\ref{fig:2dpdfb} the 2D PDF for $T$ vs $B$ is shown. The peaks in 
%   the PDFs at $T=10^2,10^4,10^6\K$ are consistent with the peaks in phase
%   PDFs in Fig.~\ref{fig:bpdf3}e. However we also see that strong order 
%   $1\mkG$ fields are present in gas of all temperatures, and the 
%   strongest $B$ is found in the warm phase near $10^4\K$.
%
%  It is evident from both plots, that a significant mean field has been 
%  generated. $\mean{B}_y$ is consistently about $3\times \mean{B}_x$ and there
%  is a strong $z$-dependence on the mean field strength. The polarity of the 
%  field is symmetric about the mid-plane and there is no evidence of mean field
%  reversals over time. 
%
%  Although there is no evidence of a mean vertical field, the fluctuations in 
%  $B_z$ are consistently stronger than the horizontal fluctuations, due to the
%  higher vertical velocity component. The fluctuations in the horizontal 
%  components are very similar to each other, despite mean $B_y$ being larger
%  than $B_x$. 

\end{chapter}

%  \clearpage                            % End the current page making sure all
%  \thispagestyle{empty}                 % tables/figures are printed.
%  \cleardoublepage                      % Necessary for correct page numbering.
%
  \part{Summary of results and future investigation}\label{part:sum}
%  \clearpage                            % End the current page making sure all
%  \thispagestyle{empty}                 % tables/figures are printed.
%  \cleardoublepage                      % Necessary for correct page numbering.
  
%-----------------------------------------------------------------------------

\begin{chapter}{\label{chap:summ}Summary of results}
%-----------------------------------------------------------------------------

\section{\label{sect:summulti} Multi-phase ISM results}

%-----------------------------------------------------------------------------

  The multi-phase gas structure obtained in these simulations appears
  robust, relatively insensitive to the physical parameters included: total gas 
  density or prescription of radiative cooling (Section~\ref{subsect:COOL}),
  SN rates, or rates of rotation and shear (Section~\ref{sect:3BB}). 
  Previous studies \citep[e.g.][]{M-LBKA05,Joung06} indicate that the 
  existence of such a multi-phase structure does not have as a prerequisite
  either rotation, shear or stratification.
  Its morphology, however, would strongly be affected by the latter. 
  Only SN rates close to the Milky Way estimates have been considered here, so
  whether there are upper and lower rates where a multi-phase structure may be
  excluded, remains to be investigated.  

  The parameterization of the radiative cooling is especially important for
  the hot phase of the ISM.
  Apart from the obvious effect that there is more hot gas with the cooling
  function WSW, which has weaker cooling than RBN at $T\ga10^3\K$
  (Fig.~\ref{fig:cool}), regions of low thermal and total pressure are more
  widespread with RBN (Fig.~\ref{fig:wsw_pdf3ph}).
  For the same reason, the WSW cooling function produces an ISM which has three
  times more thermal and kinetic energy than RBN (Fig.~\ref{fig:energetics}). 
  However, with either cooling function, thermal energy is about twice the
  kinetic energy due to perturbation velocities; more work is needed to decide if the
  relation $E_\mathrm{th}\simeq2E_\mathrm{kin}$ is of a general character.

  Examination of the 2D probability distribution of $T$ and $n$, as described in 
  Section~\ref{sect:entropy}, permits a physically motivated identification of
  natural boundaries of the major phases.
  Further refinement, distinguishing the SN active region $|z|\la200\p$ from
  the more homogeneous strata $|z|\ga200\p$  
  (Section~\ref{sect:TMPS}) improves the quality of the statistical modelling
  of each phase.
  The probability distribution of the density of each phase is approximated 
  to high quality by lognormal fits. 

  Considering that probability densities for gas temperature and number density,
  calculated for individual phases, are clearly separated, it is significant
  that probability densities for both thermal and total pressure --- the sum of
  thermal, magnetic (for MHD) and turbulent pressures --- are not segregated at
  all.
  Despite its complex thermal and dynamical structure, the gas is in 
  statistical pressure equilibrium.
  Since the SN-driven ISM is random in nature, both total and thermal pressure
  fluctuate strongly in both space and time (albeit with significantly smaller
  relative fluctuations than the gas density, temperature and perturbation
  velocity), so the pressure balance is also statistical in nature.
  These might appear to be obvious statements, since a statistically steady
  state (i.e., not involving systematic expansion or compression) must have 
  such a pressure balance.
  However, alternative conclusions may be drawn that the broadness of the
  pressure distribution in itself, irrespective of the convergence of modal
  pressures, indicates thermal pressure disequilibrium and a single phase ISM.
  Systematic deviations from pressure balance may be associated with the
  vertical outflow of the hot gas (leading to lower pressures), and with the
  compression of the cold gas by shocks and other converging flows
  (leading to somewhat increased pressures).
  If we allow for the global vertical pressure gradient
  (cf. Fig.~\ref{fig:pall4fits}) it is evident that locally phases are in total
  pressure equilibrium and a multiphase description is useful for analysis of
  the ISM.

  A direct relationship between various methods of estimating the filling 
  factor or fractional volume from observations is presented in 
  Section~\ref{sect:FF}.
  If an appropriate means of relating the observable filling factor to the
  filling factor derived from the statistical properties of the 
  distributions can be found --- perhaps by applying the simulation model fits
  --- then a more
  accurate estimate of the fractional volume of the various phases may emerge.
  This represents an improvement
  upon the assumption of locally homogeneous gas, the primary analytical tool 
  used to date in determinations of the fractional volumes of the phases.

  The modal properties of density, temperature, velocity, magnetic field, Mach
  number and pressure have been calculated for each phase separately.  
  In particular there is confirmation that the phases, with quantifiable small
  intersections in terms of thermal and density distributions, are close to
  both thermal and total statistical pressure balance.
  Statistical modelling, as used for the density, is also possible for any
  magnetohydrodynamic variable.
  This provides the opportunity to identify regions of the ISM with
  diagnostics other than temperature or density, which may help to
  characterize the dominant phase within observations.
  
  The correlation scale of the random flows is obtained in 
  Section~\ref{sect:CORR},
  from the autocorrelation functions of the velocity components. 
  Within $200\p$ of the mid-plane, the horizontal velocity components 
  have a consistent correlation scale of about $100\p$.
  In contrast, the scale of the vertical velocity 
  (which has a systematic part due to the galactic outflow of hot gas) 
  grows from about $100\p$ at the mid-plane to nearly $200\p$ at $z=200\p$,
  and further at larger heights.
  This is due to the increase of the fractional
  volume of the hot gas with distance from the mid-plane.
  Near to $|z|\simeq1\kpc$ (and beyond), most of the volume is occupied by the 
  hot gas.
  Further measurements of the correlation scale above $500\p$ are required to 
  assess to what extent such flows may reliably be modelled within a numerical 
  domain with horizontal dimensions of $1\kpc$.
  
  There is clear indication of cold gas falling back towards the mid-plane at
  speeds of a few km/s, hot gas involved in vigorous outflow away from the
  mid-plane, and some warm gas entrained in this outflow 
  (Section~\ref{subsect:GO}).
  The outflow speed of the hot gas increases up to $100\kms$ within $100\p$ of
  the mid-plane, and then slowly decreases.
  In contrast, the mean vertical velocity of the warm gas increases linearly 
  with $|z|$, up to $20\kms$ at $|z|=1\kpc$.
  
%-----------------------------------------------------------------------------
  \section{\label{sect:sumdynamo} Magnetized ISM results}
%-----------------------------------------------------------------------------

  The strongest result from the MHD models is the successful generation of a 
  mean field dynamo for all the parameters considered, including parameters
  matching the local Galaxy (Section~\ref{sect:dyn}). 
  A fluctuating dynamo may also be present, but further analysis is required. 
  In addition, an effective approach for identifying appropriately the mean 
  and random elements of the magnetic field has been identified, in terms of
  local volume averaging (Section~\ref{chap:meanB}).
  The growth rates of the mean field and the fluctuating field for each model
  have been compared with analytic estimates for the shear dynamo appropriate
  for the
  parameters in each case, and the simulation results either match or exceed 
  these estimates.
  The growth rates obtained exclude neither the possibility of a primordial
  field, nor a randomly seeded field via a Biermann battery.
  However the persistent orientation of the resultant mean field, without
  reversals in the azimuthal direction, is consistent with the hypothesis that
  galactic magnetic fields are generated through a dynamo, with the same 
  alignment above and below the mid-plane and independent of galactic 
  longitude (i.e. quadrupolar).

  In these simulations the presence of positive magnetic helicity appears to 
  limit the dynamo growth. 
  Model~$\Ompb$, with open vertical boundary conditions, had more rapid growth
  magnetic field than the otherwise equivalent Model~$\Ompa$, 
  with the vertical field condition on these boundaries.
  Although the former vents net magnetic energy through the external boundaries,
  which should slow the dynamo, it also expels significant positive magnetic
  helicity, which appears to enhance the dynamo.
  Otherwise all models generate net negative 
  magnetic helicity on the resistive time-scale during the kinematic phase; 
  this process is reversed as the
  dynamo saturates, reverting to a net relative helicity of approximately zero.
  The vertical transport of the magnetic field lines, and the associated sign 
  dependent helicity, is therefore a persistent feature of the galactic dynamo;
  and significantly, the sign of the outward helicity flux is common to both 
  sides of the mid-plane.
  
  Analysis of the multiphase structure of the magnetic field indicates that the
  strength of the magnetic field is closely connected with the density, and by
  association the warm and cold phases of the ISM. 
  Examination of the structure of this field in each phase reveals that the
  weak field in the hot gas is predominantly random. 
  There is a very strong regular component to the field in the cold gas,
  organised and amplified in alignment with the ambient warm gas.
  Most of the mean field is present within the warm gas.
  
%-----------------------------------------------------------------------------
  \section{\label{sect:future} Future investigation and experiments}
%-----------------------------------------------------------------------------

  An important technical aspect of simulations of this kind is the minimum
  numerical resolution $\Delta$ required to capture the basic physics of the
  multi-phase ISM.
  As with all other simulations of the SN-driven ISM, a host of numerical tools 
  (such as shock-capturing diffusivity) need to be employed to handle the
  extremely wide dynamical range (e.g. $10^2\la T\la10^8\K$ in terms of gas 
  temperature and $10^{-4}\la n\la10^2\cm^{-3}$ gas number density within the 
  model)
  and widespread shocks characteristic of the multi-phase ISM driven by SNe
  (detailed in Section~\ref{sect:NS}).
  $\Delta=4\p$ has been demonstrated to be sufficient with the numerical
  methods employed here, and results shown to be consistent with those obtained
  by other authors using adaptive mesh refinement with maximum resolutions of
  $2\p$ and $1.25\p$ (Section~\ref{sect:TMPS}). 
  The numerical method has been carefully tested by reproducing, quite
  accurately, the Sedov--Taylor and snowplough analytical solutions for
  individual SN remnants (Appendix~\ref{chap:EISNR}).
  Progress has been made in describing the multi-phase ISM structure and
  modelling the galactic dynamo in the vicinity close to the mid-plane.
  The relatively low vertical boundary, with carefully applied vertical
  boundary conditions, has adequately captured the critical features of the
  SN-driven ISM.
  
  From the data already available through these simulations, considerable 
  additional analysis would prove interesting.
  The distribution of the gas number density and its description in terms of
  lognormal fits has been considered, and a similar treatment of other 
  measurables, such as temperature, gas speed and pressure should also be
  possible.
  Some analysis of the total velocity correlation scales has been reported,
  but it would be useful to conduct this analysis of the gas velocity by
  individual phases, to identify the correlation scale for each.
  It would also be useful to investigate the correlation of time and position of
  the velocity field.
  The forcing by SN turbulence against a background of stratification and 
  differential rotation induces a combination of vortical and divergent 
  velocity. 
  How these are decomposed within the velocity field, and how their interaction
  is affected by changing the rates of SN injection, or galactic rotation,
  might help to understand how the galactic dynamo operates.

  Simulated line-of-sight observation from any location within a simulation,
  considering local bubbles, and taking into account changes in elevation, would
  permit us to compare simulated observational measurements with direct
  data measurements, and extend the ways in which observational
  measurements might be interpreted, with regard to underlying processes.
  This would be particularly useful in terms of Faraday rotation 
  measures for the magnetic field. 
  All of the models so far have parameters close to the solar neighbourhood, so
  the statistics could be readily identified with observational data.

  Building upon the existing model, a number of future extensions present 
  themselves.

%-----------------------------------------------------------------------------
  \subsubsection{Parameter Sweep}
%-----------------------------------------------------------------------------
  
  To understand the effects of various parameters requires each parameter to be
  considered for a range of values; at least three each, and preferably more.
  So the rates of SN injection, rotation and shear need to be varied 
  separately to identify how these affect the properties of the ISM.
  
  The model need not be restricted to galaxies similar to the Milky Way.
  Galaxies at higher redshift (young galaxies) could be investigated using 
  different stellar gravitational potentials and higher proportions of ISM to
  stars. 

%-----------------------------------------------------------------------------
  \subsubsection{Supernova Rates}
%-----------------------------------------------------------------------------
 
  Currently the models are constructed with the SN rates imposed as a parameter,
  based on a multiple of the observed SN rates.
  The Schmidt-Kennicutt Law \citep{K98} defines a relationship between the gas 
  surface density of the galactic disc with an SN rate, which is a reasonable
  fit to  the data from observations.
  Some numerical models on the scale of galaxies have successfully reproduced
  star-formation and feedback rates which reasonably match this relationship
  dynamically, rather than by imposition \citep[e.g.][]{DBP11}.
  At the local level, however, how the distribution of gas within cold clouds 
  (rather than the bulk volume of the gas in the warm ISM) is related to the 
  supernova rate is highly non-trivial.

  Nevertheless it would be preferable to derive a method of SN feedback, which 
  responds to the mass available for star formation, and the distribution of
  the gas between the cold and warm phases; i.e a system in which the SN rate is
  self-regulating, slowing down if SN activity burns aways the gas, and speeding
  up as gas cools into clouds and star formation increases. 
  Such an experiment may require a reasonably well modelled galactic fountain
  and an appropriate scheme for including realistic clustering of the SNe, to
  facilitate the venting of the hot gas from the disc as discussed in 
  Section~\ref{subsect:params}.
  Given the rarefaction of the gas density across the disc, due to spiral arms
  and other effects, it is not immediately apparent how such locally determined
  supernova rates might relate to the Schmidt-Kennicutt Law, applicable to the
  whole galaxy.
  However, experiments with a range of gas density distributions representative
  of the spiral arm (inter arm) regions could be considered, in which the 
  Schmidt-Kennicutt relations might be a lower (upper) bound against which to 
  assess the algorithm.
  An effective self-regulating method, which is reasonably consistent with the
  Schmidt-Kennicutt Law, would then be a useful analytical tool to apply in 
  galaxies with a broad range of parameters, which may differ significantly
  from the local Galaxy.

%-----------------------------------------------------------------------------
  \subsubsection{Vertical Magnetic Field}
%-----------------------------------------------------------------------------
    
  As discussed in Section~\ref{sect:dyn}, the contraint that $\average{B_z}=0$ is a construct
  of the periodic boundary conditions rather than a physical result.
  To explore the natural evolution of the vertical structure of the magnetic
  field, which is poorly understood due to difficulty in deriving 
  observational measurements, alternative horizontal boundary conditions are
  required. 
  Keeping the horizontal boundary condition on $A_z$ periodic or sheared would
  not affect $B_z$, but will contribute to the horizontal components and 
  retain the effect of shear on $B_x$. 
  With appropriate care to handle shocks crossing the boundary, setting the
  first or second derivatives for $A_x$ and $A_y$ to zero, should permit $B_z$
  to evolve naturally.
  
  Allowing the mean field to have a vertical component will also relax the 
  horizontal magnetic tension and allow greater, more physical circulation of
  gas between the disc and halo.
  Given that the objective here will be to understand the structure of the field, 
  the net flow of magnetic energy across the boundaries is of less concern, so
  the open field condition on the vertical boundary should be employed, as
  applied for Model~$\Ompb$, so as not to constrain $B_x$
  and $B_y$ nonphysically.
    
%-----------------------------------------------------------------------------
  \subsubsection{Cosmic Rays}
%-----------------------------------------------------------------------------
  
  Cosmic rays are estimated to account for about one quarter of the pressure
  in the ISM, and are strongly aligned with the magnetic field. 
  Including them effectively will allow results from the simulations to be 
  much more directly compared with observational estimates, including the
  vertical
  distribution of density, temperature and pressure, and fitted values of the
  probability distributions of these variables.
  The interaction of cosmic rays with thermodynamics and hydrodynamics has 
  not previously been explored on these scales, so subtle or even dramatic
  effects may be revealed.

  In addition, synchrotron emissions are an important tool for observations, and
  modelling these directly via simulated line-of-sight measurements will improve
  the comparison with the real observations.
  We shall also be able to compare the simulated observations with the
  direct measurements, to assess how robust the observational assumptions are.

%-----------------------------------------------------------------------------
  \subsubsection{Galactic Fountain}
%-----------------------------------------------------------------------------
  \citet{AB07} and similar models have sufficient vertical extent with
  $|z|=10\kpc$ for hot gas to cool and recycle naturally to the mid-plane.
  With their limited horizontal aspect, however, we cannot know enough about 
  the true scales of the galactic fountain. 
  With the grid scale $\Delta$ varying in the vertical direction and increasing 
  the horizontal range of the model, I intend to explore the
  physical and time scales of the galactic fountain.
  Due to the damping effect of the magnetic field, it may paradoxically be
  numerically more efficient to include magnetism for this project, and
  possibly also cosmic rays, rather than to model a purely hydrodynamic regime.  
    
%-----------------------------------------------------------------------------
  \subsubsection{Spiral Arms}
%-----------------------------------------------------------------------------
 
  The range of models included so far assume the gas is located in an inter arm
  region of the galaxy. 
  There are no density waves encountered as the numerical box orbits the 
  galactic centre.
  How the formation of stars, generation and orientation of the magnetic field
  reacts within and between the spiral arms is not well understood.
  Within this model, spiral arms could be modelled with an additional 
  gravitational 
  potential which passes through the box at set times during its orbit.  
  
%-----------------------------------------------------------------------------
  \subsubsection{Parker and Magnetorotational Instability and SN Quenching}
%-----------------------------------------------------------------------------

  Parker instabilities occur in plasma subject to gravitational forces, and 
  magnetorotational instabilities occur in differentially rotating plasma.
  These instabilities are postulated to be damped by the turbulence generated
  by SNe.
  These models could be used to test this postulation by simulating a system 
  in which either or both of these instabilities are present.
  Repeating the experiment in each case, with and without SN driven turbulence
  included, it would be possible to identify whether the instabilities are in 
  fact suppressed.
  
%-----------------------------------------------------------------------------
  \section{\label{sect:review} Review}
%-----------------------------------------------------------------------------

  In conclusion, such models are highly complex in their construction, and in
  the 
  results they generate.
  The results reported in this thesis are the product of several revised sets 
  of calculations, possible only after many sets of simulations that preceded them.
  Changes take hundreds of thousands of iterations to unfold, and it is not 
  always immediately apparent which outcomes are robust physical effects and 
  which are numerical artifacts.
  Given that the observational data and theoretical framework is far from complete,
  it is also not always obvious whether the numerical results, which appear to
  match our expectations of the ISM, are in fact true, or results which 
  appear to conflict with our expectations are nevertheless correct.

  The more the results have been analysed, the better our understanding of the
  model and the causes of various effects. 
  It would be desirable to be able to run all the simulations with exactly the
  same numerical ingredients, but such is the expense in resources and time
  that as improvements and corrections are revealed they are implemented on the
  fly, and cannot often be applied retrospectively.
  Hence there are considerable algorithmic differences between the HD models
  and the MHD
  models; and even within the sets of simulations, some numerical ingredients 
  differ, beyond the physical parameters which are under investigation.
 
  This is a new model and with each set of simulations, better experience is
  garnered on how best to interpret and report the results.
  It is now a very rich numerical resource, which has the potential to 
  unlock far more secrets of the ISM than have been listed in this thesis.
  I hope to have the opportunity to extend this model in the directions 
  described above, and in the process to shed new light on the structure and
  dynamics of galaxies.

\end{chapter}

  \appendix
%
  
%----------------------------------------------------------------------------
\begin{chapter}{\label{chap:EISNR}Evolution of an individual supernova remnant}
%----------------------------------------------------------------------------

%----------------------------------------------------------------------------
  \section{\label{sect:SNPL}The snowplough test}
%----------------------------------------------------------------------------

  The thermal and kinetic energy supplied by SNe drives, directly or 
  indirectly, all the processes discussed in this thesis.
  It is therefore crucial that the model captures correctly the energy
  conversion in the SN remnants and its transformation into the thermal and
  kinetic energies of the interstellar gas.
  As discussed in Section~\ref{sect:MSN}, the size of the region where the SN
  energy is injected corresponds to the adiabatic (Sedov--Taylor) or 
  snowplough stage.
  Given the multitude of artificial numerical effects required to model the
  extreme conditions in the multi-phase ISM, it is important to verify that the
  basic physical effects are not affected, while sufficient numerical control
  of strong shocks, rapid radiative cooling, supersonic flows, etc., is
  properly ensured.
  Another important parameter to be chosen is the numerical resolution. 
  
  Before starting the simulations of the multi-phase ISM reported in this
  thesis, I have carefully verified that the model can reproduce, to sufficient
  accuracy, the known realistic analytical solutions for the late stages of SN
  remnant expansion, until merger with the ISM.
  The minimum numerical resolution required to achieve this in this model is
  $\Delta=4\p$.
  In this Appendix, I consider a single SN remnant, initialised as described in
  Section~\ref{sect:MSN}, that expands into a homogeneous environment.
  All the numerical elements of the model are in place, but here periodic
  boundary conditions are used in all dimensions. 
  
  The parameters $\chi_1$ and $\nu_1$ are as applied in Model~\Op\ for 
  $\Delta=4\p$, but reduced here proportionally for $\Delta=2$ and $1\p$. 
  The constant $C\approx0.01$ used in Eq.~\eqref{coolxi} 
  to suppress cooling around shocks is unchanged. 
  This may allow excess cooling at higher resolution, 
  evident in the slightly reduced radii in Fig.~\ref{fig:snplb}. 
  For Model~{\OpH}, $\chi_1$ and $\nu_1$ were just as in Model~{\Op};
  for future reference, they should be appropriately adjusted, as should $C$, 
  to better optimise higher resolution performance.

%-----------------------------------------------------------------------------
  \begin{figure}[h]
  \centering
  \hspace{-1.25cm}
  \includegraphics[height=0.67\textwidth,width=0.345\textwidth]{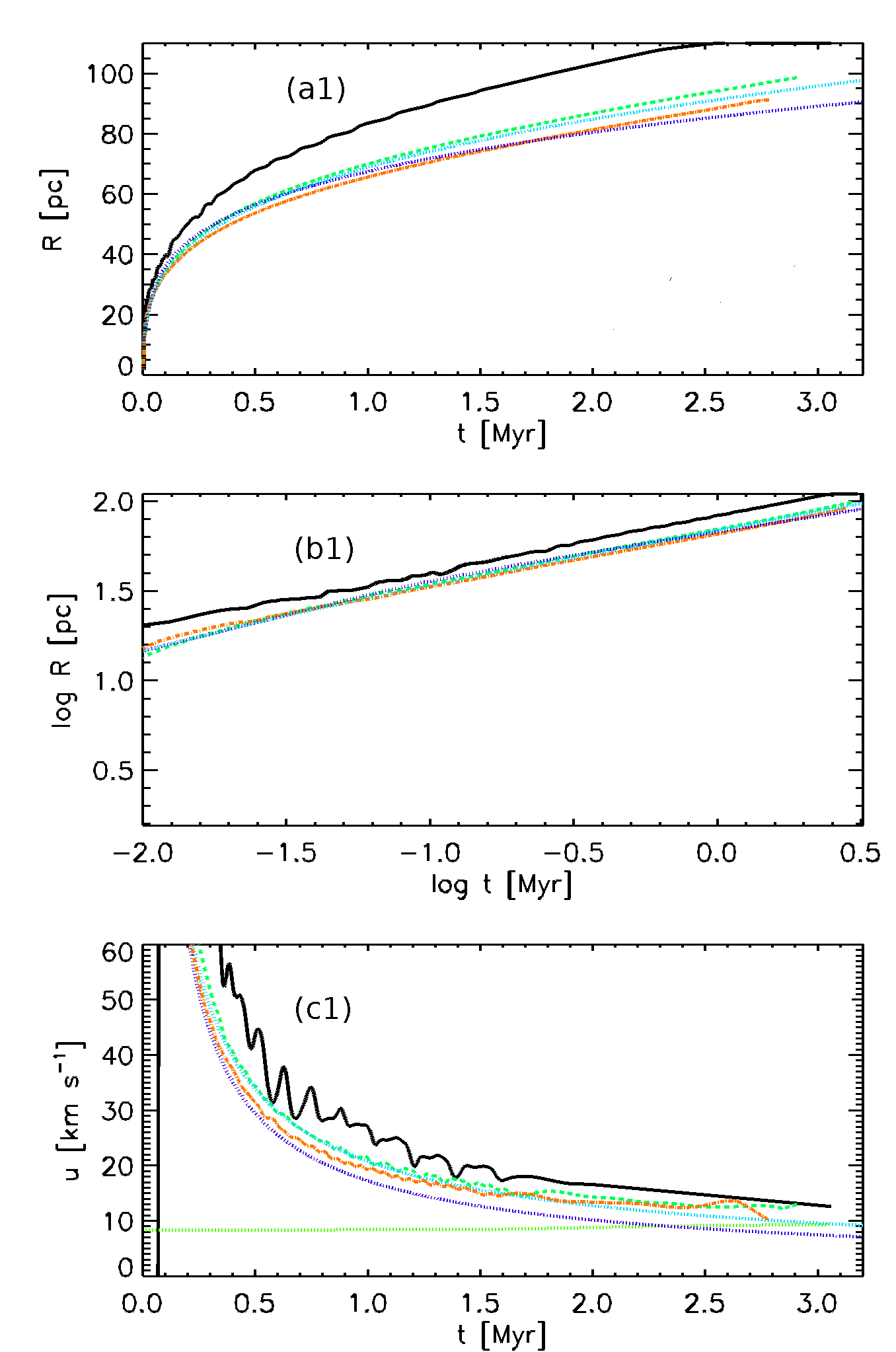}
\hfill            
  \includegraphics[height=0.67\textwidth,width=0.345\textwidth]{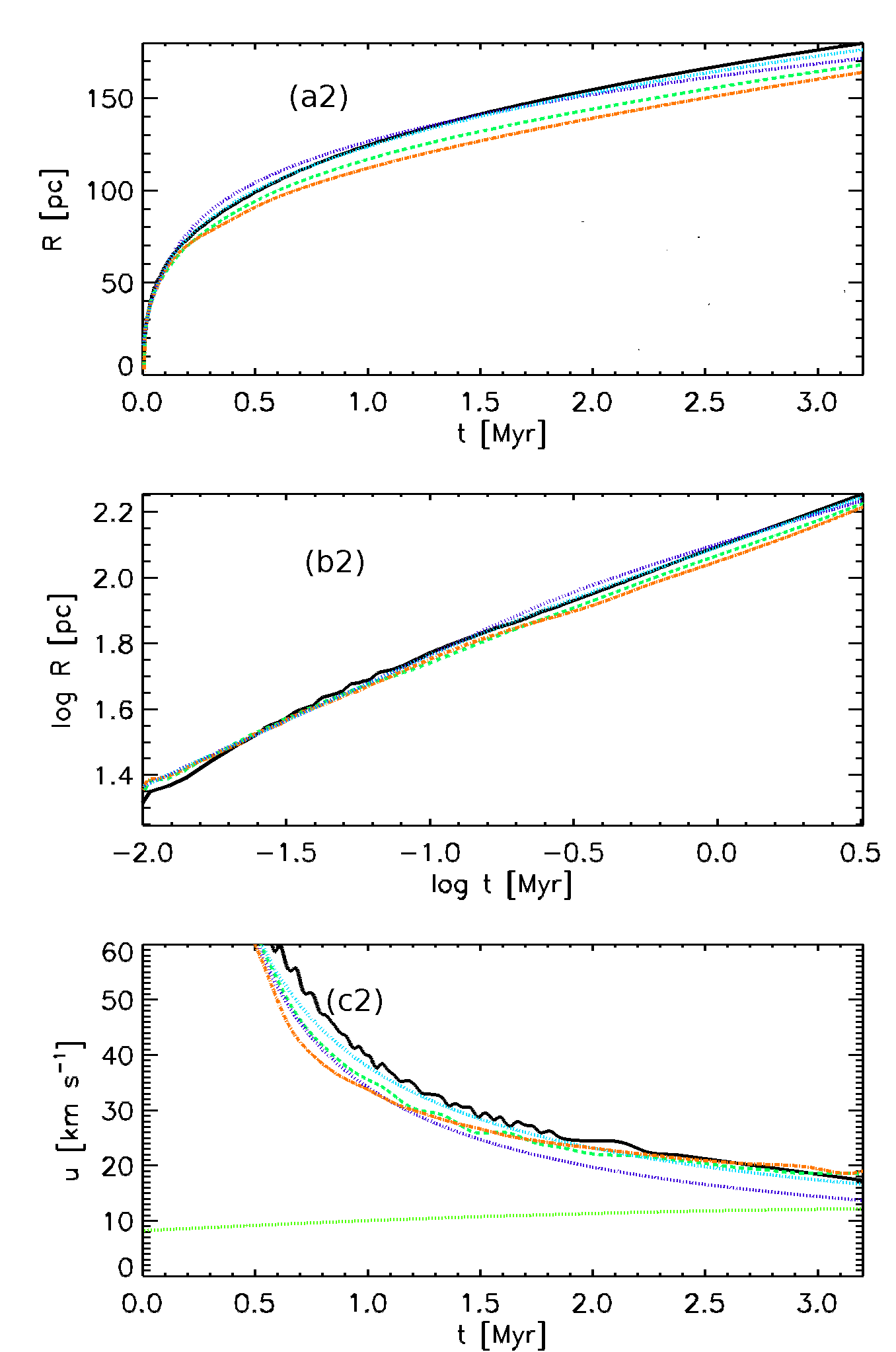}
\hfill                  
  \includegraphics[height=0.67\textwidth,width=0.345\textwidth]{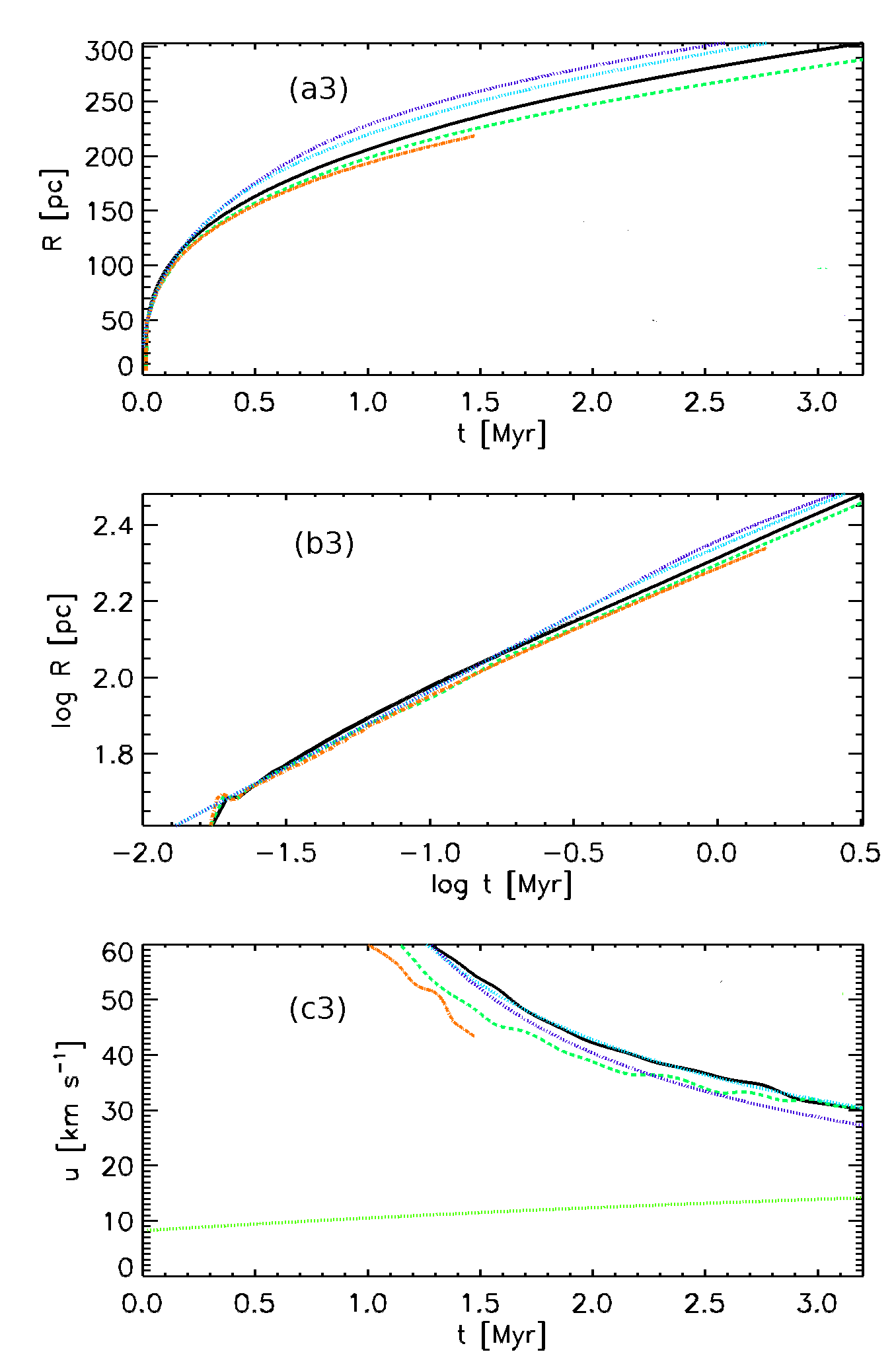}
    \caption[SN radial expansion]{
  The shell radius $R$ of an SN remnant versus time, shown in
  \textbf{(a)}~linear and \textbf{(b)}~logarithmic scales; \textbf{(c)}~the
  corresponding expansion speed $\dot{R}$.
  Frame columns 1--3 are for different ambient gas densities,
  $\rho_0/10^{24}\g\cmcube=(1.0,0.1,0.01)$ from left to right.
  Numerical results obtained under three numerical resolutions are shown:
  $\Delta=4\p$ (black, solid), 2\,pc (green, dashed) and 1\,pc (orange,
  dash-dotted).
  Dotted lines are for the standard snowplough solution (\ref{eq:snpl}) (dark
  blue) and its modification by \citet{Cioffi98} (light blue).
  The horizontal line in Panels~(\textbf{c1})--(\textbf{c3}) shows the sound
  speed in the ambient ISM.
    \label{fig:snplb}
            }
  \end{figure}

%-----------------------------------------------------------------------------

%-----------------------------------------------------------------------------
  \subsubsection{The adiabatic and snowplough stages}
%-----------------------------------------------------------------------------

  The Sedov--Taylor solution,
  \begin{equation}
    \label{eq:sed}
    R= \left(\kappa\frac{E\SN}{\rho_{0}}\right)^{{1}/{5}}t^{{2}/{5}},
  \end{equation}
  is accurately reproduced with this code at the resolution $\Delta=4\p$ or
  higher.
  Here $R$ is the remnant radius, $E\SN$ the explosion energy, $\rho_{0}$ the
  ambient gas density, and dimensionless parameter $\kappa\approx2.026$ for
  $\gamma=5/3$ \citep{Ostriker88}.
  
  Modelling even a single remnant becomes more challenging when radiative
  cooling becomes important.
  Here numerical results are compared with two analytic solutions for an SN
  remnant expanding into a perfect, homogeneous, monatomic gas at rest.
  The standard momentum-conserving snowplough solution for a radiative SN
  remnant has the form
  \begin{equation}
    \label{eq:snpl}
    R=R_{0}\left[1+4\frac{\dot{R_{0}}}{R_{0}}(t-t_{0})\right]^{{1}/{4}} ,
  \end{equation}
  where $R_{0}$ is the radius of the SN remnant at the time $t_{0}$ of the 
  transition from the adiabatic stage, and $\dot{R_{0}}$ is the shell expansion
  speed at $t_{0}$. 
  The transition time is determined by \citet{Woltjer72} 
  as that when half of the SN energy is lost to radiation; this happens when
  \begin{equation}
    \dot{R_{0}}=230\kms \left(\frac{n_{0}}{1\cmcube}\right)^{{2}/{17}}\left(\frac{E\SN}{10^{51}\erg}\right)^
  {{1}/{17}};
  \end{equation}
  the transitional expansion speed thus depends very weakly on parameters.
  
%-----------------------------------------------------------------------------
  \subsubsection{Pressure driven and momentum driven snowplough} 
%-----------------------------------------------------------------------------
  
  \citet{Cioffi98} obtained numerical and analytical solutions for an expanding
  SN remnant with special attention to the transition from the Sedov--Taylor 
  stage to the radiative stage.
  These authors adjusted an analytical solution for the pressure-driven 
  snowplough stage to fit their numerical results to an accuracy of within 2\%
  and 5\% in terms of $R$ and $\dot R$, respectively.
  (Their numerical resolution was $0.1\p$ in the interstellar gas and $0.01\p$
  within ejecta.)  They thus obtained
  \begin{equation}
    \label{eq:pds}
    R=R_{\rm{p}} \left(\frac{4}{3}\frac{t}{t_{\rm{p}}}-
    \frac{1}{3}\right)^{3/10},
  \end{equation}
  where the subscript ${\rm{p}}$ denotes the radius and time for the transition
  to the pressure driven stage.
  The estimated time of this transition is 
  \[ 
  t_{\rm{p}}\simeq13\Myr\left(\frac{E\SN}{10^{51}\erg}\right)^{3/14}
     \left(\frac{n_0}{1 \cmcube}
            \right)^{-4/7}. 
  \]
  For ambient densities of $\rho_0=(0.01,0.1,1)\times10^{-24}\g\cmcube$, 
  this yields transition times $t_\mathrm{p}\approx(25,6.6,1.8)\times10^4\yr$
  and shell radii $R_{\rm{p}}\approx(130,48,18)\p$, respectively, with speed
  $\dot{R}_\mathrm{p}=(213,296,412)\kms$.
  
  This continues into the momentum driven stage with
  \begin{equation}
    \label{eq:mcs}
    \left(\frac{R}{R_{\rm{p}}}\right)^{4}=
    \frac{3.63~\left(t-t_{\rm{m}}\right)}{t_{\rm{p}}}
    \left[1.29-\left(\frac{t_{\rm{p}}}{t_{\rm{m}}}\right)^{0.17}\right] +
    \left( \frac{R_{\rm{m}}}{R_{\rm{p}}}\right)^{4},
  \end{equation}
  where subscript ${\rm{m}}$ denotes the radius and time for this second
  transition,
  \[
  t_{\rm{m}}\simeq 61\, t_\mathrm{p}
                 \left(\frac{\dot{R}_{\rm{ej}}}{10^3\kms}\right)^3
                 \left(\frac{E\SN}{10^{51}\erg}\right)^{-3/14}
                 \left(\frac{n_0}{1\cmcube}\right)^{-3/7},
  \]
  where $\dot{R}_{\rm{ej}}\simeq5000\kms$ is the initial velocity of the
  $4M_\odot$ ejecta.
  For each $\rho_0=(0.01,0.1,1.0)\times10^{-24}\g\cmcube$, the transitions
  occur at $t_{\rm{m}}=(168,16.8,1.68)\Myr$, and $R_{\rm{m}}=(1014,281,78)\p$,
  respectively.
  The shell momentum in the latter solution tends to a constant, and the
  solution thus converges with the momentum-conserving snowplough 
  (\ref{eq:snpl}); but, depending on the ambient density, the expansion may
  become subsonic and the remnant merge with the ISM before 
  Eq.~(\ref{eq:snpl}) becomes applicable. 
  
  The simulation results are compared with the momentum-conserving snowplough
  solution and those of \citeauthor{Cioffi98} in Fig.~\ref{fig:snplb}, testing
  the model with numerical resolutions $\Delta=1$, 2 and 4\,pc for the ambient
  gas densities $\rho_0=(0.01,0.1,1.0)\times10^{-24}\g\cmcube$. 
  Shown in Fig.~\ref{fig:snplb} are: a linear plot of the remnant radius $R$
  versus time, to check if its magnitude is accurately reproduced; a double
  logarithmic plot of $R(t)$, to confirm that the scaling is right; and 
  variation of the expansion speed with time, to help assess more delicate
  properties of the solution. 
  There is good agreement with the analytical results for all the resolutions
  investigated when the ambient gas number density is below $1\cmcube$.
  For $\Delta=4\p$, the remnant radius is accurate to within about 3\% for
  $\rho_0=10^{-25}\g\cmcube$ and underestimated by up to 6\% for 
  $\rho_0=10^{-26}\g\cmcube$.
  At higher numerical resolutions, the remnant radius is underestimated by up
  to 7\% and 11\% for $\rho_0=10^{-25}\g\cmcube$ and $10^{-26}\g\cmcube$, 
  respectively.
  For $\rho_0=10^{-24}\g\cmcube$, excellent agreement is obtained for the
  higher resolutions, $\Delta=1$ and $2\p$; simulations with $\Delta=4\p$
  overestimate the remnant radius by about 20--25\% in terms of $R$ and 
  $\dot R$ at $t=2\Myr$.
  Note that a typical SN explosion site in the models described in the
  main part of the thesis has an ambient density $n_0<1\cmcube$ so that
  $\Delta=2$ or $4\p$ produce a satisfactory fit to the results, despite the
  much finer resolution of the simulations of \citeauthor{Cioffi98}.
  
  The higher than expected expansion speeds into dense gas can be explained by 
  the artificial suppression of the radiative cooling within and near to the
  shock front as described by Eq.~(\ref{coolxi}).
  Our model reproduces the low density explosions more accurately because the
  shell density is lower, and radiative cooling is therefore less important. 

%-----------------------------------------------------------------------------

\section{\label{subsect:CSHK}Cooling-heating in the shocks}

  In code units, the applied net (heating or) cooling 
  $([\Gamma-  \rho\Lambda]~T^{-1})$ has amplitudes much greater than 1, and of
  opposite sign in close proximity.
  Perturbations to $\Gamma$ or $\Lambda$ are particularly vulnerable to
  instability when acted upon by shocks, and can lead to purely numerical 
  thermal spikes, which can cause the code to crash. 
  Therefore some form of suppression of net cooling 
  within the shock profile is essential for code stability.

  In addition, the method of enhanced
  diffusion coefficients for handling shocks has the effect of broadening the
  shock profiles and increasing the density spread. High density ISM inside
  the remnants cool faster, thus diffusing the hot phase ISM more rapidly than
  would occur within the unresolvable thin shock fronts that exist physically.

  Having considered various approaches, a prescription is adopted which is
  stable, consistent with the snowplough, reduces cooling in the remnant shell
  interior, and applies cooling to the peak shell density.
  The net diffuse heating/cooling $[\Gamma-\rho\Lambda]~T^{-1}$ is multiplied by
  $\xi=e^{-C(\nabla\zeta)^2}$, where $\zeta$ is the shock diffusivity,
  defined in Eq. (\ref{shockdiff}). Therefore, $\xi$ is unity
  through most of our domain,
  but reduces towards zero in strong shocks, where $(\nabla\zeta)^2$
  is non-zero.
  $C$ is another numerical factor, the value of which is chosen to
  guarantee code stability without altering the basic physics; these
  constraints are normally fulfilled with $C\simeq0.01$.

%-----------------------------------------------------------------------------

%-----------------------------------------------------------------------------
  \section{The structure of the SN remnant}\label{subsect:RPROF}
%------------------------------------------------------------------------

%------------------------------------------------------------------------
  \begin{figure}[h]
  \centering
  \includegraphics[height=0.225\textwidth,width=.99\textwidth]{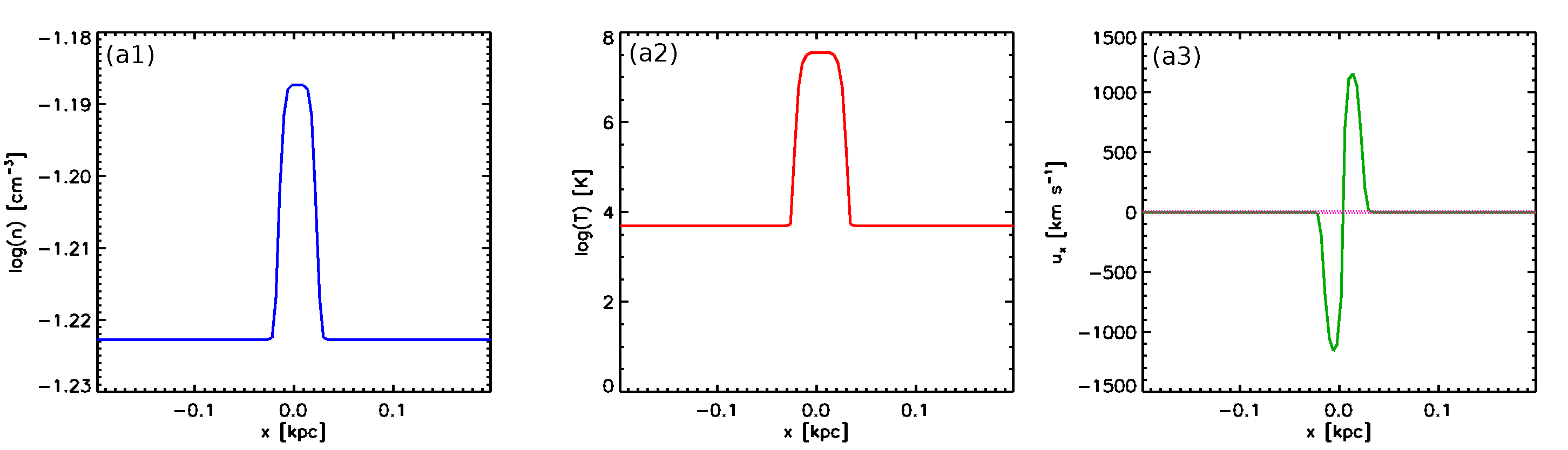}
  \hfill
  \includegraphics[height=0.225\textwidth,width=.99\textwidth]{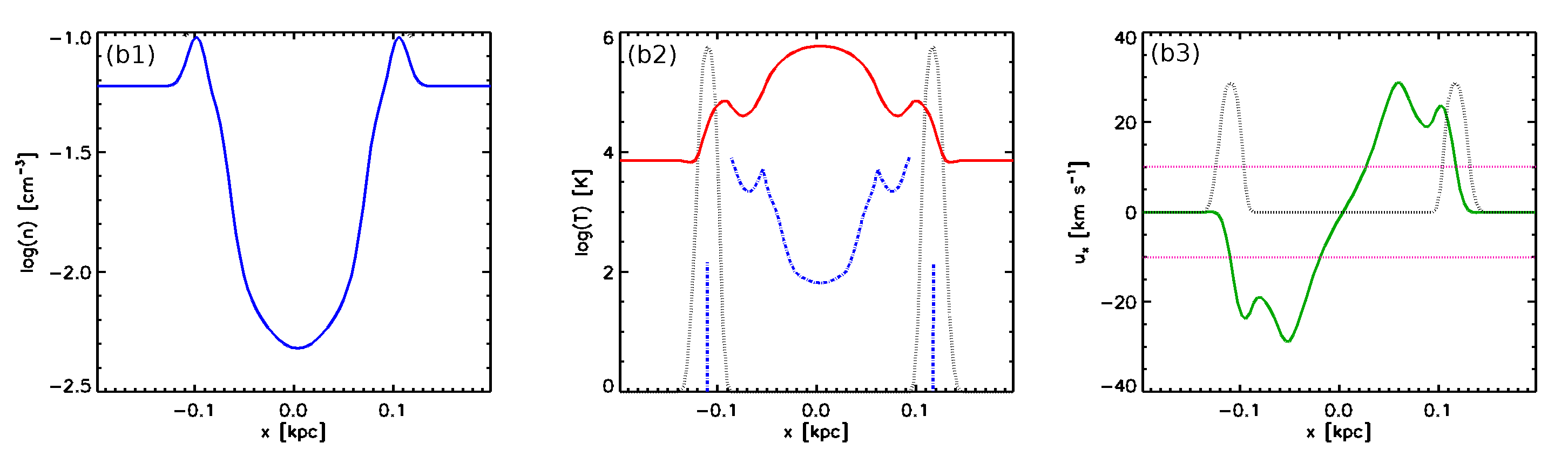}
  \hfill
  \includegraphics[height=0.225\textwidth,width=.99\textwidth]{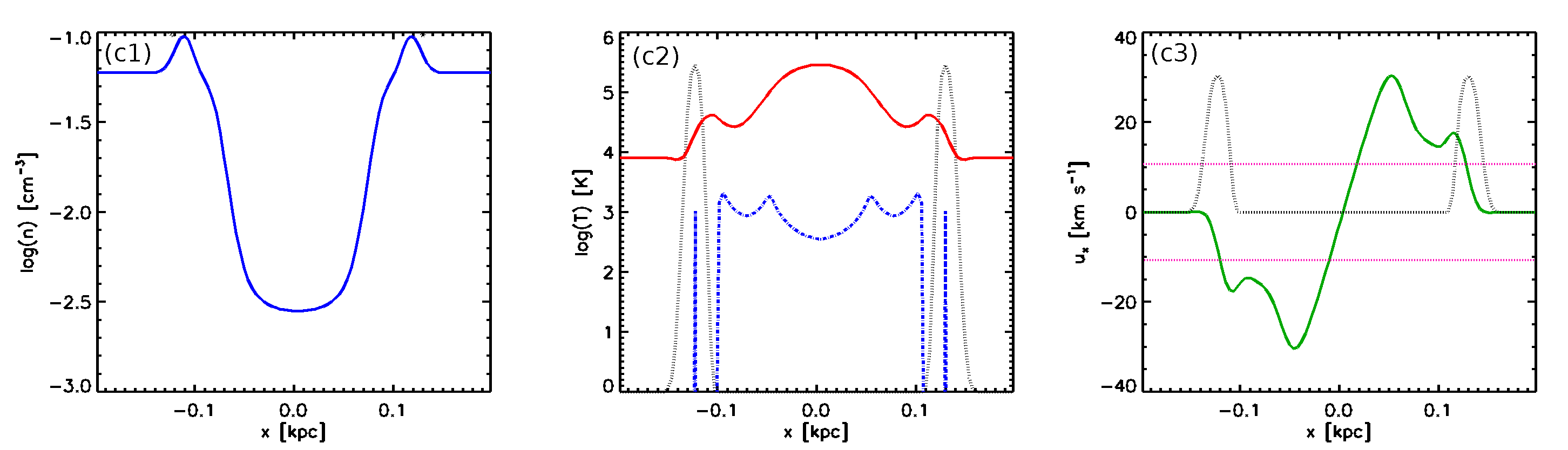}
    \caption[1D slices of remnant structure]{
  One-dimensional cuts through the origin of an SN remnant expanding into gas
  of a density $\rho_0=10^{-25}\g\cmcube$, simulated with the numerical 
  resolution $\Delta=4\p$.
  The variables shown are \textbf{(a1)--(c1)} gas density, \textbf{(a2)--(c2)}
  temperature, and \textbf{(a3)--(c3)} velocity.
  The shock viscosity profile of Eq.~(\ref{shockdiff}) (scaled to fit the
  frame, black, dotted) is shown in the temperature and velocity panels; the
  net cooling, log($-T^{-1}(\Gamma-\rho\Lambda)$), only 
  where $\rho\Lambda\ge\Gamma$, from Eq.~(\ref{eq:ent})
  (blue, dashed) is included in the temperature panel; and the ambient sound
  speed (pink, dashed) is also shown with the velocity. 
  Panels in the top row \textbf{(a)} show the injection profiles used to
  initialise the remnant at $t=0$; the lower panel rows are for the later times
  \textbf{(b)}~$t=0.72\Myr$ and \textbf{(c)}~$t=1.02\Myr$.
    \label{fig:1dtime}
            }
  \end{figure}
%-----------------------------------------------------------------------------

  Cuts through the simulated SN remnant are shown in Fig.~\ref{fig:1dtime} for
  gas density, temperature and velocity, obtained for resolution $\Delta=4\p$
  and with ambient density $\rho_0=10^{-25}\g\cm^{-3}$.
  In the temperature and velocity panels, the profile of the shock viscosity
  from Eq.~(\ref{shockdiff}) is also included (black dotted line), scaled to
  fit each plot.
  The temperature panels also show where net cooling is applied to the remnant,
  $T^{-1}(\Gamma-\rho\Lambda)<0$ from Eq.~\eqref{eq:ent} (blue dashed line), 
  while the velocity panels also show the ambient sound speed (pink dashed
  lines).
  The top panel depicts the initial distributions, at $t=0$, with which the
  mass of $4\Msol$ and $5\times10^{50}\erg$ each of thermal and kinetic energy
  are injected. 
  The other panels are for $t=0.72$ and $1.02\Myr$ after the start of the evolution, 
  from top to bottom, respectively; the actual simulation continued to
  $t=1.32\Myr$, when the remnant radius reached $130\p$.
  
  The position of the peak of the density profile is used to determine the
  shell radius shown in Fig.~\ref{fig:snplb}.
  The Rankine--Hugoniot jump conditions are not very well satisfied with the 
  numerical parameters used here.
  This is due to our numerical setup, essentially designed to control the
  shocks by spreading them sufficiently to be numerically resolvable in
  production runs that contain many interacting shocks and colliding SN shells.
  Better shock front profiles have been obtained with other choices of
  parameters and cooling control, and with better resolution.
  The density and temperature contrasts across the shock fronts are reduced by
  the shock smoothing, which inhibits the peak density and enhances gas density
  behind the shocks.
  In an isolated remnant, the peak gas number density does not exceed
  $10\cmcube$, but in the full ISM simulation densities in excess of
  $100\cmcube$ are obtained, as a result of interacting remnants and highly
  supersonic flows.
  
  The interior of the SN remnant, if more dense due to numerical smoothing 
  about the shock profile, would cool unrealistically rapidly, so that the SN
  energy would be lost to radiation rather than agitate the ambient ISM. 
  The centre panels in Fig.~\ref{fig:snplb} clarify how the cooling
  suppression described in Eq.~(\ref{coolxi}) reduces the cooling rate in the
  relatively homogeneous interior of the remnant, while still allowing rapid
  cooling in the dense shell where the gradient of the shock viscosity is
  small. 
  It is evident from the temperature cuts that the remnant still contains
  substantial amounts of hot gas  when its radius reaches 100\,pc, so it would
  be merging with the ISM in the full simulation.
  
  The panels in the right column of Fig.~\ref{fig:1dtime} demonstrate that the
  interior gas velocity can be more than twice the shell speed.
  Due to the high interior temperature, this flow is subsonic, while the
  remnant shell expands supersonically with respect to its ambient sound speed.
  The enhanced viscosity in the hotter interior (with viscosity proportional to 
  the sound speed; see Section~\ref{sect:NS}) inhibits numerical instabilities 
  that could arise from the high velocities.
  In fact, accurate modelling of the SN interiors is not essential in the
  present context (where the main interest is in a realistic description the 
  multi-phase ISM), as long as the interaction of the remnant with the ambient
  gas is well described, in terms of the energy conversion and transfer to the
  ISM, the scales and energy of turbulence, and the properties of the hot gas.

\end{chapter}

  \begin{chapter}{\label{chap:stab}Stability criteria}
%-----------------------------------------------------------------------------

\section{\label{sect:BCND}\label{subsect:courant}Courant criteria}
%--------------------------------------------------------------------------

%--------------------------------------------------------------------------
\subsubsection{Time step control}

  To achieve numerical stability with the explicit time stepping used, the
  Courant-Friedrichs-Lewy conditions have to be amply satisfied.
  For example, for advection terms, the numerical time step should be selected
  such that
  \[
    \Delta t < \kappa \frac{\Delta}{\max(c\sound, u, U)},
  \]
  where $c\sound$ is the speed of sound, $u=|\vect{u}|$ is the amplitude of the
  perturbed velocity, i.e., the deviation from the imposed azimuthal shear flow
  $U$, and $\kappa$ is a dimensionless number, determined empirically, which
  often must be significantly smaller than unity.
  Apart from the velocity field, other variables also affect the maximum time
  step, e.g., those associated with diffusion, cooling and heating, so that the
  following inequalities also have to be satisfied:
  \[
    \Delta t < \frac{\kappa_1\Delta^2}{\max(\nu, \gamma\chi, \eta)}\,,
  \quad\quad  \Delta t < \frac{\kappa_2}{H_{\max}}\,,
  \]
  where $\kappa_1$ and $\kappa_2$ are further empirical constants and
  \[
    H_{\max}=\max\left(\frac{2\nu|\mathbfss W|^2+\zeta_\nu(\nabla\cdot\vect{u})^2+\zeta_\chi(\nabla\cdot\vect{u})^2}{\cv T}\right)\,.
  \]
  $\kappa=\kappa_1=0.25$ and $\kappa_2=0.025$ are used.
  The latter, more stringent constraint has a surprisingly small impact on the
  typical time step, but a large positive effect on the numerical accuracy.
  Whilst the time step may occasionally decrease to below 0.1 or 0.01 years
  following an SN explosion, the typical time step is more than 100 years. 
  
  %-------------------------------------------------------------------------
  \subsubsection{Minimum diffusivity}
  
  Numerical stability also requires that the Reynolds and P\'eclet numbers
  defined at the resolution length $\Delta$, as well as the Field length, 
  are sufficiently small.
  These mesh P\'eclet and Reynolds numbers are defined as
  \begin{equation}
  \Pe\mesh=\frac{u \Delta}{\chi}\leq\frac{u_{\rm{max}} \Delta}{\chi}\,,
  \quad
  \Ry\mesh=\frac{u \Delta}{\nu}\leq\frac{u_{\rm{max}} \Delta}{\nu}\,, \label{mesh_re_pe}
  \end{equation}
  where $u_{\rm{max}}$ is the maximum perturbed velocity and $\Delta$
  is the mesh length.  
  For stability these must not exceed some value, typically between 1 and 10.
  
  In numerical modelling of systems with weak diffusivity, $\nu$ and $\chi$
  are usually set constant, close to the smallest value consistent with the
  numerical stability requirements.
  This level strongly depends on the maximum velocity, and hence is related to
  the local sound speed, which can exceed $1500\kms$ in the ISM models. 
  To avoid unnecessarily strong diffusion and heat conduction in the cold and
  warm phases, the corresponding diffusivity is scaled with gas temperature,
  as $T^{1/2}$. 
  As a result, the diffusive smoothing is strongest in the hot phase (where it
  is most required).
  This may cause reduced velocity and temperature inhomogeneities within the
  hot gas, and may also reduce the temperature difference between the hot gas
  and the cooler phases.
  
  The effect of thermal instability is controlled by the Field length,
  \begin{align*}
  \lambda_\mathrm{F}&\simeq\left(\frac{KT}{\rho^2\Lambda}\right)^{1/2}
  									\simeq2.4\p \left(\frac{T}{10^6\K}\right)^\frac{7}{4}\!\!
  															 \left(\frac{n}{1\cmcube}\right)^{-1}\!\!
  										           \left(\frac{\Lambda}{10^{-23}\erg\cm^3\s^{-1}}\right)^{-\frac{1}{2}},
  \end{align*}
  where any heating has been neglected. 
  To avoid unresolved density and temperature structures produced by thermal
  instability, it is required that $\lambda_\mathrm{F}>\Delta$, 
  and so the minimum value of the thermal conductivity $\chi$ follows as
  \[
  \chi_\mathrm{min}
        =\frac{1-\beta}{\gamma\tau\cool}
          \left( \frac{\Delta}{2 \upi}\right)^2,
  \]
  where $\tau\cool$ is the \textit{minimum\/} cooling time,
  and $\beta$ is the relevant exponent from the cooling function 
  (e.g.\ as in Table~\ref{table:coolSS} for WSW cooling).
  In the single remnant simulations of Appendix~\ref{chap:EISNR},
  $\tau\cool \gtrsim 0.75\Myr$. 
  In the full ISM simulations, minimum cooling times as low as $0.05\Myr$ were
  encountered.
  $\chi_\mathrm{min}$ has maxima corresponding to $\beta=0.56,-0.2,-3,
  \ldots$ for $T=313, 10^5, 2.88\times10^5\K\ldots$. 
  All of these, except for that at $T=313\K$, result in
  $\chi_\mathrm{min}<4\times10^{-4}\kms\kpc$ at $c\sound=c_1=1\kms$,
  so are satisfied by default for any $\chi_1$ sufficiently high to satisfy the
  $\Pe\mesh\leq10$ requirement.
  For $T=313\K$, at $c\sound=c_1$, then 
  $\chi_\mathrm{min}=6.6\times10^{-4}\kms\kpc>\chi_1$. 
  Thus if cooling times as short as 0.05\,Myr were to occur in the cold gas,
  $\lambda_\mathrm{F}<\Delta$, and the model would be marginally
  under-resolved. 
  Analysis of the combined distribution of density and temperature
  from Fig.~\eqref{fig:pdf2dop}, however,
  indicates that cooling times this short occur exclusively in the warm gas.
  
  With $\chi_1\approx4.1\times10^{-4}\kms\kpc$, 
  as adopted in Section~\ref{sect:NS}, 
  then $\Pe\mesh\le10$ is near the limit of numerical stability. (The choice of
  thermal diffusivity is discussed further in Appendix~\ref{sect:ti}.)
  As a result, the code may occasionally crash (notably when hot gas is 
  particularly abundant), and has to be restarted. 
  When restarting, the position or timing of the next SN explosion is 
  modified, 
  so that the particularly troublesome SN that causes the problem is avoided.
  In extreme cases, it may be necessary to increase $\chi$ temporarily 
  (for only a few hundred time steps), to reduce the value of $\Pe\mesh$
  during the period most prone to instability,
  before the model can be continued with the normal parameter values.

%----------------------------------------------------------------------------
\section{\label{sect:ti}Thermal Instability}
  
%----------------------------------------------------------------------------

  One of the two cooling functions employed in this thesis, WSW, supports
  isobaric thermal instability in the temperature range $313<T<6102\K$ where
  $\beta<1$.
  (Otherwise, for the RBN cooling function or outside this temperature range
  for WSW cooling, $\beta\geq1$ or $\Gamma\ll\rho\Lambda$, so the gas is either
  thermally stable or has no unstable equilibrium.)
  
  Under realistic conditions for the ISM, thermal instability can produce very
  small, dense gas clouds which cannot be captured with the resolution
  $\Delta=4\p$ used here. 
  Although the efficiency of thermal instability is questionable in the
  turbulent, magnetized ISM, where thermal pressure is just a part of the total
  pressure \citep[][ and references therein]{VS00,MLK04}, this instability is
  purposely suppressed in my model.
  However, this is done not by modifying the cooling function, but rather by 
  enhancing thermal diffusivity so as to avoid the growth of perturbations at
  wavelengths too short to be resolved by the numerical grid.
  
  Following \citet{Field65}, consider the characteristic wave numbers
  \[
  k_\rho=\frac{\mu(\gamma-1)\rho_0{\cal L}_\rho}{\mathcal{R}c\sound T_0},
  \quad 
  k_T=\frac{\mu(\gamma-1){\cal L}_T}{\mathcal{R}c\sound},
  \quad 
  k_K=\frac{\mathcal{R}c\sound\rho_0}{\mu(\gamma-1)K},
  \]
  where $\mathcal{R}$ is the gas constant, and the derivatives 
  $\mathcal{L}_T\equiv(\upartial\mathcal{L}/\upartial T)_\rho$ 
  and
  $\mathcal{L}_\rho\equiv(\upartial\mathcal{L}/\upartial\rho)_T$
  are calculated for constant $\rho$ and $T$, respectively. 
  The values of temperature and density in these equations,
  $T_0$ and $\rho_0$, are those at thermal equilibrium,
  $\mathcal{L}(T_0,\rho_0)=0$ with 
  $\mathcal{L}=\rho\Lambda-\Gamma$.
  Isothermal and isochoric perturbations have the characteristic wave numbers 
  $k_\rho$ and $k_T$, respectively, 
  whereas thermal conductivity $K$ is characterized by $k_K$.
  The control parameter of the instability is 
  $\varphi=k_\rho/k_K$. 
  
  The instability is suppressed by heat conduction, with the largest unstable
  wave numbers given by \citep{Field65}
  \begin{align}
  k_{\rm cc}  = & \left[k_K(k_\rho-k_T)\right]^{1/2}\,, \\
  k_{\rm cw}  = & \left[-k_K\left(k_T+\frac{k_\rho}{\gamma-1}\right)\right]^{1/2}\,,
  \end{align}
  for the condensation and wave modes, respectively, whereas the most unstable
  wave numbers are 
  \begin{align}
  k_{\rm mc}  = & \left[\frac{(1-\beta)^2}{\gamma^2}
                 +\frac{\beta(1-\beta)}{\gamma}\right]^{1/4}
                 (k_\rho k_{\rm cc})^{1/2}\,, \\
  k_{\rm mw}  = & \left|\frac{\beta-1}{\gamma}\,
                 k_\rho k_{\rm cw}\right|^{1/2}\,.
  \end{align}
   
%-----------------------------------------------------------------------------
  \begin{table}
  \caption[Thermally unstable wavelengths]{\label{table:TI}
  The unstable wavelengths of thermal instability, according to 
  \citet{Field65}, at thermally unstable equilibria $(T_0,\rho_0)$ with the WSW
  cooling function.}
  \centering
  \begin{tabular}{@{}lccccccc@{}}
  \hline
  $T_0$	&$\rho_0$			&$\varphi$ &$\lambda_\rho$	&$\lambda_\mathrm{cc}$	&$\lambda_\mathrm{mc}$	&$\lambda_\mathrm{cw}$	&$\lambda_\mathrm{mw}$\\
  $[\!\K]$&[$10^{-24}\g/\!\cm^3$]&&[pc]						&[pc]										&[pc]										&[pc]										&[pc]\\
  \hline
  \phm313&4.97 &1.91              &\phm\phm2       &\phm5									&\phm\phm5							&\phm2										&\phm\phm4\\
  4000	&1.20					&0.04				&101						&32											&\phm84		  						&14											&\phm74\\
  6102	&0.94					&0.02				&192						&44											&136										&20											&120\\
  \hline
  \end{tabular}
  \end{table}
  %-----------------------------------------------------------------
  
  Table~\ref{table:TI} contains the values of these quantities for the 
  parameters of the reference Model~\Op, presented in terms of wavelengths 
  $\lambda=2\upi/k$, rather than of wave numbers $k$.
  The unstable wavelengths of thermal instability are comfortably resolved at
  $T_0=6102\K$ and $4000\K$, with the maximum unstable wavelengths
  $\lambda_{\rm cc}=44\p$ and $32\p$, respectively, being  much larger than 
  the grid spacing $\Delta=4\p$.
  The shortest unstable wavelength of the condensation mode in the model,
  $\lambda_{\rm cc}=5\p$ at $T\approx313\K$ is marginally resolved at
  $\Delta=4\p$; gas at still lower temperatures is thermally stable.
  Unstable sound waves with $\lambda_{\rm cw}=2\p$ at $T=313\K$ are shorter
  than the numerical resolution of the reference model.
  However, for these wave modes to be unstable, the isentropic instability 
  criterion must also be satisfied, which is not the case for $\beta>0$, so
  these modes remain thermally stable.
  
  Thus, there can be confidence that the parameters of these models (most
  importantly, the thermal diffusivity) have been chosen so as to avoid any
  uncontrolled development of thermal instability, even when only the bulk 
  thermal conductivity is accounted for.  
  Since much of the cold gas, which is most unstable, has high Mach numbers, 
  thermal instability is further suppressed by the shock capturing diffusivity.

\end{chapter}
 
  \begin{chapter}{\label{subsect:MOL}\label{chap:SNosc}Mass sensitivity and mass conservation}

%-----------------------------------------------------------------------------
  \subsubsection { Effect of molecular mass}
%-----------------------------------------------------------------------------

  The two Models~RB \& \WSWa\ have about 117\% of the ISM mass of the reference
  Model~\Op, the difference corresponding to the abundance of molecular 
  hydrogen.
  The difference is apparent comparing the lower panel of
  Fig.~\ref{fig:zfill_RB_WSW} with the upper panel of Fig.~\ref{fig:zfill}.
  Including the molecular gas, the abundance of hot gas reduces with height,
  contrary to observation, while otherwise it increases with height.
  Its inclusion appears to produce unrealistically strong cooling, as evident
  in the altered profiles of velocity and temperature in the horizontal 
  averages of Fig.~\ref{fig:pav} from $t=400\Myr$, after the transition from
  Model~\WSWa\ to Model~\Op.
 
  Therefore all the other models presented in this thesis match the mass in
  Model~\Op, assuming that the cloudy molecular component is concentrated in
  very small scales, not resolved by the coarse grid.
  Comparing the lower panels of Fig.~\ref{fig:wsw_pdf3ph} for \WSWa\ with
  Fig.~\ref{fig:apdf3ph} for \Op, the higher end of all three total
  pressure distributions are very similar, but for the warm and hot gas the
  distributions extend to much longer tails of low pressure gas.
  When molecular gas is included it appears that much of the gas is
  under pressured and out of equilibrium.
  Otherwise, comparing
  Fig.~\ref{fig:wsw_pdf3ph} with Fig.~\ref{fig:pdf3ph}, the probability 
  distributions for density, velocity and Mach numbers are very similar.
  The density distributions without the molecular hydrogen are narrower, but
  the peaks match, so aside from the pressures, the phase structure appears to
  be independent of the disc surface density.
%-----------------------------------------------------------------------------
  \subsubsection { Numerical mass diffusion}
%-----------------------------------------------------------------------------

  The Pencil code is non-conservative, so that gas mass is not necessarily 
  conserved; this can be a problem due to extreme density gradients developing
  with widespread strong shocks. 
  In general, solving  Eq.~(\ref{eq:mass}) in terms of log density rather than
  linear density is numerically more efficient. 
  However solving for $\rho$, rather than $\ln\rho$ is more accurate, and
  applying a numerical diffusion to the equation of mass resolves the
  problem of numerical dissipation in the snowplough test cases described in 
  Appendix~\ref{sect:SNPL}, with mass then being conserved to within machine
  accuracy.
  Eq.~\eqref{eq:mass} thus becomes
  \begin{equation}\label{eq:massA}
  \frac{D\rho}{Dt}\,=\,- \nabla \cdot (\rho \vect{u})\,+\dot{\rho}\SN,
      +({\cal{V}}+\zeta_{\cal{V}})\nabla^2\rho
      +\nabla\zeta_{\cal{V}}\cdot\nabla\rho,
  \end{equation}
  in which constant ${\cal{V}}\approx4.1\times10^{-3}\kms\kpc$ is the
  unphysical mass diffusion coefficient and 
  $\zeta_{\cal{V}}$ is as defined in Eq.~\eqref{shockdiff}, but with 
  coefficient $c_{\cal{V}}=1$.
  A correction term is then required for mass conservation for each of 
  Eq.~\eqref{eq:mom} and \eqref{eq:ent}.
  \[\frac{D\vect{u}}{Dt}\longrightarrow\frac{D\vect{u}}{Dt}-
  \rho^{-1}\frac{{\cal{V}}\nabla^2\rho}{2}\vect{u}\,,\qquad\rho T\frac{D s}{Dt}
  \longrightarrow \rho T\frac{D s}{Dt}-\frac{{\cal{V}}\nabla^2\rho}{\cp}.
  \]
  
  However for the full model, once the ISM becomes highly turbulent, 
  there remains some numerical mass loss. 
  A comparison of mass loss through the vertical boundaries to the total mass 
  loss in the volume
  indicates that numerical dissipation accounts for 
  much less than 1\% per $\Gyr$. 
  The rate of physical loss, due to the net vertical outflow, 
  is of order $15\%$ per $\Gyr$. 
  
%-----------------------------------------------------------------------------
  \subsubsection { Boundary mass loss}
%-----------------------------------------------------------------------------
  
  In such a demanding simulation of this nature, it is inevitable that some
  heavily parameterized processes, approximations and numerical patches must 
  be employed: cooling functions, supernova injection, boundary conditions
  and shock capturing viscosity amongst others, all of which have been
  explained and hopefully satisfactorily justified.
  Perhaps the most physically compromising numerical fix, which is necessary
  to maintain the mass in the long MHD runs, relates to the issue of physical
  boundary losses.

  The general net outflow of gas through the vertical boundaries is a
  genuine physical process and consequence of sustained SN activity.
  Even if we were able to incorporate the galactic fountain and recycle this 
  gas to the mid-plane, the loss of mass from the ISM would be entirely 
  expected over a period of a few hundred Myr, never mind a Gyr. 
  The natural evolution of a galaxy as it ages would see an increasing
  proportion of the gas in the disc locked into stellar mass and more of the
  residual gas escaping the gravitational potential in galactic winds.
  As discussed in Section~\ref{sect:time}, however, we may reasonably consider
  the simulation time merely to be a statistically steady representation of
  time interval being modelled.
  Therefore, to maintain a statistically steady hydrodynamical state, it is 
  important to retain a steady density composition of the ISM, after the system
  first settles into such a state, within the first 2--4\,Myr or so.
  
  \citet{ABnei10} and \citet{Joung09}, and references within both, achieve mass 
  conservation by extending the boundary to $|z|=10\kpc$, which I prefer not to
  do to preserve numerical resources. 
  Some mass is recycled in the form of $4\Msol$ of ejecta per SN. 
  While this has useful diagnostic properties in plotting the expanding 
  remnant profiles in the single SN experiments from Appendix~\ref{sect:SNPL},
  it is negligible in relation to the ambient mass of the ISM.
  Given that mass losses from the ISM to support star formation are not included, I
  would be inclined to exclude ejecta mass in future simulations, thus making
  the SN feedback process mass neutral.
  The option of using ejecta to artificially replenish the boundary mass losses
  would require at least a ten fold increase in the current rate of ejecta, and
  this would have dramatic consequences for the accuracy of the SN energy 
  injection and would also strongly affect the local dynamics unphysically.
  
  The potentially controversial solution I have adopted is to calculate the 
  total net mass loss through the boundary and to regularly redistribute it
  everywhere in the model, proportional to the local density.
  Given the replacement rate is of order $10^{-7}$ of the ISM mass on each
  occasion, and the local ratios of gas density are preserved, there should be
  minimal impact on the local dynamics throughout the simulation.
  Similarly the local structures of energy, temperature, velocity, etc. should
  be minimally affected.
  The effect of this solution is to preserve the simulation for as long as 
  required, with no perceptible difference between the statistical properties
  of the hydrodynamics of the model before and after its implementation.
  
\subsubsection{Sensitivity to SNe location}

  \citet{K98} postulates a relationship between the surface gas density in 
  star-forming galaxies and the supernova rate.
  The estimates are very much approximations to first order on the global
  galactic parameters.
  The self-regulating mechanism, by which the amount of gas in the star forming
  regions diminishes with increasing SN activity, and the SN activity 
  decreases with diminishing gas density, is complex and not well understood.
  In this model no attempt has yet been made to probe this relationship, and
  SN rates are applied purely as a parameter with a particular rate and 
  distribution for each model, and not sensitive to the fluctuations in the
  gas density.

  The majority of Type~CC~SN are clustered in OB associations of young stars
  emerging from the star forming clouds of molecular hydrogen, which are 
  located close to the galactic mid-plane.
  It would therefore seem desirable to model these with a distribution which
  prefers locations associated with these dense regions.
  My early prescription for Type~CC location involved a quite complex 
  calculation to determine the mass distribution of the ISM each time a Type~CC
  was required, and then randomly selected, the location with a probability 
  proportional to density. 

  Two problems emerged with this method. 
  The first was numerical. 
  Given the analysis described in Appendix~\ref{sect:SNPL}, because the error
  in the approximation of cooling in the evolution of an SN remnant is higher 
  with density, these SNe lost too much of their initial energy before their
  evolution into the snowplough stage was completed. 
  Hence the ISM did not produce enough hot gas, and the turbulence was also
  inefficient at generating cold dense structures.
  The hot gas was confined to the mid-plane, where it was cooled, and 
  substantial blow outs into the lower halo could not be sustained.
  Hence the gas near the boundary remained cool at about $10^4\K$, instead of
  nearer $10^6\K$.
  The second problem was in the physics. 
  Although the stars are formed in the dense molecular clouds, after the first
  few SN in an OB association explode they sweep away the dense gas, so that 
  subsequent SN occur in the relatively diffuse gas in the wake of previous SN.
  So in fact the dependence on SN location with density is less direct.

  The method used to locate Type~I SN is to assume that $(x,y,z)$ belong to
  random distributions $(X,Y,Z)$ such that $X,Y\sim U(-0.512,0.512)$ and
  $Z\sim N(0,h_{\mathrm{I}}^2)$. 
  This simple prescription was also applied to Type~CC~SN, but with
  $h_{\mathrm{II}}$
  for the scale height, resulting in an 
  improvement in the numerical approximation of the SN remnants and also an 
  increase in the formation of both hot and cold gas, and the advection of the
  hot gas from the disc to the lower halo.

%-----------------------------------------------------------------------------
  \begin{figure}[h]
  \centering
%  \hspace{-1.25cm}
  \includegraphics[width=\linewidth]{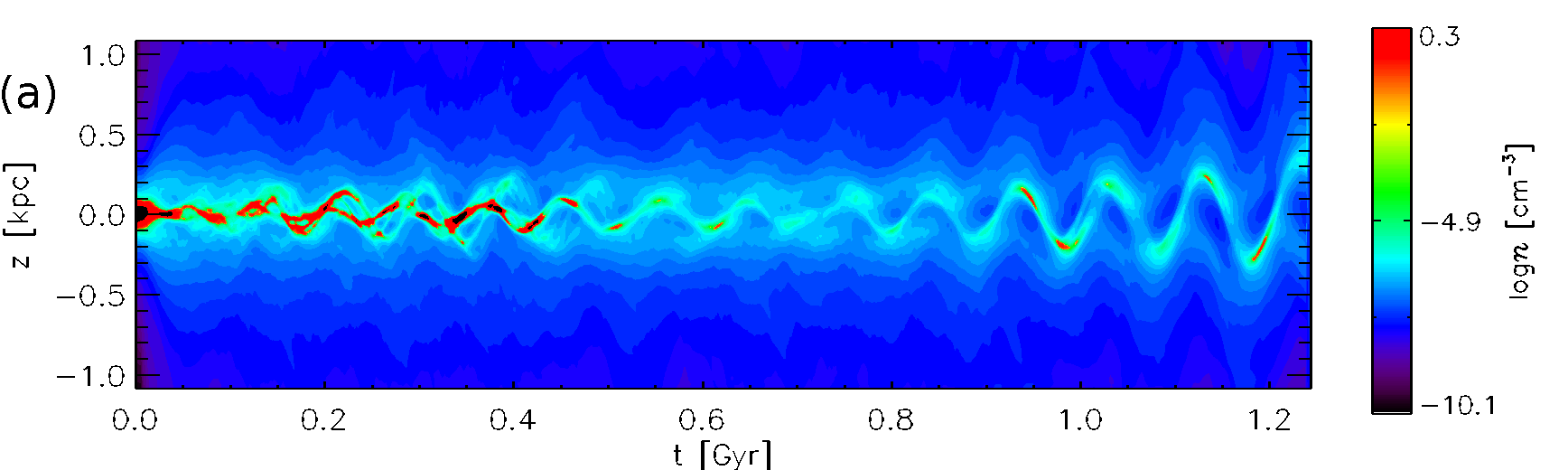}
  \includegraphics[width=\linewidth]{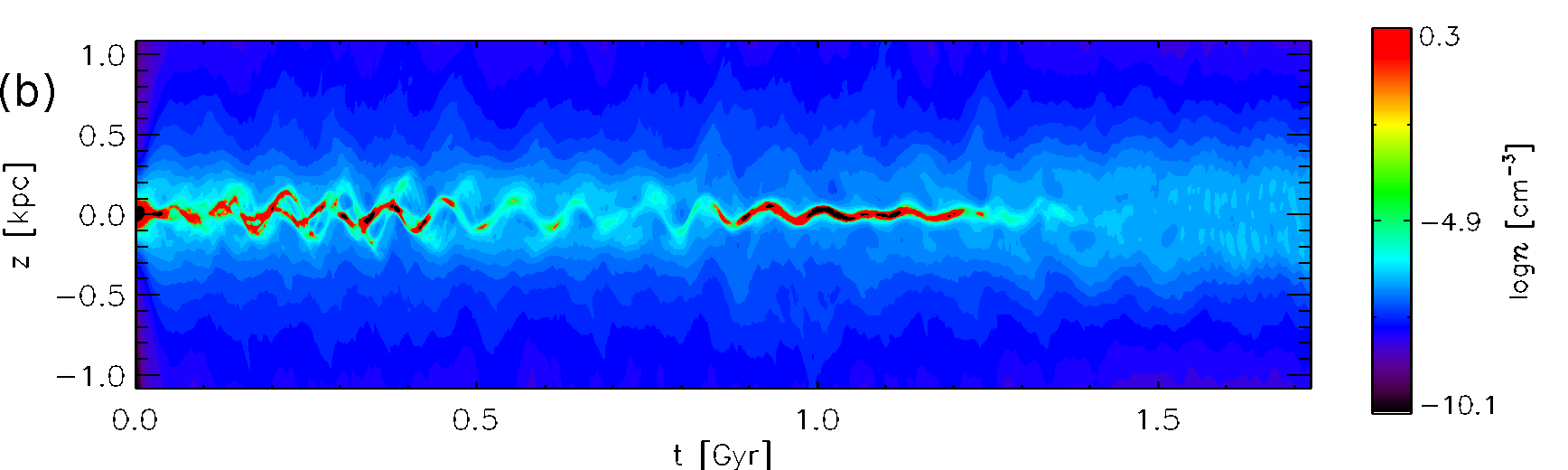}
%  \hspace{-1.5cm}
    \caption[Vertically oscillating disc]{
  Time evolution (a) of horizontal averages of gas density for Model~$\Ompd$
  with vertical distribution of Type~CC SN from $Z\sim N(0,h_{\mathrm{II}}^2)$,
  and (b) with vertical distribution of Type~CC SN from 
  $Z\sim N(z_h,h_{\mathrm{II}}^2)$, where $z_h$ corresponds to the horizontal
  location with the largest total mass at the time of the SN explosion.
    \label{fig:osc}}
  \end{figure}
%-----------------------------------------------------------------------------

  A new problem emerges with this method, however.
  There is a tendency for oscillations in the disc to be amplified over time,
  if the vertical distribution is fixed to vary about $z=0\kpc$.
  This is an unphysical effect, because one would expect the location of SNe
  in general to follow the star forming regions if they stray from the 
  mid-plane.
  This might be overlooked for the rather shorter duration hydrodynamic
  simulations, but when it becomes necessary to extend runs beyond $1\Gyr$ to
  trace the dynamo, this effect not only interferes with the physical 
  integrity of the models, it actually leads to the bulk of the gas being blown
  out of the model, and the simulation crashing.
  In Fig~\ref{fig:osc}a the horizontal averages for density from Model~$\Ompa$
  are plotted against time. 
  It can be seen that the disc oscillates from early on.
  This is natural, as intermittently one would expect the distribution of SNe
  to be greater above or below the disc. 
  It is not natural, however, that the periodicity of the oscillations should 
  be so regular nor should the amplitude be amplified. 
  From panel (a) the initial oscillations are somewhat chaotic, and the disc
  fragments, with several strands of dense gas out of phase.
  From about 500\,Myr the fragments align and the bulk of the disk mass 
  oscillates with period $\sim200\Myr$, and the displacement grows to as much
  as $500\p$ by $1.2\Gyr$.

  This is corrected by tracking the vertical position of the peak horizontal
  average gas density and adjusting the location of the SN distribution to 
  match.
  To be explicit, if the horizontal slice with the highest gas density is
  located at $z=z_h$ when a Type~CC SN site is being selected, then the 
  vertical location of the site will be selected from an $N(z_h,h_{II})$ random
  distribution, rather than as previously $N(0,h_{II})$.
  In fig.~\ref{fig:osc}b the simulation has been restarted from about 600\,Myr
  with the revised vertical distribution.
  Because the SNe are now distributed symmetrically about the disc mass, the
  systematic oscillations are damped within 300\,Myr. 
  The perturbations to the position of the disc are more natural and are no
  longer unstable. 
  In the latter stages, beyond 1.2\,Gyr, the disc density distribution 
  broadens, but this is due to the saturation of the magnetic field and the
  additional magnetic pressure supporting the gas against gravity.  
   
%-----------------------------------------------------------------------------
\end{chapter}

%-----------------------------------------------------------------------------

%-----------------------------------------------------------------------------
\begin{chapter}{Dimensional units\label{chap:units}}
%-----------------------------------------------------------------------------
  \citet[][Appendix C]{Mee07} outlines how the system of dimensionless units 
  for a simulation using the Pencil-code are uniquely specified by the 
  definition of a set of fundamental units in the centimetres, grams and
  seconds (cgs) system.
  Here I clarify how these are specified, which units are currently actually
  defined in the code and the relation to the derived units.  

  The base units are length, speed and density, which are listed in
  Table~\ref{table:spec}. 
  It is desirable in numerical modelling to satisfy the constraint of finite
  precision that variables should not differ greatly from unity. 
  By specifying two from three units of length, time and speed, the third unit
  follows. 
  The selection of Myr or Gyr for time are natural, and with a gas speed in 
  the ISM typically measured in $\kms$, it follows that unit length should 
  be of order pc or kpc, respectively.

  Because $1\Gyr=3.15570\times10^{16}\s$ and $1\kpc=3.08573\times10^{31}\cm$, to
  use unit speed as $10^5\kms$ I choose to approximate unit length
  $=3.15570\times10^{21}\cm\approx1\kpc$.
  The mean number density of atomic hydrogen near the mid-plane of the Galaxy
  is  estimated to be about $1\cmcube$.
  With the proton mass about $1.6728\times10^{-24}\g$,
  unit density $=10^{-24}\g\cm^{-3}$ is a convenient base unit to use.

%-----------------------------------------------------------------------------
  \begin{table}[h]
  \centering
  \caption{\label{table:spec}Specified units}
  \begin{tabular}{c|c|l}
  Quantity &Symbol   & Unit                                   \\[3pt]
  \hline
  \vspace{-0.25cm}   &                         &              \\ 
  Length   &$[L]$    & $3.15570\times10^{21}\cm\approx1\kpc$  \\[3pt]
  Velocity &$[u]$    & $10^5\cm \s^{-1}=1\kms$                \\[3pt]
  Density  &$[\rho]$ & $1\times10^{-24}\g\cmcube$             \\[3pt]          
  \hline 
  \end{tabular}
  \end{table}
%-----------------------------------------------------------------------------
    
  It is also possible to define separately
  units for temperature and magnetic field strength, but one should be aware
  that specifying a value which conflicts with the three fundamental units
  may lead to code instability or numerical errors. 
  These options are necessary, to permit specification where models neglect 
  physics such as density or velocity, but it is important to be aware that no 
  more than three fundamental units may be specified independently, so the
  derived determination for magnetic field strength and temperature is 
  required and not an arbitrary choice.
  
  Some of the main derived units referred to in this thesis and their
  construction in terms of the fundamental units are listed in 
  Table~\ref{table:deriv}.
  Because it is less evident how the unit of temperature and magnetic field
  are dependent on the specification of the fundamental units, their
  derivation is described explicitly below.
 
%-----------------------------------------------------------------------------
  \begin{table}[h]
  \centering
  \caption{\label{table:deriv}Derived units}
  \begin{tabular}{c|c|c|l}
  Quantity                & Symbol  & Typical Derivation                    & Unit                                    \\[3pt]
  \hline
  \vspace{-0.25cm}        &         &                                       &\\ 
  Time                    &$[t]$    & $[L][u]^{-1}$                         & $3.15570\times10^{16}\s=1\Gyr$          \\[3pt]
  Temperature             &$[T]$    & $\mu[u]^2\gamma^{-1} R^{-1}$          & $44.74127\K$                            \\[3pt]
  Mass                    &$[m]$    & $[\rho][L]^3$                         & $3.14256\times10^{40}\g$                \\[3pt]
  Energy                  &$[E]$    & $[\rho][L]^3[u]^2$                    & $3.14256\times10^{50}\erg$              \\[3pt]
  Energy density          &$[e_V]   $ & $[\rho][u]^2$                       & $1\times10^{-14}\erg\cmcube$            \\[3pt]
  Pressure                &$[p]$    & $[\rho][u]^2$                         & $1\times10^{-14}\dyn\cm^{-2}$           \\[3pt]
  Magnetic field          &$[B]$    & $\sqrt{\mu_0}[\rho]^{\frac{1}{2}}[u]$ & $3.54491\times10^{-7}\G$                \\[3pt]
  Kinematic viscosity     &$[\nu]$  & $[L][u]$                              & $3.15570\times10^{26}\erg\s\g^{-1}$     \\[3pt]
  Thermal diffusivity     &$[\chi]$ & $[L][u]$                              & $3.15570\times10^{26}\erg\s\g^{-1}$     \\[3pt]
  Magnetic diffusivity    &$[\eta]$ & $[L][u]$                              & $3.15570\times10^{26}\erg\s\g^{-1}$     \\[3pt]
  Specific entropy        &$[s]$    & $[u]^2[T]^{-1}$                       & $2.23507\times10^{8}\erg\g^{-1}\K^{-1}$ \\[3pt]
  \hline
  \end{tabular}
  \end{table}
%-----------------------------------------------------------------------------

  \subsubsection{Unit temperature and ionization}

  In cgs units the Boltzmann constant 
  $k_{\textrm {B}}\simeq1.38\times10^{-16}\erg\K^{-1}$
  and the atomic mass unit $m_u=1.66\times10^{-24}\g$. 
  From the equation of state
  \[
  c\sound^2=\gamma\frac{p}{\rho}=\gamma\frac{k_{\textrm {B}}}{m_u\mu} T,
  \]
  where the Boltzmann constant 
  $k_{\textrm {B}}=1.380658\times10^{-16}\erg\K^{-1}$, the atomic mass unit
  $m_u=1.6605402\times10^{-24} \g$, the ratio of specific heats $\gamma=5/3$
  and the mean molecular weight $\mu=0.62$.
  Hence unit temperature $T_0$ is derived from the unit sound speed $c\sound_0$,
  which is $1\kms$, as
  \[
  T_0=\frac{\mu}{\gamma R} c\sound_0^2=44.74127\K,
  \] 
  where the gas constant $R=k_{\textrm {B}}/m_u$.

  As is discussed in Section~\ref{sect:eq} in relation to Eq.~\ref{eq:eos} the
  mean molecular weight $\mu$ depends on the degree of ionization of the gas,
  due to the presence of free electrons. It could also depend on the presence of
  molecular hydrogen, but as concluded for these models, this is not resolved,
  so only the atomic particles contribute to $\mu$.

  Assuming the gas to be fully ionized then we would have one each of hydrogen
  ions and electrons for each atom and for each helium atom one ion and two
  electrons. If $X$ is the proportion of the number of hydrogen atoms and $Y$
  the proportion of helium atoms, then the total number of particles in the ISM,
  comprised purely of these two elements is given by
  \begin{align*}
    N=&\frac{X_i\rho}{m_p}+\frac{X_e\rho}{m_p}+\frac{Y_i\rho}{4m_p}+
        \frac{2Y_e\rho}{4m_p}\\
     =&\frac{\rho}{m_p}\left(\frac{4X_i+4X_e+Y_i+2Y_e}{4}\right)\\
    \mu^{-1}=&\left(\frac{8X+3(1-X)}{4}\right)=\left(\frac{5X+3}{4}\right).
  \end{align*}
  Given the estimates $X=0.908$ and $Y=0.092$ stated in 
  Section~\ref{subsect:ism} then for the fully ionized gas $\mu=0.531$. 
  For the fully neutral gas, with $X_e=Y_e=0$ then $\mu=1.074$. 
  It may be reasonable to approximate the cold gas as neutral and the hot gas
  as fully ionized, and to apply an appropriate temperature dependent transition
  between them in the warm gas, without all the additional complexity of
  modelling ions and electrons. 
  Care would then need to be taken to ensure the unit of temperature is 
  appropriate to the value of $\mu=0.531$.

  \subsubsection{Unit magnetic field strength}

  Similarly, in cgs units magnetic field energy density is defined as 
  $B^2/(2\mu_0)$
  where the magnetic field is specified in Gauss $[\G]$ and
  the magnetic vacuum permeability $\mu_0=4\upi \G^2\cm^{3}\erg^{-1}$.
  The factor of one half is required for parity with the formulation
  of the thermal and kinetic energy densities.
  So $1\G=(\mu_0\erg\cmcube)^{\frac{1}{2}}$.

\end{chapter}
                    % An appendix.

  % Uncomment the line below if you want your Bibliography to appear in the
  % table of contents.
  %\addcontentsline{toc}{chapter}{Bibliography}
  \bibliography{fred_mnras}             % References.
  \bibliographystyle{apsr}
  %\nocite{Cou90}                        % How to make a reference appear in the
                                        % bibliography when not referenced in
                                        % the text.

\end{document}